\documentclass[12pt]{article}
\makeatletter \@addtoreset{equation}{section} \makeatother
\renewcommand{\theequation}{\thesection.\arabic{equation}}
\newcommand{\ba}{\begin{array}}
\newcommand{\ea}{\end{array}}
\newcommand{\beq}{\begin{equation}}
\newcommand{\eeq}{\end{equation}}
\newcommand{\bea}{\begin{eqnarray}}
\newcommand{\eea}{\end{eqnarray}}

\def\bce{\begin{center}}
\def\ece{\end{center}}

\def\nonu{\nonumber}

\def\pa{\partial}

\def\be{\beta}

\def\la{\lambda}

\def\ps{\psi}

\def\eps6{{\displaystyle \mathop{\epsilon}^{6}}{}}
\def\g6{{\displaystyle \mathop{g}^{6}}{}}

\def\nab6{{\displaystyle \mathop{\nabla}^{6}}{}}

\def\0{{\sst{(0)}}}
\def\1{{\sst{(1)}}}
\def\2{{\sst{(2)}}}
\def\3{{\sst{(3)}}}
\def\4{{\sst{(4)}}}
\def\5{{\sst{(5)}}}
\def\6{{\sst{(6)}}}
\def\7{{\sst{(7)}}}
\def\8{{\sst{(8)}}}

\def\ba{\begin{array}}
\def\ea{\end{array}}
\def\beq{\begin{equation}}
\def\eeq{\end{equation}}
\def\be{\begin{equation}}
\def\ee{\end{equation}}

\def\la{\lambda}
\def\eps{\epsilon}

\def\ba{\begin{array}}
\def\ea{\end{array}}
\def\beq{\begin{equation}}
\def\eeq{\end{equation}}
\def\be{\begin{equation}}
\def\ee{\end{equation}}

\def\la{\lambda}
\def\eps{\epsilon}

\def\eps6{{\displaystyle \mathop{\epsilon}^{6}}{}}

\def\nab6{{\displaystyle \mathop{\nabla}^{6}}{}}

\newcommand{\bean}{\begin{eqnarray*}}
\newcommand{\eean}{\end{eqnarray*}}

\begin{document}
\thispagestyle{empty} \addtocounter{page}{-1}
   \begin{flushright}
\end{flushright}

\vspace*{1.3cm}
  
\centerline{ \Large \bf   
Higher Spin Currents} 
\vspace*{0.3cm}
\centerline{ \Large \bf 
in the ${\cal N}=2$ Stringy Coset Minimal Model } 
\vspace*{1.5cm}
\centerline{{\bf Changhyun Ahn } 
} 
\vspace*{1.0cm} 
\centerline{\it 
Department of Physics, Kyungpook National University, Taegu
41566, Korea} 
\vspace*{0.8cm} 
\centerline{\tt ahn@knu.ac.kr 
} 
\vskip2cm

\centerline{\bf Abstract}
\vspace*{0.5cm}

In the coset model based on $(A_{N-1}^{(1)} \oplus A_{N-1}^{(1)}, A_{N-1}^{(1)})$
at level $(N, N; 2N)$, it is known that 
the ${\cal N}=2$ superconformal algebra
can be realized by the two kinds of adjoint fermions.
Each Kac-Moody current of spin-$1$ is given by the product of fermions
with structure constant ($f$ symbols) as usual. 
One can construct the spin-$1$ current by combining 
the above two fermions with the structure constant and 
the spin-$1$ current by multiplying these two fermions 
with completely symmetric $SU(N)$ invariant tensor of rank $3$ ($d$ symbols). 
The lowest higher spin-$2$ current with nonzero $U(1)$ charge (corresponding 
to the zeromode eigenvalue of spin-$1$ current of ${\cal N}=2$ superconformal
algebra) can be 
obtained from these four spin-$1$ currents in quadratic form.
Similarly, the other type of lowest higher spin-$2$ current, whose $U(1)$
charge is opposite to the above one, can be obtained also.
Four higher spin-$\frac{5}{2}$ currents can be constructed from 
the operator product expansions (OPEs) between the spin-$\frac{3}{2}$ currents
of ${\cal N}=2$ superconformal algebra and the above 
two higher spin-$2$ currents.
The two higher spin-$3$ currents can be determined 
by the OPEs between the above spin-$\frac{3}{2}$ currents  
and the higher spin-$\frac{5}{2}$ currents.
Finally, the ten ${\cal N}=2$ OPEs between the four 
${\cal N}=2$ higher spin multiplets $(2, \frac{5}{2}, \frac{5}{2}, 3)$, 
$(2, \frac{5}{2}, \frac{5}{2}, 3)$, $(\frac{7}{2}, 4, 4, \frac{9}{2})$ 
and $(\frac{7}{2}, 4, 4, \frac{9}{2})$ 
are obtained explicitly for generic $N$.

\baselineskip=18pt
\newpage
\renewcommand{\theequation}
{\arabic{section}\mbox{.}\arabic{equation}}

\tableofcontents

\section{Introduction}

One of the conformal field theories (CFTs) 
in two dimensions can be described as 
the following coset model 
\bea
\frac{\hat{SU}(N)_N \oplus \hat{SU}(N)_N}{\hat{SU}(N)_{2N}}.
\label{coset}
\eea
The affine Kac-Moody algebra in the 
numerator has the levels $(N,N)$ while 
the corresponding algebra in the denominator 
has the level $2N$, which is  the sum of the levels in the numerator. 
The 
two dimensional $SU(N)$ gauge theory coupled to the adjoint Dirac fermions
associated with this coset model 
has been described in \cite{GHKSS}.
What is special feature behind the above coset model? 
This coset has ${\cal N}=(2,2)$ supersymmetry \cite{BFK} \footnote{
In this paper, we describe only the holomorphic part of the (higher spin) 
currents.
The anti-holomorphic part of the (higher spin) currents 
can be described similarly.
Then we will use the notation for the supersymmetry as ${\cal N}=2$
supersymmetry simply rather than ${\cal N}=(2,2)$ supersymmetry.}.
The Virasoro primary field has the spin-$\frac{1}{2}$ for the adjoint 
representation at the first level $N$, which is nothing but 
the dual Coxeter number
of $SU(N)$ \cite{BS}. The corresponding highest weight fields 
are given by a set of $(N^2-1)$ free fermions living in the first factor 
of the numerator.
Then the first supersymmetry generator, 
spin-$\frac{3}{2}$ current, 
can be written in terms of 
these adjoint fermions. See also $(2.74)$, $(7.49)$ and $(7.50)$ of \cite{BS}.
Furthermore, the similar analysis can be done for the other adjoint fermions
living in the second factor in the numerator
because the coset model also has the second level $N$ \cite{BFK}.
The second supersymmetry generator, spin-$\frac{3}{2}$ current, 
can be obtained from these adjoint fermions.
Then the standard ${\cal N}=2$ superconformal algebra, 
characterized by one spin-$1$, {\it two} spin-$\frac{3}{2}$ and one
spin-$2$ currents, 
in terms of
these {\it two} adjoint fermions can be realized in the stringy \footnote{
This terminology
was used in the review paper by Gaberdiel and Gopakumar \cite{GG1207}.} 
coset model \cite{BFK}.
We would like to construct the higher spin currents (and their OPEs)
in the above stringy coset minimal model beyond the currents of
${\cal N}=2$ superconformal algebra. 

As observed in \cite{GHKSS},
for $N \geq 3$, there exist higher spin currents as well as 
the ${\cal N}=2$ superconformal currents \footnote{For $N=2$, there
exists only ${\cal N}=2$ superconformal symmetry because the $d$ symbol
in $SU(2)$ vanishes identically.}.
That is, for $N=3$, the two lowest higher spin-$2$ currents 
were obtained from the Dirac fermions. Furthermore, the existence of 
two higher spin-$\frac{7}{2}$
currents and one higher spin-$4$ current has been checked from the extended vacuum character technique.
By examining these higher spin currents in details,
we would like to 
understand the higher spin symmetry algebra in the coset model (\ref{coset}),
which is much larger than the conventional $W_N$ symmetry algebra \cite{BS}.
The central charge in (\ref{coset}) is given by
\bea
c =\frac{1}{3} (N^2-1), \qquad N=3,4,5 \cdots .
\label{centralintro}
\eea
The value $c=\frac{8}{3}$ coincides with the first value in the series 
(\ref{centralintro}) \footnote{ In the ${\cal N}=2$ superconformal 
minimal models, the central charge is given by $c =\frac{3k}{(k+2)}$
where $k =1, 2, \cdots $ 
\cite{DPYZ}.
Then it is easy to see that $k=16$ case \cite{GHKSS} 
corresponds to the $c=\frac{8}{3}$.
}.  
One can check the relation (\ref{centralintro}) by obtaining the 
Sugawara construction for the stress energy tensor, spin-$2$ current,
written in terms of
two fermions in (\ref{coset}) 
and reading off the fourth-order pole of the OPE 
between the stress energy tensor and itself \footnote{
Note that the central charge term  also arises
in the OPE between the spin-$1$ currents and in the OPE
between the spin-$\frac{3}{2}$ currents. Some details on this issue 
(the normalizations of the spin-$1$ and spin-$\frac{3}{2}$ currents) 
will be described in section $2$.}.
In the large $N$ limit, 
the central charge behaves as $N^2$.
This implies that the bulk dual is presumably a string theory on 
$AdS_3$ space observed in \cite{GHKSS}. 
In the coset model with the levels $(k,1;k+1)$ in the context of 
Gaberdiel and Gopakumar's proposal \cite{GG1011,GG1205}, the central charge 
behaves as $N$ in the large $N$ limit.
The number of gauge invariant states in the former
is bigger than the ones in the latter.
The 't Hooft coupling constant in \cite{GG1011} becomes $\la =\frac{1}{2}$
in the coset model (\ref{coset}).

For the large ${\cal N}=4$ holography \cite{GG1305},  
the free symmetric product orbifold CFT   
is dual to the string theory at the tensionless limit \cite{GG1406}.
It is known that 
the stringy symmetry algebra is much bigger than the vector-like symmetry 
algebra of the Vasiliev higher spin theory \cite{PV1998}.
By studying the conformal perturbation theory of the free 
symmetric orbifold CFT (corresponding to switching on the string tension),
the additional symmetry generators of the stringy symmetry algebra seem to 
belong to different (sub-leading) Regge
trajectories \cite{GPZ1506}. Then the higher spin generators of
Vasiliev theory correspond to the leading Regge trajectory (having the 
lowest mass or anomalous dimension for a given spin).  
See also the relevant works in \cite{GG1501,GG1512}.

For the ${\cal N}=3$ holography \cite{CHR1406}, 
the deformation term breaks the higher spin symmetry 
and induces the mass to the higher spin fields \cite{HR1503,CH1506}.
The masses are not generated for the $SO(3)_R$ singlet higher spin fields 
at the leading order of $\frac{1}{c}$ where $c$ is the central charge.
However, the mass formula for the $SO(3)_R$ triplet higher spin fields 
looks like the Regge trajectory on the flat spacetime. 
Although they use the previous extended algebra (for example, 
for spins $s=3,4$), it is not clear that this extended algebra 
\cite{GG1205,BFKetal} 
coincide with
the higher spin algebra with ${\cal N}=3$ supersymmetry.
In other words, so far it is not known what is the higher spin 
symmetry algebra
for the higher spin currents together with ${\cal N}=3$ superconformal 
algebra \footnote{The currents are characterized by 
one spin-$\frac{1}{2}$, three spin-$1$, three spin-$\frac{3}{2}$ and 
one spin-$2$ currents \cite{GS,CK,Schoutens1988}. 
In ${\cal N}=2$ superspace, one can realize them 
as two ${\cal N}=2$ multiplets \cite{BO}. 
That is, $(1, \frac{3}{2}, \frac{3}{2}, 2)$ 
and $(\frac{1}{2}, 1, 1, \frac{3}{2})$.}. 
It would be interesting to see the higher spin symmetry algebra
(between the low higher spin currents)
explicitly.

Now one asks how one can make a deformation which breaks the higher spin
symmetry (and keeps the ${\cal N}=2$ superconformal symmetry) 
and see the mass formula for the higher spin fields ($SO(2)_R$ doublet
or singlet) at the leading order of $\frac{1}{c}$ (or at finite $c$). 
In order to answer this 
question, one should obtain the higher spin symmetry algebra 
for the low higher spin currents
in the stringy coset model (\ref{coset}) as a first step.  
According to the result of \cite{GHKSS}, one has 
the higher spin currents of spins 
\bea
(2, \frac{5}{2}, \frac{5}{2}, 3), \,\, (2, \frac{5}{2}, \frac{5}{2}, 3), 
\,\, (\frac{7}{2}, 4, 4, \frac{9}{2}), \,\, (\frac{7}{2}, 4, 4, \frac{9}{2}), 
  \,\, (4, \frac{9}{2}, \frac{9}{2}, 5),
\,\, \cdots \,\, ,
\label{n2mul}
\eea 
in the ${\cal N}=2$ multiplet 
notation where the five higher spin currents (two spin-$2$, 
two spin-$\frac{7}{2}$ and one spin-$4$ currents) 
are located at the first 
component of each ${\cal N}=2$ multiplet. 
We put the other three components with correct spins 
at each ${\cal N}=2$ multiplet.
Of course, the standard ${\cal N}=2$ superconformal algebra 
can be obtained from the currents of spins 
$(1, \frac{3}{2}, \frac{3}{2}, 2)$.
In previous works \cite{ASS1991,Ahn1211,Ahn1305},
the higher spin currents of spins $(\frac{5}{2},3)$, $(\frac{7}{2},4)$, 
$(4, \frac{9}{2})$, $(4, \frac{9}{2})$, $(\frac{9}{2},5)$, 
$(\frac{11}{2},6)$ and $(6, \frac{13}{2})$ are constructed together with
the currents of spins  $(\frac{3}{2}, 2)$ 
of ${\cal N}=1$ superconformal algebra. 
One can easily see that the first five ${\cal N}=1$ multiplets 
appear in the above ${\cal N}=2$ multiplets (\ref{n2mul}).
The remaining ones will also appear in the list of (\ref{n2mul})
\footnote{For the ${\cal N}=1$ supersymmetric coset model, one of the levels
is given by $k$ where $k=1, 2, \cdots$ \cite{GKO,Douglas}. See also 
the relevant works in \cite{SS,BCGG}. For the general coset model with 
arbitrary levels, see also \cite{AK1308}.}. 

In this paper, we construct the first two ${\cal N}=2$ multiplets 
(\ref{n2mul}) in terms 
of two adjoint fermions, obtain the complete OPEs between the first four
${\cal N}=2$ multiplets (in component approach and in 
${\cal N}=2$ superspace) and present how the higher spin-$\frac{7}{2}$ 
currents can be obtained from two adjoint fermions.
First of all, one should determine the lowest higher spin-$2$ current.
After this is found, then its three other component higher spin currents
can be obtained from the ${\cal N}=2$ supersymmetry. 
Similarly, the other lowest higher spin-$2$ current (and its associated 
three other component higher spin currents)
can be determined.
Now the remaining undetermined higher spin currents (third, fourth, $\cdots$,
$\cal N$=2 multiplets of (\ref{n2mul})) 
should appear in the OPEs between the known 
higher spin currents. 

As the spins increase, the right hand side of the OPE 
contains too many terms. Then how one can rewrite them in terms of
the composite fields consisting of the known (higher spin) currents?
In addition to the spin of the higher spin current (zeromode eigenvalue of 
stress energy tensor spin-$2$ current of ${\cal N}=2$ superconformal algebra), 
there exists other quantity to characterize the state corresponding to the
higher spin current.
One can use the zeromode eigenvalue of the spin-$1$ current of 
${\cal N}=2$ superconformal algebra \footnote{
In the OPE language, the spin is the coefficient of the second-order pole in 
the 
OPE between the stress energy tensor and the higher spin current 
while the 
$U(1)$ charge is the 
coefficient of the first-order pole in 
the 
OPE between the spin-$1$ current and the higher spin current.}. 

Therefore, 
it is crucial to observe the $U(1)$ charges for the higher spin currents 
in (\ref{n2mul}) because the right hand sides of any OPEs, which 
are complicated expressions  of adjoint fermions, should be reexpressed 
in terms of the known (higher spin) currents. 
In other words, once we know 
the $U(1)$ charge of the left hand side of any OPE, then 
one can figure out the algebraic structure of the right hand side
by considering the composite fields (having the correct 
$U(1)$ charge) 
appearing in the particular singular term.  
When the right hand side of the OPE cannot be written in terms of
the known (higher spin) current, then one has a new primary 
higher spin current.
Then one should check how this higher spin current can fit 
in ${\cal N}=2$ multiplet. Using the spin-$\frac{3}{2}$ currents 
of the ${\cal N}=2$ superconformal algebra, one should obtain 
the other three component higher spin currents.

In section $2$,
we review the construction of four currents of
${\cal N}=2$ superconformal algebra in terms of 
two adjoint fermions in the coset model (\ref{coset}).

In section $3$,
we construct the lowest four higher spin currents (corresponding to 
the first ${\cal N}=2$ multiplet in (\ref{n2mul})) in terms of 
two adjoint fermions which are contracted with the $f$ and $d$ symbols. 
The package by Thielemans \cite{Thielemans} is used all the times. 

In section $4$, we repeat the procedure of section $3$ for the other type of
lowest four higher spin currents (corresponding to the 
second ${\cal N}=2$ multiplet in (\ref{n2mul})) 
whose $U(1)$ charges opposite to the 
corresponding higher spin currents obtained in section $3$.

In section $5$,
we describe how we can obtain the higher spin currents beyond the lowest
higher spin currents in sections $3$ and $4$.
The third component higher spin current of ${\cal N}=2$ multiplet in 
section $3$ and the 
second component higher spin current of ${\cal N}=2$ multiplet 
in section $4$ generate 
the first component of higher spin-$s$ current of ${\cal N}=2$ multiplet.
The former increases the $U(1)$ charge of $\frac{1}{3}$
while the latter decreases the $U(1)$ charge of $\frac{1}{3}$.

In section $6$,
we describe the higher spin symmetry 
algebra between the higher spin currents obtained in 
previous sections.
We present the OPE between the lowest higher spin-$2$ current with $U(1)$ 
charge $\frac{2}{3}$ and the lowest higher spin-$2$ current with 
$U(1)$ charge $-\frac{2}{3}$ for generic $N$ (or generic central charge $c$).

In section $7$,
we consider the lowest four ${\cal N}=2$ higher spin multiplets (in component 
approach there are $16$ higher spin currents) and their OPEs in ${\cal N}=2$
superspace with the package by Krivonos and Thielemans \cite{KT}. 

In section $8$,
we summarize what we obtained in this paper and the future works are given.


In Appendices $A, B, \cdots, J$, some details appeared in previous sections 
are presented \footnote{
In particular, Appendices $H,I$ and $J$ contain the component OPEs 
corresponding to the ${\cal N}=2$ OPEs in the section $7$.}.

There are some works \cite{IKSfirst,BNZ,IKSsecond} related to the coset 
model and the higher spin theory with ${\cal N}=2$ supersymmetry 
can be found in previous works in \cite{CHR1111,CG1203,HLPR,HP,CG1207}.

\section{ The four currents of the ${\cal N}=2$ superconformal algebra
in the coset model  }

In this section, 
the four currents of spins $(1, \frac{3}{2}, \frac{3}{2}, 2)$ 
of ${\cal N}=2$ superconformal algebra 
in the coset model (\ref{coset})
will be obtained. Although they appeared in \cite{BFK} previously, we present  
the construction of those four currents  in order to understand 
how we continue to find the higher spin currents in next sections. 

Let us consider the two kinds of adjoint fermion fields,
corresponding to each $SU(N)$ factor in the coset (\ref{coset}),
which satisfy the following fundamental OPEs 
\bea
\psi^a(z) \, \psi^b(w) & = & -\frac{1}{(z-w)} \, 
\frac{1}{2} \, \delta^{ab} + \cdots,
\nonu \\
\chi^a(z) \, \chi^b(w) & = & - \frac{1}{(z-w)} \, 
\frac{1}{2} \, \delta^{ab} + \cdots, \qquad a,b = 1, 2, \cdots, (N^2-1).
\label{psichi}
\eea
Here the adjoint indices 
$a, b$ run over $a, b = 1, 2, \cdots, (N^2-1)$.
The normalization $-\frac{1}{2}$ in the first-order pole of the OPEs 
is taken.
Due to the fermionic property of these 
adjoint fields, there are extra minus signs
in the OPE when we interchange the operators.
That is, $\psi^b(w) \, \psi^a(z) = -\psi^a(z) \, \psi^b(w)$ \footnote{
Similarly the following relation for the second fermions holds
$ \chi^b(w) \, \chi^a(z) = -\chi^a(z) \, \chi^b(w)$.
The OPE $\psi^b(z) \, \psi^a(w)$ (and the OPE $\chi^b(z) \, \chi^a(w)$)
can be determined by (\ref{psichi}) using the standard Taylor expansion
\cite{BS}. 
The extra minus sign from this process can combine with the above
minus sign and leads to the same right hand sides of the OPEs in (\ref{psichi}).
In other words, the OPE 
    $\psi^b(z) \, \psi^a(w)$
can be read off from (\ref{psichi})
by replacing the index $a$ with the index $b$ 
and vice versa.}.
In the right hand side of the OPEs (\ref{psichi}), there is a  
symmetric $SU(N)$ invariant tensor
of rank $2$ denoted by $\delta^{ab}$.
Of course, there are no singular terms in the OPE
$\psi^a(z) \, \chi^b(w) = + \cdots$ because they live in different $SU(N)$
factors respectively. 

The Kac-Moody spin-$1$ $SU(N)$ adjoint currents can be defined as 
the composite of the adjoint fermion fields with totally antisymmetric 
structure constant of $SU(N)$.
That is,
\bea
J^a(z)  & \equiv &  f^{abc} \, \psi^b  \psi^c(z),
\nonu \\
K^a(z) & \equiv &  f^{abc} \, \chi^b  \chi^c(z).
\label{spin1currents}
\eea
The indices $b$ and $c$ in the right hand side of  
(\ref{spin1currents}) are summed over the $SU(N)$ adjoint indices.
The normalizations for these spin-$1$ currents are determined by the 
defining OPE for the affine Kac-Moody algebra below \footnote{
The sum of the normal ordered product $ \psi^b \, \psi^c(z)
$ and  the normal ordered product $ \psi^c \, \psi^b(z)
$ (that is, the anticommutator $ \{\psi^b, \psi^c \}(z)$) 
vanishes because there is no $w$-dependent term in the first-order pole 
in (\ref{psichi}). 
Then there is no sign change in the above spin-$1$ current 
when $\psi^b$ and $\psi^c$ are interchanged because the structure constant 
is antisymmetric in the indices $b$ and $c$. }.
The Wick theorem for the composite fields can be used 
in order to calculate the singular terms between them \cite{BS}.
Of course, the combination between the $\psi^b(z)$ and $\chi^c(w)$
can provide other type of spin-$1$ current which will be described in next 
sections.

Then 
the affine Kac-Moody algebra $\hat{SU}(N)_N \oplus \hat{SU}(N)_N$
in (\ref{coset})
is represented by the following OPEs
\bea
J^a(z) \, J^b(w) & = & - \frac{1}{(z-w)^2} \, N \, \delta^{ab}
+ \frac{1}{(z-w)} \, f^{abc} \, J^c(w) + \cdots,
\nonu \\
K^a(z) \, K^b(w) & = & -\frac{1}{(z-w)^2} \, N \, \delta^{ab}
+ \frac{1}{(z-w)} \, f^{abc} \, K^c(w) + \cdots.
\label{JKOPE}
\eea
The index $c$ in the right hand side 
of the OPEs (\ref{JKOPE})
is summed over the $SU(N)$ adjoint index $c$.
The second-order pole in (\ref{JKOPE}) 
stands for the level $(N,N)$.
It is easy to see how one obtains the level $2N$ by adding the two levels
$N$ and $N$
 \footnote{
By adding the two OPEs in (\ref{JKOPE}), 
the diagonal affine Kac-Moody algebra $\hat{SU}(N)_{2N}$
in the coset (\ref{coset}) 
can be obtained as follows:
$
(J^a+K^a)(z) \, (J^b+K^b)(w)  =  - \frac{1}{(z-w)^2} \, 2N \, \delta^{ab}
+ \frac{1}{(z-w)} \, f^{abc} \, (J^c+K^c)(w) + \cdots $,
where the trivial OPEs $J^a(z) \, K^b(w) = + \cdots$ and 
$K^a(z) \, J^b(w) = + \cdots$ 
are used.}.
There are no singular terms in the OPE $J^a(z) \, K^b(w)$
because there is no nontrivial OPE in $\psi^a(z) \, \chi^b(w) =
+ \cdots$ as described before.

We would like to construct 
the four coset currents of ${\cal N}=2$ superconformal algebra in the coset 
model.
In Appendix (\ref{appendixa}), some useful OPEs are presented.

$\bullet$ Coset spin-$1$ current

Let us consider the following spin-$1$ current 
by taking the composite of the two adjoint fermionic fields
with the $SU(N)$ invariant tensor of rank $2$ 
\footnote{ 
\label{Jcharge}
One can reepxress this current as
$J(z) = -\frac{1}{3} (\psi^a + i \chi^a)(\psi^a -i \chi^a)(z)$.
We will observe that the first factor has $U(1)$ charge $\frac{1}{3}$
while the second factor has $U(1)$ charge $-\frac{1}{3}$. Then 
the total $U(1)$ charge of $J(w)$ is zero. }
\bea
J(z) = 
 \frac{2}{3} \, i\, \delta^{ab}\, \psi^a  \chi^b(z)
=
\frac{2}{3} \, i\, \psi^a  \chi^a(z).
\label{t1}
\eea
The overall constant can be fixed by calculating the following OPE
with the help of (\ref{psichi})
\bea
J(z) \, J(w) &=& \frac{1}{(z-w)^2} \, \frac{c}{3} + \cdots,
\qquad 
c = \frac{1}{3} \, (N^2-1).
\label{JJOPE}
\eea
See also Appendix (\ref{n2scaexpression}) where the $16$
OPEs between the four currents of ${\cal N}=2$ superconformal algebra
are given.
The coset central charge will be discussed later in the context of 
stress energy coset spin-$2$ current.
We also used the fact that $\delta^{aa} = N^2-1$.

One can easily check that this coset spin-$1$ current does not have any
singular terms in the OPE $J(z) \, (J^a+K^a)(w)$.
All the coset (higher spin) currents should satisfy this requirement
\cite{BBSSfirst,BBSSsecond}.
That is,
\bea
(J^a+K^a)(z) \, J(w) & = &  + \cdots.
\label{regularcondition}
\eea 
The normalization factor $\frac{2}{3} i $ in (\ref{t1}) does not have any 
$N$-dependent factor because the $N$-dependent factor 
is canceled during this calculation. 
The numerical factor $ \frac{2}{3}$ will appear
as an $U(1)$ charge of the lowest higher spin-$2$ current in next section.  
Note that the $U(1)$ charge of $J(w)$ is zero because
there is no first-order pole in (\ref{JJOPE}). See also the footnote 
\ref{Jcharge} with (\ref{qcharges}).

Let us emphasize that 
one can check the combination 
$(\psi^a  \pm i \chi^a)(w)$ has the explicit 
$U(1)$ charges. That is,
\bea
J(z) \, (\psi^a  \pm i \chi^a)(w) & = &  \pm \frac{1}{(z-w)} \,
\frac{1}{3} (\psi^a  \pm i \chi^a)(w)
+\cdots.
\label{qcharges}
\eea
From (\ref{qcharges}),
their $U(1)$ charges are given by $\pm \frac{1}{3}$ respectively.
We will see that the higher spin currents can be written in terms of 
these two combinations with appropriate $f$ and $d$ symbols later.

$\bullet$ Coset spin-$\frac{3}{2}$ current

Let us consider one of the spin-$\frac{3}{2}$ currents
in the ${\cal N}=2$ superconformal algebra.
We obtain the following explicit form for this spin-$\frac{3}{2}$
current  
\bea
G^{+}(z) = -
\frac{1}{6\sqrt{3N}}
\Biggr[ \psi^a  J^a -3 \, \psi^a  K^a - i \, \chi^a  K^a
+ 3 \, i \, \chi^a J^a \Biggr](z).
\label{deetee}
\eea
The four terms in (\ref{deetee}) can be
determined by taking 
the spin-$\frac{1}{2}$ current and the spin-$1$ current in each 
$SU(N)$ factor in the coset model (\ref{coset}) because the spin should be
$\frac{3}{2}$. Then how one can 
determine the relative coefficients? 

One constraint is that this coset spin-$\frac{3}{2}$ current 
does not have any singular terms in the OPE with the diagonal spin-$1$
current. That is, along the line of (\ref{regularcondition}),
\bea
(J^a+K^a)(z) \, G^{+}(w) & = & +\cdots.
\label{reg}
\eea
Furthermore, the OPE between the coset spin-$1$ current and 
the coset spin-$\frac{3}{2}$ current should satisfy 
\bea
J(z) \,  G^{+}(w) & = & \frac{1}{(z-w)} \,  G^{+}(w)  +\cdots,
\label{JGOPE}
\eea
from the definition of ${\cal N}=2$ superconformal algebra.
In other words, the coset spin-$\frac{3}{2}$ current has 
$U(1)$ charge of $+1$.
See also the second equation of Appendix (\ref{n2scaexpression}).
Again from the explicit expressions in (\ref{t1}) and 
(\ref{spin1currents}) together with the coset spin-$\frac{3}{2}$ current 
with four unknown constants, the OPEs can be calculated and the two conditions 
(\ref{reg}) and (\ref{JGOPE}) are used. 
So far, the relative coefficients can be fixed and the overall factor 
can be determined later as the OPE between the two supersymmetry 
$\frac{3}{2}$ currents is obtained.
Then we obtain the expression (\ref{deetee}) except the normalization factor.

Let us rewrite the above spin-$\frac{3}{2}$ current
as follows
\bea
G^{+}(z) = -
\frac{1}{6\sqrt{3N}}
\, f^{abc} \, (\psi^a +i \chi^a)(\psi^b+i \chi^b)(\psi^c + i \chi^c)(z).
\label{g+fermions}
\eea
First of all, the first and third terms in (\ref{deetee})
can be seen from (\ref{g+fermions}) together with (\ref{spin1currents}) 
and the other two can be 
checked easily using the previous properties described before.
According to (\ref{qcharges}), this 
spin-$\frac{3}{2}$ current has $U(1)$ charge $1 (=\frac{1}{3} + 
\frac{1}{3} +\frac{1}{3})$. That is why we put the upper index $+$ in the 
spin-$\frac{3}{2}$ current $G^{+}(z)$.

$\bullet$ Coset spin-$\frac{3}{2}$ current

The second coset spin-$\frac{3}{2}$ current 
can be determined similarly.
We take four independent terms as before and
apply two conditions explained before.
It turns out that the second spin-$\frac{3}{2}$ current is  
given by
\bea
G^{-}(z) = -
\frac{1}{6\sqrt{3N}}
\Biggr[ \psi^a  J^a -3 \, \psi^a  K^a + i \, \chi^a  K^a
- 3 \, i \, \chi^a J^a \Biggr](z).
\label{deeteebar}
\eea
Two conditions are given by 
the regularity condition with the diagonal spin-$1$ current 
\bea
(J^a+K^a)(z) \, G^{-}(w) & = & +\cdots,
\label{reg1}
\eea
and the transformation with the coset spin-$1$ current
with definite $U(1)$ charge $-1$
\bea
J(z) \,  G^{-}(w) & = & -\frac{1}{(z-w)} \,  G^{-}(w)  +\cdots,
\label{jgminus}
\eea
from the ${\cal N}=2$ superconformal algebra.
The $U(1)$ charge of the spin-$\frac{3}{2}$ current $G^{-}(w)$ 
is given by $-1$ from (\ref{jgminus}).
See also the third equation of Appendix (\ref{n2scaexpression}).

Then we are left with the result (\ref{deeteebar})
except an overall factor.
Similarly,
one has the different description 
\bea
G^{-}(z) = -
\frac{1}{6\sqrt{3N}}
\, f^{abc} \, (\psi^a -i \chi^a)(\psi^b-i \chi^b)(\psi^c - i \chi^c)(z).
\label{g-fermions}
\eea
According to (\ref{qcharges}), this 
spin-$\frac{3}{2}$ current has $U(1)$ charge $-1 (=-\frac{1}{3} - 
\frac{1}{3} -\frac{1}{3})$.
It is straightforward to check the relation 
(\ref{g-fermions}) from (\ref{deeteebar}) as we did for the previous 
spin-$\frac{3}{2}$ current.
Under the change of $\chi^a(z) \rightarrow -\chi^a(z)$,
the current $G^{+}(z)$ goes to the current $G^{-}(z)$ and vice versa.

Furthermore, there exists the nontrivial OPE
between two spin-$\frac{3}{2}$ currents as follows: 
\bea
G^{+}(z) \, G^{-}(w) & = & \frac{1}{(z-w)^3} \, \frac{c}{3}+
\frac{1}{(z-w)^2} \, J(w) + \frac{1}{(z-w)} \, \left[ T +\frac{1}{2} \pa J 
\right](w) +\cdots.
\label{GG}
\eea
The first-order pole in (\ref{GG}) contains the coset 
spin-$2$ stress energy tensor 
which will appear soon. 
See also the seventh equation of Appendix (\ref{n2scaexpression}).
The highest-order pole is proportional to the central charge (\ref{JJOPE})
and the previous normalizations in the two spin-$\frac{3}{2}$ currents
can be determined from this singular term. We take the same normalization 
factor as in (\ref{deetee}) and (\ref{deeteebar})
 \footnote{
The combination $(G^{+} + G^{-})(z) =-\frac{1}{3\sqrt{3N}} \psi^a (J^a -
3 K^a)(z)$ is nothing but the ${\cal N}=1$ supersymmetry 
current where the level corresponding to the spin-$1$ current $K^a(z)$ 
is equal to $N$ \cite{Ahn1211,Ahn1305}.
The other combination $(G^{+} -G^{-})(z)= \frac{i}{3\sqrt{3N}} \chi^a (K^a -
3 J^a)(z)$
can be analyzed similarly.}.

$\bullet$ Coset spin-$2$ current

The coset spin-$2$ current can be 
obtained from the difference between the sum of each spin-$2$ current
and diagonal spin-$2$ current 
as follows 
\bea
T(z) & = & -\frac{1}{4N} \, J^a  J^a(z) -\frac{1}{4N} \, K^a  K^a(z) 
+ \frac{1}{6N}  (J^a +K^a) (J^a+K^a)(z).
\label{T}
\eea
The corresponding central charge (from the highest-order pole in the OPE
$T(z) \, T(w)$) is given by  
\bea
c = \frac{1}{3} \, (N^2-1), \qquad N =3, 4, 5, \cdots.
\label{cen}
\eea

One can easily check that this coset spin-$2$ current 
satisfies the regular condition with the diagonal spin-$1$ current
as analyzed in (\ref{regularcondition}), (\ref{reg}) and (\ref{reg1}) 
\footnote{
Moreover, the $U(1)$ charge of this coset spin-$2$ current is zero.
For the presence of higher spin currents, the minimum value of $N$ in 
(\ref{cen}) is $3$.
For $N=2$, the four currents of this section 
exist and there are no higher spin
currents as described in the introduction.}.
One can rewrite this spin-$2$ current
as 
\bea
T(z) & = & 
\frac{1}{3} (\psi^a +i \chi^a) \pa (\psi^a - i \chi^a)(z)+\frac{1}{2} 
\pa J(z) \nonu \\
& + & \frac{1}{6N} f^{abc} f^{cde} (\psi^a +i \chi^a)(\psi^b -i \chi^b)
(\psi^d +i \chi^d)(\psi^e -i \chi^e)(z),
\label{otherT}
\eea
where the spin-$1$ current $J(z)$ is given by (\ref{t1}). 
According to (\ref{qcharges}), this 
spin-$2$ current (\ref{otherT}) has $U(1)$ charge $0$.
The $N$-dependence appears in the last term.

Therefore, the four currents of 
${\cal N}=2$ superconformal algebra are given by 
(\ref{t1}), (\ref{deetee}), (\ref{deeteebar}) and (\ref{T}).
See also Appendix (\ref{n2scaexpression}).
The fundamental OPEs between the spin-$\frac{1}{2}$ currents 
are given by (\ref{psichi}) and the corresponding OPEs between 
the spin-$1$ currents 
are given in (\ref{JKOPE}).  
In next section, based on these four currents, the higher spin currents
will be constructed in the coset model explicitly. 

\section{ The lowest four higher spin currents }

In this section, the lowest higher spin currents 
will be obtained. That is, the first ${\cal N}=2$ higher spin multiplet.
Let us introduce the following 
higher spin current with the spin $h$ and $U(1)$ charge $q$
as follows:
\bea
W_q^{(h)}(z).
\label{bosonicW}
\eea
In ${\cal N}=2$ superspace, 
the ${\cal N}=2$ higher spin super current 
contains the following four higher spin currents as usual
 \footnote{We use a boldface notation for the 
${\cal N}=2$ super current. We do not use ``super'' explicitly 
in the description of 
${\cal N}=2$ 
super OPE or ${\cal N}=2$ super current.}
\bea
{\bf W}_{q}^{(h)} 
&\equiv& \left(W_{q}^{(h)}, \, W_{q+1}^{(h+\frac{1}{2})}, \,
W_{q-1}^{(h+\frac{1}{2})}, \,
W_{q}^{(h+1)} \right),
\label{superW}
\eea
where the lowest component higher spin current with spin $h$ 
and $U(1)$ charge $q$ in (\ref{superW})
corresponds to the one in (\ref{bosonicW}).
In Appendix $B$,  
the ${\cal N}=2$ superspace description for the 
${\cal N}=2$ superconformal algebra is given and 
some properties between the ${\cal N}=2$ super primary current  
and the 
${\cal N}=2$ stress energy tensor are described.
If one introduces the ${\cal N}=2$ superspace coordinates, $Z = (z, \theta, 
\bar{\theta})$, then the above becomes the relation given in 
Appendix (\ref{Wsuperexpression}) \footnote{
In other words, the first element of (\ref{superW}) corresponds to the 
$\theta, \bar{\theta}$ independent term, 
the second element corresponds to the 
$\theta$-term,
the third element corresponds to the 
$\bar{\theta}$-term,
and the fourth 
element corresponds to the 
$\theta \bar{\theta}$-term. Then one can assign 
the $U(1)$ charges for $\theta$ and $\bar{\theta}$ as $\mp 1$
respectively and their spins are given by $- \frac{1}{2}$.}.

According to the results for the extended vacuum character in \cite{GHKSS}, 
there exist the higher spin currents of spins $2, 2, \frac{7}{2}, \frac{7}{2}$
and $4$ with $U(1)$ charges $\frac{1}{3}, -\frac{1}{3}, \frac{1}{6}, 
-\frac{1}{6}$ and $0$ respectively \footnote{
Note that their $U(1)$ charge assignment is different from the $U(1)$ charge
in this paper. For example, their $U(1)$ charge for the spin-$\frac{3}{2}$
currents is given by $ \pm \frac{1}{2}$ while they are given by $\pm 1$ 
in previous section.
Then the above $U(1)$ charges are changed into 
 $\frac{2}{3}, -\frac{2}{3}, \frac{1}{3}, 
-\frac{1}{3}$ and $0$ in this paper.
Their currents are given by $W_{\frac{2}{3}}^{(2)}(z)$, $W_{-\frac{2}{3}}^{(2)}(z)$,
$W_{\frac{1}{3}}^{(\frac{7}{2})}(z)$, $W_{-\frac{1}{3}}^{(\frac{7}{2})}(z)$ and 
$W_{0}^{(4)}(z)$ in our notation.
 See also the equations $(2.28)$ and $(2.33)$ (and related descriptions) 
of \cite{GHKSS}.}.
It is straightforward to expresss the following 
${\cal N}=2$ higher spin currents by substituting the
spin $h$ and the $U(1)$ charge $q$ into (\ref{superW})
as follows: 
\bea
{\bf W}_{\frac{2}{3}}^{(2)} 
&\equiv& \left(W_{\frac{2}{3}}^{(2)}, \, W_{\frac{5}{3}}^{(\frac{5}{2})}, \,
W_{-\frac{1}{3}}^{(\frac{5}{2})}, \,
W_{\frac{2}{3}}^{(3)} \right),
\nonu \\
{\bf W}_{-\frac{2}{3}}^{(2)} &\equiv& 
\left(W_{-\frac{2}{3}}^{(2)}, \, W_{\frac{1}{3}}^{(\frac{5}{2})}, \,
W_{-\frac{5}{3}}^{(\frac{5}{2})}, \,
W_{-\frac{2}{3}}^{(3)} \right),
\nonu \\
{\bf W}_{\frac{1}{3}}^{(\frac{7}{2})} &\equiv&
\left(W_{\frac{1}{3}}^{(\frac{7}{2})}, \, W_{\frac{4}{3}}^{(4)}, \,
W_{-\frac{2}{3}}^{(4)}, \,
W_{\frac{1}{3}}^{(\frac{9}{2})} \right),
\nonu \\
{\bf W}_{-\frac{1}{3}}^{(\frac{7}{2})} &\equiv& 
\left(W_{-\frac{1}{3}}^{(\frac{7}{2})}, \, 
W_{\frac{2}{3}}^{(4)}, \, W_{-\frac{4}{3}}^{(4)}, \, 
W_{-\frac{1}{3}}^{(\frac{9}{2})} \right),
\nonu \\
{\bf W}_{0}^{(4)} &\equiv&
\left(W_{0}^{(4)}, \, 
W_{1}^{(\frac{9}{2})}, \, W_{-1}^{(\frac{9}{2})}, \, 
W_{0}^{(5)} \right),
\qquad \cdots \qquad .
\label{manyW}
\eea
The abbreviated higher spin currents in (\ref{manyW})
will be described later.
Each higher spin current possesses its own spin and $U(1)$ charge.
The 
higher spin currents with same spin have different $U(1)$ charges.
Note that the ${\cal N}=1$ 
higher spin currents with spins $(\frac{5}{2}, 3)$, 
$(\frac{7}{2}, 4)$, $(4, \frac{9}{2})$, $(4, \frac{9}{2})$, 
$(\frac{9}{2}, 5)$, $(\frac{11}{2}, 6)$ and $(6, \frac{13}{2})$ in 
\cite{Ahn1211,Ahn1305}
can be seen from the ${\cal N}=2$ version in (\ref{manyW}).

In this section, we would like to construct the 
${\cal N}=2$ higher spin multiplet (or
the lowest four higher spin currents) 
\bea
{\bf W}_{\frac{2}{3}}^{(2)} 
&\equiv& \left(W_{\frac{2}{3}}^{(2)}, \, W_{\frac{5}{3}}^{(\frac{5}{2})}, \,
W_{-\frac{1}{3}}^{(\frac{5}{2})}, \,
W_{\frac{2}{3}}^{(3)} \right).
\label{Wlowest}
\eea
One can construct the following $SU(N)$ adjoint spin-$1$ currents
by combining the two adjoint fermions with structure constant 
$f$ symbol and $d$ symbol
as follows:
\bea
L^a(z)  & \equiv &  f^{abc} \, \psi^b  \chi^c(z),
\nonu \\
M^a(z) & \equiv &  d^{abc} \, \psi^b  \chi^c(z).
\label{LMexpression}
\eea
As before, the indices $b$ and $c$ are summed over the 
$SU(N)$ adjoint indices.
Because the two fermions are independent from each other, one can construct 
them using the antisymmetric $f$ symbol and symmetric $d$ symbol.
It is easy to see that 
these are primary under the stress energy tensor (\ref{T}).
One can easily see that under the exchange of 
$\psi^a(z) \leftrightarrow \chi^a(z)$, $L^a(z)$ goes to $L^a(z)$
while $M^a(z)$ goes to $-M^a(z)$.
We will use this behavior in many OPEs appearing in Appendices.
These spin-$1$ currents will appear in the higher spin currents.

Let us introduce the intermediate $SU(N)$ adjoint spin-$2$ currents
with the symmetric $d$-symbol
as follows:
\bea
Q^a(z)  & \equiv &  d^{abc} \, J^b  J^c(z),
\nonu \\
U^a(z)  & \equiv &  d^{abc} \, J^b  L^c(z),
\nonu \\
S^a(z)  & \equiv &  d^{abc} \, K^b  K^c(z),
\nonu \\
V^a(z)  & \equiv &  d^{abc} \, K^b  L^c(z),
\nonu \\
R^a(z)  & \equiv &  d^{abc} \, J^b  K^c(z),
\nonu \\
W^a(z)  & \equiv &  d^{abc} \, L^b  L^c(z).
\label{spintwo}
\eea
The currents 
$Q^a(z)$, $S^a(z)$ and $R^a(z)$
also appeared in the ${\cal N}=1$ description in \cite{Ahn1211,Ahn1305}. 
Among these in (\ref{spintwo}), the currents 
$U^a(z)$ and $V^a(z)$ are not primary and the remaining currents are
primary under the stress energy tensor (\ref{T}) \footnote{
Under the exchange of $\psi^a(z) \leftrightarrow \chi^a(z)$,
the following relations hold:
$Q^a(z) \leftrightarrow S^a(z)$, $U^a(z) \leftrightarrow V^a(z)$,
$R^a(z) \leftrightarrow R^a(z)$ and $W^a(z) \leftrightarrow W^a(z)$.
In Appendix (\ref{appendixa}), 
we present some OPEs between the spin-$1$ currents.}.

\subsection{Higher spin-$2$ current}

Let us consider the lowest higher spin-$2$ current $W_{\frac{2}{3}}^{(2)}(z)$ 
in (\ref{Wlowest}).
How one can construct this higher spin current in terms of 
two kinds of adjoint fermions?
One can concentrate on the particular case with $N=3$.
Then the two adjoint fermions are characterized by 
$\psi^a(z)$ and $\chi^a(z)$ with $a=1, 2, \cdots, 8$.
Totally one has $16$ fermions satisfying the fundamental OPEs (\ref{psichi}).
Furthermore, there are nonzero 
$f$-symbol and $d$-symbol for the $SU(3)$ group \cite{Ahn1111}.
Then one can write down the possible spin-$2$ current
by considering all the terms, quartic terms and quadratic terms with one 
derivative. Of course, we introduce arbitrary coefficients here. 

Now we should determine these arbitrary coefficients explicitly
and express this current with the notation of a singlet under the $SU(3)$.
First of all, one has the following relation coming from 
the ${\cal N}=2$ condition for the primary. That is, the first equation of
Appendix (\ref{16TWOPE}) with $h=2$ and $q=\frac{2}{3}$.
Then one should have, together with (\ref{t1}) which consists of
eight terms, 
\bea
J(z) \, W_{\frac{2}{3}}^{(2)}(w) & = & \frac{1}{(z-w)} \, 
\frac{2}{3} \, W_{\frac{2}{3}}^{(2)}(w) + \cdots.
\label{JW}
\eea
In general, the right hand side contains the third-order and the second-order 
poles because the sum of spins in the left hand side is given by $3$.  
Then many unknown coefficients (the total number of unknown coefficients
is $2006$ for the $SU(3)$) can be fixed at this stage.
The equation of (\ref{JW}) is a linear equation among the coefficients
because the spin-$1$ current $J(z)$ in the left hand side 
is already known quantity with fixed coefficients. 
The first-order pole in (\ref{JW}) has explicit numerical factor 
$\frac{2}{3}$ which provides further constraints on the coefficients.
Recall that the $U(1)$ charge of the spin-$1$ current $J(z)$ is zero.

Furthermore, the following regularity condition should satisfy, as before,
\bea
(J^a+ K^a)(z) \, W_{\frac{2}{3}}^{(2)}(w) & = & +\cdots.
\label{JKW}
\eea
In general, there are third-order, second-order and first-order 
singular terms. However, the constraint (\ref{JKW}) allows us to insert 
more conditions on the coefficients. Moreover, 
the lowest higher spin-$2$ current should transform as a primary current 
under the stress energy tensor (\ref{T}) described in Appendix 
(\ref{16TWOPE}): the fourth equation from the bottom \footnote{
It turns out that one can determine all the coefficients except the overall 
normalization factor.
Explicitly one obtains 
$W_{\frac{2}{3}}^{(2)}(z) = -\psi^1 \chi^1 \chi^4 \chi^5(z) + \cdots +
\psi^8 \chi^6 \chi^7 \chi^8(z)$ with an overall factor.
The number of terms in $W_{\frac{2}{3}}^{(2)}(z)$ is $182$.
Note that there are no derivative terms (i.e. no quadratic terms) 
and only quartic terms survive.}.
Now we would like to express this higher spin-$2$ current for general $N$
from its $N=3$ version.
We expect that 
there should be $M^a(z)$-term because 
it contains $d$ symbol in (\ref{LMexpression}).
One can make any combinations from the other spin-$1$ currents,
$J^a(z)$, $K^a(z)$, $L^a(z)$ with $M^a(z)$.

It turns out that the lowest higher spin-$2$ current
is given by 
\bea
W_{\frac{2}{3}}^{(2)}(z) = 
\frac{1}{2\sqrt{6(N^2-4)}} \, \Biggr( J^a  M^a - K^a  M^a + 2 \, i\, 
L^a  M^a \Biggr)(z).
\label{lowestspin2}
\eea
The adjoint index $a$ is summed over $a= 1, 2, \cdots, (N^2-1)$. 
The normalization here can be fixed later 
as one computes the OPE between this higher spin-$2$ current 
and other higher spin-$2$ current which has opposite $U(1)$ charge.
As described before, 
each $SU(N)$ adjoint spin-$1$ current, $J^a(z)$, $K^a(z)$, $L^a(z)$
and $M^a(z)$ is primary and the OPEs between $J^a(z), K^a(z)$ and $L^a(z)$
and $M^a(w)$ are regular, each term appearing in (\ref{lowestspin2})
is primary under the stress energy tensor (\ref{T}). 
One can easily see that the $U(1)$ charge of $M^a(z)$ is equal to zero
and then the $U(1)$ charge of $(J^a-K^a + 2 i \, L^a)(z)$ is given by 
$\frac{2}{3}$.

One can express this higher spin-$2$ current
as 
\bea
W_{\frac{2}{3}}^{(2)}(z) = 
\frac{1}{2\sqrt{6(N^2-4)}} \,
\frac{i}{2} \, f^{abc} d^{cde} (\psi^a + i \chi^a)
(\psi^b+i \chi^b)(\psi^d+i \chi^d)(\psi^e - i \chi^e)(z),
\label{spin2otherform}
\eea
in terms of original adjoint fermions as done in (\ref{g+fermions}) 
and (\ref{g-fermions}).
According to (\ref{qcharges}), this 
higher spin-$2$ current has $U(1)$ charge $\frac{2}{3} (=\frac{1}{3} + 
\frac{1}{3} +\frac{1}{3}-\frac{1}{3})$.
Note that 
the last two factors in (\ref{spin2otherform}) can be written as
$- 2 i \, \psi^d \chi^e(z)$ which becomes $M^c(z)$ with the coefficient
$\frac{i}{2}$ and $d^{cde}$. It is obvious to observe the 
vanishing $U(1)$ charge of $M^c(z)$ from the last two factors.
Then the first two factors (of $U(1)$ charge $\frac{2}{3}$) 
with $f^{abc}$ become $(J^c -K^c +2 i \, L^c)(z)$.
One can check that the diagonal spin-$1$ current commutes with
$L^a M^a(w)$ and $(J^a-K^a)M^a(w)$ respectively \footnote{
Under the exchange of
$\psi^a(z) \leftrightarrow \chi^a(z)$, the first two terms of 
(\ref{lowestspin2}) are invariant while the last term has an extra minus 
sign. Recall that $L^a(z)$ does not change but 
$M^a(z)$ will change into $-M^a(z)$.
We will see that this is exactly the other type of lowest higher spin-$2$
current $W_{-\frac{2}{3}}^{(2)}(z)$ in next section.}.

In next two subsections, the next two higher spin currents 
can be obtained from the spin-$\frac{3}{2}$ currents (\ref{deetee}) 
and (\ref{deeteebar}) 
and the higher spin-$2$
current found in this subsection.

\subsection{Higher spin-$\frac{5}{2}$ current}

Now we would like to obtain the other three higher spin currents
living in the first ${\cal N}=2$ multiplet of (\ref{manyW}).
The higher spin-$\frac{5}{2}$ current can be determined by 
${\cal N}=2$ supersymmetry.
Recall that the fifth equation of 
Appendix (\ref{16TWOPE}) provides 
the higher spin-$\frac{5}{2}$ current 
with $q=\frac{2}{3}+1=\frac{5}{3}$.
That is, one should have 
\bea
G^{+}(z) \, W_{\frac{2}{3}}^{(2)}(w) & = & -\frac{1}{(z-w)} \, 
W_{\frac{5}{3}}^{(\frac{5}{2})}(w) +\cdots.
\label{gw}
\eea
Not that the $U(1)$ charge of $G^{+}(z)$
is given by $1$ and the sum of $U(1)$ charges in the left hand side
is preserved in the right hand side.
The spin-$\frac{3}{2}$ current $G^{+}(z)$
contains four terms in (\ref{deetee}) and the higher spin-$2$
current $W_{\frac{2}{3}}^{(2)}(w)$ contains 
three terms in (\ref{lowestspin2}).
This implies that one should calculate twelve OPEs between them.
We focus on the first-order poles in these OPEs in order to 
extract the higher spin-$\frac{5}{2}$ current in (\ref{gw}). 
Because  the explicit forms for the 
left hand side of (\ref{gw}) are completely
known, we do not have to worry about the overall factor 
of the higher spin-$\frac{5}{2}$ current.
The $N$-dependence for the higher spin-$\frac{5}{2}$ current arises 
automatically.
The explicit results at the intermediate step 
are given in Appendix (\ref{5half}).

By simplifying these, one obtains the final higher spin-$\frac{5}{2}$
current as follows: 
\bea
W_{\frac{5}{3}}^{(\frac{5}{2})}(z) &= & 
\frac{1}{6\sqrt{3N}} \, \frac{1}{2\sqrt{6(N^2-4)}} \,
\Biggr[ \frac{3}{2} \, i \, d^{abc} \, \psi^a  J^b  J^c - 5 \,
d^{abc} \, \psi^a  J^b  L^c  \nonu \\
&+& 
\frac{5}{2} \, i\, d^{abc} \, \psi^a  K^b  K^c
+ 15 \, d^{abc} \, \psi^a  K^b  L^c
-\frac{5}{2} \, d^{abc} \, \chi^a  J^b  J^c
\nonu \\
&-&   15 \, i\, d^{abc} \, \chi^a  J^b  L^c 
-\frac{3}{2} \, d^{abc} \, \chi^a  K^b  K^c
+ 5 \, i\, d^{abc} \, \chi^a  K^b  L^c
\Biggr](z). 
\label{5half1}
\eea
The $N$-dependent factors in (\ref{5half1})
originate from the overall factors of spin-$\frac{3}{2}$ current
and the higher spin-$2$ current.
Each term of (\ref{5half1}) 
is a primary under the stress energy tensor (\ref{T}).
One can express the above using the relations in (\ref{spintwo})
and then this will appear in Appendix $D$.

Let us describe how one can obtain the final result in (\ref{5half1}).
In obtaining (\ref{5half1}), we should simplify the intermediate 
results in Appendix $C$.
One has the following relations via the procedures in 
\cite{BBSSfirst,BBSSsecond,Fuchs}
\bea
d^{abc} \, \psi^a J^b K^c(z) & = &  d^{abc} f^{bde} f^{cfg} \, 
\psi^a \psi^d \psi^e \chi^f \chi^g(z),
\nonu \\
d^{abc} \, \psi^a L^b L^c(z) & = & d^{abc} f^{bde} f^{cfg} \, 
\psi^a \psi^d \chi^e \psi^f \chi^g(z) 
-\frac{N}{2} \pa \chi^a M^a(z),
\nonu \\
d^{abc} \, \chi^a J^b K^c(z) & = &  d^{abc} f^{bde} f^{cfg} \, 
\chi^a \psi^d \psi^e \chi^f \chi^g(z),
\nonu \\
d^{abc} \, \chi^a L^b L^c(z) & = &  d^{abc} f^{bde} f^{cfg} \, 
\chi^a \psi^d \chi^e \psi^f \chi^g(z) 
+\frac{N}{2} \pa \psi^a M^a(z).
\label{redundant}
\eea
The derivative terms occur in the 
second and fourth equations in (\ref{redundant}).
One has the following  
nontrivial relation (one should show)
\bea
d^{abc} \, \psi^a J^b K^c(z) +2
d^{abc} \, \psi^a L^b L^c(z) = 3
d^{abc} \, \chi^a J^b L^c(z) - N\, \pa \chi^a M^a(z).
\label{dfrelation}
\eea
Let us consider the derivative terms in (\ref{dfrelation}).
The last derivative term in (\ref{dfrelation}) comes from 
the derivative term of the 
second term in the left hand side of 
(\ref{dfrelation}) using the second equation of 
(\ref{redundant}). 
Now we 
would like to show the nonderivative terms both sides of (\ref{dfrelation}).
The nonderivative terms in (\ref{dfrelation}) 
can be checked by Jacobi identity
in the $f$ and $d$ symbols.
One way to see this relation is as follows \footnote{
The first term of the right hand side of 
(\ref{dfrelation}) is given by the nonderivative terms
$d^{abc} \, \chi^a J^b L^c(z)  =    d^{abc} f^{bde} f^{cfg} \, 
\chi^a \psi^d \psi^e \psi^f \chi^g(z)$.
Now the left hand side (nonderivative terms) of (\ref{dfrelation}) 
can be written as 
\bea
(d^{abc} f^{bde} f^{cfg} \, 
\chi^a \psi^d \psi^e \psi^f \chi^g-  d^{fac} f^{cbg} f^{bde} \, 
\psi^a \psi^d \psi^e \chi^f \chi^g + 2
 d^{abc} f^{bde} f^{cfg} \, 
\psi^a \psi^d \chi^e \psi^f \chi^g)(z),
\label{LHS}
\eea
where we reexpressed the $d^{abc} f^{cfg}$ as two other terms 
$-d^{bfc} f^{cag} - d^{fac} f^{cbg}$ and then some rearrangement between the 
fermions and relabeling the indices were used \cite{Ahn1111}.
The last term in (\ref{LHS}) comes from the nonderivative term in the 
second relation of (\ref{redundant}).
Now one can rewrite the right hand side (nonderivative terms) 
of (\ref{dfrelation}) as follows
\bea
(d^{abc} f^{bde} f^{cfg} \, 
\chi^a \psi^d \psi^e \psi^f \chi^g
+ 2
d^{abc} f^{bde} f^{cfg} \, 
\psi^a \psi^d \chi^e \psi^f \chi^g
-2 
d^{dab} f^{bce} f^{cfg} \, 
\chi^a \psi^d \psi^e \psi^f \chi^g)(z),
\label{RHS}
\eea
where the last two terms in (\ref{RHS}) are equal to
the twice of  
$d^{abc} f^{bde} f^{cfg} \, 
\chi^a \psi^d \psi^e \psi^f \chi^g(z)$ by rewriting
the factor $d^{abc} f^{bde}$ with the help of Jacobi identity as before.}.
Then we are left with the following relation we should show 
\bea
-  d^{fac} f^{cbg} f^{bde} \, 
\psi^a \psi^d \psi^e \chi^f \chi^g(z) =
-2 
d^{dab} f^{bce} f^{cfg} \, 
\chi^a \psi^d \psi^e \psi^f \chi^g(z),
\label{inter}
\eea
by comparing the nonderivative terms of the left hand side (\ref{dfrelation}) 
given by (\ref{LHS})and 
the nonderivative terms of the right hand side (\ref{dfrelation}) 
given by (\ref{RHS}).
The left hand side of (\ref{inter}) is the second term of 
(\ref{LHS}) while the right hand side of (\ref{inter})
is the last term of (\ref{RHS}). The remaining terms are 
canceled each other \footnote{
Then how one can check the above relation (\ref{inter})?
Once again by rearrangement of fermions and relabeling of the indices,
one should show that 
$(d^{fac} f^{cbg} f^{bde} + 2 d^{dfb} f^{bce} f^{cag}) 
\psi^a \psi^d \psi^e \chi^f \chi^g(z) =0$.
The left hand side of this relation  can be rewritten as
$d^{fac} ( f^{cbg} f^{bde} - 2  f^{cbe} f^{bdg}) 
\psi^a \psi^d \psi^e \chi^f \chi^g(z)$.
One can use the Jacobi identity in the first term and 
obtains  $2  f^{cbe} f^{bdg}$ exactly.
Therefore, we have checked the original relation in (\ref{dfrelation}).}.

Similarly,
one obtains the following identity
\bea
d^{abc} \, \chi^a J^b K^c(z) +2
d^{abc} \, \chi^a L^b L^c(z) = 3
d^{abc} \, \psi^a K^b L^c(z) + N\, \pa \psi^a M^a(z).
\label{otherdfrelation}
\eea
One can show this identity by following previous procedures.
Note that one sees the symmetry between $\psi^a(z)$ and $\chi^a(z)$
in (\ref{dfrelation}) and (\ref{otherdfrelation}) \footnote{
In other words, from the previous relation (\ref{dfrelation}),
one obtains (\ref{otherdfrelation}) by taking 
$\psi^a(z) \leftrightarrow \chi^a(z)$. There exists an extra 
minus sign in the last term of (\ref{otherdfrelation}) as described before.}. 

In order to express the independent terms in the higher spin-$\frac{5}{2}$
currents, one 
uses the following relations  appearing in (\ref{5half1})
\bea
d^{abc} \, \psi^a J^b J^c(z) & = &  d^{abc} f^{bde} f^{cfg} \, 
\psi^a \psi^d \psi^e \psi^f \psi^g(z), \nonu \\
d^{abc} \, \psi^a J^b L^c(z) & = &  -d^{abc} f^{bde} f^{cfg} \, 
\chi^f \psi^a \psi^d \psi^e \psi^g(z)- N \, \pa \psi^a M^a(z),
\nonu \\
d^{abc} \, \psi^a K^b K^c(z) & = &  d^{abc} f^{bde} f^{cfg} \, 
\psi^a \chi^d \chi^e \chi^f \chi^g(z)- 2N \, \pa \chi^a M^a(z), 
\nonu \\
d^{abc} \, \psi^a K^b L^c(z) & = &  d^{abc} f^{bde} f^{cfg} \, 
\psi^a \chi^d \chi^e \psi^f \chi^g(z),
\nonu \\
d^{abc} \, \chi^a J^b J^c(z) & = &  - d^{abc} f^{bde} f^{cfg} \, 
\chi^f \psi^a \psi^d \psi^e \psi^g(z) + 2N \, \pa \psi^a M^a(z),
\nonu \\
d^{abc} \, \chi^a J^b L^c(z) & = &   d^{abc} f^{bde} f^{cfg} \, 
\chi^a \psi^d \psi^e \psi^f \chi^g(z),
\nonu \\
d^{abc} \, \chi^a K^b K^c(z) & = &  d^{abc} f^{bde} f^{cfg} \, 
\chi^a \chi^d \chi^e \chi^f \chi^g(z), \nonu \\
d^{abc} \, \chi^a K^b L^c(z) & = &   - d^{abc} f^{bde} f^{cfg} \, 
\psi^f \chi^a \chi^d \chi^e \chi^g(z) + N \, \pa \chi^a M^a(z).
\label{eightexpression}
\eea
There are derivative terms in (\ref{eightexpression}).
One realizes that the higher spin-$\frac{5}{2}$ current 
does not have any derivative terms because the derivative terms 
in (\ref{eightexpression}) are canceled each other with relative coefficients
in (\ref{5half1}).   
Then we are left with the final result in (\ref{5half1})
\footnote{
Furthermore, one has the following expressions
which appear in Appendix $C$ 
\bea
f^{abc} \, \chi^a K^b M^c(z) & = & - N \, \pa \chi^a M^a(z),  
\nonu \\
f^{abc} \, \psi^a J^b M^c(z) & = & N \,  \psi^a \pa M^a(z),  
\nonu \\
f^{abc} \, \chi^a L^b M^c(z) & = & f^{abc} f^{bde} d^{cfg} \chi^a \psi^d
\chi^e \psi^f \chi^g(z)-\frac{N}{2} \pa \psi^a M^a(z),
\nonu \\
f^{abc} \, \psi^a K^b M^c(z) & = & f^{abc} f^{bde} d^{cfg} \psi^a \chi^d
\chi^e \psi^f \chi^g(z) = -2 f^{abc} f^{bde} d^{cfg} \chi^a \psi^d
\chi^e \psi^f \chi^g(z),
\nonu \\
f^{abc} \, \chi^a J^b M^c(z) & = & f^{abc} f^{bde} d^{cfg} \chi^a \psi^d
\psi^e \psi^f \chi^g(z) = -2  f^{abc} f^{bde} d^{cfg} \psi^a \psi^d
\chi^e \psi^f \chi^g(z),
\nonu \\
f^{abc} \, \psi^a L^b M^c(z) & = & f^{abc} f^{bde} d^{cfg} \psi^a \psi^d
\chi^e \psi^f \chi^g(z) -\frac{N}{2} \, \pa \chi^a M^a(z).
\label{fsymbolexp}
\eea
Also one has trivial result $\psi^a M^a(z) = 0= \chi^a M^a(z)$.
From these, one obtains $\pa \psi^a M^a(z) = - \psi^a \pa M^a(z)$
and similarly $\pa \chi^a M^a(z) = - \chi^a \pa M^a(z)$.
Therefore, via (\ref{dfrelation}) and (\ref{otherdfrelation}),
the left hand sides of these can be expressed as the right hand sides
which are some independent terms in (\ref{eightexpression}) together with
derivative terms. The $f$ symbol-dependent terms appearing in Appendix
$C$ can be repexpressed in terms of the right hand sides of 
(\ref{fsymbolexp}). The nonderivative terms in (\ref{fsymbolexp})
disappear with appropriate coefficients eventually.}. 

As long as the $U(1)$ charge is concerned, the previous expression 
(\ref{5half1}) 
for the higher spin-$\frac{5}{2}$ current is not useful.
In manifest way of $U(1)$ charge, 
one obtains 
\bea
W_{\frac{5}{3}}^{(\frac{5}{2})}(z) & = &  
\frac{1}{6\sqrt{3N}} \, \frac{1}{2\sqrt{6(N^2-4)}} \,
\frac{3}{2} i \nonu \\
 & \times & d^{abc} f^{bde} f^{cfg} (\psi^a+ i \chi^a)(\psi^d+ i \chi^d)
(\psi^e+ i \chi^e)(\psi^f+ i \chi^f)(\psi^g+ i \chi^g)(z).
\label{quintic}
\eea
One can check that the combination $(\psi^a + i \chi^a)(z)$
has $U(1)$ charge $q= \frac{1}{3}$ from (\ref{qcharges}).
It is obvious that the quintic term in (\ref{quintic})
contains the $U(1)$ charge $q=\frac{5}{3}$.
One can see the overall factor $\frac{3}{2} i$ with vanishing $\chi^a(z)$ 
in  
(\ref{quintic}) can be seen from the first term of (\ref{5half1})
with vanishing $\chi^a(z)$.
As for spin-$1$, spin-$\frac{3}{2}$, and higher spin-$2$ currents, 
the above higher spin-$\frac{5}{2}$ current has a simple  
expression contracted with $f$ and $d$ symbols.
See also \cite{ASS1991} where the similar $f$ and $d$ symbols 
appeared in the ${\cal N}=1$ higher spin-$\frac{5}{2}$ current.

\subsection{Higher spin-$\frac{5}{2}$ current}

Let us consider the second higher spin-$\frac{5}{2}$ current
with $U(1)$ charge $q=-\frac{1}{3}$. Again, from the 
${\cal N}=2$  primary condition, one has 
the following OPE, the ninth equation of Appendix (\ref{16TWOPE}),
\bea
G^{-}(z) \, W_{\frac{2}{3}}^{(2)}(w) & = & \frac{1}{(z-w)} \, 
W_{-\frac{1}{3}}^{(\frac{5}{2})}(w) +\cdots.
\label{gw1}
\eea
One can easily see the $U(1)$ charge conservation in (\ref{gw1})
where $-1+ \frac{2}{3} =-\frac{1}{3}$.
From the explicit expressions (\ref{deeteebar}) and (\ref{lowestspin2}),
one can calculate the left hand side completely 
and focus on the first-order pole.

As done before in Appendix $C$, all the first-order poles 
can be read off.
Then one obtains
\bea
W_{-\frac{1}{3}}^{(\frac{5}{2})}(z) &= & 
-\frac{1}{6\sqrt{3N}} \, \frac{1}{2\sqrt{6(N^2-4)}} \,
\Biggr[ -\frac{3}{2} \, i \, d^{abc} \, \psi^a  J^b  J^c +   
d^{abc} \, \psi^a  J^b  L^c  \nonu \\
&-& 
\frac{17}{2} \, i\, d^{abc} \, \psi^a  K^b  K^c
+ 33 \, d^{abc} \, \psi^a  K^b  L^c
-\frac{5}{2} \, d^{abc} \, \chi^a  J^b  J^c
\nonu \\
&-&  21 \, i\, d^{abc} \, \chi^a  J^b  L^c 
-\frac{3}{2} \, d^{abc} \, \chi^a  K^b  K^c
+ 7 \, i\, d^{abc} \, \chi^a  K^b  L^c
\nonu \\
&+& 
18 \, i\, d^{abc} \, \psi^a  J^b  K^c
-36 \, d^{abc} \, \chi^a  L^b  L^c
\Biggr](z). 
\label{5half2}
\eea
Compared to the previous higher spin-$\frac{5}{2}$ current in (\ref{5half1}),
there are two additional last terms which are also primary under the 
stress energy tensor (\ref{T}) respectively.  

It turns out that the corresponding higher spin-$\frac{5}{2}$ current,
showing the $U(1)$ charge manifestly, 
is given by
\bea
W_{-\frac{1}{3}}^{(\frac{5}{2})}(z) & = &  
-\frac{1}{6\sqrt{3N}} \, \frac{1}{2\sqrt{6(N^2-4)}} \, \Biggr[
  \nonu \\
 &  & 2 i \, d^{abc} f^{bde} f^{cfg} (\psi^a- i \chi^a)(\psi^d+ i \chi^d)
(\psi^e- i \chi^e)(\psi^f+ i \chi^f)(\psi^g- i \chi^g)
\nonu \\
&-& \frac{7}{2} i \, 
 d^{abc} f^{bde} f^{cfg} (\psi^a- i \chi^a)(\psi^d+ i \chi^d)
(\psi^e+ i \chi^e)(\psi^f- i \chi^f)(\psi^g- i \chi^g)
\nonu \\
&-&   12 N i\, d^{abc} \pa (\psi^a - i \chi^a) 
(\psi^b+i \chi^b)(\psi^c-i\chi^c) \Biggr](z).
\label{mani5half2}
\eea
As analyzed before (\ref{qcharges}), the factors $(\psi^a \pm i \chi^a)(z)$
have the $U(1)$ charges $\pm \frac{1}{3}$. 
It is obvious that the quintic terms  in (\ref{mani5half2})
contain two positive $U(1)$ charges and three negative charges. 
Of course, there are other possibilities where 
the two positive charges are assigned in different factors compared to 
the above expressions but they will lead to the above ones 
by using the Jacobi identities between the $f$ and $d$ symbols.
In the derivative terms, one can easily see that 
there are two negative $U(1)$ charges and one positive one which lead to 
the negative $U(1)$ charge $q=-\frac{1}{3}$.
The numerical factor $-\frac{3}{2} i$ in the first term in (\ref{5half2})
can be seen from the expressions $2i$ in the first coefficient 
and $-\frac{7}{2} i $ in the second coefficient
of (\ref{mani5half2}). Note that there are derivative terms having 
$N$-dependent coefficient in the above 
higher spin-$\frac{5}{2}$ current. This is obvious from 
(\ref{mani5half2}) rather than (\ref{5half2}).

\subsection{Higher spin-$3$ current}

Let us describe the last component higher spin-$3$ current
appearing in the first ${\cal N}=2$ higher spin multiplet in (\ref{manyW}).
One way to obtain this higher spin-$3$ current is coming from 
the tenth equation of Appendix (\ref{16TWOPE}) \footnote{
The seventh equation of Appendix (\ref{16TWOPE}) allows us to 
calculate the higher spin-$3$ current. }.
Let us consider the following OPE
\bea
G^{-}(z) \, W_{\frac{5}{3}}^{(\frac{5}{2})}(w) & = & 
-\frac{1}{(z-w)^2} \, \frac{5}{3} \, W_{\frac{2}{3}}^{(2)}(w)
+ \frac{1}{(z-w)} \, 
\left[ W_{\frac{2}{3}}^{(3)} -\frac{1}{2} \, \pa \, W_{\frac{2}{3}}^{(2)}
\right](w) +\cdots.
\label{spin3opegen}
\eea
The $U(1)$ charge is preserved in both sides.
Because one has the relations (\ref{deeteebar}) and (\ref{5half1}),
the left hand side can be computed explicitly. The detailed expressions
are given in Appendix $D$.
Due to the analysis of 
the coefficient of the descendant field (appearing in the first-order pole) 
of the higher spin-$2$
current, 
one should have $-\frac{5}{3} \times \frac{1}{4} = -\frac{5}{12}$
in the coefficient of $ \pa \, W_{\frac{2}{3}}^{(2)}(w)$.
Recall that the factor $\frac{1}{4}$ was obtained from 
the spins of $G^{-}(z)$, $W_{\frac{5}{3}}^{(\frac{5}{2})}(w)$ and 
$ W_{\frac{2}{3}}^{(2)}(w)$.
Therefore, one can add this quantity to the first-order pole and subtract the 
same quantity. Then the new higher spin-$3$ current 
can be written as the sum of $ W_{\frac{2}{3}}^{(3)}(w)$ and 
$(\frac{5}{12} -\frac{1}{2}=-\frac{1}{12}) \pa \, W_{\frac{2}{3}}^{(2)}(w)$
in (\ref{spin3opegen}).  

The higher spin-$3$ current, which is a primary, can be obtained as follows
\bea
W_{\frac{2}{3}}^{(3)} -\frac{1}{12} \, \pa \, W_{\frac{2}{3}}^{(2)}  &=& 
-\frac{1}{108N} \, 
 \frac{1}{2\sqrt{6(N^2-4)}} \Biggr[
-15 \, i\, d^{abc} \, J^a  J^b  J^c 
+ 90 \, i\,  d^{abc} \, J^a  J^b  K^c
 \nonu \\
&- &  90 \, i\, 
 d^{abc} \, J^a  K^b  K^c
-360 \,  d^{abc} \, J^a  K^b  L^c
+ 15 \, i\,  d^{abc} \, K^a  K^b  K^c
\nonu \\
&-&  \frac{45}{2} \, N\, \pa  J^a  M^a
+\frac{45}{2} \, N \,   \pa  K^a  M^a
-225N \, d^{abc} \, \psi^a  \pa \psi^b  L^c
\label{lastw3}
\\
&-&  225N \, d^{abc} \, \chi^a  \pa \chi^b  L^c 
+315N \, d^{abc} \, \pa \psi^a   \chi^b  K^c 
-315N \, d^{abc} \, \psi^a  \pa \chi^b  J^c 
\Biggr].
\nonu
\eea
There are several pairs of terms where the 
coefficients are equal up to the signs. 
The first, second, third, fifth terms in (\ref{lastw3}) 
are primary under the stress energy
tensor (\ref{T}) respectively.
Recall that $W_{\frac{2}{3}}^{(3)}(z)$, where the $U(1)$ charge is nonzero, 
is not a primary under the 
stress energy tensor according to the last equation of Appendix 
(\ref{16TWOPE}).
See also Appendix (\ref{polethree}) and Appendix (\ref{primarycombi}).
The various identities appearing in Appendix (\ref{appendixddetails})
are used frequently. 
Note that there are $N$-dependent coefficients in the derivative terms in 
(\ref{lastw3}).

One can also express the above higher spin-$3$ current in $U(1)$ manifest
way as follows: 
\bea
&& W_{\frac{2}{3}}^{(3)}(z) -\frac{1}{12} \, \pa \, W_{\frac{2}{3}}^{(2)}(z)  = 
-\frac{1}{108N} \, 
 \frac{1}{2\sqrt{6(N^2-4)}} \Biggr[
 \nonu \\
& & -  \frac{45}{2} i \, d^{abc} f^{ade} f^{bfg}
f^{chi} (\psi^d + i \chi^d )(\psi^e + i \chi^e)(\psi^f + i \chi^f)(\psi^g +
i \chi^g)(\psi^h-i \chi^h)( \psi^i - i \chi^i) 
\nonu \\
&& -45i \, N \, f^{abc} d^{cde} \pa (\psi^a+ i \chi^a) 
(\psi^b +i \chi^b)(\psi^d +i \chi^d)(\psi^e- i \chi^e)
\nonu \\
&& -\frac{45}{2} i \, N \, f^{abc} d^{cde}  (\psi^a+ i \chi^a) 
(\psi^b +i \chi^b) \pa (\psi^d +i \chi^d)(\psi^e- i \chi^e)
\nonu \\
&&  +\frac{135}{2} i \, N \, f^{abc} d^{cde}  (\psi^a+ i \chi^a) 
(\psi^b +i \chi^b)(\psi^d +i \chi^d) \pa (\psi^e- i \chi^e) \Biggr](z).
\label{manispin3}
\eea
Now it is easy to see that 
the sextic (nonderivative) terms in (\ref{manispin3}) have 
four $U(1)$ charges with $q=\frac{1}{3}$ and two 
ones with $q=-\frac{1}{3}$, leading to the total $U(1)$ charge of 
$q=\frac{2}{3}$ (\ref{qcharges}). For the derivative terms, 
the derivative acts on each factor appearing in 
the higher spin-$2$ current.
The relative coefficients are different from each other. 
The derivative term acting on the $(\psi^b +i \chi^b)$ is not independent 
term and can be absorbed in the second line of (\ref{manispin3}).
Of course, the higher spin-$3$ current, 
$W_{\frac{2}{3}}^{(3)}(z)$, which is not a primary,
can be written explicitly by using (\ref{lastw3}) or (\ref{manispin3})
with the help of the higher spin-$2$ current (\ref{lowestspin2}) or 
(\ref{spin2otherform}). 

Therefore, in this section, the four higher spin currents appearing in the
first ${\cal N}=2$ multiplet in (\ref{manyW}) are determined completely.
There are either (\ref{lowestspin2}), (\ref{5half1}), (\ref{5half2}) and (\ref{lastw3}) 
or 
(\ref{spin2otherform}), (\ref{quintic}), (\ref{mani5half2}) and 
(\ref{manispin3}) in terms of two adjoint fermions. 
The derivative terms appear in the higher spin currents 
$W_{-\frac{1}{3}}^{(\frac{5}{2})}(z)$ and $W_{\frac{2}{3}}^{(3)}(z)$.

\section{The other lowest four higher spin currents }

In this section, the other lowest higher spin currents
in the second higher spin ${\cal N}=2$ multiplet in (\ref{manyW}) 
will be obtained by following the procedures in previous section.

\subsection{Higher spin-$2$ current}

By substituting $q=-\frac{2}{3}$ with $h=2$ into the first equation 
of Appendix (\ref{16TWOPE}), one obtains
\bea
J(z) \, W_{-\frac{2}{3}}^{(2)}(w) & = & -\frac{1}{(z-w)} \, 
\frac{2}{3} \, W_{-\frac{2}{3}}^{(2)}(w) + \cdots.
\label{jwsection4}
\eea
As done in the higher spin-$2$ current with $q=\frac{2}{3}$,
one can take the same ansatz for all the possible terms
having the spin-$2$ (quartic terms and quadratic terms)
with arbitrary coefficients which will be determined later.  
Due to the minus sign in the right hand side of (\ref{jwsection4}),
in general, one has different relations between these coefficients
compared to the previous section.
One also has the regularity condition as follows:
\bea
(J^a+ K^a)(z) \, W_{-\frac{2}{3}}^{(2)}(w) & = & +\cdots.
\label{regsection4}
\eea

Then one obtains the following higher spin-$2$ current, satisfying 
(\ref{jwsection4}) and (\ref{regsection4}),
\bea
W_{-\frac{2}{3}}^{(2)}(z) = 
\frac{1}{2\sqrt{6(N^2-4)}} \, \Biggr( J^a  M^a - K^a  M^a - 2 \, i\, 
L^a  M^a \Biggr)(z).
\label{spin2other}
\eea
This looks similar to the previous higher spin-$2$ current 
with $q=\frac{2}{3}$. The only difference appears in the coefficient in the last
term of (\ref{spin2other}). 
As observed previously, 
the $M^a(z)$ has $U(1)$ charge zero.
Then the combination $(J^a-K^a-2\, i\, L^a)(z)$ should have $U(1)$ charge 
$q=-\frac{2}{3}$ which can be checked explicitly.

One can express this higher spin-$2$ current
as 
\bea
W_{-\frac{2}{3}}^{(2)}(z) = 
-\frac{1}{2\sqrt{6(N^2-4)}} \,
\frac{i}{2} \, f^{abc} d^{cde} (\psi^a - i \chi^a)
(\psi^b-i \chi^b)(\psi^d-i \chi^d)(\psi^e + i \chi^e)(z),
\label{spin2othersection4}
\eea
in terms of original adjoint fermions.
By exchanging $\psi^a(z) \leftrightarrow \chi^a(z)$, the previous 
result (\ref{spin2otherform}) leads to (\ref{spin2othersection4}).
Under this transformation, one has
$(\psi^a \pm i \chi^a)(z) \rightarrow \pm i (\psi^a \mp i \chi^a)(z)$.
According to (\ref{qcharges}), this 
higher spin-$2$ current has $U(1)$ charge $-\frac{2}{3} (=-\frac{1}{3} - 
\frac{1}{3} -\frac{1}{3} +\frac{1}{3})$.
Note that 
the last two factors in (\ref{spin2othersection4}) can be written as
$ 2 i \, \psi^d \chi^e(z)$ which becomes $M^c(z)$ with the coefficient
$-\frac{i}{2}$ and $d^{cde}$. It is obvious to observe the 
vanishing $U(1)$ charge of $M^c(z)$ from the last two factors
as before.
Then the first two factors (of $U(1)$ charge $-\frac{2}{3}$) 
with $f^{abc}$ become $(J^c -K^c - 2 i \, L^c)(z)$.

\subsection{Higher spin-$\frac{5}{2}$ current}

Let us move on the second component higher spin-$\frac{5}{2}$ current
in the second ${\cal N}=2$ higher spin multiplet in (\ref{manyW}).
Again, from the defining equation of the fifth equation in 
Appendix (\ref{16TWOPE})
by substituting $h=2$ and $q=-\frac{2}{3}$, one obtains
\bea
G^{+}(z) \, W_{-\frac{2}{3}}^{(2)}(w) & = & -\frac{1}{(z-w)} \, 
W_{\frac{1}{3}}^{(\frac{5}{2})}(w) +\cdots.
\label{gw5half}
\eea
One can calculate the left hand side of (\ref{gw5half})
with (\ref{deetee}) and (\ref{spin2other}) 
or (\ref{spin2othersection4}) and focus on 
the first-order pole.
From Appendix $C$, one can collect the corresponding 
expressions.

It turns out that 
the corresponding higher spin-$\frac{5}{2}$ current is given by
\bea
W_{\frac{1}{3}}^{(\frac{5}{2})}(z) &= & 
\frac{1}{6\sqrt{3N}} \, \frac{1}{2\sqrt{6(N^2-4)}} \,
\Biggr[ \frac{3}{2} \, i \, d^{abc} \, \psi^a  J^b  J^c +   
d^{abc} \, \psi^a  J^b  L^c  \nonu \\
&+& 
\frac{17}{2} \, i\, d^{abc} \, \psi^a  K^b  K^c
+ 33 \, d^{abc} \, \psi^a  K^b  L^c
-\frac{5}{2} \, d^{abc} \, \chi^a  J^b  J^c
\nonu \\
&+&  21 \, i\, d^{abc} \, \chi^a  J^b  L^c 
-\frac{3}{2} \, d^{abc} \, \chi^a  K^b  K^c
- 7 \, i\, d^{abc} \, \chi^a  K^b  L^c
\nonu \\
&-& 
18 \, i\, d^{abc} \, \psi^a  J^b  K^c
-36 \, d^{abc} \, \chi^a  L^b  L^c
\Biggr](z). 
\label{aboveexpression}
\eea
Now this (\ref{aboveexpression}) 
looks similar to the previous higher spin-$\frac{5}{2}$
current (\ref{5half2}) with $q=-\frac{1}{3}$. Only the signs of 
the numerical coefficients
are different.

Furthermore, one can reexpress the above higher spin-$\frac{5}{2}$ current 
in terms of the original adjoint fermions as follows:
\bea
W_{\frac{1}{3}}^{(\frac{5}{2})}(z) & = &  
\frac{1}{6\sqrt{3N}} \, \frac{1}{2\sqrt{6(N^2-4)}} \, \Biggr[
  \nonu \\
 & - & 2 i \, d^{abc} f^{bde} f^{cfg} (\psi^a+ i \chi^a)(\psi^d- i \chi^d)
(\psi^e+ i \chi^e)(\psi^f- i \chi^f)(\psi^g+ i \chi^g)
\nonu \\
&+& \frac{7}{2} i \, 
 d^{abc} f^{bde} f^{cfg} (\psi^a+ i \chi^a)(\psi^d- i \chi^d)
(\psi^e- i \chi^e)(\psi^f+ i \chi^f)(\psi^g+ i \chi^g)
\nonu \\
&+&   12 N i\, d^{abc} \pa (\psi^a + i \chi^a) 
(\psi^b-i \chi^b)(\psi^c+i\chi^c) \Biggr](z).
\label{5halfsecond1}
\eea
Under the transformation $\psi^a(z) \leftrightarrow \chi^a(z)$, 
the previous higher spin-$\frac{5}{2}$ current with $q=-\frac{1}{3}$
(\ref{mani5half2}) 
goes to (\ref{5halfsecond1}) with an exception of overall factor $-i$.
One can check the $U(1)$ charge of this higher spin-$\frac{5}{2}$ 
current using (\ref{qcharges}).

\subsection{Higher spin-$\frac{5}{2}$ current}

The ninth equation of Appendix 
(\ref{16TWOPE}) with $h=2$ and $q=-\frac{2}{3}$
implies 
the following OPE
\bea
G^{-}(z) \, W_{-\frac{2}{3}}^{(2)}(w) & = & \frac{1}{(z-w)} \, 
W_{-\frac{5}{3}}^{(\frac{5}{2})}(w) +\cdots.
\label{gwothercase}
\eea
Then as we did before, 
by calculating the left hand side, one obtains the explicit form for the 
higher spin-$\frac{5}{2}$ current with $q=-\frac{5}{3}$.
From the explicit forms in (\ref{deeteebar}) and (\ref{spin2other}), 
the complete structures in the first-order pole are 
determined and 
the details are again in Appendix $C$.

Therefore, one obtains, by reading off the first-order pole in 
(\ref{gwothercase}),
\bea
W_{-\frac{5}{3}}^{(\frac{5}{2})}(z) &= & 
-\frac{1}{6\sqrt{3N}} \, \frac{1}{2\sqrt{6(N^2-4)}} \,
\Biggr[ -\frac{3}{2} \, i \, d^{abc} \, \psi^a  J^b  J^c -5\,
d^{abc} \, \psi^a  J^b  L^c  \nonu \\
&-& 
\frac{5}{2} \, i\, d^{abc} \, \psi^a  K^b  K^c
+ 15 \, d^{abc} \, \psi^a  K^b  L^c
-\frac{5}{2} \, d^{abc} \, \chi^a  J^b  J^c
\nonu \\
&+&   15 \, i\, d^{abc} \, \chi^a  J^b  L^c 
-\frac{3}{2} \, d^{abc} \, \chi^a  K^b  K^c
- 5 \, i\, d^{abc} \, \chi^a  K^b  L^c
\Biggr](z). 
\label{5halflast}
\eea
In this case, the field contents are the same as the ones in (\ref{5half1}).
The signs of the numerical factors are different.
As in previous cases, one can rewrite (\ref{5halflast}) as   
\bea
W_{-\frac{5}{3}}^{(\frac{5}{2})}(z) & = &  
\frac{1}{6\sqrt{3N}} \, \frac{1}{2\sqrt{6(N^2-4)}} \,
\frac{3}{2} i \nonu \\
 & \times & d^{abc} f^{bde} f^{cfg} (\psi^a- i \chi^a)(\psi^d- i \chi^d)
(\psi^e- i \chi^e)(\psi^f- i \chi^f)(\psi^g- i \chi^g)(z).
\label{second5half2}
\eea
One obtains (\ref{second5half2}) by taking the transformation
$\psi^a(z) \leftrightarrow \chi^a(z)$ on (\ref{quintic}) except of an overall 
factor $i$.  
One can check the $U(1)$ charge of this higher spin-$\frac{5}{2}$ 
current using (\ref{qcharges}).
The above higher spin-$\frac{5}{2}$ current has very simple form
contracted with $f$ and $d$ symbols
\footnote{One can construct the following combination
$( \frac{7}{5} W_{\frac{5}{3}}^{(\frac{5}{2})} + \frac{7}{5} 
W_{-\frac{5}{3}}^{(\frac{5}{2})}  + W_{-\frac{1}{3}}^{(\frac{5}{2})}+
W_{\frac{1}{3}}^{(\frac{5}{2})})(z)$ which is proportional to $ d^{abc} (3 
\psi^a J^b J^c -15 \psi^a J^b K^c + 10 \psi^a K^b K^c)(z)$. 
This is exactly the higher spin-$\frac{5}{2}$ current of ${\cal N}=1$ version in 
\cite{Ahn1211,Ahn1305}. The relative coefficient $\frac{7}{5}$
removes the terms of $d^{abc} \chi^a J^b L^c(z)$ and $d^{abc} 
\chi^a K^b L^c(z)$.}.

\subsection{Higher spin-$3$ current}

The seventh equation of Appendix 
(\ref{16TWOPE}) with $h=2$ and $q=-\frac{2}{3}$ 
implies 
\bea
G^{+}(z) \, W_{-\frac{5}{3}}^{(\frac{5}{2})}(w) & = & 
\frac{1}{(z-w)^2} \, \frac{5}{3} \, W_{-\frac{2}{3}}^{(2)}(w)
+ \frac{1}{(z-w)} \, 
\left[ W_{-\frac{2}{3}}^{(3)}(z) +\frac{1}{2} \, \pa \, W_{-\frac{2}{3}}^{(2)}
\right](w) 
\nonu \\
& + & \cdots.
\label{opegw}
\eea
The corresponding coefficient of the descendant field $\pa \, 
W_{-\frac{2}{3}}^{(2)}(w)$ in the first-order pole of (\ref{opegw}) 
is equal to $\frac{5}{3} \times \frac{1}{4}=
\frac{5}{12}$.
Then by subtracting $\frac{5}{12}$ and adding $\frac{1}{2}$ ($-\frac{5}{12}+
\frac{1}{2}= \frac{1}{12}$), one can 
obtain the new higher spin-$3$ current, which is  a primary, 
as follows:
\bea
W_{-\frac{2}{3}}^{(3)} 
 +\frac{1}{12} \, \pa \, W_{-\frac{2}{3}}^{(2)}
&=& -\frac{1}{108N} \, 
 \frac{1}{2\sqrt{6(N^2-4)}} \Biggr[
-15 \, i\, d^{abc} \, J^a  J^b  J^c 
+ 90 \, i\,  d^{abc} \, J^a  J^b  K^c
 \nonu \\
&- &  90 \, i\, 
 d^{abc} \, J^a  K^b  K^c
+360 \,  d^{abc} \, J^a  K^b  L^c
+ 15 \, i\,  d^{abc} \, K^a  K^b  K^c
\nonu \\
&+&  \frac{45}{2} \, N\, \pa  J^a  M^a
-\frac{45}{2} \, N \,   \pa  K^a  M^a
+225N \, d^{abc} \, \psi^a  \pa \psi^b  L^c
\label{lastspin3}
 \\
&+&  225N \, d^{abc} \, \chi^a  \pa \chi^b  L^c 
-315N \, d^{abc} \, \pa \psi^a   \chi^b  K^c 
+315N \, d^{abc} \, \psi^a  \pa \chi^b  J^c 
\Biggr].
\nonu
\eea
See also Appendix (\ref{primarycombi}).
One realizes that the field contents of (\ref{lastspin3})
are the same as the ones in (\ref{lastw3}).
The signs of the numerical coefficients are different from each other.

Furthermore, the $U(1)$ manifest way to describe this higher spin-$3$ current
can be written as 
\bea
&& W_{-\frac{2}{3}}^{(3)}(z) +\frac{1}{12} \, \pa \, W_{-\frac{2}{3}}^{(2)}(z)  = 
-\frac{1}{108N} \, 
 \frac{1}{2\sqrt{6(N^2-4)}} \Biggr[
 \nonu \\
& & -  \frac{45}{2} i \, d^{abc} f^{ade} f^{bfg}
f^{chi} (\psi^d - i \chi^d )(\psi^e - i \chi^e)(\psi^f - i \chi^f)(\psi^g -
i \chi^g)(\psi^h+i \chi^h)( \psi^i + i \chi^i) 
\nonu \\
&& -45i \, N \, f^{abc} d^{cde} \pa (\psi^a- i \chi^a) 
(\psi^b -i \chi^b)(\psi^d -i \chi^d)(\psi^e+ i \chi^e)
\nonu \\
&& -\frac{45}{2} i \, N \, f^{abc} d^{cde}  (\psi^a- i \chi^a) 
(\psi^b -i \chi^b) \pa (\psi^d -i \chi^d)(\psi^e+ i \chi^e)
\nonu \\
&&  +\frac{135}{2} i \, N \, f^{abc} d^{cde}  (\psi^a- i \chi^a) 
(\psi^b -i \chi^b)(\psi^d -i \chi^d) \pa (\psi^e+ i \chi^e) \Biggr](z).
\label{otherspinthreeexp}
\eea
One can also write down the higher spin-$3$ current $ W_{-\frac{2}{3}}^{(3)}(z)$ 
from 
(\ref{otherspinthreeexp}) by writing the derivative terms of the higher 
spin-$2$ current. 
One can check the $U(1)$ charge of this higher spin-$3$ 
current using (\ref{qcharges}) 
\footnote{One can add the two expressions (\ref{lastw3}) and 
(\ref{lastspin3}). Then this is proportional to the following
result $ d^{abc} ( J^a  J^b  J^c 
-6    J^a  J^b  K^c
+6 
   J^a  K^b  K^c
-    K^a  K^b  K^c)(z)$ which is nothing but the higher spin-$3$ current
with the condition $k=N$ 
in the ${\cal N}=1$ version in \cite{Ahn1211,Ahn1305}. }.

Therefore, in this section, the four higher spin currents appearing in the
second ${\cal N}=2$ multiplet in (\ref{manyW}) are determined explicitly.
They are given by (\ref{spin2other}), (\ref{aboveexpression}), 
(\ref{5halflast}) and (\ref{lastspin3}).

\section{The next higher spin currents }

So far, the two higher spin ${\cal N}=2$ multiplets 
in (\ref{manyW}) are obtained in previous sections.
The next question is how one can determine the next higher spin ${\cal N}=2$ 
multiplet.
First of all, because one can calculate the following OPE  between 
the first component 
and the third component of the first ${\cal N}=2$ higher spin multiplet
(\ref{manyW}), 
\bea
W_{\frac{2}{3}}^{(2)}(z) \, 
W_{-\frac{1}{3}}^{(\frac{5}{2})}(w),
\label{onethreeOPE}
\eea
for $N=3$, one can examine the right hand side of this OPE (\ref{onethreeOPE}).
It turns out that one obtains the following new higher spin-$\frac{7}{2}$
current with $q=\frac{1}{3}$ at the first-order pole in (\ref{onethreeOPE}) 
as follows:
\bea
W_{\frac{1}{3}}^{(\frac{7}{2})}(z).
\label{7half}
\eea
In other words, one observes the multiple product of
two adjoint fermions with possible derivatives 
at the first-order pole of (\ref{onethreeOPE})
and these cannot be written in terms of any combinations of 
previously known (higher spin) currents found so far.  
Then it is straightforward to obtain the other three component currents 
living in the third ${\cal N}=2$ multiplet in (\ref{manyW}) using the 
spin-$\frac{3}{2}$ currents of ${\cal N}=2$ superconformal algebra,
as done in sections $3$ and $4$. 
See also Appendix $(\ref{h1operesult})$.

Now one can consider the following OPE
between the first component and the second component 
of the second ${\cal N}=2$ higher spin multiplet in (\ref{manyW}),
\bea
W_{-\frac{2}{3}}^{(2)}(z) \, 
W_{\frac{1}{3}}^{(\frac{5}{2})}(w)
\label{onetwodual}
\eea
in order to see the other higher spin-$\frac{7}{2}$ current with 
$q = -\frac{1}{3}$.
In this case, the first-order pole of (\ref{onetwodual})
provides the following higher spin-$\frac{7}{2}$ current for $N=3$ 
\bea
W_{-\frac{1}{3}}^{(\frac{7}{2})}(z),
\label{7halfdual}
\eea
which has opposite $U(1)$ charge to the previous one (\ref{7half}).
Then again, from the ${\cal N}=2$ supersymmetry, 
one obtains the other three higher spin currents belonging to 
the fourth ${\cal N}=2$ multiplet (\ref{manyW}) from (\ref{7halfdual})
\footnote{In Appendix $E$, we present some
details for the  construction of higher spin-$\frac{7}{2}$ currents.
We can obtain the final forms for these higher spin-$\frac{7}{2}$ currents
via the simplifications of the normal ordered products.}.
See also Appendix (\ref{h3ope}).

We generalize the ${\cal N}=2$ multiplets in (\ref{manyW})
to the higher spin case as follows (See also Appendix 
(\ref{Wsuperexpression})):
\bea
{\bf W}_{\frac{2}{3}}^{(2)} 
&\equiv& \left(W_{\frac{2}{3}}^{(2)}, \, W_{\frac{5}{3}}^{(\frac{5}{2})}, \,
W_{-\frac{1}{3}}^{(\frac{5}{2})}, \,
W_{\frac{2}{3}}^{(3)} \right),
\nonu \\
{\bf W}_{-\frac{2}{3}}^{(2)} &\equiv& 
\left(W_{-\frac{2}{3}}^{(2)}, \, W_{\frac{1}{3}}^{(\frac{5}{2})}, \,
W_{-\frac{5}{3}}^{(\frac{5}{2})}, \,
W_{-\frac{2}{3}}^{(3)} \right),
\nonu \\
{\bf W}_{\frac{1}{3}}^{(\frac{7}{2})} &\equiv&
\left(W_{\frac{1}{3}}^{(\frac{7}{2})}, \, W_{\frac{4}{3}}^{(4)}, \,
W_{-\frac{2}{3}}^{(4)}, \,
W_{\frac{1}{3}}^{(\frac{9}{2})} \right),
\nonu \\
{\bf W}_{-\frac{1}{3}}^{(\frac{7}{2})} &\equiv& 
\left(W_{-\frac{1}{3}}^{(\frac{7}{2})}, \, 
W_{\frac{2}{3}}^{(4)}, \, W_{-\frac{4}{3}}^{(4)}, \, 
W_{-\frac{1}{3}}^{(\frac{9}{2})} \right),
\nonu \\
{\bf W}_{0}^{(4)} &\equiv&
\left(W_{0}^{(4)}, \, 
W_{1}^{(\frac{9}{2})}, \, W_{-1}^{(\frac{9}{2})}, \, 
W_{0}^{(5)} \right),
\nonu \\
{\bf W}_{-\frac{1}{3}}^{(\frac{11}{2})} &\equiv&
\left(W_{-\frac{1}{3}}^{(\frac{11}{2})}, \, 
W_{\frac{2}{3}}^{(6)}, \, W_{-\frac{4}{3}}^{(6)}, \, 
W_{-\frac{1}{3}}^{(\frac{13}{2})} \right),
\nonu \\
{\bf W}_{\frac{1}{3}}^{(\frac{11}{2})} &\equiv&
\left(W_{\frac{1}{3}}^{(\frac{11}{2})}, \, 
W_{\frac{4}{3}}^{(6)}, \, W_{-\frac{2}{3}}^{(6)}, \, 
W_{\frac{1}{3}}^{(\frac{13}{2})} \right),
\nonu \\
{\bf W}_{-\frac{2}{3}}^{(6)} &\equiv&
\left(W_{-\frac{2}{3}}^{(6)}, \, W_{\frac{1}{3}}^{(\frac{13}{2})}, \,
W_{-\frac{5}{3}}^{(\frac{13}{2})}, \,
W_{-\frac{2}{3}}^{(7)} \right),
\nonu \\
{\bf W}_{\frac{2}{3}}^{(6)} &\equiv&
\left(W_{\frac{2}{3}}^{(6)}, \, W_{\frac{5}{3}}^{(\frac{13}{2})}, \,
W_{-\frac{1}{3}}^{(\frac{13}{2})}, \,
W_{\frac{2}{3}}^{(7)}\right), 
\nonu \\
 & \vdots & 
\nonu \\
{\bf W}_{-\frac{2n}{3}+\frac{5}{3}}^{(2n-\frac{1}{2})} &\equiv&
\left(W_{-\frac{2n}{3}+\frac{5}{3}}^{(2n-\frac{1}{2})}, \, 
W_{-\frac{2n}{3}+\frac{8}{3}}^{(2n)}, \, W_{-\frac{2n}{3}+\frac{2}{3}}^{(2n)}, \, 
W_{-\frac{2n}{3}+\frac{5}{3}}^{(2n+\frac{1}{2})} \right),
\nonu \\
{\bf W}_{\frac{2n}{3}-\frac{5}{3}}^{(2n-\frac{1}{2})} &\equiv&
\left(W_{\frac{2n}{3}-\frac{5}{3}}^{(2n-\frac{1}{2})}, \, 
W_{\frac{2n}{3}-\frac{2}{3}}^{(2n)}, \, W_{\frac{2n}{3}-\frac{8}{3}}^{(2n)}, \, 
W_{\frac{2n}{3}-\frac{5}{3}}^{(2n+\frac{1}{2})} \right),
\nonu \\
{\bf W}_{-\frac{2n}{3}+\frac{4}{3}}^{(2n)} &\equiv&
\left(W_{-\frac{2n}{3}+\frac{4}{3}}^{(2n)}, \, 
W_{-\frac{2n}{3}+\frac{7}{3}}^{(2n+\frac{1}{2})}, \,
W_{-\frac{2n}{3}+\frac{1}{3}}^{(2n+\frac{1}{2})}, \,
W_{-\frac{2n}{3}+\frac{4}{3}}^{(2n+1)} \right),
\nonu \\
{\bf W}_{\frac{2n}{3}-\frac{4}{3}}^{(2n)} &\equiv&
\left(W_{\frac{2n}{3}-\frac{4}{3}}^{(2n)}, \, 
W_{\frac{2n}{3}-\frac{1}{3}}^{(2n+\frac{1}{2})}, \,
W_{\frac{2n}{3}-\frac{7}{3}}^{(2n+\frac{1}{2})}, \,
W_{\frac{2n}{3}-\frac{4}{3}}^{(2n+1)}\right),
\nonu \\
& \vdots &
\label{allhigher}
\eea

Then how one can obtain the next higher spin-$4$ current $W_0^{(4)}(z)$
living in the lowest component current in the fifth ${\cal N}=2$ 
multiplet in (\ref{allhigher})? 
One way to obtain this higher spin current is to calculate 
the OPE $W_{-\frac{1}{3}}^{(\frac{5}{2})}(z) \, W_{\frac{1}{3}}^{(\frac{7}{2})}(w)$
between the third component in the first ${\cal N}=2$ multiplet 
and the first component  in the third ${\cal N}=2$ multiplet.
It turns out that the second-order pole of this OPE 
leads to the above higher spin-$4$ current $W_0^{(4)}(w)$ belonging to
the fifth ${\cal N}=2$ multiplet for $N=3$.
The remaining three higher spin currents living in the 
fifth ${\cal N}=2$ multiplet can be obtained 
from the ${\cal N}=2$ supersymmetry on this ${\cal N}=2$ multiplet
as done in sections $3$ and $4$.
See also Appendix (\ref{i1ope}).

We can further analyze the next higher spin-$\frac{11}{2}$ current
 $W_{-\frac{1}{3}}^{(\frac{11}{2})}(z)$
living in the lowest component current in the sixth ${\cal N}=2$ 
multiplet in (\ref{allhigher}) \footnote{For $N=3$, we have determined 
the first five ${\cal N}=2$ higher spin currents in (\ref{allhigher}). }.
In this case, we expect that 
the OPE $W_{-\frac{1}{3}}^{(\frac{5}{2})}(z) \, W_{0}^{(4)}(w)$
will allow us to obtain this higher spin current 
because the $U(1)$ charge of the left hand side 
of this OPE gives the correct $q=-\frac{1}{3}$. Due to the spin counting, 
the first-order pole can have this higher spin-$\frac{11}{2}$ current with
$q=-\frac{1}{3}$.
What happens for  the higher spin-$\frac{11}{2}$ current
 $W_{\frac{1}{3}}^{(\frac{11}{2})}(z)$
 in the lowest component current in the seventh ${\cal N}=2$ 
multiplet in (\ref{allhigher})?
Similarly, 
we expect that the OPE $W_{\frac{1}{3}}^{(\frac{5}{2})}(z) \, W_{0}^{(4)}(w)$
can allow us to obtain this higher spin current 
because the $U(1)$ charge of the left hand side 
of this OPE provides the correct $q=\frac{1}{3}$
\footnote{
Then the two higher spin currents 
of  $W_{\mp \frac{1}{3}}^{(\frac{5}{2})}(z)$ living in the 
first and second ${\cal N}=2$ multiplet of (\ref{manyW})
play the role of the ``generators'' of the new higher spin currents 
with the known higher spin currents. In order to obtain 
the higher spin current    $W_{-\frac{2}{3}}^{(6)}(w)$,
one can consider 
the the OPE $W_{-\frac{1}{3}}^{(\frac{5}{2})}(z) \, 
W_{-\frac{1}{3}}^{(\frac{11}{2})}(w)$ and focus on the second-order pole.
For the higher spin current    $W_{\frac{2}{3}}^{(6)}(w)$,
we expect that
 the  second-order pole in the OPE $W_{\frac{1}{3}}^{(\frac{5}{2})}(z) \, 
W_{\frac{1}{3}}^{(\frac{11}{2})}(w)$ will provide the right structure on 
the higher spin current.}.

Let us focus on the last four ${\cal N}=2$ multiplets in (\ref{allhigher}).
One obtains the higher spin current 
$ W_{-\frac{2n}{3}+\frac{4}{3}}^{(2n)}(w)$ from the second-order pole 
in the OPE $W_{-\frac{1}{3}}^{(\frac{5}{2})}(z) \, 
W_{-\frac{2n}{3}+\frac{5}{3}}^{(2n-\frac{1}{2})}(w)$.
The $U(1)$ charge and spin countings give this result.
Similarly, one obtains the higher spin current 
$ W_{\frac{2n}{3}-\frac{4}{3}}^{(2n)}(w)$ from the second-order pole 
in the OPE $W_{\frac{1}{3}}^{(\frac{5}{2})}(z) \, 
W_{\frac{2n}{3}-\frac{5}{3}}^{(2n-\frac{1}{2})}(w)$. 
Furthermore, 
one obtains the higher spin current 
$ W_{-\frac{2n}{3}+\frac{5}{3}}^{(2n-\frac{1}{2})}(w)$ from the first-order pole 
in the OPE $W_{-\frac{1}{3}}^{(\frac{5}{2})}(z) \, 
W_{-\frac{2n}{3}+2}^{(2n-2)}(w)$.
Similarly, one obtains the higher spin current 
$ W_{\frac{2n}{3}-\frac{5}{3}}^{(2n-\frac{1}{2})}(w)$ from the first-order pole 
in the OPE $W_{\frac{1}{3}}^{(\frac{5}{2})}(z) \, 
W_{\frac{2n}{3}-2}^{(2n-2)}(w)$. 
The second, third and fourth components higher spin currents
can be obtained from the ${\cal N}=2$ supersymmetry on 
the (determined) lowest higher spin current
described before. 

Therefore, one can generate the new higher spin currents
living in the lowest component higher spin currents of
${\cal N}=2$ multiplets in (\ref{allhigher})
by using the two higher spin-$\frac{5}{2}$ currents 
$W_{\mp \frac{1}{3}}^{(\frac{5}{2})}(z)$ together with the known 
higher spin currents systematically.

\section{The OPEs between the higher spin currents }

In order to obtain the higher spin symmetry algebra
between the higher spin currents, we should
calculate the OPEs between the higher spin currents.
In this section, we present the OPE between the higher spin-$2$ currents 
living in the first components of first and second ${\cal N}=2$ multiplets 
in (\ref{manyW}).

There exist the trivial OPEs 
$W_{\frac{2}{3}}^{(2)}(z) \, W_{\frac{2}{3}}^{(2)}(w)= + \cdots
$ and $
W_{\frac{2}{3}}^{(2)}(z) \, W_{\frac{5}{3}}^{(\frac{5}{2})}(w)
= + \cdots
$ for $N=3$.
The regularity of these OPEs can be checked by the counting of 
spin and $U(1)$ charge \footnote{
First of all we classify the possible composite fields for given spin. Then 
we should check whether these will satisfy the definite $U(1)$ charge.
For example, for the former, the possible spin in the right hand side
is given by $0, 1, 2$ and $3$. Then the composite fields should 
have the $U(1)$ charge as $\frac{4}{3}$. For the spin-$1$ current, one has
$J(w)$. For the spin-$2$ current, one has $J J(w), \pa J(w), T(w), 
W_{\frac{2}{3}}^{(2)}(w)$ and $W_{-\frac{2}{3}}^{(2)}(w)$. Similarly for the spin-$3$
current, one has $J J J(w)$, $\pa J J(w)$, $\pa^2 J(w)$, $J T(w)$, 
$G^{-} G^{+}$, $\pa T(w)$, $ J W_{\frac{2}{3}}^{(2)}(w)$, 
$ J W_{-\frac{2}{3}}^{(2)}(w)$, $\pa W_{\frac{2}{3}}^{(2)}(w)$ and 
$ \pa W_{-\frac{2}{3}}^{(2)}(w)$. Then, one sees that there is no possible 
composite field having the $U(1)$ charge $\frac{4}{3}$.  }.
The next OPE
$
W_{\frac{2}{3}}^{(2)}(z) \, W_{-\frac{1}{3}}^{(\frac{5}{2})}(w)
$ has nontrivial OPE and the new higher spin-$\frac{7}{2}$ current 
$ W_{\frac{1}{3}}^{(\frac{7}{2})}(w)$
can be found at the first-order pole, as observed in previous section.
One sees that the first-order pole of next OPE
$W_{\frac{2}{3}}^{(2)}(z) \, W_{\frac{2}{3}}^{(3)}(w)$
gives the higher spin-$4$ current $ W_{\frac{4}{3}}^{(4)}(w)$
which is the second component of the third ${\cal N}=2$ 
higher spin multiplet.
See also Appendix (\ref{h1operesult}).

One can continue to calculate the remaining six OPEs between the 
lowest higher spin currents in the first ${\cal N}=2$ multiplet. 
The OPE 
$W_{\frac{5}{3}}^{(\frac{5}{2})}(z) \, W_{\frac{5}{3}}^{(\frac{5}{2})}(w)$
does not have any singular terms for $N=3$ and 
the first-order pole of the OPE
$W_{\frac{5}{3}}^{(\frac{5}{2})}(z) \, W_{-\frac{1}{3}}^{(\frac{5}{2})}(w)$
leads to the higher spin-$4$ current $ W_{\frac{4}{3}}^{(4)}(w)$ appeared 
as above for $N=3$.
The next OPE
$W_{\frac{5}{3}}^{(\frac{5}{2})}(z) \, W_{\frac{2}{3}}^{(3)}(w)$
does not give any singular terms
\footnote{
For the remaining three OPEs, one can analyze further.
The first-order pole of the OPE
$W_{-\frac{1}{3}}^{(\frac{5}{2})}(z) \, W_{-\frac{1}{3}}^{(\frac{5}{2})}(w)$
gives the higher spin current 
$ W_{-\frac{2}{3}}^{(4)}(w)$ which is the third component of 
the third ${\cal N}=2$ multiplet in (\ref{manyW}).
The next OPE
$W_{-\frac{1}{3}}^{(\frac{5}{2})}(z) \, W_{\frac{2}{3}}^{(3)}(w)$
contains the information on the two higher spin currents,
$ W_{\frac{1}{3}}^{(\frac{7}{2})}(w)$ and $ W_{\frac{1}{3}}^{(\frac{9}{2})}(w)$
corresponding to the first and the last components of the third ${\cal N}=2$
multiplet in (\ref{manyW}).
The former appear in the second-order pole while the latter 
appears in the first-order pole.
Therefore, the presence of third ${\cal N}=2$ multiplet in (\ref{manyW})
can be seen from the above fifteen OPEs considered so far.
Now one can calculate the last OPE
$W_{\frac{2}{3}}^{(3)}(z) \, W_{\frac{2}{3}}^{(3)}(w)$
and the second-order pole  provides the higher spin current 
$ W_{\frac{4}{3}}^{(4)}(w)$ which is the second component of 
the third ${\cal N}=2$ multiplet.
See also Appendix (\ref{h1operesult}).
}.

The next OPEs we can consider are 
the ones between the first and second ${\cal N}=2$ multiplets
in (\ref{manyW}). 
Then let us calculate the OPE 
$W_{\frac{2}{3}}^{(2)}(z) \, W_{-\frac{2}{3}}^{(2)}(w)$.
Note that this OPE has zero $U(1)$ charge $q=0$.
The sum of spins in the left hand side 
is given by four. 
Then the right hand of this OPE can start with the fourth-order pole 
where the central charge term can appear.
The other singular term can appear in general.
One should determine the possible
composite fields with correct spin and $U(1)$ charge
at the specific pole.
The third-order pole contains 
the spin-$1$ field with $q=0$.
The second-order pole has
the spin-$2$ field with vanishing $U(1)$ charge
and the first-order pole 
possesses the spin-$3$ field where the $U(1)$ charge is zero. 
Recall that the ${\cal N}=2$ superconformal algebra currents 
have their spins and $U(1)$ charges as follows:
$(1_0, \frac{3}{2}_{+}, \frac{3}{2}_{-}, 2_0)$ from Appendix (\ref{stressN2}).

The normalization we take here is that the fourth-order pole is given by
$\frac{c}{2}$. Some of the singular terms for this OPE
are presented in
Appendix $F$. The fourth-order pole contribution comes from 
the OPEs $J^a M^a(z) \, J^b M^b(w)$,
$K^a M^a(z) \, K^b M^b(w)$,
and $L^a M^a(z) \, L^b M^b(w)$.
We can introduce the same normalization factor 
for the higher spin-$2$ currents. 
Then the choice in (\ref{lowestspin2}) and (\ref{spin2other})
together with the results in Appendix $F$ gives the correct 
$\frac{c}{2}$. 

Now let us move on the third-order pole.
According to the results in Appendix $F$, the third-order pole
can appear in the OPEs $J^aM^a(z) \, L^b M^b(w)$ and 
$K^a M^a(z) \, L^b M^b(w)$(and  $L^a M^a(z) \, J^b M^b(w)$ and 
$L^a M^a(z) \, K^b M^b(w)$).
By calculating the third-order poles with correct 
coefficients, one realizes that the third-order pole is given by
$J(w)$. 

The next second-order pole can be analyzed similarly.
Again from Appendix $F$, the second-order poles can appear in 
all possible nine terms. We should express them in terms of 
the currents of ${\cal N}=2$ superconformal algebra.
The higher spin currents  are given in (\ref{manyW}) with the spins and 
$U(1)$ charges.  
One can easily check that the only possible spin-$2$ current with vanishing 
$U(1)$ charge consists of the currents of ${\cal N}=2$ superconformal 
algebra without any higher spin currents.
One can check that 
the following identities are useful to simplify the OPEs  
\bea
M^a M^a(w) & = & -\frac{1}{2} J^a K^a(w) -\frac{4}{N} \psi^a \psi^b
\chi^b \chi^a(w) +\frac{(N^2-2)}{4N^2} J^a J^a(w) + \frac{(N^2-2)}{4N^2} 
K^a K^a(w),
\nonu \\ 
L^a L^a(w) & = & -\frac{1}{2} J^aK^a(w) +\frac{1}{4} K^a K^a(w) +
\frac{1}{4} J^a J^a(w), 
\nonu \\
J J(w) & = & -\frac{4}{9}  \left( \psi^a \chi^a
\psi^b \chi^b +\frac{1}{8N} K^a K^a +\frac{1}{8N} J^a J^a \right)(w).
\label{quadraticexpression}
\eea

In principle, we can also analyze the first-order pole. In this case, 
the spin of the composite field with vanishing $U(1)$ charge is given by
three. Then one should rewrite all the first-order pole in terms of 
fully normal ordered product \cite{BBSSfirst,BBSSsecond} 
in order to express in terms of the known
spin-$3$ field consisting of the currents of the ${\cal N}=2$ superconformal 
algebra.

It turns out that the OPE between 
the lowest higher spin-$2$ current with $q=\frac{2}{3}$
and the lowest higher spin-$2$ current with 
$q=-\frac{2}{3}$ is given by 
\bea
W_{\frac{2}{3}}^{(2)}(z) \, W_{-\frac{2}{3}}^{(2)}(w) & = & 
\frac{1}{(z-w)^4} \, \frac{c}{2} + \frac{1}{(z-w)^3} \, J(w)
\nonu \\
& + & \frac{1}{(z-w)^2} \left[ -\frac{2}{(c-1)} 
\, J  J + \frac{1}{2}\, \pa  J + \frac{2(3c-1)}{3(c-1)} \, T \right](w)
\nonu \\
&+& \frac{1}{(z-w)} \, \frac{1}{(c-1) (c+6) (2 c-3)} \,
 \left[ \frac{1}{3} \, (3 c+8) (8 c+9) 
\, J  T \right. \nonu \\
& - &  \frac{1}{3} \, (32 c+127) \, J J J 
 \nonu \\
& - &   (c^2-36 c+8) \, G^{-}  G^{+}
+\frac{1}{6} \, (12 c^3+53 c^2-234 c+60)\, 
\pa  T \nonu \\
&+ & \left.  \frac{1}{6} \, (c+2) (2 c^2+2 c+5) \, 
\pa^2  J - 2 (c+6) (2c-3)\, \pa  J  J \right](w)
+ \cdots.
\label{twospin2ope}
\eea
We use the relations (\ref{quadraticexpression}).
In the large $c$ limit (or large $N$ limit),
all the nonlinear terms in (\ref{twospin2ope}) disappear. 
Furthermore, one can analyze the $\frac{1}{c}$- and $ \frac{1}{c^2}$-terms
\cite{Ahn1206}.
Compared to the ${\cal N}=2$ $W_3$ algebra \cite{Romans} where one of the OPEs
is given by the spin-$2$ current and itself,
the additional field contents in (\ref{twospin2ope})
arise in the right hand side.
That is, the 
central term, $JJ(w)$, $T(w)$, $\pa J J(w)$ and $\pa T(w)$
terms in (\ref{twospin2ope}) appeared in \cite{Romans,Ahn1208}.
Let us emphasize that the above OPE is given by two different currents
although their spins are the same but the $U(1)$ charges are different.
Therefore, the nonderivative terms in the third and first-order poles 
can appear in the above OPE in general.
Furthermore, we expect that there should be the descendant fields
$\pa J(w)$ and $\pa^2 J(w)$ 
in the second and first-order poles. 
Due to the $U(1)$ charge conservation, either $W_{ \frac{2}{3}}^{(2)}(w)$ or
$W_{-\frac{2}{3}}^{(2)}(w)$ cannot appear in the second-order pole.
One can reexpress the first-order pole as the sum of (quasi) primary fields
and other descendant fields, as usual \footnote{
One can check that $J J(w)$, $T(w)$, 
$J T(w)$, and $J J J(w)$ are quasi primary fields.}.

One can repeat the remaining fifteen 
OPEs in the two ${\cal N}=2$ lowest higher spin currents.
It is rather nontrivial and complicated to complete these OPEs for generic
$N$. For $N=3$, the explicit forms in (\ref{manyW}) are determined completely.
The question is how to generalize them for generic $N$.
In order to obtain the higher spin current algebra, 
the Jacobi identity method in next section will be used.
  

\section{The OPEs between the higher spin currents in ${\cal N}=2$ 
superspace }

In this section, by using the package by Krivonos and Thielemans \cite{KT},
some OPEs
of the higher spin currents are determined in the ${\cal N}=2$
superspace.
That is, the ten OPEs between the first four ${\cal N}=2$ higher spin multiplets
in (\ref{manyW}).

\subsection{The OPEs between the higher spin-$2$ currents 
${\bf W}_{\pm \frac{2}{3}}^{(2)}(Z)$
} 

Let us consider the OPE between 
$ {\bf W}_{\frac{2}{3}}^{(2)}(Z_1)$ and $ {\bf W}_{\frac{2}{3}}^{(2)}(Z_2)$.
That is, the OPE between the first ${\cal N}=2$ multiplet and itself 
in (\ref{manyW}).
The corresponding component results for $N=3$ are obtained from section $6$.
Now one can introduce the arbitrary coefficients in the right hand side of the
OPE.  
Inside of the package \cite{KT}, one introduces the OPE 
in Appendix (\ref{SuperTT}), the OPEs in Appendix (\ref{TWOPEseveral})
for  ${\bf W}_{\frac{2}{3}}^{(2)}(Z_2)$ and 
${\bf W}_{\frac{1}{3}}^{(\frac{7}{2})}(Z_2)$ which is an 
${\cal N}=2$  extension of the 
$W_{\frac{1}{3}}^{(\frac{7}{2})}(w)$.
Then one can write down  $
  {\bf W}_{\frac{2}{3}}^{(2)}(Z_1) \, {\bf W}_{\frac{2}{3}}^{(2)}(Z_2)$
with arbitrary coefficients.

By using the Jacobi identity 
\footnote{The outcome of ${\tt OPEJacobi}$ is a double list of operators 
\cite{Thielemansphd}. The higher spin currents with large spin 
appear in the beginning of
this list while the higher spin currents with small spin 
appear at the end of this list. It is better to analyze the elements at the 
end of the list first.}
between the three (higher spin) currents 
$({\bf T},  {\bf W}_{\frac{2}{3}}^{(2)}, {\bf W}_{\frac{2}{3}}^{(2)})$,
we obtain the following result
\bea
&& {\bf W}_{\frac{2}{3}}^{(2)}(Z_1) \, {\bf W}_{\frac{2}{3}}^{(2)}(Z_2)  = 
C_{(2) \, (2)}^{(\frac{7}{2}) \, +} \, \left(
\frac{\bar{\theta}_{12}}{z_{12}} \, {\bf W}_{\frac{1}{3}}^{(\frac{7}{2})}(Z_2)
- \frac{\theta_{12} \, \bar{\theta}_{12}}{z_{12}} \, \frac{1}{2} \,
D \, {\bf W}_{\frac{1}{3}}^{(\frac{7}{2})}(Z_2) \right) + \cdots.
\label{oneone}
\eea
At this level, the unknown structure constant appearing in the 
first term of (\ref{oneone}) is present \footnote{
Recall that the $U(1)$ charge of $\bar{\theta}_{12}$ is $+1$ ($\theta_{12}$
has $-1$)
while the covariant derivative $D_2$ is $+1$. Then it is easy to see that 
the $U(1)$ charge is preserved in both sides of (\ref{oneone}).}.
See also Appendix (\ref{h1operesult}) where the component results are given.

The next is the OPE between 
the higher spin-$2$ current with $q=\frac{2}{3}$
and the higher spin-$2$ current with $q=-\frac{2}{3}$. That is, 
$ {\bf W}_{\frac{2}{3}}^{(2)}(Z_1) \, {\bf W}_{-\frac{2}{3}}^{(2)}(Z_2)$.
From the component results in section $6$,
one expects that the right hand side of this OPE
can consist of the composite fields between the currents 
of the ${\cal N}=2$ superconformal algebra.
The nontrivial thing is to write down the right hand side 
with arbitrary coefficients.
In order to obtain the consistent solution for the Jacobi identity, 
it is necessary to write down all the possible terms in the right 
hand side. Otherwise, the outcome for the Jacobi identity
will give us the inconsistent solutions.  
Of course, 
 the OPE in Appendix (\ref{TWOPEseveral})
for  ${\bf W}_{-\frac{2}{3}}^{(2)}(Z_2)$ should be included inside the package.

After using the Jacobi identity 
 between the three (higher spin) currents 
$({\bf T},  {\bf W}_{\frac{2}{3}}^{(2)}, {\bf W}_{-\frac{2}{3}}^{(2)})$,
one obtains the following OPE
\bea
&& {\bf W}_{\frac{2}{3}}^{(2)}(Z_1) \, {\bf W}_{-\frac{2}{3}}^{(2)}(Z_2) = 
-\frac{\theta_{12} \, \bar{\theta}_{12}}{z_{12}^5} \, \frac{c}{6}
+\frac{1}{z_{12}^4} \, \frac{c}{2} + 
 \frac{\theta_{12} \, \bar{\theta}_{12}}{z_{12}^4} \, \frac{8}{3} \, {\bf T}(Z_2)
 + \frac{1}{z_{12}^3} \, {\bf T}(Z_2)- \frac{\theta_{12}}{z_{12}^3} 
\, \frac{5}{2} \, D {\bf T}(Z_2) 
\nonu \\
&& + 
\frac{\bar{\theta}_{12}}{z_{12}^3} \, \frac{7}{2} \, \overline{D} {\bf T}(Z_2)
 + \frac{\theta_{12} \, \bar{\theta}_{12}}{z_{12}^3} \, \frac{1}{(c-1)} \, 
\left[ \frac{1}{18} \, (3 c+34) \, 
[D, \overline{D}]{\bf T} +\frac{37}{6} \, {\bf T} {\bf T} +
\frac{17}{6}\, (c-1) \, \pa {\bf T}\right](Z_2) 
\nonu \\
& & +  \frac{1}{z_{12}^2} \, 
\frac{1}{(c-1)}
\, \left[ -\frac{1}{3} (3 c-1) \,
[D, \overline{D}]{\bf T} -2 \, {\bf T} {\bf T} 
+\frac{1}{2} \, (c-1)\, \pa {\bf T}
\right](Z_2) 
\nonu \\
&& +  
\frac{\theta_{12}}{z_{12}^2} \, \frac{1}{(c-1)}\, 
\left[ -\frac{5}{6} \, (2 c-5) \, \pa D {\bf T} 
-\frac{15}{2 } \, {\bf T} D {\bf T}\right](Z_2)  \nonu \\
&& +   
\frac{\bar{\theta}_{12}}{z_{12}^2} \, \frac{1}{(c-1)}\,  
\left[ \frac{7}{6} \, (2 c-1) \,
\pa \overline{D} {\bf T} +\frac{7}{2 } \,
{\bf T} \overline{D} {\bf T}\right](Z_2)
\nonu \\
&& +   \frac{\theta_{12} \, \bar{\theta}_{12}}{z_{12}^2}
\, \frac{1}{(c-1) (c+6) (2 c-3)} \,
\left[ \frac{1}{72} \, (6 c^3-551 c^2+2718 c-3120)\,
\pa [D, \overline{D}]{\bf T} \right. \nonu \\
&  & -   \frac{1}{18} \, (204 c^2-469 c-18)\,{\bf T} \,
[D, \overline{D}] {\bf T}  
 -\frac{2}{9} \, (101 c-239)\, 
{\bf T} {\bf T} {\bf T}  
\nonu \\
&& +    \frac{1}{3} \, (131 c^2-306 c+208)\,
  \overline{D} {\bf T} D {\bf T}
 \nonu \\
&& + \left.
\frac{107}{12} \, (c+6) (2 c-3)\, \pa {\bf T} {\bf T } 
+    
\frac{1}{18} \, (52 c^3+51 c^2-396 c+260) \, \pa^2 {\bf T} \right](Z_2) 
\nonu \\
& & +   \frac{1}{z_{12}} \, \frac{1}{ (c-1) (c+6) (2 c-3)} \, 
\left[ -\frac{1}{12} \, (12 c^3+53 c^2-234 c+60)\,
\pa [D, \overline{D}]{\bf T} \right. \nonu \\
& & -  \frac{1}{6} \, (3 c+8) (8 c+9)\,
{\bf T} [D, \overline{D}] {\bf T}   
-   \frac{1}{3} \, (32 c+127) \,
 {\bf T} {\bf T} {\bf T}  +   (c^2-36 c+8)\,
\overline{D} {\bf T} D {\bf T} 
 \nonu \\
&& \left. -2 \,(c+6)(2c-3)\, \pa {\bf T} {\bf T } 
 +   \frac{1}{6} \, (c+2) (2 c^2+2 c+5)\,
 \pa^2 {\bf T} \right](Z_2) 
\nonu \\
& &
+  \frac{\theta_{12}}{z_{12}}
\, \frac{1}{ (c-1) (c+6) (2 c-3)} \, 
\left[ -\frac{5}{4} \, (c^3-2 c^2+12 c-30)\,
 \pa^2 D {\bf T}
+\frac{5}{2} \,(4 c-41)\,
{\bf T} {\bf T} D {\bf T}
\right. 
\nonu \\
& & +  \frac{5}{12} \, (27 c^2-38 c-30)\,
[D, \overline{D}]{\bf T} D {\bf T}
  -     \frac{5}{6} \,(11 c^2+91 c-240)\,
\pa 
D {\bf T}  {\bf T}
\nonu \\
&& \left. -\frac{5}{4} \, (7 c^2-2 c-4)\, 
\pa {\bf T}  D {\bf T}
\right](Z_2)
\nonu \\
 & & +   
\frac{\bar{\theta}_{12}}{z_{12}} \, \frac{1}{(c-1) (c+6) (2 c-3)} \, \left[ 
\frac{7}{24} \, (6 c^3+3 c^2+4 c+42)\,
 \pa^2 \overline{D} {\bf T}
-\frac{7}{2} \, (12 c+7)\,
{\bf T} {\bf T} \overline{D} {\bf T}
\right.
\nonu \\
& &   -    \frac{7}{12} \, (27 c^2-32 c+6)\,
 \overline{D} {\bf T} [D, \overline{D}]{\bf T}
+ \frac{7}{6} \, (5 c^2-37 c-12)\,  \pa 
\overline{D} {\bf T} {\bf T}
 \nonu \\
&& \left. +   
\frac{7}{4} \, (c^2+40 c-56)\,
\pa {\bf T}  \overline{D} {\bf T} 
\right](Z_2)
\nonu \\
& & +   
\frac{\theta_{12} \, \bar{\theta}_{12}}{z_{12}}
\, \frac{1}{(c-1) (c+1) (c+6) (2 c-3) (5 c-9)}\,
\left[
\right. \nonu \\
&& \frac{35}{72} \, (12 c^4-57 c^3+70 c^2+9 c+90)\, 
\pa^2 [D, \overline{D}] {\bf T} 
 \nonu \\
& & -  \frac{35}{9} \, (30 c^3-23 c^2+59 c+57) \,  
  {\bf T} {\bf T} [D, \overline{D}] {\bf T}
  - \frac{175}{18} \,(2 c-1) (16 c+5)\, 
{\bf T} {\bf T} {\bf T} {\bf T} 
\nonu \\
& & +  
\frac{35}{3} \,(40 c^3-131 c^2+23 c-26)\,
 {\bf T} \overline{D}
{\bf T}  D {\bf T} 
+   \frac{35}{18} \, (75 c^4-159 c^3-58 c-72)\,
\pa \overline{D} {\bf T}  D {\bf T} \nonu \\
& & -  
\frac{35}{216} \, (264 c^3-827 c^2+759 c+90)\,
 [D, \overline{D}] {\bf T} [D, \overline{D}] {\bf T}
\nonu \\
& & -  
\frac{35}{36} \, (30 c^4-34 c^3-253 c^2+231 c-240) \,
    \pa  [D, \overline{D}]  {\bf T}  {\bf T}
\nonu \\
 & & +  
\frac{35}{18} \, (75 c^4-321 c^3+258 c^2+514 c-360)\,
 \pa D {\bf T} \overline{D} {\bf T}
\nonu \\
 & & - 
\frac{35}{36} \,(12 c^2+c-36)(c+1)(5c-9)\,
\pa {\bf T} [D, \overline{D}] {\bf T}
-\frac{35}{6} \,(8 c-17)(c+1)(5c-9)\,
\pa {\bf T} {\bf T} 
{\bf T}  \nonu \\
& & +   
\frac{35}{24} \, (c-2) (20 c^3+118 c^2-159 c-37)\,
\pa {\bf T} \pa {\bf T} 
\nonu \\
&& +  \frac{35}{36} \, (40 c^4+64 c^3-207 c^2-381 c+950)\,
\pa^2 {\bf T} {\bf T} 
\nonu \\
&  & +   \left.
\frac{35}{216} \, (30 c^5-54 c^4-235 c^3+782 c^2-177 c+210)\,
   \pa^3 {\bf T} 
\right](Z_2) +\cdots.
\label{onetwo}
\eea
In the large $c$ limit,
all the nonlinear terms in (\ref{onetwo}) disappear. 
One can also analyze the subleading 
$\frac{1}{c}, \cdots, \frac{1}{c^3}$-terms. 
Note that the normalization for the higher spin currents 
${\bf W}_{\pm \frac{2}{3}}^{(2)}(Z)$ can be seen from the 
singular term $\frac{1}{z_{12}^4}$ with the coefficient $\frac{c}{2}$.
All the structure constants in (\ref{onetwo}) are completely
fixed and can be written in terms of the function of the central charge
$c$ \footnote{
The covariant derivative $\overline{D}_2$ has $U(1)$ charge 
$-1$. One can check the $U(1)$ charge of the  
right hand side of (\ref{onetwo}) is zero.}. See also Appendix 
(\ref{h2ope}) for the component result.

Inside of the package \cite{KT}, one also introduces the OPE 
in Appendix (\ref{SuperTT}), the OPE in Appendix (\ref{TWOPEseveral})
for   
${\bf W}_{-\frac{1}{3}}^{(\frac{7}{2})}(Z_2)$ which is an 
${\cal N}=2$  extension of the 
$W_{-\frac{1}{3}}^{(\frac{7}{2})}(w)$.
Then one can write down  the OPE $
  {\bf W}_{-\frac{2}{3}}^{(2)}(Z_1) \, {\bf W}_{-\frac{2}{3}}^{(2)}(Z_2)$
with arbitrary coefficients as done before.
By using the Jacobi identity between 
the three (higher spin) currents 
$({\bf T},  {\bf W}_{-\frac{2}{3}}^{(2)}, {\bf W}_{-\frac{2}{3}}^{(2)})$,
we obtain the following result
\bea
{\bf W}_{-\frac{2}{3}}^{(2)}(Z_1) \, {\bf W}_{-\frac{2}{3}}^{(2)}(Z_2) & = & 
C_{(2) \, (2)}^{(\frac{7}{2}) \, -} \, \left(
\frac{{\theta}_{12}}{z_{12}} \, {\bf W}_{-\frac{1}{3}}^{(\frac{7}{2})}(Z_2)
+ \frac{\theta_{12} \, \bar{\theta}_{12}}{z_{12}} \, 
\frac{1}{2} \, \overline{D} \, 
{\bf W}_{-\frac{1}{3}}^{(\frac{7}{2})}(Z_2) \right)+\cdots.
\label{twotwo}
\eea
The unknown structure constant appearing in the 
first term of (\ref{twotwo}) is present.
As before Appendix (\ref{h3ope}) describes the component result.

The above three OPEs (\ref{oneone}), (\ref{onetwo}) 
and (\ref{twotwo}) can be summarized by
\bea
\Biggr[{\bf W}_{\pm \frac{2}{3}}^{(2)} \Biggr] \cdot \Biggr[{\bf W}_{\pm\frac{2}{3}}^{(2)}\Biggr]
= \Biggr[{\bf W}_{\pm\frac{1}{3}}^{(\frac{7}{2})}\Biggr], \qquad
\Biggr[{\bf W}_{ \frac{2}{3}}^{(2)} \Biggr] \cdot \Biggr[{\bf W}_{-\frac{2}{3}}^{(2)}
\Biggr]
=\Biggr[I \Biggr].
\label{fusion}
\eea
Here $[I]$ stands for the ${\cal N}=2$ superconformal family
of the identity operator.  
Therefore, the third and fourth ${\cal N}=2$ multiplet in (\ref{manyW})
can be obtained from the first equation of (\ref{fusion}).

\subsection{The OPEs between the higher spin-$2$ current and the 
higher spin-$\frac{7}{2}$ current 
}

Let us consider the next OPE 
between $
 {\bf W}_{\frac{2}{3}}^{(2)}(Z_1)$ and $ {\bf W}_{\frac{1}{3}}^{(\frac{7}{2})}(Z_2)$.
That is, the OPE between the first and the third ${\cal N}=2$ multiplets 
in (\ref{manyW}).
The corresponding component results for $N=3$ can be obtained.
For example, the lowest component higher spin current 
$W_0^{(4)}(w)$, living in the fifth ${\cal N}=2$ multiplet in (\ref{manyW}),  
appears in the second-order pole of the OPE
$W_{\frac{2}{3}}^{(2)}(z) \, W_{-\frac{2}{3}}^{(4)}(w)$ for $N=3$.
See also Appendix (\ref{i1ope}).
It is obvious to see that 
the $U(1)$ charge counting in the OPE 
implies this particular singular term if one takes the first operator 
as the lowest higher spin current $W_{\frac{2}{3}}^{(2)}(z)$.
Note that the above  higher spin current $W_0^{(4)}(w)$
is the first element of vanishing $U(1)$ charge in the list of 
(\ref{allhigher}).
Now one can introduce the arbitrary coefficients in the right hand side of the
${\cal N}=2$ OPE.  
Inside of the package \cite{KT}, one introduces 
the OPE in Appendix (\ref{TWOPEseveral})
for ${\bf W}_{0}^{(4)}(Z_2)$.
Then one can write down  the OPE $
  {\bf W}_{\frac{2}{3}}^{(2)}(Z_1) \, {\bf W}_{\frac{1}{3}}^{(\frac{7}{2})}(Z_2)$
with arbitrary coefficients.

By using the Jacobi identity between the three (higher spin) currents 
$({\bf T},  {\bf W}_{\frac{2}{3}}^{(2)}, {\bf W}_{\frac{1}{3}}^{(\frac{7}{2})})$,
we obtain the following result
\bea
&& {\bf W}_{\frac{2}{3}}^{(2)}(Z_1) \, {\bf W}_{\frac{1}{3}}^{(\frac{7}{2})}(Z_2) = 
C_{(2) \, (\frac{7}{2})}^{(4) \, +} \, \left(
\frac{\bar{\theta}_{12}}{z_{12}^2} \,
 {\bf W}_{0}^{(4)}(Z_2)
-   \frac{\theta_{12} \, \bar{\theta}_{12}}{z_{12}^2}
\, \frac{7}{24} \, D  {\bf W}_{0}^{(4)}(Z_2)
\right. \nonu \\
&& -\frac{1}{z_{12}} \, \frac{1}{4} \, D {\bf W}_{0}^{(4)}(Z_2)
 +   
\frac{\bar{\theta}_{12}}{z_{12}} \left[ \frac{1}{24} \, [D, \overline{D}]
 {\bf W}_{0}^{(4)} + \frac{3}{8} \, \pa  {\bf W}_{0}^{(4)}
\right](Z_2)
\label{11opeope}
 \\
& & + \left.    
\frac{\theta_{12} \, \bar{\theta}_{12}}{z_{12}} \left[
 -\frac{5(c-3)}{36(c+9)} \, \pa    D  
{\bf W}_{0}^{(4)}  + \frac{20}{3(c+9)} \, D {\bf T} {\bf W}_{0}^{(4)} 
 -   
\frac{5}{3(c+9)} {\bf T}
D {\bf W}_{0}^{(4)} \right](Z_2) \right) +\cdots.
\nonu
\eea
In the large $c$ limit,
all the nonlinear terms in (\ref{11opeope}) disappear.
One can analyze the subleading $\frac{1}{c}$-term. 
Except an overall structure constant factor in (\ref{11opeope}), 
all the relative coefficients are determined during this calculation.

Let us move on the next OPE between the first and the fourth 
${\cal N}=2$ multiplets in (\ref{manyW}).
The lowest component higher spin current 
$W_{-\frac{2}{3}}^{(2)}(w)$, living in the second ${\cal N}=2$ multiplet 
in (\ref{manyW}),  
appears in the fourth-order pole of the OPE
$W_{\frac{2}{3}}^{(2)}(z) \, W_{-\frac{4}{3}}^{(4)}(w)$ for $N=3$.
It is obvious to see that 
the $U(1)$ charge counting in the OPE 
implies this particular OPE if one takes the first operator 
as the lowest higher spin current $W_{\frac{2}{3}}^{(2)}(z)$ \footnote{ 
One can calculate the OPE 
$W_{\frac{2}{3}}^{(2)}(z) \, W_{-\frac{1}{3}}^{(\frac{7}{2})}(w)$
and obtain all the singular terms where the third-order pole contains
 $ W_{\frac{1}{3}}^{(\frac{5}{2})}(w)$. 
Furthermore, one obtains the OPE
$W_{\frac{2}{3}}^{(2)}(z) \, W_{\frac{2}{3}}^{(4)}(w)$
where the highest singular term is the second-order pole.
Similarly, one can calculate 
the OPE  $W_{\frac{2}{3}}^{(2)}(z) \, W_{-\frac{1}{3}}^{(\frac{9}{2})}(w)$ where
the fourth-order pole has $ W_{\frac{1}{3}}^{(\frac{5}{2})}(w)$ for $N=3$.
See also Appendix (\ref{i2ope}).
}.
Note that the above  higher spin current $W_{-\frac{2}{3}}^{(2)}(w)$
is the first element of having $U(1)$ charge $-\frac{2}{3}$ in the list of 
(\ref{allhigher}).
Then by reading off the 
possible terms in the right hand sides of the above
four OPEs in the component approach and generalizing them in ${\cal N}=2$
superspace,
one can write down the right hand side of this OPE 
with arbitrary coefficients as follows: 
\bea
&& {\bf W}_{\frac{2}{3}}^{(2)}(Z_1) \, {\bf W}_{-\frac{1}{3}}^{(\frac{7}{2})}(Z_2) = 
C_{(2) \, (\frac{7}{2})}^{(2) \, +} \, \left(
\frac{\bar{\theta}_{12}}{z_{12}^4} \, {\bf W}_{-\frac{2}{3}}^{(2)}(Z_2)
+\frac{\theta_{12} \, \bar{\theta}_{12}}{z_{12}^4} \,
c_2 \, D {\bf W}_{-\frac{2}{3}}^{(2)} +
\frac{1}{z_{12}^3} \,
c_3 \, D {\bf W}_{-\frac{2}{3}}^{(2)} \right. 
\nonu \\
&& +
\frac{ \bar{\theta}_{12}}{z_{12}^3} \, \left[
c_4 \, [D, \overline{D}]  {\bf W}_{-\frac{2}{3}}^{(2)} 
+ c_5 {\bf T}  {\bf W}_{-\frac{2}{3}}^{(2)}  
+  c_6 \, \pa {\bf W}_{-\frac{2}{3}}^{(2)} 
\right] 
\nonu \\ 
&& +
\frac{\theta_{12} \, \bar{\theta}_{12}}{z_{12}^3} 
\, 
\left[ 
c_7 \, \pa D  {\bf W}_{-\frac{2}{3}}^{(2)} 
+ c_8 \, {\bf T}  D  {\bf W}_{-\frac{2}{3}}^{(2)} 
+ c_9 \, D {\bf T}    {\bf W}_{-\frac{2}{3}}^{(2)} 
\right]
\nonu \\
&& + \frac{1}{z_{12}^2} \, 
\left[ 
c_{10} \, \pa D  {\bf W}_{-\frac{2}{3}}^{(2)} 
+ c_{11} \, {\bf T}  D  {\bf W}_{-\frac{2}{3}}^{(2)} 
+ c_{12} \, D {\bf T}    {\bf W}_{-\frac{2}{3}}^{(2)} 
\right]
 +  \frac{\theta_{12}}{z_{12}^2} \, c_{13} \, D  {\bf T} D
 {\bf W}_{-\frac{2}{3}}^{(2)}
\nonu \\
&& + \frac{\bar{\theta}_{12}}{z_{12}^2} \, \left[ 
c_{14} \, \pa [D, \overline{D}]  {\bf W}_{-\frac{2}{3}}^{(2)}
+ \cdots 
+ c_{22} \, \pa^2  {\bf W}_{-\frac{2}{3}}^{(2)}
\right]
\nonu \\
&& +   \frac{\theta_{12} \, \bar{\theta}_{12}}{z_{12}^2}
\left[ c_{23} \,
\pa^2 D    {\bf W}_{-\frac{2}{3}}^{(2)}
+\cdots 
+   c_{31} \, {\bf T} D {\bf T}  {\bf W}_{-\frac{2}{3}}^{(2)}
\right]
\nonu \\
&& +   \frac{1}{z_{12}}
\left[ c_{32} \, \pa^2 D    {\bf W}_{-\frac{2}{3}}^{(2)}
+\cdots 
+ c_{40} \, \pa  {\bf T} D {\bf W}_{-\frac{2}{3}}^{(2)} 
\right]
\nonu \\
 && +  
\frac{\theta_{12}}{z_{12}} \left[ 
c_{41} \, {\bf T} D {\bf T} D {\bf W}_{-\frac{2}{3}}^{(2)}
+\cdots 
+ c_{43} \, \pa D {\bf T} D {\bf W}_{-\frac{2}{3}}^{(2)}
\right] 
\nonu \\
&& +  
\frac{ \bar{\theta}_{12}}{z_{12}} 
 \left[ c_{44} \,
\overline{D} {\bf T} D {\bf T}  {\bf W}_{-\frac{2}{3}}^{(2)}
+ \cdots
+ c_{65} \, \pa^3  {\bf W}_{-\frac{2}{3}}^{(2)}
\right]
\nonu \\
& & \left. +\frac{\theta_{12} \, \bar{\theta}_{12}}{z_{12}} 
\left[ c_{66}\, 
\pa^3 D  {\bf W}_{-\frac{2}{3}}^{(2)}
+ \cdots +
c_{87} \, {\bf T} {\bf T} D {\bf T} {\bf W}_{-\frac{2}{3}}^{(2)}
  \right]  \right)(Z_2)+\cdots.
\label{12opeope}
\eea
Using the Jacobi identity 
 between 
the three (higher spin) currents 
$({\bf T},  {\bf W}_{\frac{2}{3}}^{(2)}, {\bf W}_{-\frac{1}{3}}^{(\frac{7}{2})})$,
the coefficients (except an overall coefficient factor) 
are fixed and their explicit results 
are given in Appendix (\ref{14OPE}). 
By $U(1)$ charge counting, the nonlinear term ${\bf W}_{\frac{2}{3}}^{(2)} 
{\bf W}_{\frac{2}{3}}^{(2)}(Z_2)$ in the singular term 
$\frac{\theta_{12}}{z_{12}^2}$ (and the descendant terms in other 
singular terms) can arise but this cannot happen
because they become identically zero after using the Jacobi identity. 
The nonlinear higher spin currents do not appear in (\ref{12opeope}).
Furthermore, the higher spin current ${\bf W}_{\frac{1}{3}}^{(\frac{7}{2})}(Z_2)$
can also appear in the $\frac{1}{z_{12}^2}$ term but 
this cannot happen.

Now one can go to the other OPE between 
the second and the fourth ${\cal N}=2$ multiplets in (\ref{manyW}).
As before,
the lowest component higher spin current 
$W_{\frac{2}{3}}^{(2)}(w)$, living in the first ${\cal N}=2$ multiplet 
in (\ref{manyW}),  
appears in the fourth-order pole in the OPE
$W_{-\frac{2}{3}}^{(2)}(z) \, W_{\frac{4}{3}}^{(4)}(w)$ for $N=3$.
See also Appendix (\ref{i3ope}).
The $U(1)$ charge counting in the OPE 
implies this particular OPE if one takes the first operator 
as the lowest higher spin current $W_{-\frac{2}{3}}^{(2)}(z)$. 
Note that the above  higher spin current $W_{\frac{2}{3}}^{(2)}(w)$
is the first element of having $U(1)$ charge $\frac{2}{3}$ in the list of 
(\ref{allhigher}).
Then the right hand side of this OPE 
with arbitrary coefficients can be written as follows: 
\bea
&& {\bf W}_{-\frac{2}{3}}^{(2)}(Z_1) \, {\bf W}_{\frac{1}{3}}^{(\frac{7}{2})}(Z_2) = 
C_{(2) \, (\frac{7}{2})}^{(2) \, -} \, 
 \nonu \\
&& \times \left( \frac{1}{C_{(2) \, (\frac{7}{2})}^{(2) \, +}} \, 
{\bf W}_{\frac{2}{3}}^{(2)}(Z_1) \, 
{\bf W}_{-\frac{1}{3}}^{(\frac{7}{2})}(Z_2)\Biggr|_{\theta_{12} \leftrightarrow \bar{\theta}_{12}, D \leftrightarrow \overline{D},{\bf W}_{-\frac{2}{3}}^{(2)}(Z_2) \rightarrow {\bf W}_{\frac{2}{3}}^{(2)}(Z_2), c_i \rightarrow d_i}  \right).
\label{21opeope}
\eea
Here the previous OPE result (\ref{12opeope})
was used. The explicit form for (\ref{21opeope}) is given in 
Appendix (\ref{23OPE}).
The Jacobi identity 
 between 
the three (higher spin) currents 
$({\bf T},  {\bf W}_{-\frac{2}{3}}^{(2)}, {\bf W}_{\frac{1}{3}}^{(\frac{7}{2})})$
determines
the coefficients (except an overall coefficient factor) 
and their explicit results 
are given in Appendix (\ref{dcoeff}). 

Now the last OPE in this subsection we consider 
is the OPE between the 
second and the fourth ${\cal N}=2$ multiplets in (\ref{manyW}).
The lowest component higher spin current 
$W_0^{(4)}(w)$, 
living in the fifth ${\cal N}=2$ multiplet in (\ref{manyW}),  
appears in the second-order pole OPE
$W_{-\frac{2}{3}}^{(2)}(z) \, W_{\frac{2}{3}}^{(4)}(w)$ for $N=3$.
See also Appendix (\ref{i4ope}).
The Jacobi identity between the three (higher spin) currents 
$({\bf T},  {\bf W}_{-\frac{2}{3}}^{(2)}, {\bf W}_{-\frac{1}{3}}^{(\frac{7}{2})})$
determines  the following result
\bea
&& {\bf W}_{-\frac{2}{3}}^{(2)}(Z_1) \, {\bf W}_{-\frac{1}{3}}^{(\frac{7}{2})}(Z_2) = 
C_{(2) \, (\frac{7}{2})}^{(4) \, -} \, \left( \frac{\theta_{12}}{z_{12}^2} \,
 {\bf W}_{0}^{(4)}(Z_2)
+   \frac{\theta_{12} \, \bar{\theta}_{12}}{z_{12}^2}
\, \frac{7}{24} \, \overline{D}  {\bf W}_{0}^{(4)}(Z_2)
\right. \nonu \\
&& -\frac{1}{z_{12}} \,  \frac{1}{4} \, \overline{D}  {\bf W}_{0}^{(4)}(Z_2) 
 +  
\frac{\theta_{12}}{z_{12}} \left[ -\frac{1}{24} \, [D, \overline{D}]
 {\bf W}_{0}^{(4)} + \frac{3}{8} \, \pa  {\bf W}_{0}^{(4)}
\right](Z_2)
\label{22opeope}
 \\
&& + \left.  
\frac{\theta_{12} \, \bar{\theta}_{12}}{z_{12}} \left[
 \frac{5(c-3)}{36(c+9)} \, \pa    \overline{D}  
{\bf W}_{0}^{(4)}  + \frac{20}{3(c+9)} \, 
\overline{D} {\bf T}  {\bf W}_{0}^{(4)} 
 -  \frac{5}{3(c+9)} \,
{\bf T} \overline{D}  {\bf W}_{0}^{(4)} \right](Z_2) \right)
+\cdots.
\nonu 
\eea
The large $c$ limit can be analyzed before.

The above four OPEs (\ref{11opeope}), (\ref{12opeope}), 
(\ref{21opeope}), and 
(\ref{22opeope}) can be simplified as
\bea
\Biggr[{\bf W}_{\pm \frac{2}{3}}^{(2)} \Biggr] \cdot \Biggr[
{\bf W}_{\pm\frac{1}{3}}^{(\frac{7}{2})}\Biggr]
= \Biggr[{\bf W}_{0}^{(4)} \Biggr], \qquad
\Biggr[{\bf W}_{ \mp \frac{2}{3}}^{(2)} \Biggr] 
\cdot \Biggr[{\bf W}_{\pm \frac{1}{3}}^{(\frac{7}{2})} \Biggr]
= \Biggr[{\bf W}_{\pm \frac{2}{3}}^{(2)} \Biggr].
\label{fusion1}
\eea
Therefore, the fifth ${\cal N}=2$ multiplet in (\ref{manyW})
can be obtained from the first relations in (\ref{fusion1}) \footnote{
In this paper, because the OPE between the higher spin current 
${\bf W}_{0}^{(4)}(Z)$ and the other higher spin current is not known,
one cannot use the Jacobi identity including both 
the higher spin current and  ${\bf W}_{0}^{(4)}(Z)$.
For example, in order to obtain the Jacobi identity 
between the higher spin currents $({\bf W}_{\frac{2}{3}}^{(2)}, 
{\bf W}_{\frac{2}{3}}^{(2)}, {\bf W}_{\frac{1}{3}}^{(\frac{7}{2})})$, we should 
calculate more OPEs.}. 

\subsection{The OPEs between the higher spin-$\frac{7}{2}$ currents 
}

Let us consider the OPE
between the third ${\cal N}=2$ multiplet and itself in (\ref{manyW}).
As before,
the lowest component higher spin current 
$W_{-\frac{1}{3}}^{(\frac{7}{2})}(w)$, living in the fourth ${\cal N}=2$ multiplet 
in (\ref{manyW}),  
appears in the fourth-order pole of the OPE
$W_{\frac{1}{3}}^{(\frac{7}{2})}(z) \, W_{-\frac{2}{3}}^{(4)}(w)$ for $N=3$.
See also Appendix (\ref{j1ope}).
Similarly, 
OPE
$W_{\frac{1}{3}}^{(\frac{7}{2})}(z) \, W_{\frac{1}{3}}^{(\frac{7}{2})}(w)$
contains the third-order pole having 
$W_{\frac{2}{3}}^{(4)}(w)$ for $N=3$.

Then one can write down the following OPE with arbitrary 
coefficients
\bea
&& {\bf W}_{\frac{1}{3}}^{(\frac{7}{2})}(Z_1) \, {\bf W}_{\frac{1}{3}}^{(\frac{7}{2})}(Z_2) =  
C_{(\frac{7}{2}) \, (\frac{7}{2})}^{(\frac{7}{2}) \, +} \, 
\nonu \\
&& \times \left( 
\frac{1}{C_{(2) \, (\frac{7}{2})}^{(2) \, +}} \, 
{\bf W}_{\frac{2}{3}}^{(2)}(Z_1) \, 
{\bf W}_{-\frac{1}{3}}^{(\frac{7}{2})}(Z_2)\Biggr|_{
{\bf W}_{-\frac{2}{3}}^{(2)}(Z_2) 
\rightarrow {\bf W}_{-\frac{1}{3}}^{(\frac{7}{2})}(Z_2), \, \,\, c_i \rightarrow e_i}
\right),
\label{opeope11}
\eea
where the previous expression (\ref{12opeope}) is used.
The three (higher spin) currents 
$({\bf T},  {\bf W}_{\frac{1}{3}}^{(\frac{7}{2})}, {\bf W}_{\frac{1}{3}}^{(\frac{7}{2})})$
determines
the coefficients (except an overall coefficient factor) 
and their explicit results 
are given in Appendix (\ref{ecoeff}). 
As long as the $U(1)$ charge is concerned, the higher spin current 
${\bf W}_{\frac{2}{3}}^{(2)}(Z_2)$ can appear in the above OPE but this 
cannot happen.

Furthermore, the OPE between the third and fourth ${\cal N}=2$ multiplets
can be described as follows. 
Due to the $U(1)$ charge conservation, one can have 
${\bf W}_0^{(4)}(Z_2)$ in the right hand side but it has been checked that 
the third-order pole 
in the OPE between  $ W_{\frac{1}{3}}^{(\frac{7}{2})}(z)\, 
W_{-\frac{1}{3}}^{(\frac{7}{2})}(w)$ for $N=3$
does not contain the above $W_0^{(4)}(w)$ in the component approach.
See also Appendix (\ref{j2ope}).
Only spin-$4$ composite fields coming from the currents 
of ${\cal N}=2$ superconformal algebra arise in the above 
third-order pole. 
This implies that by considering the 
general terms $ (\pa^{l_1} D^{l_2} \overline{D}^{l_3} {\bf T}^{l_4} \cdots
 \pa^{m_1} D^{m_2} \overline{D}^{m_3} {\bf T}^{m_4})(Z_2)$, 
one can write down the possible ansatz with 
various arbitrary coefficients more than four hundreds
as follows:
\bea
&& {\bf W}_{\frac{1}{3}}^{(\frac{7}{2})}(Z_1) \, 
{\bf W}_{-\frac{1}{3}}^{(\frac{7}{2})}(Z_2) = 
\frac{\theta_{12} \, \bar{\theta}_{12}}{z_{12}^8} \, g_{379} +
\frac{1}{z_{12}^7} \, \frac{2c}{7}+
\frac{\theta_{12} \, \bar{\theta}_{12}}{z_{12}^7} \, g_{380} \, {\bf T}(Z_2)
+ \frac{1}{z_{12}^6} \, g_{381} \, {\bf T}(Z_2)
\nonu \\
&& +\frac{\theta_{12}}{z_{12}^6} \, g_{382} \, D {\bf T}(Z_2)
+ \frac{\bar{\theta}_{12}}{z_{12}^6} \, g_{383} \, \overline{D} {\bf T}(Z_2)
+ \frac{\theta_{12} \, \bar{\theta}_{12}}{z_{12}^6} \, 
\left[ g_{384} \, [ D, \overline{D} ] {\bf T}
+ g_{385} \, {\bf T} {\bf T} + g_{386} \, \pa {\bf T} \right]
(Z_2)
\nonu \\
&&+ \frac{1}{z_{12}^5} \, \left[ g_{387} \,
[D, \overline{D} ] {\bf T} + g_{388} \, {\bf T} 
{\bf T} + g_{399} \, \pa {\bf T} \right](Z_2) 
 + \frac{\theta_{12}}{z_{12}^5} \, \left[ 
g_{389} \, \pa D {\bf T} + g_{390} \, {\bf T} D {\bf T}
\right](Z_2)
\nonu \\
&&+ \frac{\bar{\theta}_{12}}{z_{12}^5} \, \left[ 
g_{391}\, \pa \overline{D} {\bf T} 
+ g_{392} \, {\bf T} \overline{D} {\bf T}
\right](Z_2)
 +  \frac{\theta_{12} \, \bar{\theta}_{12}}{z_{12}^5} \, \left[  
g_{393} \, \pa [ D,  \overline{D} ] {\bf T} + 
\cdots + 
g_{398} \, \pa^2 {\bf T}
\right](Z_2)
\nonu \\
&&+ \frac{1}{z_{12}^4} \, \left[ g_{400} \, \pa [D, \overline{D}] {\bf T}
+\cdots 
+  g_{404} \, \pa {\bf T} {\bf T}
+ g_{428} \, \pa^2 {\bf T} \right](Z_2)
\nonu \\
&&+ \frac{\theta_{12}}{z_{12}^4} \, \left[ 
g_{405} \, \pa^2 D {\bf T} 
+ \cdots 
+   g_{409} \, \pa {\bf T} D {\bf T}
\right](Z_2)
+ 
 \frac{\bar{\theta}_{12}}{z_{12}^4} \, \left[ 
g_{410} \, \pa^2 \overline{D} {\bf T} 
+\cdots
+   g_{414} \, \pa {\bf T} \overline{D} {\bf T}
\right](Z_2)
\nonu \\
&&+ \frac{\theta_{12} \, \bar{\theta}_{12}}{z_{12}^4}
\left[ g_{415} \, \pa^2 [ D, \overline{D} ] {\bf T}
\cdots + 
g_{427} \, \pa^3 {\bf T}
\right](Z_2)
\label{opeope12}
\\
&&+  \frac{1}{z_{12}^3}
\left[ g_1 \, \pa^2 [ D, \overline{D} ] {\bf T}
+\cdots
+ g_{57} \, \pa^3 {\bf T}
\right](Z_2)
+ \frac{\theta_{12}}{z_{12}^3}
\, \left[ 
g_{13} \, \pa^3 D {\bf T} 
+\cdots
+g_{22} \, \pa^2 {\bf T} D {\bf T}
\right](Z_2)
\nonu \\
&&+ \frac{\bar{\theta}_{12}}{z_{12}^3}
\, \left[ g_{23} \,
\pa^3 \overline{D} {\bf T} 
+ \cdots
+ g_{32} \, \pa^2 {\bf T} \overline{D} {\bf T}
\right](Z_2)
+ \frac{\theta_{12} \, \bar{\theta}_{12}}{z_{12}^3}
\left[ g_{33} \, \pa^3 [ D, \overline{D} ] {\bf T} 
+ \cdots
+  g_{56} \, \pa^4 {\bf T}
\right](Z_2)
\nonu \\
&&+  \frac{1}{z_{12}^2}
\left[ g_{58}\, \pa^3 [ D, \overline{D} ] {\bf T} 
+ \cdots
+  g_{170} \, \pa^4 {\bf T}
\right](Z_2)
+ \frac{\theta_{12}}{z_{12}^2}
\left[ g_{81} \,
\pa^4 D {\bf T}
+\cdots
+ g_{101} \, \pa^3 {\bf T} D {\bf T}
\right](Z_2)
\nonu \\
&&+ \frac{\bar{\theta}_{12}}{z_{12}^2}
\left[ 
g_{102} \, \pa^4 \overline{D} {\bf T}
+\cdots
+ g_{122} \, \pa^3 {\bf T} \overline{D} {\bf T}
\right](Z_2)
\nonu \\
&& + \frac{\theta_{12} \, \bar{\theta}_{12}}{z_{12}^2}
\left[ g_{123} \, \pa^4  [ D, \overline{D} ] {\bf T}
+ \cdots
+
g_{169} \, \pa^5 {\bf T}
\right](Z_2)
\nonu \\
&&+ \frac{1}{z_{12}}
\left[ g_{171} \, \pa^4  [ D, \overline{D} ] {\bf T}
+\cdots
 + g_{378} \, \pa^5 {\bf T}
\right](Z_2)
+ \frac{\theta_{12}}{z_{12}} \left[ 
 g_{217} \, \pa^5 D {\bf T}
+\cdots
+  g_{255} \, \pa^4 {\bf T}  D {\bf T} \right](Z_2)
\nonu \\
&&+ \frac{\bar{\theta}_{12}}{z_{12}} \left[ 
 g_{256} \, \pa^5 \overline{D} {\bf T}
+\cdots
+  g_{294} \, \pa^4 {\bf T} \overline{D} {\bf T}
\right](Z_2)
\nonu \\
&& + \frac{\theta_{12} \, \bar{\theta}_{12}}{z_{12}} \,
\left[  g_{295} \, \pa^5 [ D, \overline{D} ] {\bf T}
+  \cdots 
+ g_{377} \,  \pa^6 {\bf T}
\right](Z_2).
\nonu
\eea
Note that 
the undetermined coefficients $g_i$ where $i =1,2, \cdots, 428$
do not appear in order unfortunately. 
Again, 
three (higher spin) currents 
$({\bf T},  {\bf W}_{\frac{1}{3}}^{(\frac{7}{2})}, {\bf W}_{-\frac{1}{3}}^{(\frac{7}{2})})$
determines
the coefficients  completely, where the normalization 
for the higher spin currents 
$ {\bf W}_{\pm \frac{1}{3}}^{(\frac{7}{2})}(Z)$
are fixed through the singular term $\frac{1}{z_{12}^7} \, \frac{2c}{7}$, 
and their explicit results 
are given in Appendices (\ref{34OPE}) and (\ref{gcoeff}). 
Although we present the OPE (\ref{opeope12}), the complete 
expression is given in Appendices (\ref{34OPE}) and (\ref{gcoeff}).


The final OPE between the fourth ${\cal N}=2$
multiplet and itself 
can be summarized by the following expression
\bea
&& {\bf W}_{-\frac{1}{3}}^{(\frac{7}{2})}(Z_1) \, 
{\bf W}_{-\frac{1}{3}}^{(\frac{7}{2})}(Z_2) = 
C_{(\frac{7}{2}) \, (\frac{7}{2})}^{(\frac{7}{2}) \, -} \, 
\nonu \\
&& \times \left(
 \frac{1}{C_{(\frac{7}{2}) \, (\frac{7}{2})}^{(\frac{7}{2}) \, +}} \, 
{\bf W}_{\frac{1}{3}}^{(\frac{7}{2})}(Z_1) \, 
{\bf W}_{\frac{1}{3}}^{(\frac{7}{2})}(Z_2)\Biggr|_{\theta_{12} \leftrightarrow \bar{\theta}_{12}, 
\,\,\, D \leftrightarrow \overline{D},
\,\,\, {\bf W}_{-\frac{1}{3}}^{(\frac{7}{2})}(Z_2) 
\rightarrow {\bf W}_{\frac{1}{3}}^{(\frac{7}{2})}(Z_2), 
\,\,\,
e_i \rightarrow f_i} 
\right),
\label{opeope22} 
\eea
where the OPE (\ref{opeope11}) is used.
In this case, the lowest component higher spin current 
$W_{\frac{1}{3}}^{(\frac{7}{2})}(w)$, living in the third ${\cal N}=2$ multiplet 
in (\ref{manyW}),  
appears in the fourth-order pole in the OPE
$W_{-\frac{1}{3}}^{(\frac{7}{2})}(z) \, W_{\frac{2}{3}}^{(4)}(w)$ for $N=3$.
See also Appendix (\ref{j3ope}).
The detailed expression for the coefficients, after using 
the Jacobi identity between the three (higher spin) currents 
$({\bf T},  {\bf W}_{-\frac{1}{3}}^{(\frac{7}{2})}, 
{\bf W}_{-\frac{1}{3}}^{(\frac{7}{2})})$,
 is given by
Appendix (\ref{fcoeff}).

The above OPEs (\ref{opeope11}), (\ref{opeope12}) and (\ref{opeope22})
can be described as
\bea
\Biggr[{\bf W}_{\pm \frac{1}{3}}^{(\frac{7}{2})} \Biggr] 
\cdot \Biggr[{\bf W}_{\pm\frac{1}{3}}^{(\frac{7}{2})} \Biggr]
= \Biggr[{\bf W}_{\mp \frac{1}{3}}^{(\frac{7}{2})} \Biggr], \qquad
\Biggr[{\bf W}_{ \frac{1}{3}}^{(\frac{7}{2})} \Biggr] 
\cdot \Biggr[{\bf W}_{-\frac{1}{3}}^{(\frac{7}{2})} \Biggr]
=\Biggr[I \Biggr].
\label{fusionexpexp}
\eea
In other words, the higher spin current algebra
for ${\bf W}_{\pm \frac{1}{3}}^{(\frac{7}{2})}(Z)$ is closed from (\ref{fusionexpexp}). 

From the 
the Jacobi identity between the higher spin currents 
$({\bf W}_{\frac{2}{3}}^{(2)}, {\bf W}_{\frac{2}{3}}^{(2)},
{\bf W}_{-\frac{2}{3}}^{(2)})$,
one obtains the relation
\bea
C_{(2)\,(\frac{7}{2})}^{(2) \, -} \,  C_{(2)\,(2)}^{(\frac{7}{2}) \, +} 
& =& 
\frac{70 (810 c^5+15354 c^4-76443 c^3+75177 c^2+55324 c-80590)}{243   (c-1) (c+1) (c+6) (2 c-3) (5 c-9)}.
\label{ccproduct}
\eea
From the three OPEs between the above higher spin currents, one realizes 
that the structure constant $ C_{(2)\,(2)}^{(\frac{7}{2}) \, +}$
appears in the OPE (\ref{oneone}) and 
the structure constant $C_{(2)\,(\frac{7}{2})}^{(2) \, -}$
appears in the OPE (\ref{21opeope}) by selecting the first and second 
higher spin currents in the Jacobi identity first and then 
selecting the higher spin current ${\bf W}_{ \frac{1}{3}}^{(\frac{7}{2})}(Z)$
and the third higher spin current ${\bf W}_{-\frac{2}{3}}^{(2)}(Z)$.
Furthermore, the OPE (\ref{onetwo}) contains the stress energy tensor (and its
descendant fields) with known $c$-dependent coefficients after
selecting the first and the third higher spin currents in the Jacobi 
identity first and then selecting the second higher spin current 
${\bf W}_{-\frac{2}{3}}^{(2)}(Z)$ and the above stress energy tensor.   
This implies that 
one can rewrite $C_{(2)\,(\frac{7}{2})}^{(2) \, -}$ in terms of
$C_{(2)\,(2)}^{(\frac{7}{2}) \, +} $ with the help of (\ref{ccproduct}).

Similarly,
the Jacobi identity between the higher spin currents 
$({\bf W}_{-\frac{2}{3}}^{(2)}, {\bf W}_{-\frac{2}{3}}^{(2)},
{\bf W}_{\frac{2}{3}}^{(2)})$ implies the following relation
\bea
C_{(2)\,(\frac{7}{2})}^{(2) \, +} \,  C_{(2)\,(2)}^{(\frac{7}{2}) \, -} & = & 
\frac{70 (810 c^5+15354 c^4-76443 c^3+75177 c^2+55324 c-80590)}
{243  (c-1) (c+1) (c+6) (2 c-3) (5 c-9)}.
\label{ccproduct1}
\eea
In this case, from the OPE between the first and second higher spin currents
in the Jacobi identity, one observes the OPE (\ref{twotwo}) has
the structure constant  $ C_{(2)\,(2)}^{(\frac{7}{2}) \, -}$
and the OPE between the third higher spin current and the higher spin current
appearing in the right hand side of (\ref{twotwo}) leads to
the OPE (\ref{12opeope}) where the structure constant 
$C_{(2)\,(\frac{7}{2})}^{(2) \, +} $ appears. 
Furthermore, the combination between the first and third higher spin currents
in the Jacobi identity
implies the OPE (\ref{onetwo}) with the stress energy tensor 
and the OPE between the remaining (second) higher spin current and the 
stress energy tensor gives the higher spin itself with known $c$-dependent 
coefficients. Therefore, it turns out that one obtains the  
above result (\ref{ccproduct1}).
Then
one can rewrite $C_{(2)\,(\frac{7}{2})}^{(2) \, +}$ in terms of
$C_{(2)\,(2)}^{(\frac{7}{2}) \, -} $ with the help of (\ref{ccproduct1}).

By using 
the Jacobi identity between the higher spin currents 
$({\bf W}_{\frac{2}{3}}^{(2)}, {\bf W}_{\frac{2}{3}}^{(2)},
{\bf W}_{-\frac{1}{3}}^{(\frac{7}{2})})$,
one has the simple relation
\bea
C_{(2)\,(\frac{7}{2})}^{(2) \, +} = -\frac{4}{7} \,  
C_{(2)\,(2)}^{(\frac{7}{2}) \, +}. 
\label{ccproduct2}
\eea
According to previous description, 
 the structure constant $ C_{(2)\,(2)}^{(\frac{7}{2}) \, +}$
appears in the OPE (\ref{oneone}).
Then after doing the OPE between the higher spin current 
${\bf W}_{\frac{1}{3}}^{(\frac{7}{2})}(Z_1)$ appearing in the 
OPE (\ref{oneone}) and the third higher spin current in the Jacobi identity
leads to the OPE (\ref{opeope12}) where one can see the fixed coefficients
which depend on the central charge. 
Furthermore, the other combination in the Jacobi identity 
provides the OPE between the first and third higher spin currents.
This is given by the OPE (\ref{12opeope}) where the structure constant 
$C_{(2)\,(2)}^{(\frac{7}{2}) \, +}$ appears. After that, the OPE between the 
remaining (second) higher spin current and the higher spin current 
${\bf W}_{-\frac{2}{3}}^{(2)}(Z_2)$ gives the stress energy tensor with fixed 
coefficients via the OPE (\ref{onetwo}). Then we are left with the 
above result (\ref{ccproduct2}). 

Also 
the Jacobi identity between the higher spin currents 
$({\bf W}_{-\frac{2}{3}}^{(2)}, {\bf W}_{-\frac{2}{3}}^{(2)},
{\bf W}_{\frac{1}{3}}^{(\frac{7}{2})})$ determines the relation 
\bea
C_{(2)\,(\frac{7}{2})}^{(2) \, -} = -\frac{4}{7} \,  
C_{(2)\,(2)}^{(\frac{7}{2}) \, -}. 
\label{ccproduct3}
\eea
From the OPE between the first and second higher spin currents,
one sees the structure constant  
$C_{(2)\,(2)}^{(\frac{7}{2}) \, -}$ in the OPE (\ref{twotwo}).
Then 
 after using the OPE between the higher spin current 
${\bf W}_{-\frac{1}{3}}^{(\frac{7}{2})}(Z_1)$ appearing in the 
OPE (\ref{twotwo}) and the third higher spin current in the Jacobi identity
leads to the OPE (\ref{opeope12}) where one can see the fixed coefficients
which depend on the central charge. 
The other combination in the Jacobi identity 
provides the OPE between the first and third higher spin currents.
This is given by the OPE (\ref{21opeope}) where the structure constant 
$C_{(2)\,(2)}^{(\frac{7}{2}) \, -}$ appears. After that, the OPE between the 
remaining (second) higher spin current and the higher spin current 
${\bf W}_{\frac{2}{3}}^{(2)}(Z_2)$ gives the stress energy tensor (and its
descendant fields) with fixed 
coefficients via the OPE (\ref{onetwo}).
Then one arrives at the above result (\ref{ccproduct3}).

It is obvious that by combining the two equations (\ref{ccproduct1}) 
and (\ref{ccproduct2}),
one obtains
\bea
 C_{(2)\,(2)}^{(\frac{7}{2}) \, +} \,
C_{(2)\,(2)}^{(\frac{7}{2}) \, -}  & = &
-\frac{245 (810 c^5+15354 c^4-76443 c^3+75177 c^2+55324 c-80590)}
{486  (c-1) (c+1) (c+6) (2 c-3) (5 c-9)}.
\label{22result}
\eea
One can also obtain (\ref{22result}) from the two relations
(\ref{ccproduct}) and (\ref{ccproduct3}).

Finally, 
the Jacobi identity between the higher spin currents 
$({\bf W}_{\frac{1}{3}}^{(\frac{7}{2})}, {\bf W}_{-\frac{1}{3}}^{(\frac{7}{2})},
{\bf W}_{\frac{1}{3}}^{(\frac{7}{2})})$ (or 
the Jacobi identity between the higher spin currents 
$({\bf W}_{\frac{1}{3}}^{(\frac{7}{2})}, {\bf W}_{-\frac{1}{3}}^{(\frac{7}{2})},
{\bf W}_{-\frac{1}{3}}^{(\frac{7}{2})})$) determines the relation 
{\small
\bea
& & C_{(\frac{7}{2})\,(\frac{7}{2})}^{(\frac{7}{2}) \, +} \,
C_{(\frac{7}{2})\,(\frac{7}{2})}^{(\frac{7}{2}) \, -} \,
  \nonu \\
& & =  
-\frac{968 (c+9) (3 c-2) (3 c+4) (27 c-46) (3 c^2+90 c-265) 
}{
45927 (c-2) (c-1) (c+1) (c+6) (c+12) (c+18) (2 c-3) (4 c-9) (5 c-9) (7 c-15) 
} \times
\nonu \\
& &   
[\frac{1}{(25515 c^6+2447010 c^5+37120599 c^4-159264468 c^3+23829036 c^2+286911248 c-147772320)} ]
\nonu \\
& & \times  
(3214890 c^{10}+456897105 c^9+15491804931 c^8+80626717305 c^7
 \nonu \\
& & -  1337882375511 c^6
+  
4266884659422 c^5-3477197652650 c^4-4929653958916 c^3 \nonu \\
& & +   8674839058952 c^2-1802006074448 c-1456307369280).
\label{lastcc}
\eea}
In this case, by choosing the first and second higher spin currents,
one has the stress energy tensor (and its descendant fields) via 
the OPE (\ref{opeope12}) and then by considering the OPE 
between the stress energy tensor (with fixed $c$-dependent coefficients) 
and the third higher spin current
will provide the higher spin itself.  
Moreover, by taking the first and third higher spin currents,
one has the OPE (\ref{opeope11}) with structure constant
$ C_{(\frac{7}{2})\,(\frac{7}{2})}^{(\frac{7}{2}) \, +}$
and then considering the OPE between the higher spin current 
${\bf W}_{-\frac{1}{3}}^{(\frac{7}{2})}(Z_1)$ and the second higher spin current,
one obtains the OPE (\ref{opeope22}) where one sees the structure constant 
$C_{(\frac{7}{2})\,(\frac{7}{2})}^{(\frac{7}{2}) \, -}$. Then we are left with 
the above result (\ref{lastcc}).

By collecting the above relations, we are left with the undetermined 
structure constants $ C_{(2)\,(2)}^{(\frac{7}{2}) \, +}$, 
$ C_{(\frac{7}{2})\,(\frac{7}{2})}^{(\frac{7}{2}) \, +}$, 
$ C_{(2)\,(\frac{7}{2})}^{(4) \, +}$ and  $C_{(2)\,(\frac{7}{2})}^{(4) \, -}$
\footnote{
If one makes the rescalings 
$
{\bf W}_{\frac{1}{3}}^{(\frac{7}{2})}(Z)  \rightarrow  
\left( C_{(\frac{7}{2})\,(\frac{7}{2})}^{(\frac{7}{2}) \, +} \,
 C_{(\frac{7}{2})\,(\frac{7}{2})}^{(\frac{7}{2}) \, +} \,
C_{(\frac{7}{2})\,(\frac{7}{2})}^{(\frac{7}{2}) \, -} \right)^{\frac{1}{3}}
{\bf W}_{\frac{1}{3}}^{(\frac{7}{2})}(Z)$ and
$
{\bf W}_{-\frac{1}{3}}^{(\frac{7}{2})}(Z)  \rightarrow  
\left( C_{(\frac{7}{2})\,(\frac{7}{2})}^{(\frac{7}{2}) \, -} \,
 C_{(\frac{7}{2})\,(\frac{7}{2})}^{(\frac{7}{2}) \,-} \,
C_{(\frac{7}{2})\,(\frac{7}{2})}^{(\frac{7}{2}) \, +} \right)^{\frac{1}{3}}
{\bf W}_{-\frac{1}{3}}^{(\frac{7}{2})}(Z)$,
then the overall factors in (\ref{opeope11}) and (\ref{opeope22}) 
are given by 
the right hand side of (\ref{lastcc}).
Moreover the right hand side of (\ref{opeope12}) has 
the extra factor which is equal to the right hand side of (\ref{lastcc}) also.
Then 
the following rescalings hold
$
{\bf W}_{\frac{2}{3}}^{(2)}(Z)  \rightarrow  
\left( C_{(\frac{7}{2})\,(\frac{7}{2})}^{(\frac{7}{2}) \, +} \,
 C_{(\frac{7}{2})\,(\frac{7}{2})}^{(\frac{7}{2}) \, +} \,
C_{(\frac{7}{2})\,(\frac{7}{2})}^{(\frac{7}{2}) \, -} \right)^{\frac{1}{6}}
\,  \left(C_{(2)\,(2)}^{(\frac{7}{2}) \, -}\right)^{\frac{1}{2}}
{\bf W}_{\frac{2}{3}}^{(2)}(Z)$,
and 
$
{\bf W}_{-\frac{2}{3}}^{(2)}(Z)  \rightarrow  
\left( C_{(\frac{7}{2})\,(\frac{7}{2})}^{(\frac{7}{2}) \, -} \,
 C_{(\frac{7}{2})\,(\frac{7}{2})}^{(\frac{7}{2}) \,-} \,
C_{(\frac{7}{2})\,(\frac{7}{2})}^{(\frac{7}{2}) \, +} \right)^{\frac{1}{6}}
\,
\left(C_{(2)\,(2)}^{(\frac{7}{2}) \, +}\right)^{\frac{1}{2}}
{\bf W}_{-\frac{2}{3}}^{(2)}(Z)$.
Then the right hand sides of (\ref{11opeope}) 
and (\ref{22opeope}) contain the
right hand side of (\ref{22result}).
The right hand side of (\ref{12opeope}) can be changed also.}.

For the Jacobi identity 
 between the higher spin currents 
$({\bf W}_{\frac{2}{3}}^{(2)}, {\bf W}_{\frac{2}{3}}^{(2)},
{\bf W}_{-\frac{2}{3}}^{(2)})$,
one has
\bea
\frac{5 (1440 c^4+34131 c^3-74910 c^2-14858 c+73493)  }
{{243 (c-1) (c+1) (c+6) (2 c-3) (5 c-9)}}
(3 c-8)
\, {\bf W}_{\frac{2}{3}}^{(2)}(Z).
\label{someres}
\eea
Then for $c=\frac{8}{3}$ which is the minimum value in (\ref{cen}), 
this term (\ref{someres}) vanishes.
Furthermore, the descendant field of $ {\bf W}_{\frac{2}{3}}^{(2)}(Z)$,
\bea
&& \left(\frac{93}{10} \,  {\bf T} D \pa  {\bf W}_{\frac{2}{3}}^{(2)}
-\frac{43}{10} D {\bf T} \pa  {\bf W}_{\frac{2}{3}}^{(2)}
+\frac{4}{3} \pa D {\bf T} {\bf W}_{\frac{2}{3}}^{(2)}
-\frac{7}{2} \pa {\bf T} D {\bf W}_{\frac{2}{3}}^{(2)}
-\frac{209}{60} \pa^2 D  {\bf W}_{\frac{2}{3}}^{(2)} \right. \nonu \\
&& \left.
-\frac{13}{4} [ D, \overline{D} ] {\bf T} D  {\bf W}_{\frac{2}{3}}^{(2)}
+\frac{6}{5} D {\bf T} [D, \overline{D} ]  {\bf W}_{\frac{2}{3}}^{(2)}
-\frac{51}{4} {\bf T} {\bf T} D {\bf W}_{\frac{2}{3}}^{(2)}
+3 {\bf T} D {\bf T} {\bf W}_{\frac{2}{3}}^{(2)}\right)(Z),
\label{nullfield}
\eea
appears. However, this expression (\ref{nullfield})
becomes a null field at $c= \frac{8}{3}$.
One way to see this feature, following the procedure in \cite{ASS1991}, 
is to calculate 
the OPE
${\bf T}(Z_1)$ and the field (\ref{nullfield}) at $Z_2$.
Then the highest-order pole contains 
$ \frac{\bar{\theta}_{12}}{z_{12}^4} \, \frac{7}{9} (3c-8)  \, 
{\bf W}_{\frac{2}{3}}^{(2)}(Z_2)$ plus other singular terms.
This implies that for $c=\frac{8}{3}$, the above field 
(\ref{nullfield}) is a null field.
Similar analysis for 
 the Jacobi identity 
 between the higher spin currents 
$({\bf W}_{-\frac{2}{3}}^{(2)}, {\bf W}_{-\frac{2}{3}}^{(2)},
{\bf W}_{\frac{2}{3}}^{(2)})$ can be done \footnote{
As observed in 
\cite{Hornfeck1992}, the Zamolodchikov's extended conformal algebra
\cite{Zamol1985}
consisting of the spin-$\frac{5}{2}$ current as well as the spin-$2$ stress
energy tensor contains the null field.
That is, the spin-$\frac{5}{2}$ current appears in 
the Jacobi identity with the coefficient $(14c+13)$.
Then for $c=-\frac{13}{14}$, this term vanishes. Moreover, the descendant 
field of this current appears in the Jacobi identity and becomes a null 
field for $c =-\frac{13}{14}$ because the OPE between the stress energy tensor
with this descendant field leads to the highest-order pole with 
the spin-$\frac{5}{2}$ current with the coefficient $(14c+13)$.  
The
 Jacobi identity 
 between the higher spin currents 
$({\bf W}_{\frac{1}{3}}^{(\frac{7}{2})}, {\bf W}_{\frac{1}{3}}^{(\frac{7}{2})},
{\bf W}_{-\frac{1}{3}}^{(\frac{7}{2})})$ leads to 
$
(3c-8) f(c)
\, 
{\bf W}_{\frac{1}{3}}^{(\frac{7}{2})}(Z)$,
 where $f(c)$ is a 
complicated fractional expression in $c$.
Then for $c=\frac{8}{3}$, this term  vanishes.
Furthermore, there exists 
the descendant field of $ {\bf W}_{\frac{1}{3}}^{(\frac{7}{2})}(Z)$,
$
\left(
\frac{26703}{637} {\bf T} {\bf T} \pa^2 [D, \overline{D}]
{\bf W}_{\frac{1}{3}}^{(\frac{7}{2})}
-\frac{8933}{1274} {\bf T} \pa^3  [D, \overline{D}]
{\bf W}_{\frac{1}{3}}^{(\frac{7}{2})}
+\frac{180}{637} \pa^4 [D, \overline{D}]
{\bf W}_{\frac{1}{3}}^{(\frac{7}{2})} + \mbox{other 109-terms}\right)$.
Then the highest-order pole in the OPE between the 
stress energy tensor and the above descendant field contains 
$ \frac{\theta_{12}\, \bar{\theta}_{12}}{z_{12}^7} \, g(c) \, (3c-8)  \, 
{\bf W}_{\frac{1}{3}}^{(\frac{7}{2})}(Z_2)$ where $g(c)$ is a 
complicated fractional expression in $c$ (plus other singular terms).
This implies that for $c=\frac{8}{3}$, the above field 
 is a null field.
The analysis for  the
 Jacobi identity  
$({\bf W}_{-\frac{1}{3}}^{(\frac{7}{2})}, {\bf W}_{-\frac{1}{3}}^{(\frac{7}{2})},
{\bf W}_{\frac{1}{3}}^{(\frac{7}{2})})$
can be done.}.

One expects that
the sixth-ninth higher spin currents in (\ref{allhigher})
can be obtained the following  OPEs 
\bea
&& 
\Biggr[{\bf W}_{\pm \frac{2}{3}}^{(2)} \Biggr] 
\cdot \Biggr[{\bf W}_{0}^{(4)} \Biggr] =   \Biggr[
{\bf W}_{\mp \frac{1}{3}}^{(\frac{11}{2})} \Biggr], \qquad
\Biggr[{\bf W}_{ \pm \frac{2}{3}}^{(2)} \Biggr] \cdot 
\Biggr[{\bf W}_{\mp \frac{1}{3}}^{(\frac{11}{2})}\Biggr]
=\Biggr[{\bf W}_{\mp \frac{2}{3}}^{(6)}\Biggr],
\nonu \\
&& \Biggr[{\bf W}_{ \pm \frac{2}{3}}^{(2)}\Biggr] 
\cdot \Biggr[{\bf W}_{\pm \frac{1}{3}}^{(\frac{11}{2})}\Biggr]
 =  \Biggr[{\bf W}_{0}^{(4)}\Biggr].
\label{extraope}
\eea
Of course, in the right hand sides of (\ref{extraope}), 
the previous known higher spin currents can appear.
It would be interesting to see these extra OPEs in details in the future.


\section{Conclusions and outlook }

In this paper, 
the first two ${\cal N}=2$ higher spin multiplets in (\ref{manyW})
are obtained from the two adjoint fermions living in the stringy coset minimal 
model (\ref{coset}).
We also obtain the corresponding OPEs between the first four ${\cal N}=2$ 
higher spin multiplets in (\ref{manyW}) in ${\cal N}=2$ superspace
for generic $N$ (or central charge $c$) 
by using the Jacobi identity.

Now we present the future directions as follows.

$\bullet$
The ${\cal N}=3$ (or ${\cal N}=4$) supersymmetric coset minimal model

One can think of the following coset models
\bea
\frac{\hat{SU}(N)_N \oplus \hat{SU}(N)_N \oplus 
\hat{SU}(N)_N }{\hat{SU}(N)_{3N}}, \qquad
\frac{\hat{SU}(N)_N \oplus \hat{SU}(N)_N \oplus 
\hat{SU}(N)_N \oplus \hat{SU}(N)_N}{\hat{SU}(N)_{4N}}.
\nonu
\eea
Then it is an open problem to obtain 
three (four) spin-$\frac{3}{2}$ currents of ${\cal N}=3$ (large ${\cal N}=4$)
superconformal algebra from 
 the three (four) kinds of adjoint fermions. 
One expects that the spin-$2$ stress energy tensor 
can be determined by the Sugawara construction.
The nontrivial thing is to obtain the correct OPEs between the 
spin-$\frac{3}{2}$ currents.
Moreover, it is an open problem to check whether there are 
higher sin currents in the context of large ${\cal N}=4$ holography
\cite{GG1305,Ahn1311,Ahn1408,AK1411,Ahn1504,AK1506,AK1509}.

$\bullet$
The explicit Casimir higher spin-$\frac{7}{2}, 4, \frac{9}{2}$ currents 

So far, the Casimir higher spin-$2, \frac{5}{2}, 3$ currents are found.
It would be interesting to obtain the third and fourth ${\cal N}=2$ 
multiplets in (\ref{manyW}) in terms of two adjoint fermions.
From the $U(1)$ charge counting, one can have 
\bea 
&& W_{ \pm \frac{1}{3}}^{(\frac{7}{2})}(z)  = 
d^{abc} d^{cde} f^{bfg} f^{dhi} f^{ejk} \nonu \\
&& \times  (\psi^a \pm  i \chi^a)(
\psi^f + i \chi^f)( \psi^g -i \chi^g)(\psi^h + i \chi^h)(\psi^i-i \chi^i)
(\psi^j + i \chi^j)(\psi^k-i \chi^k)(z) + \cdots.
\nonu 
\eea
Here the abbreviated parts come from the derivative terms and 
the nonderivative terms with different choices of signs in the $\chi^a(z)$
which preserves the corresponding $U(1)$ charge. 
Furthermore, 
the higher spin current
$W_{\pm \frac{4}{3}}^{(4)}(z)$ can be written in terms of
$
d^{abc} d^{cde} f^{afg} f^{bhi} f^{djk} f^{elm} (
\psi^f \pm i \chi^f)( \psi^g \pm i \chi^g)(\psi^h \pm i \chi^h)(\psi^i 
\pm i \chi^i)
(\psi^j + i \chi^j)(\psi^k-i \chi^k) (\psi^l +  i \chi^l)(
\psi^m - i \chi^m)(z) + \cdots$.
Similarly, one can have 
$W_{\pm \frac{2}{3}}^{(4)}(z) =
d^{abc} d^{cde} f^{afg} f^{bhi} f^{djk} f^{elm} (
\psi^f \pm i \chi^f)( \psi^g \pm i \chi^g)(\psi^h + i \chi^h)(\psi^i 
- i \chi^i)
(\psi^j + i \chi^j)(\psi^k-i \chi^k) (\psi^l +  i \chi^l)(
\psi^m - i \chi^m)(z) + \cdots$.
For the higher spin current with vanishing
$U(1)$ charge, one expects 
 $W_{0}^{(4)}(z) =
d^{abc} d^{cde} f^{afg} f^{bhi} f^{djk} f^{elm} (
\psi^f + i \chi^f)( \psi^g - i \chi^g)(\psi^h + i \chi^h)(\psi^i 
- i \chi^i)
(\psi^j + i \chi^j)(\psi^k-i \chi^k) (\psi^l +  i \chi^l)(
\psi^m - i \chi^m)(z) + \cdots$.

$\bullet$
Marginal deformation

One of the motivations of this paper is to 
describe the marginal deformation which breaks the higher spin symmetry 
and obtain the mass for the higher spin currents in the large $N$ limit.
The coset model we consider here has ${\cal N}=2$ supersymmetry and there 
should be a marginal deformation. 
It would be interesting to determine the mass for the higher spin currents
with the help of the explicit symmetry algebra found in this paper. 
According to the observation of  \cite{CHR1406,HR1503,CH1506},
the $SO(2)_R$ doublet rather than $SO(2)_R$ singlet can have nonzero 
mass contribution. For the integer spin $SO(2)_R$ doublet, there are
the higher spin currents $W_{\pm \frac{4}{3}}^{(4)}(z)$ and 
$W_{\mp \frac{2}{3}}^{(4)}(z)$. Their OPEs can be found from  the ${\cal N}=2$
version,
(\ref{opeope11}), (\ref{opeope12}) or (\ref{opeope22}).
Unfortunately we did not present them in this paper. As emphasized in Appendix
$J$, one can read off them from the above ${\cal N}=2$ version using the 
command $\tt N2OPEToComponents$ inside of \cite{KT}.     
For the integer spin $SO(2)_R$ singlet, one has 
$W_{\pm \frac{2}{3}}^{(2)}(z)$ and $W_{\pm \frac{2}{3}}^{(3)}(z)$.
Their OPEs can be found from Appendix $H$ in the large $N$ limit 
(or in the finite $N$).
For the half integer case (either $SO(2)_R$ doublet or $SO(2)_R$ singlet)
one can also analyze the mass contribution. 
From Appendix $I$, one has the necessary OPEs between the 
$SO(2)_R$ singlets and the $SO(2)_R$ doublets.

$\bullet$ Possible bulk theory computation

It is an open problem to obtain the $AdS_3$ dual string theory
(or extension of $AdS_3$ higher spin theory).
First of all, in the context of type IIB string theory, 
the ${\cal N}=2$ supersymmetry should be maintained.
The $7$-dimensional space may contain the two sphere ${\bf S}^2$
having the $SO(2)_R$ symmetry. 
See also the relevant work \cite{BNZ}.

$\bullet$ The orthogonal version

It is natural to ask whether the present description can apply to 
the orthogonal coset minimal model 
\cite{Ahn1106,GV,Ahn1202,AP1301,AP1310,AP1410,AKP}.
The first thing to do is to obtain 
the realization of ${\cal N}=2$ superconformal algebra in the coset
$\frac{\hat{SO}(2N)_{2N-2} \oplus \hat{SO}(2N)_{2N-2} }{\hat{SO}(2N)_{4N-4}}$
model where the two levels are given by 
the dual Coxeter number of $SO(2N)$ and the level in the denominator
is the sum of these two levels. The central charge is given by 
$c =\frac{1}{3} N(2N-1)$ which behaves as $ \frac{2}{3} N^2$ if the large 
$N$ limit is taken. 
One expects that there exist $N(2N-1)$
free fermions transforming in the adjoint representation 
of $SO(2N)$ and the quadratic expression with the structure 
constant realizes the usual affine Kac-Moody algebra.
The adjoint index for the fermions is given by 
either a single notation or double notation. 
One can also analyze the $SO(2N+1)$ case. 

$\bullet$ The ${\cal N}=2$ description of adjoint fermions 

It is interesting to see whether there exists an ${\cal N}=2$ supersymmetric 
extension of the two adjoint fermions. How one can write down 
$ \Psi^a(Z) = \psi^a(z) + \cdots $ (and $\Xi^a(Z) = \chi^a(z) + \cdots$)?
Then how one can express the (higher spin) currents in terms of 
these ${\cal N}=2$ adjoint fermions? 
Can we introduce any ${\cal N}=2$ constraints on these
${\cal N}=2$ currents?

$\bullet$
The ${\cal N}=2$ OPE 
$
\Biggr[{\bf W}_{ q}^{(h)} \Biggr] 
\cdot \Biggr[{\bf W}_{-q}^{(h)} \Biggr]
=\Biggr[I \Biggr]$

In this paper, we have obtained this relation for 
$(h=2, q=\frac{2}{3})$ and $(h=\frac{7}{2}, q=\frac{1}{3})$.
It would be interesting to observe whether the above relation
satisfies for any $(h,q)$ or not. The point is 
whether the right hand side of the above OPE contains 
any other combinations among 
the higher spin currents with vanishing $U(1)$ charge or not 
\cite{Blumenhagen}. 

\vspace{.7cm}

\centerline{\bf Acknowledgments}

CA would like to thank H. Kim for discussions. 
This research was supported by Basic Science Research Program through
the National Research Foundation of Korea (NRF)  
funded by the Ministry of Education  
(No. 2015R1D1A1A01059064).
CA acknowledges warm hospitality from 
the School of  Liberal Arts (and Institute of Convergence Fundamental
Studies), Seoul National University of Science and Technology.

\newpage

\appendix

\renewcommand{\theequation}{\Alph{section}\mbox{.}\arabic{equation}}

\section{ The fundamental OPEs between the adjoint 
spin-$\frac{1}{2}$ and spin-$1$ currents }

We summarize the 
fundamental OPEs between 
 the adjoint 
spin-$\frac{1}{2}$ currents
and the adjoint spin-$1$ currents as follows:
\bea
\psi^a(z) \, J^b(w) &= & \frac{1}{(z-w)} \, f^{abc} \, \psi^c(w)
+ \cdots,
\nonu \\
\psi^a(z) \, L^b(w) &= & \frac{1}{(z-w)} \, \frac{1}{2} \, f^{abc} \, 
\chi^c(w)
+ \cdots,
\nonu \\
\psi^a(z) \, M^b(w) &= & -\frac{1}{(z-w)} \, \frac{1}{2} \,
d^{abc} \, \chi^c(w)
+ \cdots,
\nonu \\
\chi^a(z) \, K^b(w) &= & \frac{1}{(z-w)} \, f^{abc} \, \chi^c(w)
+ \cdots,
\nonu \\
\chi^a(z) \, L^b(w) &= & \frac{1}{(z-w)} \, \frac{1}{2} \, f^{abc} \, 
\psi^c(w)
+ \cdots,
\nonu \\
\chi^a(z) \, M^b(w) &= & \frac{1}{(z-w)} \, \frac{1}{2} \,
d^{abc} \, \psi^c(w)
+ \cdots,
\nonu \\
J^a(z) \, J^b(w) &= & -\frac{1}{(z-w)^2} \, N \, \delta^{ab} +
\frac{1}{(z-w)}  \,
f^{abc} \, J^c(w)
+ \cdots,
\nonu \\
J^a(z) \, L^b(w) &= & 
\frac{1}{(z-w)}  \,
f^{ace}  \, f^{bde} \, \psi^c  \chi^d(w)
+ \cdots,
\nonu \\
J^a(z) \, M^b(w) &= & 
-\frac{1}{(z-w)}  \,
f^{ace}  \, d^{bde} \, \psi^c  \chi^d(w)
+ \cdots,
\nonu \\
K^a(z) \, K^b(w) &= & -\frac{1}{(z-w)^2} \, N \, \delta^{ab} +
\frac{1}{(z-w)}  \,
f^{abc} \, K^c(w)
+ \cdots,
\nonu \\
K^a(z) \, L^b(w) &= & 
-\frac{1}{(z-w)}  \,
f^{ade}  \, f^{bce} \, \psi^c  \chi^d(w)
+ \cdots,
\nonu \\
K^a(z) \, M^b(w) &= & 
-\frac{1}{(z-w)}  \,
f^{ade}  \, d^{bce} \, \psi^c  \chi^d(w)
+ \cdots,
\nonu \\
L^a(z) \, L^b(w) &= & -\frac{1}{(z-w)^2} \, \frac{N}{2} \, \delta^{ab} +
\frac{1}{(z-w)}  \,
\frac{1}{4} \, f^{abc} \, (J^c+ K^c)(w)
+ \cdots,
\nonu \\
L^a(z) \, M^b(w) &= & 
-\frac{1}{(z-w)}  \,
\frac{1}{2} \, f^{ace} \, d^{bde} \, (\chi^c \chi^d
- \psi^c \psi^d)(w)
+ \cdots,
\label{appendixa}
\\
M^a(z) \, M^b(w) &= & -\frac{1}{(z-w)^2} \, \frac{(N^2-4)}{2N} \, \delta^{ab} 
+
\frac{1}{(z-w)}  \,
\frac{1}{2} \, d^{ace} \, d^{bde} (\chi^c  \chi^d +
 \psi^c  \psi^d)(w)
+ \cdots.
\nonu
\eea
The defining relations are given in (\ref{spin1currents}) and 
(\ref{LMexpression}).

\section{ The OPEs between the ${\cal N}=2$ stress energy tensor 
and the 
${\cal N}=2$ primary current}

The standard OPE between ${\cal N}=2$ stress energy tensor and itself in 
${\cal N}=2$ superspace is given by
\bea
{\bf T}(Z_1) \, {\bf T}(Z_2) = \frac{1}{z_{12}^2} \, \frac{c}{3} 
+\frac{\theta_{12} \, \bar{\theta}_{12}}
{z_{12}^2} \, {\bf T}(Z_2)-
\frac{\theta_{12}}{z_{12}} \, D {\bf T}(Z_2)
+\frac{\bar{\theta}_{12}}{z_{12}} \, \overline{D} {\bf T}(Z_2)
+ \frac{\theta_{12} \, \bar{\theta}_{12}}
{z_{12}} \, \pa \, {\bf T}(Z_2)
\label{SuperTT}
\eea
where we have the explicit component expression for the stress 
energy tensor
\bea
{\bf T}(Z) = J(z) + \theta \, G^{+}(z) + \bar{\theta} \, G^{-}(z) +
\theta \, \bar{\theta} \, T(z).
\label{stressN2}
\eea

The equivalent $16$ component OPEs corresponding to Appendix (\ref{SuperTT}) 
are given by 
\bea
J(z) \, J(w) & = & \frac{1}{(z-w)^2} \, \frac{c}{3} + \cdots,
\nonu \\
J(z) \, G^{+}(w) & = & \frac{1}{(z-w)} \, G^{+}(w) + \cdots,
\nonu \\
J(z) \, G^{-}(w) & = & -\frac{1}{(z-w)} \, G^{-}(w) + \cdots,
\nonu \\
J(z) \, T(w) & = & \frac{1}{(z-w)^2} \, T(w) + \cdots,
\nonu \\
G^{+}(z) \, J(w) & = & -\frac{1}{(z-w)} \, G^{+}(w) +\cdots,
\nonu \\
G^{+}(z) \, G^{+}(w) & = &  + \cdots,
\nonu \\ 
G^{+}(z) \, G^{-}(w) & = & \frac{1}{(z-w)^3} \, \frac{c}{3} + 
\frac{1}{(z-w)^2} \, J(w) + 
\frac{1}{(z-w)} \, \left[ T + \frac{1}{2} \pa J \right](w) + \cdots,
\nonu \\
G^{+}(z) \, T(w) & = & 
\frac{1}{(z-w)^2} \, \frac{3}{2} \, G^{+}(w) +
\frac{1}{(z-w)} \, \frac{1}{2} \, \pa G^{+}(w) +
\cdots,
\nonu \\
G^{-}(z) \, J(w) & = & \frac{1}{(z-w)} \, G^{-}(w) +\cdots,
\nonu \\
G^{-}(z) \, G^{+}(w) & = & \frac{1}{(z-w)^3} \, \frac{c}{3} - 
\frac{1}{(z-w)^2} \, J(w) + 
\frac{1}{(z-w)} \, \left[ T - \frac{1}{2} \pa J \right](w) + \cdots,
\nonu \\
G^{-}(z) \, G^{-}(w) & = & + \cdots,
\nonu \\
G^{-}(z) \, T(w) & = & 
\frac{1}{(z-w)^2} \, \frac{3}{2} \, G^{-}(w) +
\frac{1}{(z-w)} \, \frac{1}{2} \, \pa G^{-}(w) +
\cdots,
\nonu \\
T(z) \, J(w) &=& \frac{1}{(z-w)^2} \, J(w) + \frac{1}{(z-w)} \, \pa J(w) +
\cdots,
\nonu \\
T(z) \, G^{+}(w) & = & \frac{1}{(z-w)^2} \, \frac{3}{2} \, 
G^{+}(w) + \frac{1}{(z-w)} \, \pa G^{+}(w) + \cdots, 
\nonu \\
T(z) \, G^{-}(w) & = & \frac{1}{(z-w)^2} \, \frac{3}{2} \, 
G^{-}(w) + \frac{1}{(z-w)} \, \pa G^{-}(w) + \cdots, 
\nonu \\
T(z) \, T(w) & = & 
\frac{1}{(z-w)^4} \, \frac{c}{2} +\frac{1}{(z-w)^2} \, 2 T(w) +
\frac{1}{(z-w)} \, \pa T(w) +\cdots.
\label{n2scaexpression}
\eea

Let us introduce the ${\cal N}=2$ primary current of 
spin $h$ 
with nonzero $U(1)$ charge $q$  as follows:
\bea
{\bf T}(Z_1) \, {\bf W}_q^{(h)}(Z_2) & = & 
\left( \frac{\theta_{12} \, \bar{\theta}_{12}}
{z_{12}^2} \, h +  \frac{1}{z_{12}} \, q \right){\bf W}_q^{(h)}(Z_2)-
\frac{\theta_{12}}{z_{12}} \, D {\bf W}_q^{(h)}(Z_2)
+\frac{\bar{\theta}_{12}}{z_{12}} \, \overline{D} {\bf W}_q^{(h)}(Z_2)
\nonu \\
&& +  \frac{\theta_{12} \, \bar{\theta}_{12}}
{z_{12}} \, \pa \, {\bf W}_q^{(h)}(Z_2).
\label{TWOPEseveral}
\eea
As before, the component result for the  ${\cal N}=2$ primary current
is given by
\bea
{\bf W}_q^{(h)}(Z) = W_q^{(h)}(z) + \theta \, W_{q+1}^{(h+\frac{1}{2})}(z) + 
\bar{\theta} \, W_{q-1}^{(h+\frac{1}{2})}(z) +
\theta \, \bar{\theta} \, W_q^{(h+1)}(z).
\label{Wsuperexpression}
\eea
In this classification, the above ${\cal N}=2$ stress energy tensor
is characterized by $h=1$ and $q=0$.
More explicitly, $W_0^{(1)}(z) =J(z)$, $W_1^{(\frac{3}{2})}(z) =G^{+}(z)$,
$W_{-1}^{(\frac{3}{2})}(z) =G^{-}(z)$ and $W_0^{(2)}(z) =T(z)$.

In component approach, we have the following $16$ OPEs corresponding to
Appendix (\ref{TWOPEseveral})
\bea
J(z) \, W_q^{(h)}(w) & = & \frac{1}{(z-w)} \, q \, W_q^{(h)}(w) + \cdots,
\nonu \\
J(z) \, W_{q+1}^{(h+\frac{1}{2})}(w) & = & \frac{1}{(z-w)} \, \left(q+1\right) \, 
W_{q+1}^{(h+\frac{1}{2})}(w) + \cdots,
\nonu \\
J(z) \, W_{q-1}^{(h+\frac{1}{2})}(w) & = & \frac{1}{(z-w)} \, \left(q-1\right) \, 
W_{q-1}^{(h+\frac{1}{2})}(w) + \cdots,
\nonu \\
J(z) \, W_q^{(h+1)}(w) & = & \frac{1}{(z-w)^2} \, h \, W_q^{(h)}(w)+
\frac{1}{(z-w)} \, q \, W_q^{(h+1)}(w) + \cdots,
\nonu \\
G^{+}(z) \, W_q^{(h)}(w) & = & -\frac{1}{(z-w)} \,  W_{q+1}^{(h+\frac{1}{2})}(w) 
+ \cdots,
\nonu \\
G^{+}(z) \, W_{q+1}^{(h+\frac{1}{2})}(w) & = &  + \cdots,
\nonu \\
G^{+}(z) \, W_{q-1}^{(h+\frac{1}{2})}(w) & = & \frac{1}{(z-w)^2} \, 
\left(h + \frac{q}{2} \right) W_q^{(h)} + \frac{1}{(z-w)} \left[ 
W_{q}^{(h+1)} +\frac{1}{2} \pa W_q^{(h)} \right](w) + \cdots,
\nonu \\
G^{+}(z) \, W_q^{(h+1)}(w) & = & \frac{1}{(z-w)^2} \, 
\left[h+\frac{1}{2} (q+1) \right] \, 
W_{q+1}^{(h+\frac{1}{2})}(w)+
\frac{1}{(z-w)} \, \frac{1}{2}\, \pa  \, W_{q+1}^{(h+\frac{1}{2})}(w) + \cdots,
\nonu \\
G^{-}(z) \, W_q^{(h)}(w) & = & \frac{1}{(z-w)} \,  W_{q-1}^{(h+\frac{1}{2})}(w) 
+ \cdots,
\nonu \\
G^{-}(z) \, W_{q+1}^{(h+\frac{1}{2})}(w) & = & \frac{1}{(z-w)^2} \, 
\left(-h + \frac{q}{2} \right) W_q^{(h)} + \frac{1}{(z-w)} \left[ 
W_{q}^{(h+1)} -\frac{1}{2} \pa W_q^{(h)} \right](w) + \cdots,
\nonu \\
G^{-}(z) \, W_{q-1}^{(h+\frac{1}{2})}(w) & = &  + \cdots,
\nonu \\
G^{-}(z) \, W_q^{(h+1)}(w) & = & \frac{1}{(z-w)^2} \, 
\left[h-\frac{1}{2} (q-1) \right] \, 
W_{q-1}^{(h+\frac{1}{2})}(w)+
\frac{1}{(z-w)} \, \frac{1}{2}\, \pa  \, W_{q-1}^{(h+\frac{1}{2})}(w) + \cdots,
\nonu \\
T(z) \, W_q^{(h)}(w) & = & \frac{1}{(z-w)^2} \, h \, W_q^{(h)}(w) +
 \frac{1}{(z-w)} \, \pa \, W_q^{(h)}(w) + \cdots,
\nonu \\
T(z) \, W_{q+1}^{(h+\frac{1}{2})}(w) & = & \frac{1}{(z-w)^2} \, (h+\frac{1}{2}) \, 
W_{q+1}^{(h+\frac{1}{2})}(w) +
 \frac{1}{(z-w)} \, \pa \, W_{q+1}^{(h+\frac{1}{2})}(w) + \cdots,
\nonu \\
T(z) \, W_{q-1}^{(h+\frac{1}{2})}(w) & = & \frac{1}{(z-w)^2} \, (h+\frac{1}{2}) \, 
W_{q-1}^{(h+\frac{1}{2})}(w) +
 \frac{1}{(z-w)} \, \pa \, W_{q-1}^{(h+\frac{1}{2})}(w) + \cdots,
\label{16TWOPE} 
\\
T(z) \, W_q^{(h+1)}(w) & = & \frac{1}{(z-w)^3} \, \frac{q}{2} \, W_q^{(h)} +
 \frac{1}{(z-w)^2} \, (h+1) \, W_q^{(h+1)}(w) +
 \frac{1}{(z-w)} \, \pa \, W_q^{(h+1)}(w) \nonu \\
& + & \cdots.
\nonu 
\eea
For $q=0$, the above relations reproduce the previous relations (for example
in 
\cite{Ahn1208}).
For the nonzero $q$, the last component current
$ W_q^{(h+1)}(w)$ in Appendix (\ref{Wsuperexpression}) 
is not a primary current because 
there exists the third-order pole in the OPE between the 
stress energy tensor of spin-$2$: the last equation of Appendix 
(\ref{16TWOPE}).
In order to obtain the primary current, one should 
consider the following OPE which can be obtained from the fourth equation from 
the bottom of Appendix (\ref{16TWOPE})
\bea
T(z) \, \pa \, W_q^{(h)}(w) & = & \frac{1}{(z-w)^3} \, 2h \, W_q^{(h)}(w)+
\frac{1}{(z-w)^2} \, (h+1) \, \pa \, W_q^{(h)}(w) \nonu \\
& + & 
 \frac{1}{(z-w)} \, \pa^2 \, W_q^{(h)}(w) +  \cdots,
\label{polethree}
\eea
where the third-order pole is nonvanishing.
Then it is easy to see that 
the combination of 
\bea
W_q^{(h+1)}(w)- \frac{q}{4h} \, \pa\,  W_q^{(h)}(w)
\label{primarycombi}
\eea
is primary current because 
the third-order pole with the stress energy tensor vanishes
\footnote{
The defining OPEs in Appendix (\ref{16TWOPE})
can provide how to obtain the remaining three component currents 
for given lowest higher spin current $W_q^{(h)}(z)$.
The fifth relation gives the second component current 
by reading off the first-order pole.
Similarly the third component current can be 
determined from the ninth relation of Appendix (\ref{16TWOPE}). 
Finally, the last component current can be obtained either 
seventh relation 
or tenth relation.
The first several ${\cal N}=2$ primary currents in the coset model
are presented by
$
{\bf W}_{\pm \frac{2}{3}}^{(2)}(Z)$, 
$ {\bf W}_{\pm \frac{1}{3}}^{(\frac{7}{2})}(Z)$,
$ 
{\bf W}_{0}^{(4)}(Z)$, 
${\bf W}_{\mp \frac{1}{3}}^{(\frac{11}{2})}(Z)$, 
${\bf W}_{\mp \frac{2}{3}}^{(6)}(Z), \cdots $ as in (\ref{manyW}) or 
(\ref{allhigher}). }.

\section{The first-order poles in the OPEs $G^{\pm}(z)
\, W_{\pm \frac{2}{3}}^{(2)}(w)$ }

In subsection $3.2$, we have seen the construction of 
higher spin-$\frac{5}{2}$ current. We present here 
some details for the first-order poles as follows with 
(\ref{spin1currents}), (\ref{LMexpression}) 
and (\ref{spintwo}):
\bea
\psi^a  J^a(z) \, \, J^b  M^b(w) \mid_{\frac{1}{(z-w)}} & = &
-3 N \, \pa \psi^a M^a(w) -\frac{3}{2} \, \chi^a Q^a(w),
\nonu \\
\psi^a  J^a(z) \, \, K^b  M^b(w) \mid_{\frac{1}{(z-w)}} & = & -\frac{3}{2} 
\, \chi^a R^a(w),
\nonu \\
\psi^a  J^a(z) \, \, L^b  M^b(w) \mid_{\frac{1}{(z-w)}} & = &
\frac{3}{2} \,  f^{abc} \, \chi^a J^b M^c(w) -\frac{3}{2} \, \chi^a U^a(w),
\nonu \\
\psi^a  K^a(z) \, \, J^b  M^b(w) \mid_{\frac{1}{(z-w)}} & = &
f^{abc} \, \psi^a K^b M^c(w)- N \, \psi^a \pa M^a(w) 
-\frac{1}{2} \, \chi^a R^a(w) + \psi^a U^a(w),
\nonu \\
\psi^a  K^a(z) \, \, K^b  M^b(w) \mid_{\frac{1}{(z-w)}} & = &
-2 N \, \pa \psi^a M^a(w) - f^{abc} \, \psi^a K^b M^c(w) -\frac{1}{2}\,
\chi^a S^a(w) + \psi^a V^a(w),
\nonu \\
\psi^a  K^a(z) \, \, L^b  M^b(w) \mid_{\frac{1}{(z-w)}} & = &
\frac{1}{2} \,f^{abc} \, \chi^a K^b M^c(w)+ f^{abc} \, \psi^a L^b M^c(w) 
-\frac{1}{2} \, \chi^a V^a(w) \nonu \\
& + & \psi^a W^a(w),
\nonu \\
\chi^a  K^a(z) \, \, J^b  M^b(w) \mid_{\frac{1}{(z-w)}} & = &
\frac{3}{2} 
\, \psi^a R^a(w),
\label{5half}
 \\
\chi^a  K^a(z) \, \, K^b  M^b(w) \mid_{\frac{1}{(z-w)}} & = &
-3 N \, \pa \chi^a M^a(w) +\frac{3}{2} \, \psi^a S^a(w),
\nonu \\
\chi^a  K^a(z) \, \, L^b  M^b(w) \mid_{\frac{1}{(z-w)}} & = &
\frac{3}{2} \,  f^{abc} \, \psi^a K^b M^c(w) +\frac{3}{2} \, \psi^a V^a(w),
\nonu \\
\chi^a  J^a(z) \, \, J^b  M^b(w) \mid_{\frac{1}{(z-w)}} & = &
-2 N \, \pa \chi^a M^a(w) - f^{abc} \, \chi^a J^b M^c(w) +\frac{1}{2}\,
\psi^a Q^a(w) - \chi^a U^a(w),
\nonu \\
\chi^a  J^a(z) \, \, K^b  M^b(w) \mid_{\frac{1}{(z-w)}} & = & 
f^{abc} \, \chi^a J^b M^c(w)+ N \, \pa \chi^a  M^a(w) 
+\frac{1}{2} \, \psi^a R^a(w) - \chi^a V^a(w),
\nonu \\
\chi^a  J^a(z) \, \, L^b  M^b(w) \mid_{\frac{1}{(z-w)}} & = &
\frac{1}{2} \,f^{abc} \, \psi^a J^b M^c(w)+ f^{abc} \, \chi^a L^b M^c(w) 
+\frac{1}{2} \, \psi^a U^a(w) \nonu \\
& - & \chi^a W^a(w).
\nonu
\eea
As explained before, the last half of the OPEs in Appendix 
(\ref{5half})
can be obtained from the first half of the OPEs 
using the symmetry under the exchange of 
$\psi^a(z) \leftrightarrow \chi^a(z)$. 

\section{The first-order poles in the OPEs $G^{\mp}(z)
\, W_{\pm \frac{5}{3}}^{(\frac{5}{2})}(w)$ }

In subsection $3.4$, the higher spin-$3$ current was obtained.
The first-order poles (the total number is $4 \times 8=32$) together with 
(\ref{spintwo}) are summarized by 
\bea
\psi^a  J^a(z) \, \, \psi^b  Q^b(w) \mid_{\frac{1}{(z-w)}} & = &
15 N \, d^{abc} \, \psi^a \pa \psi^b J^c(w),
\nonu \\
\psi^a  J^a(z) \, \, \psi^b  U^b(w) \mid_{\frac{1}{(z-w)}} & = &
-\frac{3}{2}  \, J^a U^a(w) +3 N \, d^{abc} \, \psi^a \pa \psi^b L^c(w)
+\frac{3}{2} \, d^{abc} f^{cde} \, \psi^a J^b J^d \chi^e(w),
\nonu \\
\psi^a  J^a(z) \, \, \psi^b  S^b(w) \mid_{\frac{1}{(z-w)}} & = &
-\frac{3}{2}  \, J^a S^a(w),
\nonu \\
\psi^a  J^a(z) \, \, \psi^b  V^b(w) \mid_{\frac{1}{(z-w)}} & = &
-\frac{3}{2} \, J^a V^a(w) + \frac{3}{2} \, d^{abc} f^{cde} \, 
\psi^a K^b J^d \chi^e(w),
\nonu \\
\psi^a  J^a(z) \, \, \chi^b  Q^b(w) \mid_{\frac{1}{(z-w)}} & = &
3 N \, d^{abc} \, \chi^a \pa \psi^b J^c(w) + 3 N \, d^{abc} \, \chi^a J^b \pa 
\psi^c(w),
\nonu \\
\psi^a J^a(z) \, \, \chi^b  U^b(w) \mid_{\frac{1}{(z-w)}}  & = &
3 N \, d^{abc} \, \chi^a \pa \psi^b L^c(w) +\frac{3}{2}\, d^{abc} f^{cde} \,
\chi^a J^b J^d \chi^e(w), 
\nonu \\
\psi^a  J^a(z) \, \, \chi^b  S^b(w) \mid_{\frac{1}{(z-w)}} & = &
+ \cdots,
\nonu \\
\psi^a  J^a(z) \, \, \chi^b  V^b(w) \mid_{\frac{1}{(z-w)}} & = &
\frac{3}{2} \, d^{abc} f^{cde}\, \chi^a K^b J^d \chi^e(w),
\nonu \\
\psi^a  K^a(z) \, \, \psi^b  Q^b(w) \mid_{\frac{1}{(z-w)}} & = &
-\frac{1}{2} \, K^a Q^a(w) - d^{abc} f^{cde} \, \psi^a J^b \psi^d K^e(w)
+ N \, d^{abc} \, \psi^a \pa \psi^b K^c(w)
\nonu \\
& - & d^{abc} f^{cde} \, \psi^a J^b \psi^d K^e(w),
\nonu \\
\psi^a  K^a(z) \, \, \psi^b  U^b(w) \mid_{\frac{1}{(z-w)}} & = &
-\frac{1}{2} \, K^a U^a(w) - d^{abc} f^{cde} \, \psi^a (\psi^d K^e) L^b(w)
\nonu \\
& + &  \frac{1}{2} \, d^{abc} f^{cde} \, \psi^a J^b K^d \chi^e(w)
-d^{abc} f^{cde} f^{efg} \, \psi^a J^b \psi^d \psi^f \chi^g(w),
\nonu \\
\psi^a  K^a(z) \, \, \psi^b  S^b(w) \mid_{\frac{1}{(z-w)}} & = &
-\frac{1}{2} \, K^a S^a(w) + N \, d^{abc} \, \psi^a \pa \psi^b K^c(w)
+ d^{abc} f^{cde}\, \psi^a(\psi^d K^e) K^b(w)
\nonu \\
& +  & N \, d^{abc} \, \psi^a K^b \pa \psi^c(w)
+ d^{abc} f^{cde} \, \psi^a K^b \psi^d K^e(w),
\nonu \\
\psi^a  K^a(z) \, \, \psi^b  V^b(w) \mid_{\frac{1}{(z-w)}} & = &
-\frac{1}{2} \, K^a V^a(w) + N \, d^{abc} \, \psi^a \pa \psi^b L^c(w)
+ d^{abc} f^{cde} \, \psi^a(\psi^d K^e)L^b(w)
\nonu \\
& + & \frac{1}{2} d^{abc} f^{cde} \psi^a K^b K^d \chi^e(w)
- d^{abc} f^{cde} f^{efg} \psi^a K^b \psi^d \psi^f \chi^g(w),
\nonu \\
\psi^a  K^a(z) \, \, \chi^b  Q^b(w) \mid_{\frac{1}{(z-w)}} & = &
-f^{abc} \, (\psi^a \chi^b)Q^c(w)
- d^{abc} f^{cde} \, \chi^a J^b \psi^d K^e(w)
\nonu \\
& + & N \, d^{abc} \, \chi^a \pa \psi^b K^c(w)
\nonu \\
&-& N \, \pa K^a M^a(w) - d^{abc}f^{cde} \, \chi^a J^b \psi^d K^e(w),
\nonu \\
\psi^a  K^a(z) \, \, \chi^b  U^b(w) \mid_{\frac{1}{(z-w)}} & = &
-f^{abc} \, (\psi^a \chi^b) U^c(w) - d^{abc} f^{cde} \,
\chi^a (\psi^d K^e) L^b(w)
\nonu \\
&+ & \frac{1}{2} d^{abc} f^{cde} \, \chi^a J^b K^d \chi^e(w)
-d^{abc} f^{cde} f^{efg} \, \chi^a J^b \psi^d \psi^f \chi^g(w),
\nonu \\
\psi^a  K^a(z) \, \, \chi^b  S^b(w) \mid_{\frac{1}{(z-w)}} & = &
- f^{abc} \, (\psi^a \chi^b) S^c(w)
+ N \, d^{abc} \, \chi^a \pa \psi^b K^c(w)
\nonu \\
& + &  d^{abc} f^{cde} \, \chi^a (\psi^d K^e) K^b(w)
\nonu \\
&+ & N \, d^{abc} \, \chi^a K^b \pa \psi^c(w)
+ d^{abc} f^{cde} \, \chi^a K^b \psi^d K^e(w),
\nonu \\
\psi^a  K^a(z) \, \, \chi^b  V^b(w) \mid_{\frac{1}{(z-w)}} & = &
-f^{abc} \, (\psi^a \chi^b) V^c(w) + N \, d^{abc} \, \chi^a \pa \psi^b L^c(w)
\nonu \\
& + &  d^{abc} f^{cde} \, \chi^a (\psi^d K^e) L^b(w)
\nonu \\
&+& \frac{1}{2} d^{abc} f^{cde} \, \chi^a K^b K^d \chi^e(w)
- d^{abc} f^{cde} f^{efg} \, \chi^a K^b \psi^d \psi^f \chi^g(w),
\nonu \\
\chi^a  K^a(z) \, \, \psi^b  Q^b(w) \mid_{\frac{1}{(z-w)}} & = &
+ \cdots,
\nonu \\
\chi^a  K^a(z) \, \, \psi^b  U^b(w) \mid_{\frac{1}{(z-w)}} & = &
\frac{3}{2} \, d^{abc} f^{cde} \psi^a J^b K^d \psi^e(w),
\nonu \\
\chi^a  K^a(z) \, \, \psi^b  S^b(w) \mid_{\frac{1}{(z-w)}} & = &
3 N \, d^{abc} \, \psi^a \pa \chi^b K^c(w) + 3 N \, d^{abc} \, \psi^a K^b \pa 
\chi^c(w),
\nonu \\
\chi^a  K^a(z) \, \, \psi^b  V^b(w) \mid_{\frac{1}{(z-w)}} & = &
3 N \, d^{abc} \, \psi^a \pa \chi^b L^c(w) +\frac{3}{2} \, d^{abc} f^{cde} \,
\psi^a K^b K^d \psi^e(w), 
\nonu \\
\chi^a  K^a(z) \, \, \chi^b  Q^b(w) \mid_{\frac{1}{(z-w)}} & = &
-\frac{3}{2}  \, K^a Q^a(w),
\nonu \\
\chi^a  K^a(z) \, \, \chi^b  U^b(w) \mid_{\frac{1}{(z-w)}} & = &
-\frac{3}{2} \, K^a U^a(w) + \frac{3}{2} \, d^{abc} f^{cde} \, 
\chi^a J^b K^d \psi^e(w),
\nonu \\
\chi^a  K^a(z) \, \, \chi^b  S^b(w) \mid_{\frac{1}{(z-w)}} & = &
15 N \, d^{abc} \, \chi^a \pa \chi^b K^c(w),
\nonu \\
\chi^a  K^a(z) \, \, \chi^b  V^b(w) \mid_{\frac{1}{(z-w)}} & = &
-\frac{3}{2}  \, K^a V^a(w) +3 N \, d^{abc} \, \chi^a \pa \chi^b L^c(w)
\nonu \\
& + & \frac{3}{2} \, d^{abc} f^{cde} \, \chi^a K^b K^d \psi^e(w),
\nonu \\
\chi^a  J^a(z) \, \, \psi^b  Q^b(w) \mid_{\frac{1}{(z-w)}} & = &
- f^{abc} \, (\chi^a \psi^b) Q^c(w)
+ N \, d^{abc} \, \psi^a \pa \chi^b J^c(w)
\nonu \\
& + & d^{abc} f^{cde} \, \psi^a (\chi^d J^e) J^b(w)
\nonu \\
&+& N \, d^{abc} \, \psi^a J^b \pa \chi^c(w)
+ d^{abc} f^{cde} \, \psi^a J^b \chi^d J^e(w),
\nonu \\
\chi^a  J^a(z) \, \, \psi^b  U^b(w) \mid_{\frac{1}{(z-w)}} & = &
-f^{abc} \, (\chi^a \psi^b) U^c(w) + N \, d^{abc} \, \psi^a \pa \chi^b L^c(w)
\nonu \\
& + & d^{abc} f^{cde} \, \psi^a (\chi^d J^e) L^b(w)
\nonu \\
&+& \frac{1}{2} \, d^{abc} f^{cde} \, \psi^a J^b J^d \psi^e(w)
- d^{abc} f^{cde} f^{efg} \, \psi^a J^b \chi^d \chi^f \psi^g(w),
\nonu \\
\chi^a  J^a(z) \, \, \psi^b  S^b(w) \mid_{\frac{1}{(z-w)}} & = &
-f^{abc} \, (\chi^a \psi^b)S^c(w)
- d^{abc} f^{cde} \, \psi^a K^b \chi^d J^e(w)
\nonu \\
& + &  N \, d^{abc} \, \psi^a \pa \chi^b J^c(w)
+ N \, \pa J^a M^a(w) - d^{abc}f^{cde} \, \psi^a K^b \chi^d J^e(w),
\nonu \\
\chi^a  J^a(z) \, \, \psi^b  V^b(w) \mid_{\frac{1}{(z-w)}} & = &
-f^{abc} \, (\chi^a \psi^b) V^c(w) - d^{abc} f^{cde} \,
\psi^a (\chi^d J^e) L^b(w)
\nonu \\
&+& \frac{1}{2} \, d^{abc} f^{cde} \, \psi^a K^b J^d \psi^e(w)
-d^{abc} f^{cde} f^{efg} \, \psi^a K^b \chi^d \chi^f \psi^g(w),
\nonu \\
\chi^a  J^a(z) \, \, \chi^b  Q^b(w) \mid_{\frac{1}{(z-w)}} & = &
-\frac{1}{2} \, J^a Q^a(w) + N \, d^{abc} \, \chi^a \pa \chi^b J^c(w)
+ d^{abc} f^{cde} \chi^a(\chi^d J^e) J^b(w)
\nonu \\
& +& N \, d^{abc} \, \chi^a J^b \pa \chi^c(w)
+ d^{abc} f^{cde} \, \chi^a J^b \chi^d J^e(w),
\nonu \\
\chi^a  J^a(z) \, \, \chi^b  U^b(w) \mid_{\frac{1}{(z-w)}} & = &
-\frac{1}{2} \, J^a U^a(w) + N \, d^{abc} \, \chi^a \pa \chi^b L^c(w)
+ d^{abc} f^{cde} \, \chi^a(\chi^d J^e)L^b(w)
\nonu \\
&+ & \frac{1}{2} d^{abc} f^{cde} \chi^a J^b J^d \psi^e(w)
- d^{abc} f^{cde} f^{efg} \, \chi^a J^b \chi^d \chi^f \psi^g(w),
\nonu \\
\chi^a  J^a(z) \, \, \chi^b  S^b(w) \mid_{\frac{1}{(z-w)}} & = &
-\frac{1}{2} \, J^a S^a(w) - d^{abc} f^{cde} \, \chi^a K^b \chi^d J^e(w)
+ N \, d^{abc} \, \chi^a \pa \chi^b J^c(w)
\nonu \\
& - & d^{abc} f^{cde} \, \chi^a K^b \chi^d J^e(w),
\nonu \\
\chi^a  J^a(z) \, \, \chi^b  V^b(w) \mid_{\frac{1}{(z-w)}} & = &
-\frac{1}{2} \, J^a V^a(w) - d^{abc} f^{cde} \, \chi^a (\chi^d J^e) L^b(w)
\nonu \\
& + & \frac{1}{2} \, d^{abc} f^{cde} \, \chi^a K^b J^d \psi^e(w)
- d^{abc} f^{cde} f^{efg} \, \chi^a K^b \chi^d \chi^f \psi^g(w).
\label{spin3exp}
\eea
In this case, one can use the symmetry between $\psi^a(z)$ and $\chi^a(z)$.
That is, the last half of these OPEs in Appendix (\ref{spin3exp}) 
can be obtained from the 
first half of those OPEs.
In order to obtain the final higher spin-$3$ current, we should simplify 
these expressions in terms of fully normal ordered product 
\cite{BBSSfirst,BBSSsecond}.
We present some useful identities (which can be checked using 
the Jacobi identities between the $f$ and $d$ symbols) as follows:
\bea
d^{bcd} f^{def} \, \psi^b J^c J^e \chi^f(w) & = & -\frac{1}{2} d^{abc} J^a J^b L^c(w)
+\frac{N}{2} \pa J^a M^a(w) + N d^{abc} J^a \pa \psi^b \chi^c(w),
\nonu \\
d^{bcd} f^{def} \, \psi^b K^c J^e \chi^f(w) & = &
-d^{abc} K^a J^b L^c(w)+ N d^{abc} K^a \pa \psi^b \chi^c(w) \nonu \\
& - & 2N d^{abc} \psi^a J^b
\pa \chi^c(w),
\nonu \\
d^{bcd} f^{def} \, \chi^b J^c J^e \chi^f(w) & = & -\frac{1}{2} d^{abc} 
J^a J^b K^c(w),
\nonu \\
d^{bcd} f^{def} \, \chi^b K^c J^e \chi^f(w) & = &
-3N d^{abc} \chi^a \pa \chi^b J^c(w) - d^{abc} J^a K^b K^c(w),
\nonu \\
d^{bcd} f^{cae} \, \psi^b J^d \psi^e K^a(w) & = &
-d^{abc} J^a J^b K^c(w) -3N d^{abc} \psi^a \pa \psi^b K^c(w),
\nonu \\
d^{bcd} f^{cae} f^{def} \psi^b \pa (\psi^f K^a)(w) & = & 
- N d^{abc} \psi^a \pa \psi^b K^c(w),
\nonu \\
d^{bcd} f^{dae} \, \psi^b J^c \psi^e K^a(w) & = &
-d^{abc} J^a J^b K^c(w) -3N d^{abc} \psi^a \pa \psi^b K^c(w),
\nonu \\
d^{bcd} f^{def} \, \psi^b J^c K^e \chi^f(w) & = &
2N d^{gbc} \psi^b J^c \ps \chi^g(w),
\nonu \\
d^{bcd} f^{def} f^{fag} \psi^b J^c \psi^e \psi^a \chi^g(w) & = &
-d^{bcd} d^{daf} \psi^b J^c \psi^a M^f(w),
\nonu \\
d^{bcd} f^{cae} \psi^b (\psi^a K^e) K^d(w) & = & 
\frac{1}{2} d^{abc} J^a K^b K^c(w) + N d^{abc} \psi^a \pa \psi^b K^c(w),
\nonu \\
d^{bcd} f^{dae} \psi^b K^c \psi^a K^e(w) & = & \frac{1}{2} d^{abc} J^a K^b K^c(w),
\nonu \\
d^{bcd} f^{cef} \psi^b (\psi^e K^f) L^d(w) & = & -\frac{N}{2} d^{abc} \psi^a 
\pa \chi^b K^c(w) -\frac{N}{2} \pa K^a M^a(w) -\frac{N}{2} \pa M^a J^a(w)
\nonu \\
& - & \frac{N}{4} M^a \pa J^a(w) +\frac{N}{2} d^{abc} \psi^a \pa \psi^b L^c(w) 
\nonu \\
& + &
d^{bcd} f^{cef} \psi^b L^d \psi^e K^f(w),
\nonu \\
d^{bcd} f^{def} \psi^b K^c K^e \chi^f(w) & = & 2N d^{abc} \psi^a \pa \chi^b K^c(w),
\nonu \\
d^{bcd} f^{cae} \chi^b J^d \psi^e K^a(w) & = & 
-N d^{dge} J^d \psi^e \pa \chi^g - d^{abc}J^a K^b L^c(w) -2N d^{abc} \chi^a \pa
\psi^b K^c(w), 
\nonu \\
f^{cae} f^{def} d^{bcd} \chi^b \pa (\psi^f K^a)(w) & = & 
-N d^{abc} \chi^a \pa \psi^b K^c(w) 
+N \pa K^a M^a(w),
\nonu \\
f^{dae} d^{bcd} \chi^b J^c \psi^e K^a(w) & = & 
-N d^{abc} \psi^a \pa \chi^b J^c(w) - d^{abc} J^a K^b L^c(w)
\nonu \\
& - & 2N d^{abc} \chi^a \pa \psi^b K^c(w),
\nonu \\
f^{bfg} d^{bcd} (\psi^f \chi^g)(J^c L^d) & = & 
d^{abc} L^a J^b L^c(w),
\nonu \\
f^{def} d^{bcd} \chi^b J^c K^e \chi^f(w) & = & 
2N d^{abc} \chi^a \pa \chi^b J^c(w),
\nonu \\
d^{bcd} f^{cfg} \chi^b (\psi^f K^g) K^d(w) & = & 
d^{bcd} f^{cfg} \chi^b K^d \psi^f K^g(w)-N d^{abc} \pa \psi^a \chi^b K^c(w)
\nonu \\
& - & N \pa K^a M^a(w),
\nonu \\
d^{bcd} f^{dfg} \chi^b K^c \psi^f K^g(w) & = & 
-N d^{abc} \psi^a \pa \chi^b K^c(w) + N \pa K^a M^a(w),
\nonu \\
f^{cef} d^{bcd} \chi^b (\psi^f K^e) L^d(w) & = &
f^{cef} d^{bcd} \chi^b L^d \psi^f K^e(w) +\frac{N}{2} d^{abc} \chi^a \pa 
\chi^b K^c(w)
+\frac{N}{2} \pa M^a L^a(w) \nonu \\
& + & \frac{N}{2} f^{abc} \pa \psi^a 
\chi^b M^c(w)
+\frac{N}{2} \pa L^a M^a(w) +\frac{N}{2} d^{abc} \pa \psi^a \chi^b L^c(w), 
\nonu \\
d^{bcd} f^{bfg} (\psi^f \chi^g)(K^c L^d)(w) &= &
d^{abc} L^a K^b L^c(w),
\nonu \\
d^{bcd} f^{cef} \chi^b (\psi^e K^f) L^d(w) & = &
-f^{cef} d^{bcd} \chi^b L^d \psi^f K^e(w) -\frac{N}{2} d^{abc} \chi^a \pa \chi^b
K^c(w) \nonu \\
& - & \frac{N}{2} \pa M^a L^a (w) 
\nonu \\
&-& \frac{N}{2} f^{abc} \pa \psi^a \chi^b M^c(w)
-\frac{N}{2} \pa L^a M^a(w)
-\frac{N}{2} d^{abc} \pa \psi^a \chi^b L^c(w),
\nonu \\
d^{bcd} f^{def} \chi^b K^c K^e \chi^f(w) & = & 
2N d^{abc} \chi^a \pa \chi^b K^c(w),
\nonu \\
d^{bcd} f^{def} \chi^b J^c \psi^e L^f(w) & = & 
-\frac{1}{4} d^{abc} J^a J^b K^c(w),
\nonu \\
d^{bcd} f^{def} \chi^b K^c \psi^e L^f(w) & = & 
-\frac{1}{2} d^{abc} J^a K^b K^c(w)
-\frac{3N}{2} d^{abc} \psi^a \pa \chi^b L^c(w) \nonu \\
& + & \frac{3N}{2} f^{abc} \psi^a \pa 
\chi^b M^c(w),
\nonu \\
d^{bcd} f^{def} \psi^b K^c \psi^e L^f(w) & = & 
-\frac{1}{2} d^{abc} J^a K^b L^c(w) +\frac{N}{2} d^{abc} \pa \psi^a \chi^b K^c(w)
\nonu \\
& - & N d^{abc} \psi^a \pa \chi^b J^c(w), 
\nonu \\
d^{bcd} f^{cef} \psi^b L^d \psi^f K^e(w) & = & 
-\frac{3}{2} d^{abc} J^a K^b L^c(w) +\frac{N}{2} d^{abc} \psi^a \pa \psi^b L^c(w)
\nonu \\
& + & \frac{3N}{2} d^{abc} \pa \psi^a \chi^b K^c(w) \nonu \\
&-& \frac{3N}{2} d^{abc} \psi^a \pa \chi^b J^c(w) -\frac{N}{2} d^{abc}
\psi^a \pa \chi^b K^c(w) -\frac{N}{4} \pa J^a M^a(w),
\nonu \\
d^{bcd} f^{cef} \chi^b L^d \psi^f K^e(w) & = & 
-d^{abc} K^a L^b L^c(w) -\frac{1}{2} d^{abc} J^a K^b K^c(w) +\frac{1}{6} d^{abc}
K^a K^b K^c(w)
\nonu \\
&-& 3N d^{abc} \psi^a \pa \chi^b L^c(w)
-N f^{abc} \pa \psi^a \chi^b M^c(w) \nonu \\
& + & N f^{abc} \psi^a \pa \chi^b M^c(w),
\nonu \\
d^{abc} K^a K^b L^c(w) & = & -N \pa K^a M^a(w) -4N d^{abc} \chi^a \pa 
\chi^b L^c(w),
\nonu \\
\pa M^a J^a(w) & = & 
\frac{1}{2} \pa J^a M^a(w) - d^{abc} \psi^a \pa \psi^b L^c(w) + d^{abc} \psi^a 
\pa \chi^b J^c(w),
\nonu \\
\pa M^a L^a(w) & = & 
d^{abc} \pa \psi^a \chi^b L^c(w) + d^{abc} \psi^a \pa \chi^b L^c(w),
\nonu \\
\pa L^a M^a(w) & = & 
-d^{abc} \chi^a \pa \chi^b  J^c(w) + d^{abc} \psi^a \pa \chi^b L^c(w)+
d^{abc} \psi^a \pa \psi^b K^c(w) \nonu \\
&+ & d^{abc} \pa \psi^a \chi^b L^c(w),
\nonu \\
\pa M^a K^a(w) & = &
d^{abc} \pa \psi^a \chi^b K^c(w) + d^{abc} \chi^a \pa \chi^b L^c(w) +\frac{1}{2}
\pa K^a M^a(w),
\nonu \\
d^{abc} \chi^a \pa \chi^b L^c(w) & = & 
d^{abc} \psi^a \pa \chi^b K^c(w) -\frac{1}{2} M^a \pa K^a(w), 
\nonu \\
f^{abc} \psi^a \pa \chi^b M^c(w) & = & 
-d^{abc} J^a \chi^b \pa \chi^c(w) + d^{abc} \psi^a \pa \chi^b L^c(w),
\nonu \\
d^{abc} J^a \pa \psi^b \chi^c(w) & = & 
\frac{1}{2} \pa J^a M^a(w) - d^{abc} \psi^a \pa \psi^b L^c(w),
\nonu \\
f^{abc} \pa \psi^a \chi^b M^c(w) & = & 
d^{abc} \psi^a \pa \psi^b K^c(w) + d^{abc} \pa \psi^a  \chi^b L^c(w),
\nonu \\
d^{abc} \pa \psi^a  \chi^b J^c(w) & = & 
 \frac{1}{2} \pa J^a M^a(w) -d^{abc} \psi^a \pa \psi^b   L^c(w),
\nonu \\
d^{abc} L^a J^b L^c(w) & = & d^{abc} J^a L^b L^c(w) - N d^{abc} \pa \psi^a \chi^b
L^c(w) - N d^{abc} \psi^a \pa \chi^b L^c(w),
\nonu \\
d^{abc} L^a K^b L^c(w) & = & 
d^{abc} K^a L^b L^c(w) + N d^{abc} \psi^a \pa \chi^b L^c(w) +N d^{abc} \pa \psi^a
\chi^b L^c(w),
\nonu \\
d^{abc} L^a J^b J^c(w) & = & 
d^{abc} J^a J^b L^c(w) - N \pa J^a M^a(w) + 2N d^{abc} \psi^a \pa \psi^b
 L^c(w) \nonu \\
& - & 2N d^{abc} \psi^a \pa \chi^b J^c(w),
\nonu \\
d^{abc} L^a K^b K^c(w) & = & 
d^{abc} K^a K^b L^c(w) + N \pa K^a M^a(w) + 2N d^{abc} \chi^a \pa \chi^b
 L^c(w) \nonu \\
& - & 2N d^{abc} \chi^a \pa \psi^b K^c(w).
\label{appendixddetails}
\eea

\section{The first-order poles in the OPEs  
$W_{ \pm \frac{2}{3}}^{(2)}(z) 
\, W_{\mp \frac{1}{3}}^{(\frac{5}{2})}(w)$ }

Let us consider the higher spin-$\frac{7}{2}$ current with $q=\frac{1}{3}$
or $q=-\frac{1}{3}$. According to the section $5$,
one should calculate the OPE between  $W_{ \pm \frac{2}{3}}^{(2)}(z)$ and 
$W_{\mp \frac{1}{3}}^{(\frac{5}{2})}(w)$.
The first-order poles of $3 \times 10=30$ OPEs are given by 
\bea
&& J^a  M^a(z) \, \, \psi^b  Q^b(w) \mid_{\frac{1}{(z-w)}}  = 
f^{abc} \, (\psi^a M^b)Q^c(w) + \frac{1}{2} \, d^{abc} \, (J^a \chi^b)Q^c(w)
\nonu \\
&& - 2 N \, d^{abc} \, \psi^a \pa M^b J^c(w)
\nonu \\
& & + d^{abc} f^{cde} \, \psi^a (J^d M^e) J^b(w)
 + d^{abc} d^{def} f^{cdg} \,
\psi^e (J^a \psi^g \chi^b) J^f(w)
\nonu \\
& & - 2 N \, d^{abc} \, \psi^a J^b \pa M^c(w)
+ d^{abc} f^{cde} \, \psi^a J^b J^d M^e(w)
\nonu \\
& & +  d^{abc} d^{def} f^{cdg} \, \psi^e J^f J^a \psi^g \chi^b(w),
\nonu \\
&& J^a  M^a(z) \, \, \psi^b  U^b(w) \mid_{\frac{1}{(z-w)}}  = 
f^{abc} \, (\psi^a M^b)U^c(w) + \frac{1}{2} \, d^{abc} \, (J^a \chi^b)U^c(w)
\nonu \\
&& - 2 N \, d^{abc} \, \psi^a \pa M^b L^c(w)
\nonu \\
&& +  d^{abc} f^{cde} \, \psi^a (J^d M^e) L^b(w)
 + d^{abc} d^{def} f^{cdg} \,
\psi^e (J^a \psi^g \chi^b) L^f(w) 
\nonu \\
&& + d^{abc} f^{cde} f^{efg} \, \psi^a J^b (\psi^g \chi^d) M^f(w)
+\frac{1}{2}  d^{abc} d^{def} f^{cdg} \, \psi^e J^f J^a \chi^g \chi^b(w),
\nonu \\
&& - \frac{1}{2}  d^{abc} d^{def} f^{cdg} \, \psi^d J^e J^a \psi^g \psi^b(w),
\nonu \\
&& J^a  M^a(z) \, \, \psi^b  S^b(w) \mid_{\frac{1}{(z-w)}}  = 
f^{abc} \, (\psi^a M^b)S^c(w) + \frac{1}{2} \, d^{abc} \, (J^a \chi^b)S^c(w)
\nonu \\
& & - d^{abc} d^{def} f^{cdg} \,
\psi^a (J^e \psi^f \chi^g) K^b(w)
- d^{abc} d^{def} f^{cdg} \, \psi^a J^e K^b \psi^f \chi^g(w),
\nonu \\
&& J^a  M^a(z) \, \, \psi^b  V^b(w) \mid_{\frac{1}{(z-w)}}  = 
f^{abc} \, (\psi^a M^b)V^c(w) + \frac{1}{2} \, d^{abc} \, (J^a \chi^b)V^c(w)
\nonu \\
&& - d^{abc} d^{def} f^{cdg} \,
\psi^a (J^e \psi^f \chi^g) L^b(w)
+ d^{abc} f^{cde} f^{efg} \,
\psi^a K^b ( \psi^g \chi^d) M^f(w)
\nonu \\
& & -  \frac{1}{2} d^{abc} d^{def} f^{cdg} \, \psi^a K^b J^e \chi^g \chi^f(w)
+ \frac{1}{2} d^{abc} d^{def} f^{cdg} \, \psi^a K^b J^e \psi^g \psi^f(w),
\nonu \\
&& J^a  M^a(z) \, \, \chi^b  Q^b(w) \mid_{\frac{1}{(z-w)}}  = 
- \frac{1}{2} \, d^{abc} \, (J^a \psi^b)Q^c(w)
- 2 N \, d^{abc} \, \chi^a \pa M^b J^c(w)
\nonu \\
& & + d^{abc} f^{cde} \, \chi^a (J^d M^e) J^b(w)
 + d^{abc} d^{def} f^{cdg} \,
\chi^e (J^a \psi^g \chi^b) J^f(w)
\nonu \\
& & - 2 N \, d^{abc} \, \chi^a J^b \pa M^c(w)
+ d^{abc} f^{cde} \, \chi^a J^b J^d M^e(w)
\nonu \\
& & +   d^{abc} d^{def} f^{cdg} \, \chi^e J^f J^a \psi^g \chi^b(w),
\nonu \\
&& J^a  M^a(z) \, \, \chi^b  U^b(w) \mid_{\frac{1}{(z-w)}}  = 
- \frac{1}{2} \, d^{abc} \, (J^a \psi^b)U^c(w)
- 2 N \, d^{abc} \, \chi^a \pa M^b L^c(w)
\nonu \\
& & +  d^{abc} f^{cde} \, \chi^a (J^d M^e) L^b(w)
 + d^{abc} d^{def} f^{cdg} \,
\chi^e (J^a \psi^g \chi^b) L^f(w) 
\nonu \\
& & +  d^{abc} f^{cde} f^{efg} \, \chi^a J^b (\psi^g \chi^d) M^f(w)
+\frac{1}{2}  d^{abc} d^{def} f^{cdg} \, \chi^e J^f J^a \chi^g \chi^b(w)
\nonu \\
& & - \frac{1}{2}  d^{abc} d^{def} f^{cdg} \, \chi^e J^f J^a \psi^g \psi^b(w),
\nonu \\
&& J^a  M^a(z) \, \, \chi^b  S^b(w) \mid_{\frac{1}{(z-w)}}  = 
 -\frac{1}{2} \, d^{abc} \, (J^a \psi^b)S^c(w)
\nonu \\
& & - d^{abc} d^{def} f^{cdg} \,
\chi^a (J^e \psi^f \chi^g) K^b(w)
- d^{abc} d^{def} f^{cdg} \, \chi^a J^e K^b \psi^f \chi^g(w),
\nonu \\
&& J^a  M^a(z) \, \, \chi^b  V^b(w) \mid_{\frac{1}{(z-w)}}  = 
- \frac{1}{2} \, d^{abc} \, (J^a \psi^b)V^c(w)
\nonu \\
&& - d^{abc} d^{def} f^{cdg} \,
\chi^a (J^e \psi^f \chi^g) L^b(w)
+ d^{abc} f^{cde} f^{efg} \,
\chi^a K^b ( \psi^g \chi^d) M^f(w)
\nonu \\
& & -  \frac{1}{2} d^{abc} d^{def} f^{cdg} \, \chi^a K^b J^e \chi^g \chi^f(w)
+ \frac{1}{2} d^{abc} d^{def} f^{cdg} \, \chi^a K^b J^e \psi^g \psi^f(w),
\nonu \\
 & & J^a  M^a(z) \, \, \psi^b  R^b(w) \mid_{\frac{1}{(z-w)}}  = 
f^{abc} \, (\psi^a M^b) R^c(w)+ \frac{1}{2} \, d^{abc} \, (J^a \chi^b)R^c(w)
\nonu \\
&& -2 N d^{abc} \, \psi^a \pa M^b K^c(w)
\nonu \\
& & + d^{abc} f^{cde} \,
\psi^a (J^d M^e) K^b(w)
+ d^{abc} d^{def} f^{cdg} \,
\psi^e (J^a  \psi^g \chi^b) K^f(w)
\nonu \\
& & -   d^{abc} d^{def} f^{cdg} \, \psi^a J^b J^e \psi^f \chi^g(w),
\nonu \\
&& J^a  M^a(z) \, \, \chi^b  W^b(w) \mid_{\frac{1}{(z-w)}}  = 
- \frac{1}{2} \, d^{abc} \, (J^a \psi^b)W^c(w)
\nonu \\
& & +  d^{abc} f^{cde} f^{efg} \,
\chi^a ((\psi^g \chi^d) M^f) L^b(w)
\nonu \\
&& +  \frac{1}{2} d^{abc} d^{def} f^{cdg} \, \chi^e (J^a \chi^g \chi^b) L^f(w)
\nonu \\
& & -   \frac{1}{2} d^{abc} d^{def} f^{cdg} \, \chi^e (J^a \psi^g \psi^b) L^f(w)
+ d^{abc} f^{cde} f^{efg} \, \chi^a L^b (\psi^g \chi^d) M^f(w)
\nonu \\
& & + \frac{1}{2} d^{abc} d^{def} f^{cdg} \chi^e L^f J^a \chi^g \chi^b(w) 
-\frac{1}{2}
d^{abc} d^{def} f^{cdg}   \chi^e L^f J^a \psi^g \psi^b(w),
\nonu \\
&& K^a  M^a(z) \, \, \psi^b  Q^b(w) \mid_{\frac{1}{(z-w)}}  = 
\frac{1}{2} \, d^{abc} \, (K^a \chi^b)Q^c(w)
\nonu \\
& & +  d^{abc} d^{def} f^{cdg} \,
\psi^a (K^e \chi^f \psi^g) J^b(w)
+ d^{abc} d^{def} f^{cdg} \, \psi^a K^e J^b \chi^f \psi^g(w),
\nonu \\
&& K^a  M^a(z) \, \, \psi^b  U^b(w) \mid_{\frac{1}{(z-w)}}  = 
 \frac{1}{2} \, d^{abc} \, (K^a \chi^b)U^c(w)
\nonu \\
& & + d^{abc} d^{def} f^{cdg} \,
\psi^a (K^e \chi^f \psi^g) L^b(w)
+ d^{abc} f^{cde} f^{efg} \,
\psi^a J^b ( \chi^g \psi^d) M^f(w)
\nonu \\
& & +  \frac{1}{2} d^{abc} d^{def} f^{cdg} \, \psi^a J^b K^e \psi^g \psi^f(w)
- \frac{1}{2} d^{abc} d^{def} f^{cdg} \, \psi^a J^b K^e \chi^g \chi^f(w),
\nonu \\
&& K^a  M^a(z) \, \, \psi^b  S^b(w) \mid_{\frac{1}{(z-w)}}  = 
 \frac{1}{2} \, d^{abc} \, (K^a \chi^b)S^c(w)
- 2 N \, d^{abc} \, \psi^a \pa M^b K^c(w)
\nonu \\
& & + d^{abc} f^{cde} \, \psi^a (K^d M^e) K^b(w)
 - d^{abc} d^{def} f^{cdg} \,
\psi^e (K^a \chi^g \psi^b) K^f(w)
\nonu \\
& & - 2 N \, d^{abc} \, \psi^a K^b \pa M^c(w)
+ d^{abc} f^{cde} \, \psi^a K^b K^d M^e(w)
\nonu \\
& & -  d^{abc} d^{def} f^{cdg} \, \psi^e K^f K^a \chi^g \psi^b(w),
\nonu \\
&& K^a  M^a(z) \, \, \psi^b  V^b(w) \mid_{\frac{1}{(z-w)}}  = 
- \frac{1}{2} \, d^{abc} \, (K^a \chi^b)V^c(w)
- 2 N \, d^{abc} \, \psi^a \pa M^b L^c(w)
\nonu \\
&& +  d^{abc} f^{cde} \, \psi^a (K^d M^e) L^b(w)
 + d^{abc} d^{def} f^{cdg} \,
\psi^e (K^a \chi^g \psi^b) L^f(w) 
\nonu \\
& & +  d^{abc} f^{cde} f^{efg} \, \psi^a K^b (\chi^g \psi^d) M^f(w)
\nonu \\
&& +  \frac{1}{2}  d^{abc} d^{def} f^{cdg} \, \psi^e K^f K^a \psi^g \psi^b(w),
\nonu \\
&& K^a  M^a(z) \, \, \chi^b  Q^b(w) \mid_{\frac{1}{(z-w)}}  = 
f^{abc} \, (\chi^a M^b)Q^c(w) - \frac{1}{2} \, d^{abc} \, (K^a \psi^b)Q^c(w)
\nonu \\
& & +  d^{abc} d^{def} f^{cdg} \,
\chi^a (K^e \chi^f \psi^g) J^b(w)
+ d^{abc} d^{def} f^{cdg} \, \chi^a K^e J^b \chi^f \psi^g(w),
\nonu \\
&& K^a  M^a(z) \, \, \chi^b  U^b(w) \mid_{\frac{1}{(z-w)}}  = 
f^{abc} \, (\chi^a M^b)U^c(w) - \frac{1}{2} \, d^{abc} \, (K^a \psi^b)U^c(w)
\nonu \\
& & +  d^{abc} d^{def} f^{cdg} \,
\chi^a (K^e \chi^f \psi^g) L^b(w)
+ d^{abc} f^{cde} f^{efg} \,
\chi^a J^b ( \chi^g \psi^d) M^f(w)
\nonu \\
& & +   \frac{1}{2} d^{abc} d^{def} f^{cdg} \, \chi^a J^b K^e \psi^g \psi^f(w)
- \frac{1}{2} d^{abc} d^{def} f^{cdg} \, \chi^a J^b K^e \chi^g \chi^f(w),
\nonu \\
&& K^a  M^a(z) \, \, \chi^b  S^b(w) \mid_{\frac{1}{(z-w)}}  = 
f^{abc} \, (\chi^a M^b)S^c(w) - \frac{1}{2} \, d^{abc} \, (K^a \psi^b)S^c(w)
\nonu \\
&& - 2 N \, d^{abc} \, \chi^a \pa M^b K^c(w)
\nonu \\
& & +  d^{abc} f^{cde} \, \chi^a (K^d M^e) K^b(w)
 - d^{abc} d^{def} f^{cdg} \,
\chi^e (K^a \chi^g \psi^b) K^f(w)
\nonu \\
& & - 2 N \, d^{abc} \, \chi^a K^b \pa M^c(w)
+ d^{abc} f^{cde} \, \chi^a K^b K^d M^e(w)
\nonu \\
& & -  d^{abc} d^{def} f^{cdg} \, \chi^e K^f K^a \chi^g \psi^b(w),
\nonu \\
&& K^a  M^a(z) \, \, \chi^b  V^b(w) \mid_{\frac{1}{(z-w)}}  = 
f^{abc} \, (\chi^a M^b)V^c(w) - \frac{1}{2} \, d^{abc} \, (K^a \psi^b)V^c(w)
\nonu \\
&& - 2 N \, d^{abc} \, \chi^a \pa M^b L^c(w)
\nonu \\
&& +   d^{abc} f^{cde} \, \chi^a (K^d M^e) L^b(w)
 - d^{abc} d^{def} f^{cdg} \,
\chi^e (K^a \chi^g \psi^b) L^f(w) 
\nonu \\
& & +  d^{abc} f^{cde} f^{efg} \, \chi^a K^b (\chi^g \psi^d) M^f(w)
-\frac{1}{2}  d^{abc} d^{def} f^{cdg} \, \chi^e K^f K^a \psi^g \psi^b(w),
\nonu \\
&& K^a  M^a(z) \, \, \psi^b  R^b(w) \mid_{\frac{1}{(z-w)}}  = 
\frac{1}{2} \, d^{abc} \, (K^a \chi^b)R^c(w) 
-2 N\, d^{abc} \, \psi^a \pa M^b J^c(w)
\nonu \\
& & -  d^{abc} f^{cde} \psi^a (K^e M^d) J^b(w) 
\nonu \\
& & +  d^{abc} d^{def} f^{cdg} \, \psi^a (K^e \chi^g \psi^f)J^b(w)
+ d^{abc} d^{def} f^{cdg} \, \psi^a K^b K^e \chi^f \psi^g(w),
\nonu \\
&& K^a  M^a(z) \, \, \chi^b  W^b(w) \mid_{\frac{1}{(z-w)}}  = 
f^{abc} \, (\chi^a M^b)W^c(w)-\frac{1}{2} \, d^{abc} \, (K^a\psi^b)W^c(w)
\nonu \\
& & +   d^{abc} f^{cde} f^{efg} \chi^a((\chi^g \psi^d)M^f)L^b(w)
\nonu \\
& & +  \frac{1}{2} \, d^{abc} d^{def} f^{cdg} \chi^a (K^e \psi^g \psi^f) L^b(w)
\nonu \\
& & +  \frac{1}{2} \,  d^{abc} d^{def} f^{cdg} \chi^a (K^e \chi^g \chi^f) L^b(w)
+  d^{abc} f^{cde} f^{efg} \, \chi^a L^b (\chi^g \psi^d) M^f(w) 
\nonu \\
& & +  \frac{1}{2} \, d^{abc} d^{def} f^{cdg} \chi^a L^b K^e \psi^g \psi^f(w)
- \frac{1}{2}  \, d^{abc} d^{def} f^{cdg} \chi^a L^b K^e \chi^g \chi^f(w),
\nonu \\
&& L^a  M^a(z) \, \, \psi^b  Q^b(w) \mid_{\frac{1}{(z-w)}}  = 
-\frac{1}{2} \, f^{abc} \, (\chi^a M^b)Q^c(w) +\frac{1}{2} \, d^{abc}\,
(L^a \chi^b)Q^c(w) \nonu \\
&& + d^{abc} f^{def} f^{cdg} \, \psi^a (M^e \psi^g \chi^f) J^b(w)
- d^{abc} d^{def} f^{cdg} \, \psi^a (L^e \psi^g \chi^f) J^b(w)
\nonu \\
&& +  d^{abc} f^{def} f^{cdg} \, \psi^a J^b M^e \psi^g \chi^f(w) -
d^{abc} f^{def} f^{cdg} \, \psi^a J^b L^e \psi^g \chi^f(w),
\nonu \\
&& L^a  M^a(z) \, \, \psi^b  U^b(w) \mid_{\frac{1}{(z-w)}}  = 
\frac{1}{2} \, f^{abc} (\chi^a M^b) U^b(w)+\frac{1}{2} \, d^{abc}
(L^a \chi^b) U^c(w) \nonu \\
& & +   d^{abc} f^{def} f^{cdg} \, \psi^a 
((\psi^g \chi^f) M^e)L^b(w)  
- d^{abc} d^{def} f^{cdg} \, \psi^a (L^e \psi^g \chi^f) L^b(w)
\nonu \\
& & - N \, d^{abc} \, \psi^a J^b \pa M^c(w)
+\frac{1}{4} d^{abc} f^{cde} \psi^a J^b J^d M^e(w)
\nonu \\
& & +  \frac{1}{4} \, d^{abc} f^{cde} \, \psi^a J^b K^d M^e(w)
- \frac{1}{2} \, d^{abc} d^{def} f^{cdg} \, \psi^a J^b L^e \chi^g \chi^f(w) 
\nonu \\
& & +  \frac{1}{2} \, d^{abc} d^{def} f^{cdg} \, \psi^a J^b L^e \psi^g \psi^f(w),
\nonu \\
&& L^a  M^a(z) \, \, \psi^b  S^b(w) \mid_{\frac{1}{(z-w)}}  = 
\frac{1}{2}\, f^{abc} \, (\chi^a M^b)S^c(w)
+\frac{1}{2} \, d^{abc} \, (L^a \chi^b)S^c(w)
\nonu \\
& & - d^{abc} f^{def} f^{cdg} \, \psi^a ((\psi^f \chi^g)M^e) K^b(w)
- d^{abc} d^{def} f^{cdg} \, \psi^a (L^e \psi^f \chi^g) K^b(w) 
\nonu \\
& & -  d^{abc} f^{def} f^{cdg}  \psi^a K^b (\psi^f \chi^g) M^e(w)
- d^{abc} d^{def} f^{cdg} \, \psi^a K^b L^e \psi^f \chi^g(w),
\nonu \\
&& L^a  M^a(z) \, \, \psi^b  V^b(w) \mid_{\frac{1}{(z-w)}}  = 
\frac{1}{2}\, f^{abc} \, (\chi^a M^b)V^c(w)
+\frac{1}{2} \, d^{abc} \, (L^a \chi^b)V^c(w)
\nonu \\
& & - d^{abc} f^{def} f^{cdg} \, \psi^a ((\psi^f \chi^g)M^e) L^b(w)
- d^{abc} d^{def} f^{cdg} \, \psi^a (L^e \psi^f \chi^g) L^b(w)  
\nonu \\
& & - N \, d^{abc} \, \psi^a K^b \pa M^c(w)
+\frac{1}{4} \, d^{abc} f^{cde} \, \psi^a K^b J^d M^e(w)
\nonu \\
&& +  \frac{1}{4} \, d^{abc} f^{cde} \, \psi^a K^b K^d M^e(w)
-\frac{1}{2} \, d^{abc} d^{def} f^{cdg} \, 
\psi^a K^b L^e \chi^g \chi^f(w)
\nonu \\
& & +  \frac{1}{2} \, d^{abc} d^{def} f^{cdg} \, 
\psi^a K^b L^e \psi^g \psi^f(w),
\nonu \\
&& L^a  M^a(z) \, \, \chi^b  Q^b(w) \mid_{\frac{1}{(z-w)}}  = 
\frac{1}{2}\, f^{abc} \, (\psi^a M^b)Q^c(w)
-\frac{1}{2} \, d^{abc} \, (L^a \psi^b)Q^c(w)
\nonu \\
& & - d^{abc} f^{def} f^{cdg} \, \chi^a ((\chi^f \psi^g)M^e) J^b(w)
+ d^{abc} d^{def} f^{cdg} \, \chi^a (L^e \chi^f \psi^g) J^b(w) 
\nonu \\
& & -  d^{abc} f^{def} f^{cdg}  \chi^a J^b (\chi^f \psi^g) M^e(w)
+ d^{abc} d^{def} f^{cdg} \, \chi^a J^b L^e \chi^f \psi^g(w),
\nonu \\
&& L^a  M^a(z) \, \, \chi^b  U^b(w) \mid_{\frac{1}{(z-w)}}  = 
\frac{1}{2}\, f^{abc} \, (\psi^a M^b)U^c(w)
-\frac{1}{2} \, d^{abc} \, (L^a \psi^b)U^c(w)
\nonu \\
& & - d^{abc} f^{def} f^{cdg} \, \chi^a ((\chi^f \psi^g)M^e) L^b(w)
+ d^{abc} d^{def} f^{cdg} \, \chi^a (L^e \chi^f \psi^g) L^b(w)  
\nonu \\
& & - N \, d^{abc} \, \chi^a J^b \pa M^c(w)
+\frac{1}{4} \, d^{abc} f^{cde} \, \chi^a J^b K^d M^e(w)
\nonu \\
& & +  \frac{1}{4} \, d^{abc} f^{cde} \, \chi^a J^b J^d M^e(w)
+\frac{1}{2} \, d^{abc} d^{def} f^{cdg} \, 
\chi^a J^b L^e \psi^g \psi^f(w)
\nonu \\
& & - \frac{1}{2} \, d^{abc} d^{def} f^{cdg} \, 
\chi^a J^b L^e \chi^g \chi^f(w),
\nonu \\
&& L^a  M^a(z) \, \, \chi^b  S^b(w) \mid_{\frac{1}{(z-w)}}  = 
-\frac{1}{2} \, f^{abc} \, (\psi^a M^b)S^c(w) -\frac{1}{2} \, d^{abc}\,
(L^a \psi^b)S^c(w) \nonu \\
& & +  d^{abc} f^{def} f^{cdg} \, \chi^a (M^e \chi^g \psi^f) K^b(w)
+ d^{abc} d^{def} f^{cdg} \, \chi^a (L^e \chi^g \psi^f) K^b(w)
\nonu \\
& & +  d^{abc} f^{def} f^{cdg} \, \chi^a K^b M^e \chi^g \psi^f(w) +
d^{abc} f^{def} f^{cdg} \, \chi^a K^b L^e \chi^g \psi^f(w),
\nonu \\
&& L^a  M^a(z) \, \, \chi^b  V^b(w) \mid_{\frac{1}{(z-w)}}  = 
\frac{1}{2} \, f^{abc} (\psi^a M^b) V^b(w)-\frac{1}{2} \, d^{abc}
(L^a \psi^b) V^c(w) \nonu \\
&& +   d^{abc} f^{def} f^{cdg} \, \chi^a 
((\chi^g \psi^f) M^e)L^b(w)  
+ d^{abc} d^{def} f^{cdg} \, \chi^a (L^e \chi^g \psi^f) L^b(w)
\nonu \\
& & - N \, d^{abc} \, \chi^a K^b \pa M^c(w)
+\frac{1}{4} d^{abc} f^{cde} \chi^a K^b K^d M^e(w)
\nonu \\
& & +  \frac{1}{4} \, d^{abc} f^{cde} \, \chi^a K^b J^d M^e(w)
+ \frac{1}{2} \, d^{abc} d^{def} f^{cdg} \, \chi^a K^b L^e \psi^g \psi^f(w) 
\nonu \\
& & - \frac{1}{2} \, d^{abc} d^{def} f^{cdg} \, \chi^a K^b L^e \chi^g \chi^f(w),
\nonu \\
&& L^a  M^a(z) \, \, \psi^b  R^b(w) \mid_{\frac{1}{(z-w)}}  = 
\frac{1}{2} \, f^{abc} \, (\chi^a M^b)R^c(w)
+ \frac{1}{2} \, d^{abc} \, (L^a \chi^b)R^c(w)
\nonu \\
& & +  d^{abc} f^{def} f^{cdg} \, \psi^a ((\psi^g \chi^f)M^e) K^b(w)
- d^{abc} d^{def} f^{cdg} \, \psi^a (L^e \psi^g \chi^f) K^b(w)
\nonu \\
& & -  d^{abc} f^{def} f^{cdg} \, \psi^a J^b (\psi^f \chi^g)M^e(w)  
- d^{abc} d^{def} f^{cdg} \, \psi^a J^b L^e \psi^f \chi^g(w),
\nonu \\
&& L^a  M^a(z) \, \, \chi^b  W^b(w) \mid_{\frac{1}{(z-w)}}  = 
\frac{1}{2} \, f^{abc} \, (\psi^a M^b)W^c(w)
-\frac{1}{2} \, d^{abc} \, (L^a \psi^b)W^c(w)
\nonu \\
&& - N \, d^{abc} \, \chi^a \pa M^b L^c(w)
\nonu \\
& & +  \frac{1}{4} d^{abc} f^{cde} \chi^a (J^d M^e) L^b(w)
+\frac{1}{4} d^{abc} f^{cde} \chi^a (K^d M^e) L^b(w)
\nonu \\
& & - \frac{1}{2} \, d^{abc} d^{def} f^{cdg} \, \chi^a (L^e \chi^g \chi^f) L^b(w)
 + \frac{1}{2} \, d^{abc} d^{def} f^{cdg} \, \chi^a (L^e \psi^g \psi^f) L^b(w)
\nonu \\
& & - N \, d^{abc} \, \chi^a L^b \pa M^c(w)
+\frac{1}{4} \, d^{abc} f^{cde} \, \chi^a L^b J^d M^e(w)
\nonu \\
& & +  \frac{1}{4} \, d^{abc} f^{cde} \, \chi^a L^b K^d M^e(w)
- \frac{1}{2} \, d^{abc} d^{def} f^{cdg} \, \chi^a L^b L^e \chi^g \chi^f(w)
\nonu \\
& & + \frac{1}{2} \, d^{abc} d^{def} f^{cdg} \, \chi^a L^b L^e \psi^g \psi^f(w).
\label{spin7halfdetails}
\eea
As noticed before, the half of them in Appendix (\ref{spin7halfdetails})
are obtained from the remaining ones using the symmetry under 
$\psi^a(z) \leftrightarrow \chi^a(z)$.
In order to obtain the complete form for the higher spin-$\frac{7}{2}$
currents, the fully normal ordered products from the intermediate 
expressions in
Appendix (\ref{spin7halfdetails}) are needed.  

For example, $f^{abc} (\psi^a M^b) Q^c(w)$ should be simplified further.
That is, this can be written as
$f^{abc} (\psi^a M^b) Q^c(w) = f^{abc} Q^c (\psi^a M^b) (w) -f^{abc}
[Q^c, \psi^a M^b](w)$.
In order to simplify the second term, one should calculate 
the OPE $
Q^c(z) \, \psi^a M^b(w)$.
This 
becomes
\bea
Q^c(z) \, \psi^a M^b(w) & = & - \frac{1}{(z-w)^2} \, 3 (N^2-4)
\chi^a J^a(w) \nonu \\
& + & \frac{1}{(z-w)} \, \left[ 
 -2 (N^2-4) \, \pa \chi^a J^a -4
(N^2-4) \,  \chi^a \pa J^a \right. \nonu \\
& + & \left. 
 2N \, d^{abc}  \psi^a M^b J^c +2(N^2-4) \,  \pa \psi^a L^a
\right](w) +\cdots. 
\label{interinter}
\eea
From Appendix (\ref{interinter}), the above commutator can be obtained 
and the final result can be written as
\bea
f^{abc} (\psi^a M^b) Q^c(w) & = &
f^{abc} Q^c \psi^a M^b (w) +\frac{1}{2}(N^2-4) \, \pa^2 \chi^a J^a(w) +
3(N^2-4) \, \pa \chi^a \pa J^a(w) 
\nonu \\
& + & \frac{5}{2}(N^2-4) \, \chi^a \pa^2 J^a(w)
-2N \, d^{abc} \pa (\psi^a M^b J^c)(w) 
\nonu \\
& - & 2(N^2-4) \, \pa (\pa \psi^a L^a)(w).
\label{psimq}
\eea

The fourth term of the first equation of Appendix (\ref{spin7halfdetails})
should be simplified further as done in Appendix (\ref{psimq}). 
This can be written as 
$d^{abc} f^{cde} \, \psi^a (J^d M^e) J^b(w) = d^{abc} f^{cde} \, 
\psi^a J^b J^d M^e(w) - d^{abc} f^{cde} \, \psi^a [J^b, J^d M^e](w)$.
For the commutator in this relation, the following OPE 
should be calculated as follows:
\bea
J^b(z) \, J^d M^e(w) & = & \frac{1}{(z-w)^2} \, \left[ - N \delta^{bd} \, M^e
- f^{bde'} f^{e'c''e''} d^{ed''e''} \psi^{c''} \chi^{d''}\right](w)
\nonu \\
& + & \frac{1}{(z-w)} \, \left[ f^{bde'} J^{e'} M^{e} - f^{bc'e'} d^{ed'e'} 
J^d \psi^{c'} \chi^{d'} \right](w) + \cdots. 
\label{JJMOPE}
\eea
In order to obtain the above commutator, one should 
use Appendices 
$(A.6), (A.7)$ and $(A.15)$ of \cite{Ahn1111} in Appendix (\ref{JJMOPE}).
Then one can easily see that this becomes 
$ \frac{1}{2}(N^2-4) \, \psi^a \pa^2 L^a + \frac{N}{2} 
d^{abc} \psi^a \pa (J^b M^c) +
\frac{N}{2} d^{abc} d^{cde} \psi^a \pa (J^d \psi^e \chi^b) -\frac{N}{2} d^{abc}
d^{cde} \psi^a \pa (J^d \psi^b \chi^e)$
\footnote{
For the fifth term of the first equation of Appendix 
(\ref{spin7halfdetails}),
the property of Appendix $(A.11)$ of \cite{Ahn1111} can be used.
When we simplify the last term of the first equation of 
Appendix 
(\ref{spin7halfdetails}), Appendix $(A.10)$ of \cite{Ahn1111} is used.}. 

One can analyze the second term of the second equation of
Appendix (\ref{spin7halfdetails}) which is given by  
$ d^{abc} \, (J^a \chi^b)U^c(w)$.
This can be written as 
$ d^{abc} \, (J^a \chi^b)U^c(w) =  d^{abc} \, U^a J^b \chi^c(w) -  
d^{abc} \, [U^c, J^a \chi^b](w)$.
As before, one should know the OPE
$U^c(z) \, (J^a \chi^b)(w)$ in order to calculate the above commutator.
It turns out that
\bea
U^c(z) \, d^{abc} \, J^a \chi^b(w) & = &
-\frac{1}{(z-w)^2} \, 3(N^2-4) \chi^a L^a(w) 
\nonu \\
& + &
\frac{1}{(z-w)} \, \left[ -(N^2-4) \, \pa \psi^a J^a -3 (N^2-4)\,
\chi^a \pa L^a \right. \nonu \\
& + & \frac{(N^2-4)}{N} f^{abc} \chi^a J^b L^c +\frac{4(4-N^2)}{N^2}
\chi^a J^a \psi^b \chi^b 
\label{simple}
\\
& +& \left. \frac{(8-N^2)}{2N} \, d^{abc} d^{cde} \, \chi^a J^b \psi^d \chi^e
-\frac{N}{2} \, d^{abc} d^{cde} \chi^a J^d \psi^b \chi^e
\right](w) +\cdots. 
\nonu
\eea
Here we used the fact that the expression $f^{ac'e'} f^{e'd'e} d^{edc} d^{cba} 
\chi^b J^d \psi^{c'} \chi^{d'}(w)$ is equal to the last three terms of the 
first-order pole in Appendix 
(\ref{simple}) with the help of Appendix $(A.14)$ of
\cite{Ahn1111}. 
Then it is straightforward to express the above commutator using the 
OPE in Appendix (\ref{simple}).

\section{The singular terms in the OPE 
$W_{  \frac{2}{3}}^{(2)}(z)
\, W_{  -\frac{2}{3}}^{(2)}(w)$ }

In section $6$, one of the OPEs between the higher spin currents 
is given. We would like to present the OPE 
between 
the higher spin-$2$ currents $W_{  \pm \frac{2}{3}}^{(2)}(z)$.
The six OPEs (rather than nine OPEs) are given by
\bea
&& J^a  M^a(z) \, \, J^b  M^b(w)  =  
\frac{1}{(z-w)^4} \, (N^2-4)(N^2-1) \nonu \\
& & +  \frac{1}{(z-w)^2} \, 
\left[ -3 N \, M^a M^a +\frac{(N^2-4)}{2N} J^a K^a - 
\frac{3(N^2-4)}{4N} J^a J^a + \frac{(N^2-4)}{4N} K^a K^a
\right](w) \nonu \\
&& +
{\cal O} (\frac{1}{(z-w)}) + \cdots,
\nonu \\
&& J^a  M^a(z) \, \, K^b  M^b(w)  =  
\nonu \\
&& \frac{1}{(z-w)^2} \, 
\left[  N \, M^a M^a -\frac{3(N^2-4)}{2N} J^a K^a - 
\frac{(N^2-4)}{4N} J^a J^a - \frac{(N^2-4)}{4N} K^a K^a
\right](w) 
\nonu \\
&& +
{\cal O} (\frac{1}{(z-w)})
+\cdots,
\nonu \\
&& J^a  M^a(z) \, \, L^b  M^b(w)  =  
-\frac{1}{(z-w)^3} \, 2 (N^2-4) \, \psi^a \chi^a(w) 
-\frac{1}{(z-w)^2} \, 2 (N^2-4) \pa \psi^a \chi^a(w) \nonu \\
&&  +
{\cal O} (\frac{1}{(z-w)}) + \cdots,   
\nonu \\
&& K^a  M^a(z) \, \, K^b  M^b(w)  =  
\frac{1}{(z-w)^4} \, (N^2-4)(N^2-1) \nonu \\
& & +  \frac{1}{(z-w)^2} \, 
\left[ -3 N \, M^a M^a +\frac{(N^2-4)}{2N} J^a K^a - 
\frac{3(N^2-4)}{4N} K^a K^a + \frac{(N^2-4)}{4N} J^a J^a
\right](w) \nonu \\
&&  +
{\cal O} (\frac{1}{(z-w)})+\cdots,
\nonu \\
&& K^a  M^a(z) \, \, L^b  M^b(w)  =  
\frac{1}{(z-w)^3} \, 2 (N^2-4) \, \psi^a \chi^a(w) 
+\frac{1}{(z-w)^2} \, 2 (N^2-4)  \psi^a \pa \chi^a(w) \nonu \\
&&  +
{\cal O} (\frac{1}{(z-w)}) + \cdots,
\nonu \\
&& L^a  M^a(z) \, \, L^b  M^b(w)  =  \frac{1}{(z-w)^4} \,  \frac{1}{2}\, (N^2-4)(N^2-1)
\nonu \\
&& + \frac{1}{(z-w)^2} \, \left[ 
-2N \, M^a M^a 
+\frac{(N^2-4)}{4N} \, J^a K^a +\frac{3(N^2-4)}{8N} \, J^a J^a 
\right. \nonu \\
&& \left. +  \frac{3(N^2-4)}{8N} \, K^a K^a -\frac{3(N^2-4)}{2N} \, L^a L^a 
\right](w)  +
{\cal O} (\frac{1}{(z-w)}) + \cdots.
\label{spin2spin2OPE}
\eea
We did not present the first-order poles in Appendix 
(\ref{spin2spin2OPE}).
One can use the symmetry under the transformation $\psi^a(z) \leftrightarrow 
\chi^a(z)$.
Then one can check (\ref{twospin2ope}) by using the above 
results in Appendix 
(\ref{spin2spin2OPE}) with correct coefficients.
For the first-order poles, one resorts to the description of section $7$.

\section{The details for the OPEs 
between the higher spin currents in ${\cal N}=2$
superspace }

In this Appendix, the ${\cal N}=2$ description for the 
OPEs between the higher spin currents are given based on the section $7$.

\subsection{ The OPEs ${\bf W}_{\pm \frac{2}{3}}^{(2)}(Z_1) \,
{\bf W}_{ \pm \frac{1}{3}}^{(\frac{7}{2})}(Z_2)
$ }

The OPE between the first higher spin ${\cal N}=2$ multiplet and 
the third higher 
spin ${\cal N}=2$ multiplet in (\ref{manyW}) can be summarized by
{\small
\bea
&& {\bf W}_{\frac{2}{3}}^{(2)}(Z_1) \, {\bf W}_{\frac{1}{3}}^{(\frac{7}{2})}(Z_2) = 
C_{(2) \, (\frac{7}{2})}^{(4) \, +} \, \left(
\frac{\bar{\theta}_{12}}{z_{12}^2} \,
 {\bf W}_{0}^{(4)}(Z_2)
-   \frac{\theta_{12} \, \bar{\theta}_{12}}{z_{12}^2}
\, \frac{7}{24} \, D  {\bf W}_{0}^{(4)}(Z_2)
\right. \nonu \\
&& -\frac{1}{z_{12}} \, \frac{1}{4} \, D {\bf W}_{0}^{(4)}(Z_2)
 +   
\frac{\bar{\theta}_{12}}{z_{12}} \left[ \frac{1}{24} \, [D, \overline{D}]
 {\bf W}_{0}^{(4)} + \frac{3}{8} \, \pa  {\bf W}_{0}^{(4)}
\right](Z_2)
\nonu \\
& & + \left.    
\frac{\theta_{12} \, \bar{\theta}_{12}}{z_{12}} \left[
 -\frac{5(c-3)}{36(c+9)} \, \pa    D  
{\bf W}_{0}^{(4)}  + \frac{20}{3(c+9)} \, D {\bf T} {\bf W}_{0}^{(4)} 
 -   
\frac{5}{3(c+9)} {\bf T}
D {\bf W}_{0}^{(4)} \right](Z_2) \right) +\cdots.
\label{13OPE} 
\eea}
Similarly, 
the OPE between the first higher spin ${\cal N}=2$ multiplet and 
the fourth higher 
spin ${\cal N}=2$ multiplet in (\ref{manyW}) can be described by
{\small
\bea
&& {\bf W}_{\frac{2}{3}}^{(2)}(Z_1) \, {\bf W}_{-\frac{1}{3}}^{(\frac{7}{2})}(Z_2) = 
C_{(2) \, (\frac{7}{2})}^{(2) \, +} \, \left(
\frac{\bar{\theta}_{12}}{z_{12}^4} \, {\bf W}_{-\frac{2}{3}}^{(2)}(Z_2)
+\frac{\theta_{12} \, \bar{\theta}_{12}}{z_{12}^4} \,
c_2 \, D {\bf W}_{-\frac{2}{3}}^{(2)}(Z_2) +
\frac{1}{z_{12}^3} \,
c_3 \, D {\bf W}_{-\frac{2}{3}}^{(2)}(Z_2) \right. 
\nonu \\
&& +
\frac{ \bar{\theta}_{12}}{z_{12}^3} \, \left[
c_4 \, [D, \overline{D}]  {\bf W}_{-\frac{2}{3}}^{(2)} 
+ c_5 {\bf T}  {\bf W}_{-\frac{2}{3}}^{(2)}  
+  c_6 \, \pa {\bf W}_{-\frac{2}{3}}^{(2)} 
\right](Z_2) 
\nonu \\ 
&& +
\frac{\theta_{12} \, \bar{\theta}_{12}}{z_{12}^3} 
\, 
\left[ 
c_7 \, \pa D  {\bf W}_{-\frac{2}{3}}^{(2)} 
+ c_8 \, {\bf T}  D  {\bf W}_{-\frac{2}{3}}^{(2)} 
+ c_9 \, D {\bf T}    {\bf W}_{-\frac{2}{3}}^{(2)} 
\right](Z_2)
\nonu \\
&& + \frac{1}{z_{12}^2} \, 
\left[ 
c_{10} \, \pa D  {\bf W}_{-\frac{2}{3}}^{(2)} 
+ c_{11} \, {\bf T}  D  {\bf W}_{-\frac{2}{3}}^{(2)} 
+ c_{12} \, D {\bf T}    {\bf W}_{-\frac{2}{3}}^{(2)} 
\right]
(Z_2)
 +  \frac{\theta_{12}}{z_{12}^2} \, c_{13} \, D  {\bf T} D
 {\bf W}_{-\frac{2}{3}}^{(2)}(Z_2)
\nonu \\
&& + \frac{\bar{\theta}_{12}}{z_{12}^2} \, \left[ 
c_{14} \, \pa [D, \overline{D}]  {\bf W}_{-\frac{2}{3}}^{(2)}
+ c_{15} \, {\bf T} [D, \overline{D}]  {\bf W}_{-\frac{2}{3}}^{(2)}
+ c_{16} \, {\bf T} {\bf T}  {\bf W}_{-\frac{2}{3}}^{(2)}
 + 
c_{17} \, {\bf T} \pa {\bf W}_{-\frac{2}{3}}^{(2)} 
+ c_{18} \, \overline{D} {\bf T} D  {\bf W}_{-\frac{2}{3}}^{(2)} 
\right.
\nonu \\
&&
+ \left. 
c_{19} \,  [D, \overline{D}] {\bf T}  {\bf W}_{-\frac{2}{3}}^{(2)} 
 + 
c_{20} \, D {\bf T} \overline{D}  {\bf W}_{-\frac{2}{3}}^{(2)}
+ c_{21} \, \pa {\bf T}  {\bf W}_{-\frac{2}{3}}^{(2)}
+ c_{22} \, \pa^2  {\bf W}_{-\frac{2}{3}}^{(2)}
\right](Z_2)
\nonu \\
&& +   \frac{\theta_{12} \, \bar{\theta}_{12}}{z_{12}^2}
\left[ c_{23} \,
\pa^2 D    {\bf W}_{-\frac{2}{3}}^{(2)}
+ c_{24} \, {\bf T} \pa D  {\bf W}_{-\frac{2}{3}}^{(2)}
+ c_{25} \, {\bf T} {\bf T} D  {\bf W}_{-\frac{2}{3}}^{(2)}
 +
 c_{26} \, [D, \overline{D}] {\bf T} D  {\bf W}_{-\frac{2}{3}}^{(2)}
\right. \nonu \\
&& + 
\left.
c_{27} \, D {\bf T}  [D, \overline{D}] {\bf W}_{-\frac{2}{3}}^{(2)}
 + c_{28} \, D {\bf T} \pa  {\bf W}_{-\frac{2}{3}}^{(2)}
 +  
 c_{29} \, \pa {\bf T} D  {\bf W}_{-\frac{2}{3}}^{(2)}
+ c_{30} \, \pa D {\bf T} {\bf W}_{-\frac{2}{3}}^{(2)}
+   c_{31} \, {\bf T} D {\bf T}  {\bf W}_{-\frac{2}{3}}^{(2)}
\right](Z_2)
\nonu \\
&& +   \frac{1}{z_{12}}
\left[ c_{32} \, \pa^2 D    {\bf W}_{-\frac{2}{3}}^{(2)}
+ c_{33} \, {\bf T} \pa D  {\bf W}_{-\frac{2}{3}}^{(2)}
+ c_{34} \, {\bf T} {\bf T} D  {\bf W}_{-\frac{2}{3}}^{(2)}
+ c_{35} \,  {\bf T} D {\bf T}   {\bf W}_{-\frac{2}{3}}^{(2)}
 + c_{36} \,
 [D, \overline{D}] {\bf T} D  {\bf W}_{-\frac{2}{3}}^{(2)}
\right. \nonu \\
&& + \left.
c_{37} \, D {\bf T}  [D, \overline{D}] {\bf W}_{-\frac{2}{3}}^{(2)}
+ c_{38} \, D {\bf T} \pa  {\bf W}_{-\frac{2}{3}}^{(2)}
 +  
 c_{39} \, \pa D {\bf T}   {\bf W}_{-\frac{2}{3}}^{(2)}
+ c_{40} \, \pa  {\bf T} D {\bf W}_{-\frac{2}{3}}^{(2)} 
\right](Z_2)
\nonu \\
 && +  
\frac{\theta_{12}}{z_{12}} \left[ 
c_{41} \, {\bf T} D {\bf T} D {\bf W}_{-\frac{2}{3}}^{(2)}
+ c_{42} \,  D {\bf T} \pa D {\bf W}_{-\frac{2}{3}}^{(2)}
+ c_{43} \, \pa D {\bf T} D {\bf W}_{-\frac{2}{3}}^{(2)}
\right](Z_2) 
\nonu \\
&& +  
\frac{ \bar{\theta}_{12}}{z_{12}} 
 \left[ c_{44} \,
\overline{D} {\bf T} D {\bf T}  {\bf W}_{-\frac{2}{3}}^{(2)}
+  c_{45} \, \pa^2 [D, \overline{D}] {\bf W}_{-\frac{2}{3}}^{(2)}
+  c_{46} \, {\bf T} \pa [D, \overline{D}] {\bf W}_{-\frac{2}{3}}^{(2)}
 + c_{47} \, {\bf T} {\bf T}  [D, \overline{D}] {\bf W}_{-\frac{2}{3}}^{(2)}
\right. \nonu \\
&& + c_{48} \, {\bf T} {\bf T} {\bf T}  {\bf W}_{-\frac{2}{3}}^{(2)}
+ c_{49} \,{\bf T} {\bf T}  \pa {\bf W}_{-\frac{2}{3}}^{(2)}
 + c_{50} \, {\bf T} \overline{D} {\bf T} D   {\bf W}_{-\frac{2}{3}}^{(2)}
+ c_{51} \, {\bf T}  [D, \overline{D}]  {\bf T}  {\bf W}_{-\frac{2}{3}}^{(2)}
+ c_{52} \, {\bf T} D {\bf T} \overline{D}  {\bf W}_{-\frac{2}{3}}^{(2)}
\nonu \\
& & + c_{53} \, {\bf T} \pa^2  {\bf W}_{-\frac{2}{3}}^{(2)}
+  c_{54}\, \overline{D} {\bf T} \pa D   {\bf W}_{-\frac{2}{3}}^{(2)}
+ c_{55} \, \pa \overline{D} {\bf T}  D   {\bf W}_{-\frac{2}{3}}^{(2)}
 +   c_{56}\,
[D, \overline{D}]  {\bf T}  [D, \overline{D}]  {\bf W}_{-\frac{2}{3}}^{(2)} 
\nonu \\
&& + c_{57} \, [D, \overline{D}]  {\bf T} \pa  {\bf W}_{-\frac{2}{3}}^{(2)}
+  c_{58} \, \pa [D, \overline{D}]  {\bf T} {\bf W}_{-\frac{2}{3}}^{(2)}
 + c_{59} \, D  {\bf T} \pa  \overline{D}  {\bf W}_{-\frac{2}{3}}^{(2)}
+ c_{60} \, \pa D  {\bf T}   \overline{D}  {\bf W}_{-\frac{2}{3}}^{(2)}
\nonu \\
&& \left.
+ c_{61} \, \pa {\bf T}  [D, \overline{D}]  {\bf W}_{-\frac{2}{3}}^{(2)}
 +  c_{62} \, \pa {\bf T} {\bf T}  {\bf W}_{-\frac{2}{3}}^{(2)}
+  c_{63} \, \pa {\bf T} \pa  {\bf W}_{-\frac{2}{3}}^{(2)}
+ c_{64} \, \pa^2 {\bf T}  {\bf W}_{-\frac{2}{3}}^{(2)}
+ c_{65} \, \pa^3  {\bf W}_{-\frac{2}{3}}^{(2)}
\right](Z_2) 
\nonu \\
& & +\frac{\theta_{12} \, \bar{\theta}_{12}}{z_{12}} 
\left[ c_{66}\, 
\pa^3 D  {\bf W}_{-\frac{2}{3}}^{(2)}
+ c_{67} \, {\bf T} \pa^2 D  {\bf W}_{-\frac{2}{3}}^{(2)}
+ c_{68} \, {\bf T} {\bf T} \pa D  {\bf W}_{-\frac{2}{3}}^{(2)} 
 + c_{69} \, {\bf T} {\bf T}  {\bf T} D  {\bf W}_{-\frac{2}{3}}^{(2)} 
\right. \nonu \\
&& + c_{70} \, {\bf T}  [D, \overline{D}] {\bf T} D {\bf W}_{-\frac{2}{3}}^{(2)} 
+ c_{71} \, {\bf T} D {\bf T}  [D, \overline{D}] {\bf W}_{-\frac{2}{3}}^{(2)} 
 + c_{72} \, {\bf T} D {\bf T} \pa  {\bf W}_{-\frac{2}{3}}^{(2)} 
+ c_{73} \, \overline{D} {\bf T} D {\bf T} D  {\bf W}_{-\frac{2}{3}}^{(2)} 
\nonu \\
&& + c_{74} \,
[D, \overline{D}]  {\bf T} \pa D {\bf W}_{-\frac{2}{3}}^{(2)} 
 +  c_{75} \, \pa [D, \overline{D}]  {\bf T} D  {\bf W}_{-\frac{2}{3}}^{(2)}
+ c_{76} \, D {\bf T} \pa  [D, \overline{D}]   {\bf W}_{-\frac{2}{3}}^{(2)}
+ c_{77} \, D {\bf T} \pa^2  {\bf W}_{-\frac{2}{3}}^{(2)}
\nonu \\
& & + c_{78} \, \pa D {\bf T}  [D, \overline{D}]   {\bf W}_{-\frac{2}{3}}^{(2)}
+ c_{79} \, \pa D {\bf T} \pa  {\bf W}_{-\frac{2}{3}}^{(2)}
+ c_{80} \, \pa {\bf T} \pa D  {\bf W}_{-\frac{2}{3}}^{(2)}
 + c_{81}\, \pa {\bf T} {\bf T} D {\bf W}_{-\frac{2}{3}}^{(2)}
\nonu \\
&& + c_{82} \, \pa^2 {\bf T} D {\bf W}_{-\frac{2}{3}}^{(2)}
+ c_{83} \, \pa^2 D {\bf T} {\bf W}_{-\frac{2}{3}}^{(2)}
 + c_{84} \, \pa {\bf T} D {\bf T}  {\bf W}_{-\frac{2}{3}}^{(2)}
+ c_{85} \, \pa D {\bf T} {\bf T}  {\bf W}_{-\frac{2}{3}}^{(2)}
+  c_{86} \, [D, \overline{D}] {\bf T} D {\bf T} {\bf W}_{-\frac{2}{3}}^{(2)}
\nonu \\
& & +  \left. \left.
c_{87} \, {\bf T} {\bf T} D {\bf T} {\bf W}_{-\frac{2}{3}}^{(2)}
  \right](Z_2)  \right)+\cdots,
\label{14OPE}
\eea
}
where the coefficients appearing in 
Appendix (\ref{14OPE}) are given by
{\small 
\bea
c_2& = &-\frac{1}{14}, \qquad
c_3 = -\frac{3}{7}, \qquad
c_4 = \frac{3 }{70}, \qquad
c_5 = 0, \qquad
c_6 = \frac{17 }{70}, \qquad
c_7 = -\frac{(2 c-81) }{35 (c+6)}, 
\nonu \\
c_8 & = &  -\frac{93 }{35 (c+6)}, \qquad
c_9 = \frac{31 }{5 (c+6)}, \qquad
c_{10} = -\frac{3 (c-3) }{35 (c+6)}, \qquad
c_{11} = -\frac{27 }{35 (c+6)}, 
\nonu \\
c_{12} & = &  \frac{9 }{5 (c+6)}, \nonu \\
c_{13}& = &\frac{15 }{7 (c+6)}, \qquad
c_{14} = \frac{3 (9 c^4-156 c^3-8130 c^2+11704 c+24069) }{70 (c+6) (9 c-11) (3 c^2+54 c-169)}, \nonu \\
c_{15}& = &-\frac{9 (3 c^3-897 c^2+10517 c-1783) }{70 (c+6) (9 c-11) (3 c^2+54 c-169)}, \qquad
c_{16} = -\frac{234 (c+1) }{(9 c-11) (3 c^2+54 c-169)}, \nonu \\
c_{17}& = &\frac{3 (75 c^3-4617 c^2-55519 c+142773) }{70 (c+6) (9 c-11) (3 c^2+54 c-169)}, \qquad
c_{18} = -\frac{9 (381 c^3+3499 c^2-11015 c-11933) }{35 (c+6) (9 c-11) (3 c^2+54 c-169)}, \nonu \\
c_{19}& = &-\frac{(87 c^2+896 c-2271) }{10 (c+6) (3 c^2+54 c-169)}, \qquad
c_{20} = \frac{3 (9 c^3-2199 c^2+3345 c+3353) }{5 (c+6) (9 c-11) (3 c^2+54 c-169)}, \nonu \\
c_{21}& = &-\frac{3 (21 c^3-671 c^2-13725 c+44167) }{10 (c+6) (9 c-11) (3 c^2+54 c-169)}, \nonu \\
c_{22}& = &\frac{(90 c^4-1875 c^3-24159 c^2+82747 c-35859) }{70 (c+6) (9 c-11) (3 c^2+54 c-169)}, \nonu \\
c_{23}& = &-\frac{(189 c^5-10089 c^4-162501 c^3+1083579 c^2-2319700 c+1852098) 
}{140 (c+6) (3 c+2) (9 c-11) (3 c^2+54 c-169)}, \nonu \\
c_{24}& = &-\frac{3 (1125 c^4+1866 c^3-400176 c^2+1289780 c-925103) 
}{70 (c+6) (3 c+2) (9 c-11) (3 c^2+54 c-169)}, \nonu \\
c_{25}& = &-\frac{9 (138 c^3+3272 c^2-20155 c+16311)
 }{5 (c+6) (3 c+2) (9 c-11) (3 c^2+54 c-169)}, \nonu \\
c_{26}& = &\frac{5 (243 c^4-1872 c^3-22806 c^2+118666 c-129483) 
}{28 (c+6) (3 c+2) (9 c-11) (3 c^2+54 c-169)}, \nonu \\
c_{27}& = &\frac{3 (1188 c^4-5337 c^3+82971 c^2-293445 c+420059) 
}{140 (c+6) (3 c+2) (9 c-11) (3 c^2+54 c-169)}, \nonu \\
c_{28}& = &\frac{(17550 c^4+358209 c^3-1944915 c^2+3998133 c-3565693) 
}{140 (c+6) (3 c+2) (9 c-11) (3 c^2+54 c-169)}, \nonu \\
c_{29}& = &-\frac{3 (9927 c^4+192834 c^3-69356 c^2-2083236 c+1532827) 
}{140 (c+6) (3 c+2) (9 c-11) (3 c^2+54 c-169)}, \nonu \\
c_{30}& = &\frac{(3456 c^4+55017 c^3-189063 c^2-196243 c+477781) 
}{10 (c+6) (3 c+2) (9 c-11) (3 c^2+54 c-169)}, \nonu \\
c_{31}& = &\frac{63 (111 c^3+1589 c^2-6025 c+5697)
 }{5 (c+6) (3 c+2) (9 c-11) (3 c^2+54 c-169)}, \nonu \\
c_{32}& = &-\frac{3 (27 c^5-747 c^4-15963 c^3+26637 c^2+60860 c-38166) 
}{70 (c+6) (3 c+2) (9 c-11) (3 c^2+54 c-169)}, \nonu \\
c_{33}& = &-\frac{9 (45 c^4+678 c^3-8628 c^2-8020 c+29841)
 }{35 (c+6) (3 c+2) (9 c-11) (3 c^2+54 c-169)}, \nonu \\
c_{34}& = &\frac{54 (132 c^3+1258 c^2-995 c+2279)
 }{35 (c+6) (3 c+2) (9 c-11) (3 c^2+54 c-169)}, \nonu \\
c_{35}& = &\frac{18 (63 c^3+237 c^2+3075 c-1499)
 }{5 (c+6) (3 c+2) (9 c-11) (3 c^2+54 c-169)}, \nonu \\
c_{36}& = &\frac{15 (81 c^4+810 c^3-1836 c^2-3428 c+2217) 
}{14 (c+6) (3 c+2) (9 c-11) (3 c^2+54 c-169)}, \nonu \\
c_{37}& = &\frac{9 (54 c^4-3591 c^3+6033 c^2+19605 c-1073)
 }{70 (c+6) (3 c+2) (9 c-11) (3 c^2+54 c-169)}, \nonu \\
c_{38}& = &\frac{9 (180 c^4+8589 c^3+1245 c^2-53167 c+4597)
 }{70 (c+6) (3 c+2) (9 c-11) (3 c^2+54 c-169)}, \nonu \\
c_{39}& = &\frac{9 (36 c^4+167 c^3-3893 c^2+4967 c+2391) 
}{5 (c+6) (3 c+2) (9 c-11) (3 c^2+54 c-169)}, \nonu \\
c_{40}& = &-\frac{9 (261 c^4+3852 c^3+8902 c^2-41898 c-75809) 
}{70 (c+6) (3 c+2) (9 c-11) (3 c^2+54 c-169)}, \nonu \\
c_{41}& = &\frac{18 }{(c+6) (3 c+2)}, \qquad
c_{42} = \frac{3 (3 c-19) }{7 (c+6) (3 c+2)}, \qquad
c_{43} = \frac{3 (10 c+23) }{7 (c+6) (3 c+2)}, \nonu \\
c_{44}& = &\frac{6 
}{35 d_1(c)}
(6c-5)(12636 c^5+213291 c^4-869547 c^3-1688193 c^2+9155809 c-7487400), \nonu \\
c_{45}& = &\frac{3 
}{980 d_1(c)}
(6c-5)(162 c^7-7785 c^6-221163 c^5+372513 c^4 \nonu \\
& + & 2956895 c^3-15133452 c^2+45663254 c-56921328), \nonu \\
c_{46}& = &-\frac{9 
}{490 d_1(c) }
(6c-5)(18 c^6-8007 c^5+199476 c^4 \nonu \\
&+ & 723368 c^3-5758220 c^2+13579913 c-17090600), \nonu \\
c_{47}& = &-\frac{27 
}{245 d_1(c)}
(6c-5)(600 c^5-920 c^4-68870 c^3+499227 c^2-563431 c+810392), \nonu \\
c_{48}& = &-\frac{36 
}{7 d_1(c) }
(c+6)(6c-5)(24 c^3-1413 c^2+7051 c-312), \nonu \\
c_{49}& = &-\frac{9 
}{245 d_1(c)}
(6c-5)(9948 c^5+69578 c^4+6344 c^3+2045165 c^2-3013125 c-9608376), \nonu \\
c_{50}& = &-\frac{36 
}{245 d_1(c)}
(6c-5)(2736 c^5+35961 c^4-367677 c^3-88833 c^2+3519449 c-5809320), \nonu \\
c_{51}& = &-\frac{3 
}{35 d_1(c)}
(6c-5)(486 c^5-12429 c^4+202083 c^3-592713 c^2+1289849 c-929640), \nonu \\
c_{52}& = &\frac{18 
}{35 d_1(c)}
(6c-5)(18 c^5-6807 c^4-46011 c^3+502931 c^2-1728763 c+1726520), \nonu \\
c_{53}& = &\frac{3(6c-5) 
}{490 d_1(c)}
(954 c^6-65019 c^5-954840 c^4 \nonu \\
& - & 1317980 c^3+20872044 c^2-16574843 c-26680200), 
 \nonu \\
c_{54}& = &-\frac{9(6c-5) 
}{245 d_1(c)}
(3510 c^6+4647 c^5-560158 c^4 \nonu \\
& + & 1409686 c^3+5284684 c^2-25272657 c+28763240), \nonu \\
c_{55}& = &-\frac{3 
}{245 d_1(c)}
(6c-5)(30456 c^6+555030 c^5+979827 c^4 \nonu \\
& - & 16722391 c^3-956413 c^2+94359951 c-69377880), \nonu \\
c_{56}& = &-\frac{3 
}{980 d_1(c)}
(c+6)(6c-5)(9c-11)(810 c^4-3699 c^3-69146 c^2+419403 c-575480), \nonu \\
c_{57}& = &-\frac{1
}{980 d_1(c)}
(6c-5)(119394 c^6+1136979 c^5-5384853 c^4 \nonu \\
& - & 11413555 c^3+114851475 c^2-255103168 c+161816640), \nonu \\
c_{58}& = &-\frac{1
}{70 d_1(c) }
(6c-5)(15876 c^6+232614 c^5-454779 c^4\nonu \\
& - & 4689629 c^3+12949785 c^2-6736983 c-1874280), \nonu \\
c_{59}& = &\frac{3 
}{35 d_1(c)}
(6c-5)(162 c^6-27603 c^5+45714 c^4 \nonu \\
& + & 386502 c^3-2695994 c^2+7793319 c-7547864), \nonu \\
c_{60}& = &\frac{3 
}{35 d_1(c)}
(6c-5)(108 c^6-27528 c^5-143997 c^4 \nonu \\
& + & 566823 c^3+2483685 c^2-7353379 c+3020920), \nonu \\
c_{61}& = &-\frac{9 
}{980 d_1(c)}
(6c-5)(270 c^6-36843 c^5+8837 c^4 \nonu \\
& + & 1993967 c^3+1569217 c^2-23275344 c+17205056), \nonu \\
c_{62}& = &-\frac{9 
}{35 d_1(c)}
(6c-5)(5262 c^5+80377 c^4+188761 c^3-1514111 c^2+181483 c+1936840), \nonu \\
c_{63}& = &-\frac{3(6c-5) 
}{980 d_1(c)}
(54 c^6+86325 c^5+18223 c^4 \nonu \\
& - & 1900127 c^3+7301039 c^2+43132590 c-142114928), \nonu \\
c_{64}& = &-\frac{(6c-5)
}{70 d_1(c)}
(1296 c^6-59886 c^5-1374639 c^4 \nonu \\
& + & 1261671 c^3+28832545 c^2-58221783 c+821400), \nonu \\
c_{65}& = &\frac{
 1}{2940 d_1(c)}
\nonu \\
& \times & (6c-5)
(3726 c^7-198711 c^6-2233857 c^5+6866295 c^4 \nonu \\
& + &  
30429673 c^3-143190204 c^2+250626670 c-207457968), \nonu \\
c_{66}& = &-\frac{
1
 }{294 d_1(c)}
(972 c^8-79272 c^7+7731 c^6+10446411 c^5-21084303 c^4 \nonu \\
& - & 
208264183 c^3+891248796 c^2-1200942160 c+537945600), \nonu \\
c_{67}& = &-\frac{6 
}{49 d_1(c)}
(864 c^7-32760 c^6-535152 c^5+1699326 c^4 \nonu \\
& + & 14157958 c^3-72609371 c^2+112633825 c-57153800), \nonu \\
c_{68}& = &-\frac{18 
}{49 d_1(c)}
(396 c^6+93084 c^5-338545 c^4 \nonu \\
& - & 4056553 c^3+23472490 c^2-42123390 c+24287600), \nonu \\
c_{69}& = &\frac{36 
}{7 d_1(c)}
(3312 c^5-4632 c^4-92505 c^3+603674 c^2-1237965 c+784200), \nonu \\
c_{70}& = &\frac{15 
}{49 d_1(c)}
(23004 c^6-11124 c^5-989991 c^4 \nonu \\
& + & 2728689 c^3+3856415 c^2-18890897 c+14478440), \nonu \\
c_{71}& = &\frac{9 
}{49 d_1(c)}
(3996 c^6-54540 c^5-552621 c^4 \nonu \\
& + & 3168369 c^3-7758539 c^2+17550647 c-12700760), \nonu \\
c_{72}& = &\frac{3 
}{49 d_1(c)}
(66420 c^6+903852 c^5-5656347 c^4\nonu \\
& - & 27245901 c^3+212618587 c^2-391025811 c+209729080), \nonu \\
c_{73} & = &\frac{6 
 }{49 d_1(c)}
(182088 c^6+69984 c^5-3586797 c^4 \nonu \\
& + & 2890611 c^3-4721361 c^2+49900535 c-42855000), \nonu \\
c_{74}& = &\frac{5 
}{98 d_1(c)}
(4860 c^7-173124 c^6-1050993 c^5+8755464 c^4 \nonu \\
& + & 907370 c^3-81713274 c^2+149730149 c-78979080), \nonu \\
c_{75}& = &\frac{
1 }{98 d_1(c) }
 (29160 c^7+501228 c^6-2858976 c^5-8735367 c^4
 \nonu \\
& + &  85139311 c^3-272136831 c^2+442206735 c-210327000), \nonu \\
c_{76}& = &\frac{3 
}{98 d_1(c)}
(3564 c^7-150444 c^6-113949 c^5+2761866 c^4 \nonu \\
& - & 11202750 c^3+54951740 c^2-120084663 c+72084440), \nonu \\
c_{77}& = &\frac{
1 }{98 d_1(c) }
 (28836 c^7-502740 c^6-7227999 c^5+24981018 c^4
 \nonu \\
& + & 
124984842 c^3-579844400 c^2+719963079 c-286824920), \nonu \\
c_{78}& = &\frac{3 
}{98 d_1(c)}
(6480 c^7+25452 c^6+517908 c^5+2105547 c^4 \nonu \\
& - & 14714569 c^3+10292323 c^2-2604737 c-250840), \nonu \\
c_{79}& = &\frac{
 1}{98 d_1(c)}
 (101088 c^7+1972404 c^6+3464640 c^5-39604647 c^4
 \nonu \\
& - &  44660867 c^3+246743165 c^2-54701219 c-84179080), \nonu \\
c_{80}& = &-\frac{15 
}{98 d_1(c)}
(3564 c^7+30636 c^6-582069 c^5-569956 c^4 \nonu \\
& + & 16379738 c^3-42067358 c^2+31022649 c-2553640), \nonu \\
c_{81}& = &-\frac{9 
}{49 d_1(c)}
(28260 c^6+649932 c^5-1011487 c^4 \nonu \\
& - & 17746275 c^3+57026779 c^2-46205105 c+3639400), \nonu \\
c_{82}& = &-\frac{
1 }{98 d_1(c)}
 (124416 c^7+2668572 c^6+4393764 c^5-86497545 c^4
 \nonu \\
& + & 
20413401 c^3+667925879 c^2-1085835075 c+458100600), \nonu \\
c_{83}& = &\frac{2 
}{7 d_1(c)}
(5508 c^7+67923 c^6-86064 c^5-1427376 c^4 \nonu \\
& + & 1333962 c^3+13913834 c^2-34980625 c+18308200), \nonu \\
c_{84}& = &\frac{6 
}{7 d_1(c)}
(15984 c^6+320931 c^5-444090 c^4 \nonu \\
& - & 8473867 c^3+25850560 c^2-19328210 c+2659600), \nonu \\
c_{85}& = &\frac{12 
}{7 d_1(c)}
(4644 c^6+92313 c^5+105393 c^4-2228802 c^3+4284334 c^2-4512340 c+1734400), \nonu \\
c_{86}& = &-\frac{2 
}{7 d_1(c)}
(9c-11)(7776 c^5+26487 c^4-282471 c^3+347798 c^2+376350 c-583600), \nonu \\
c_{87}& = &-\frac{36 
}{7 d_1(c)}
(7488 c^5+2082 c^4-212150 c^3+976741 c^2-2519915 c+1814200), \nonu \\
d_1(c) &\equiv &  (c+6) (3 c+2) (6 c-5) (9 c-11) (2 c^2+9 c-40) 
(3 c^2+54 c-169). 
\label{coeff}
\eea }
In the large $c$ limit,
all the nonlinear terms associated with the coefficients in 
Appendix (\ref{coeff}) disappear.
One can also analyze the $\frac{1}{c}, \cdots, \frac{1}{c^3}$-terms. 
We introduce $d_1(c)$ in Appendix (\ref{coeff}) 
which appears in the denominators of the 
coefficients. 
The OPE between the second higher spin ${\cal N}=2$ multiplet 
and the third higher 
spin ${\cal N}=2$ multiplet in (\ref{manyW}) can be written as
{\small
\bea
&& {\bf W}_{-\frac{2}{3}}^{(2)}(Z_1) \, {\bf W}_{\frac{1}{3}}^{(\frac{7}{2})}(Z_2) = 
C_{(2) \, (\frac{7}{2})}^{(2) \, -} \, \left(
\frac{\theta_{12}}{z_{12}^4} \, {\bf W}_{\frac{2}{3}}^{(2)}(Z_2)
+\frac{\theta_{12} \, \bar{\theta}_{12}}{z_{12}^4} \,
d_2 \, \overline{D} {\bf W}_{\frac{2}{3}}^{(2)}(Z_2) +
+\frac{1}{z_{12}^3} \,
d_3 \, \overline{D} {\bf W}_{\frac{2}{3}}^{(2)}(Z_2) \right.
\nonu \\
&& +
\frac{ \theta_{12}}{z_{12}^3} \, \left[
d_4 \, [D, \overline{D}]  {\bf W}_{\frac{2}{3}}^{(2)} 
+ d_5 \, {\bf T}  {\bf W}_{\frac{2}{3}}^{(2)}  
+  d_6 \, \pa {\bf W}_{\frac{2}{3}}^{(2)} 
\right](Z_2) 
\nonu \\
& & +
\frac{\theta_{12} \, \bar{\theta}_{12}}{z_{12}^3} 
\, 
\left[ d_7 \,
\pa \overline{D}  {\bf W}_{\frac{2}{3}}^{(2)} 
+ d_8 \,{\bf T}  \overline{D}  {\bf W}_{\frac{2}{3}}^{(2)} 
+ d_9 \, \overline{D} {\bf T}    {\bf W}_{\frac{2}{3}}^{(2)} 
\right](Z_2)
\nonu \\
&& + \frac{1}{z_{12}^2} \, 
\left[ d_{10} \,
\pa \overline{D}  {\bf W}_{\frac{2}{3}}^{(2)} 
+ d_{11} \, {\bf T}  \overline{D}  {\bf W}_{\frac{2}{3}}^{(2)} 
+ d_{12} \, \overline{D} {\bf T}    {\bf W}_{\frac{2}{3}}^{(2)} 
\right]
(Z_2)
 +  \frac{\bar{\theta}_{12}}{z_{12}^2} \, 
d_{13} \, \overline{D}  {\bf T} \overline{D}
 {\bf W}_{\frac{2}{3}}^{(2)}(Z_2)
\nonu \\
&& + \frac{\theta_{12}}{z_{12}^2} \, \left[ 
d_{14} \, \pa [D, \overline{D}]  {\bf W}_{\frac{2}{3}}^{(2)}
+ d_{15} \, {\bf T} [D, \overline{D}]  {\bf W}_{\frac{2}{3}}^{(2)}
+ d_{16} \, {\bf T} {\bf T}  {\bf W}_{\frac{2}{3}}^{(2)}
 + 
d_{17} \, {\bf T} \pa {\bf W}_{\frac{2}{3}}^{(2)} 
+ d_{18} \, D {\bf T} \overline{D}  {\bf W}_{\frac{2}{3}}^{(2)} 
\right. \nonu \\
&& +  \left. d_{19} \, [D, \overline{D}] {\bf T}  {\bf W}_{\frac{2}{3}}^{(2)} 
 + 
d_{20} \, \overline{D} {\bf T} D  {\bf W}_{\frac{2}{3}}^{(2)}
+ d_{21} \, \pa {\bf T}  {\bf W}_{\frac{2}{3}}^{(2)}
+ d_{22} \, \pa^2  {\bf W}_{\frac{2}{3}}^{(2)}
\right](Z_2)
\nonu \\
&& +   \frac{\theta_{12} \, \bar{\theta}_{12}}{z_{12}^2}
\left[ d_{23} \,
\pa^2 \overline{D}    {\bf W}_{\frac{2}{3}}^{(2)}
+ d_{24} \, {\bf T} \pa  \overline{D}  {\bf W}_{\frac{2}{3}}^{(2)}
+ d_{25} \, {\bf T} {\bf T}  \overline{D}  {\bf W}_{\frac{2}{3}}^{(2)}
 +
 d_{26} \, [D, \overline{D}] {\bf T}  \overline{D}  {\bf W}_{\frac{2}{3}}^{(2)}
\right. \nonu \\
&& + \left.
d_{27} \, \overline{D} {\bf T}  [D, \overline{D}] {\bf W}_{\frac{2}{3}}^{(2)}
+ d_{28} \, \overline{D} {\bf T} \pa  {\bf W}_{\frac{2}{3}}^{(2)}
+  
 d_{29} \, \pa {\bf T} \overline{D}  {\bf W}_{\frac{2}{3}}^{(2)}
+ d_{30}\, \pa \overline{D} {\bf T} {\bf W}_{\frac{2}{3}}^{(2)}
+ d_{31} \, {\bf T} \overline{D} {\bf T}  {\bf W}_{\frac{2}{3}}^{(2)}
\right](Z_2)
\nonu \\
&& +   \frac{1}{z_{12}}
\left[  d_{32} \, \pa^2  \overline{D}    {\bf W}_{\frac{2}{3}}^{(2)}
+ d_{33} \, {\bf T} \pa  \overline{D}  {\bf W}_{\frac{2}{3}}^{(2)}
+ d_{34} \, {\bf T} {\bf T}  \overline{D}  {\bf W}_{\frac{2}{3}}^{(2)}
 + d_{35} \,
 {\bf T} \overline{D}{\bf T}   {\bf W}_{\frac{2}{3}}^{(2)}
+ d_{36} \,  [D, \overline{D}] {\bf T} \overline{D} {\bf W}_{\frac{2}{3}}^{(2)}
\right. \nonu \\
&& \left.
+ d_{37} \, \overline{D} {\bf T}  [D, \overline{D}] {\bf W}_{\frac{2}{3}}^{(2)}
+ d_{38} \, \overline{D} {\bf T} \pa  {\bf W}_{\frac{2}{3}}^{(2)}
 +  
 d_{39} \, \pa  \overline{D} {\bf T}   {\bf W}_{\frac{2}{3}}^{(2)}
+ d_{40} \, \pa  {\bf T}  \overline{D} {\bf W}_{\frac{2}{3}}^{(2)}
\right](Z_2)
\nonu \\
 && +  
\frac{\bar{\theta}_{12}}{z_{12}} \left[ 
d_{41} \, {\bf T} \overline{D} {\bf T} \overline{D} {\bf W}_{\frac{2}{3}}^{(2)}
+ d_{42} \, \overline{D} {\bf T} \pa \overline{D} {\bf W}_{\frac{2}{3}}^{(2)}
+ d_{43} \, \pa  \overline{D} {\bf T} \overline{D} {\bf W}_{\frac{2}{3}}^{(2)}
\right](Z_2) 
\nonu \\
&& +  
\frac{ \theta_{12}}{z_{12}} 
 \left[ d_{44} \,
 \overline{D} {\bf T} D {\bf T} {\bf W}_{\frac{2}{3}}^{(2)}
+  d_{45}\, \pa^2 [D, \overline{D}] {\bf W}_{\frac{2}{3}}^{(2)}
+  d_{46} \, {\bf T} \pa [D, \overline{D}] {\bf W}_{\frac{2}{3}}^{(2)}
 + d_{47} \, {\bf T} {\bf T}  [D, \overline{D}] {\bf W}_{\frac{2}{3}}^{(2)}
\right. \nonu \\
&& + d_{48} \, {\bf T} {\bf T} {\bf T}  {\bf W}_{\frac{2}{3}}^{(2)}
+ d_{49} \, {\bf T} {\bf T}  \pa {\bf W}_{\frac{2}{3}}^{(2)}
 + d_{50} \, {\bf T} D {\bf T} \overline{D}   {\bf W}_{\frac{2}{3}}^{(2)}
+ d_{51} \, {\bf T}  [D, \overline{D}]  {\bf T}  {\bf W}_{\frac{2}{3}}^{(2)}
+ d_{52} \, {\bf T} \overline{D} {\bf T} D  {\bf W}_{\frac{2}{3}}^{(2)}
\nonu \\
& & + d_{53} \, {\bf T} \pa^2  {\bf W}_{\frac{2}{3}}^{(2)}
+  d_{54}\, D {\bf T} \pa \overline{D}   {\bf W}_{\frac{2}{3}}^{(2)}
+ d_{55} \, \pa D {\bf T}  \overline{D}   {\bf W}_{\frac{2}{3}}^{(2)}
 +   d_{56}\, 
[D, \overline{D}]  {\bf T}  [D, \overline{D}]  {\bf W}_{\frac{2}{3}}^{(2)} 
\nonu \\
&& + d_{57} \, [D, \overline{D}]  {\bf T} \pa  {\bf W}_{\frac{2}{3}}^{(2)}
+  d_{58} \, \pa [D, \overline{D}]  {\bf T} {\bf W}_{\frac{2}{3}}^{(2)}
+ d_{59} \, \overline{D}  {\bf T} \pa  D  {\bf W}_{\frac{2}{3}}^{(2)}
+ d_{60} \, \pa  \overline{D}  {\bf T}   D  {\bf W}_{\frac{2}{3}}^{(2)}
\nonu \\
&& + \left. 
d_{61} \, \pa {\bf T}  [D, \overline{D}]  {\bf W}_{\frac{2}{3}}^{(2)}
 +   d_{62} \, \pa {\bf T} {\bf T}  {\bf W}_{\frac{2}{3}}^{(2)}
+ d_{63} \, \pa {\bf T} \pa  {\bf W}_{\frac{2}{3}}^{(2)}
+ d_{64} \, \pa^2 {\bf T}  {\bf W}_{\frac{2}{3}}^{(2)}
+ d_{65} \, \pa^3  {\bf W}_{\frac{2}{3}}^{(2)}
\right](Z_2) 
\nonu \\
& & +\frac{\theta_{12} \, \bar{\theta}_{12}}{z_{12}} 
\left[ d_{66} \,
\pa^3 \overline{D}  {\bf W}_{\frac{2}{3}}^{(2)}
+ d_{67} \, {\bf T} \pa^2  \overline{D}  {\bf W}_{\frac{2}{3}}^{(2)}
+ d_{68} \, {\bf T} {\bf T} \pa  \overline{D}  {\bf W}_{\frac{2}{3}}^{(2)} 
 + d_{69} \, {\bf T} {\bf T} {\bf T} \overline{D}  {\bf W}_{\frac{2}{3}}^{(2)} 
\right. \nonu \\
&& + d_{70} \,
{\bf T}  [D, \overline{D}] {\bf T} \overline{D} {\bf W}_{\frac{2}{3}}^{(2)} 
+ d_{71} \, {\bf T} \overline{D} {\bf T}  [D, \overline{D}] {\bf W}_{\frac{2}{3}}^{(2)} 
 + d_{72} \, {\bf T} \overline{D} {\bf T} \pa  {\bf W}_{\frac{2}{3}}^{(2)} 
+  d_{73} \, \overline{D} {\bf T}  D {\bf T} \overline{D}  {\bf W}_{\frac{2}{3}}^{(2)} 
\nonu \\
&& + d_{74} \,
[D, \overline{D}]  {\bf T} \pa \overline{D} {\bf W}_{\frac{2}{3}}^{(2)} 
+  d_{75} \, \pa 
[D, \overline{D}]  {\bf T} \overline{D}  {\bf W}_{\frac{2}{3}}^{(2)}
+ d_{76} \, \overline{D} {\bf T} \pa  [D, \overline{D}]   {\bf W}_{\frac{2}{3}}^{(2)}
+ d_{77} \, \overline{D} {\bf T} \pa^2  {\bf W}_{\frac{2}{3}}^{(2)}
\nonu \\
& & + d_{78} \, 
\pa \overline{D} {\bf T}  [D, \overline{D}]   {\bf W}_{\frac{2}{3}}^{(2)}
+ d_{79} \, \pa \overline{D} {\bf T} \pa  {\bf W}_{\frac{2}{3}}^{(2)}
+ d_{80} \, \pa {\bf T} \pa  \overline{D}  {\bf W}_{\frac{2}{3}}^{(2)}
 + d_{81} \, \pa {\bf T} {\bf T}  \overline{D} {\bf W}_{\frac{2}{3}}^{(2)}
\nonu \\
&& + d_{82} \, \pa^2 {\bf T}  \overline{D} {\bf W}_{\frac{2}{3}}^{(2)}
+ d_{83} \, \pa^2  \overline{D} {\bf T} {\bf W}_{\frac{2}{3}}^{(2)}
 + d_{84} \, \pa {\bf T}  \overline{D} {\bf T}  {\bf W}_{\frac{2}{3}}^{(2)}
+ d_{85} \, \pa \overline{D} {\bf T} {\bf T}  {\bf W}_{\frac{2}{3}}^{(2)}
+  d_{86} \, \overline{D} {\bf T} [D, \overline{D}] {\bf T} {\bf W}_{\frac{2}{3}}^{(2)}
\nonu \\
& & +  \left. \left.
d_{87} \, {\bf T} {\bf T} \overline{D} {\bf T} {\bf W}_{\frac{2}{3}}^{(2)}
  \right](Z_2) \right) +\cdots,
\label{23OPE}
\eea
}
where the coefficients in Appendix (\ref{23OPE}) 
can be written in terms of previous ones $c_i$ in 
Appendix (\ref{coeff}) 
and are given by  
{\small
\bea
d_2 & = & - c_2, \qquad
d_{3}   =   c_3, \qquad d_4 =- c_4, \qquad d_5 =0, \qquad
d_6 = c_6, \qquad d_7 = - c_7, \nonu \\
d_8  & = & c_8, \qquad d_9 = c_{9},
\qquad d_{10} =c_{10}, \qquad
d_{11} = -c_{11}, \qquad d_{12} = -c_{12}, \qquad
d_{13} = -c_{13}, \nonu \\
d_{14} & = & -c_{14},
\qquad
d_{15} = c_{15}, \qquad
d_{16} = c_{16}, \qquad
d_{17} =- c_{17}, \qquad
d_{18} = - c_{18}, \qquad
d_{19} = c_{19},
\nonu \\
d_{20} & = & - c_{20}, \qquad
d_{21}   =   -c_{21}, \qquad d_{22} = c_{22}, \qquad d_{23} = - c_{23}, \qquad
d_{24} = c_{24}, \qquad d_{25} = - c_{25}, \nonu \\
d_{26}  & = & -c_{26}, \qquad d_{27} = -c_{27},
\qquad d_{28} =c_{28}, \qquad
d_{29} = c_{29}, \qquad d_{30} = c_{30}, \qquad
d_{31} = -c_{31}, \nonu \\
d_{32} & = & c_{32},
\qquad
d_{33} = -c_{33}, \qquad
d_{34} = c_{34}, \qquad
d_{35} = c_{35}, \qquad
d_{36} =  c_{36}, \qquad
d_{37} = c_{37},
\nonu \\
d_{38} & = & - c_{38}, \qquad
d_{39}   =   -c_{39}, \qquad d_{40} =- c_{40}, 
\qquad d_{41} =  c_{41}, \qquad
d_{42} = -c_{42}, \qquad d_{43} = - c_{43}, \nonu \\
d_{44}  & = & -c_{44}, \qquad d_{45} = -c_{45},
\qquad d_{46} =c_{46}, \qquad
d_{47} = -c_{47}, \qquad d_{48} = -c_{48}, \qquad
d_{49} = c_{49}, \nonu \\
d_{50} & = & c_{50},
\qquad
d_{51} = -c_{51}, \qquad
d_{52} = c_{52}, \qquad
d_{53} =- c_{53}, \qquad
d_{54} = - c_{54}, \qquad
d_{55} = -c_{55},
\nonu \\
d_{56} & = & - c_{56}, \qquad
d_{57}   =   c_{57}, 
\qquad
 d_{59}   =   - c_{59}, \qquad
d_{60} = -c_{60}, \qquad d_{61} =  c_{61}, \nonu \\
d_{62}  & = & c_{62}, \qquad d_{63} = -c_{63},
\qquad d_{64} = -c_{64}, \qquad
d_{65} = c_{65}, \qquad 
d_{66} = -c_{66} \qquad
d_{67} = c_{67}, \nonu \\
d_{68} & = & -c_{68},
\qquad
d_{69} = c_{69}, \qquad
d_{70} = c_{70}, \qquad
d_{71} = c_{71}, \qquad
d_{72} = - c_{72}, \qquad
d_{73} = c_{73},
\nonu  \\
d_{74} & = & -c_{74},
\qquad
d_{76}  =   -c_{76}, \qquad
d_{77} = c_{77}, \qquad
d_{78} = - c_{78}, \qquad
d_{79} = c_{79},
\qquad
d_{80}   =   c_{80},
\nonu \\
d_{81} & = & - c_{81}, 
\qquad
d_{82}  =   c_{82}, 
\qquad
d_{84}  =   - c_{84}, \qquad
d_{85} = -c_{85},
\qquad
d_{86}   =   c_{86}, \qquad
d_{87} = c_{87},
\nonu \\
d_{58} & = & 
-\frac{1}{70 d_1(c)}
(6c-5)
(15876 c^6+308430 c^5+824967 c^4 \nonu \\
& - & 9906911 c^3+2820627 c^2+48197871 c-46798680), 
\nonu \\
d_{75} & = & 
-\frac{1
 }{98 d_1(c)}
   5 (5832 c^7-118260 c^6-655776 c^5+2557083 c^4
 \nonu \\
& + & 
13559129 c^3-48761733 c^2+28560705 c+9360600), 
\nonu \\
d_{83}  & = &  \frac{
1 }{7 d_1(c)} 
(11016 c^7+205830 c^6-19281 c^5-5688348 c^4
 \nonu \\
& + & 
8905287 c^3+27389040 c^2-79353500 c+43036000).
\label{dcoeff}
\eea
}
Note that the last three terms in Appendix (\ref{dcoeff}) 
are different from the corresponding 
coefficients in Appendix (\ref{coeff}).
Furthermore, 
the OPE between the second higher spin ${\cal N}=2$ multiplet 
and the fourth higher 
spin ${\cal N}=2$ multiplet in (\ref{manyW}) can be described as
{\small
\bea
&& {\bf W}_{-\frac{2}{3}}^{(2)}(Z_1) \, {\bf W}_{-\frac{1}{3}}^{(\frac{7}{2})}(Z_2) = 
C_{(2) \, (\frac{7}{2})}^{(4) \, -} \, \left( \frac{\theta_{12}}{z_{12}^2} \,
 {\bf W}_{0}^{(4)}(Z_2)
+   \frac{\theta_{12} \, \bar{\theta}_{12}}{z_{12}^2}
\, \frac{7}{24} \, \overline{D}  {\bf W}_{0}^{(4)}(Z_2)
\right. \nonu \\
&& -\frac{1}{z_{12}} \,  \frac{1}{4} \, \overline{D}  {\bf W}_{0}^{(4)}(Z_2) 
 +  
\frac{\theta_{12}}{z_{12}} \left[ -\frac{1}{24} \, [D, \overline{D}]
 {\bf W}_{0}^{(4)} + \frac{3}{8} \, \pa  {\bf W}_{0}^{(4)}
\right](Z_2)
\nonu \\
&& + \left.  
\frac{\theta_{12} \, \bar{\theta}_{12}}{z_{12}} \left[
 \frac{5(c-3)}{36(c+9)} \, \pa    \overline{D}  
{\bf W}_{0}^{(4)}  + \frac{20}{3(c+9)} \, 
\overline{D} {\bf T}  {\bf W}_{0}^{(4)} 
 -  \frac{5}{3(c+9)} \,
{\bf T} \overline{D}  {\bf W}_{0}^{(4)} \right](Z_2) \right)
+\cdots,
\label{24OPE}
\eea
}
which looks like as Appendix (\ref{13OPE}).
Again,
in the large $c$ limit,
all the nonlinear terms in Appendix (\ref{24OPE}) disappear.
One can also analyze the $\frac{1}{c}$-terms. 

\subsection{ The OPEs ${\bf W}_{\pm \frac{1}{3}}^{(\frac{7}{2})}(Z_1) \,
{\bf W}_{ \pm \frac{1}{3}}^{(\frac{7}{2})}(Z_2)
$ }

Now let us consider the OPE 
between the third higher spin ${\cal N}=2$ multiplet and itself 
in (\ref{manyW}).
{\small
\bea
&& {\bf W}_{\frac{1}{3}}^{(\frac{7}{2})}(Z_1) \, {\bf W}_{\frac{1}{3}}^{(\frac{7}{2})}(Z_2) =  
C_{(\frac{7}{2}) \, (\frac{7}{2})}^{(\frac{7}{2}) \, +} \, \left(
\frac{\bar{\theta}_{12}}{z_{12}^4} \, {\bf W}_{-\frac{1}{3}}^{(\frac{7}{2})}(Z_2)
+\frac{\theta_{12} \, \bar{\theta}_{12}}{z_{12}^4} \,
e_2 \, D {\bf W}_{-\frac{1}{3}}^{(\frac{7}{2})}(Z_2) +
+\frac{1}{z_{12}^3} \,
e_3 \, D {\bf W}_{-\frac{1}{3}}^{(\frac{7}{2})}(Z_2) \right.
\nonu \\
&& +
\frac{ \bar{\theta}_{12}}{z_{12}^3} \, \left[
e_4 \, [D, \overline{D}]  {\bf W}_{-\frac{1}{3}}^{(\frac{7}{2})} 
+ e_5 \, {\bf T}  {\bf W}_{-\frac{1}{3}}^{(\frac{7}{2})}  
+ e_6 \,  \pa {\bf W}_{-\frac{1}{3}}^{(\frac{7}{2})} 
\right](Z_2) 
\nonu \\
&& +
\frac{\theta_{12} \, \bar{\theta}_{12}}{z_{12}^3} 
\, 
\left[ e_7 \, \pa D  {\bf W}_{-\frac{1}{3}}^{(\frac{7}{2})} 
+ e_8 \, {\bf T}  D  {\bf W}_{-\frac{1}{3}}^{(\frac{7}{2})} 
+ e_9 \, D {\bf T}    {\bf W}_{-\frac{1}{3}}^{(\frac{7}{2})} 
\right](Z_2)
\nonu \\
&& + \frac{1}{z_{12}^2} \, 
\left[ e_{10} \,
\pa D  {\bf W}_{-\frac{1}{3}}^{(\frac{7}{2})} 
+ e_{11} \, {\bf T}  D  {\bf W}_{-\frac{1}{3}}^{(\frac{7}{2})} 
+ e_{12} \, D {\bf T}    {\bf W}_{-\frac{1}{3}}^{(\frac{7}{2})} 
\right]
(Z_2)
 +   \frac{\theta_{12}}{z_{12}^2} \, e_{13} \, D  {\bf T} D
 {\bf W}_{-\frac{1}{3}}^{(\frac{7}{2})}(Z_2)
\nonu \\
&& + \frac{\bar{\theta}_{12}}{z_{12}^2} \, \left[ 
e_{14} \, \pa [D, \overline{D}]  {\bf W}_{-\frac{1}{3}}^{(\frac{7}{2})}
+ e_{15} \, {\bf T} [D, \overline{D}]  {\bf W}_{-\frac{1}{3}}^{(\frac{7}{2})}
+ e_{16} \, {\bf T} {\bf T}  {\bf W}_{-\frac{1}{3}}^{(\frac{7}{2})}
 + 
e_{17} \, {\bf T} \pa {\bf W}_{-\frac{1}{3}}^{(\frac{7}{2})} 
+ e_{18} \, \overline{D} {\bf T} D  {\bf W}_{-\frac{1}{3}}^{(\frac{7}{2})} 
\right. \nonu \\&&
+ \left. e_{19} \, [D, \overline{D}] {\bf T}  {\bf W}_{-\frac{1}{3}}^{(\frac{7}{2})} 
+ 
e_{20} \, D {\bf T} \overline{D}  {\bf W}_{-\frac{1}{3}}^{(\frac{7}{2})}
+ e_{21} \, \pa {\bf T}  {\bf W}_{-\frac{1}{3}}^{(\frac{7}{2})}
+e_{22} \, \pa^2  {\bf W}_{-\frac{1}{3}}^{(\frac{7}{2})}
\right](Z_2)
\nonu \\
&& +   \frac{\theta_{12} \, \bar{\theta}_{12}}{z_{12}^2}
\left[ e_{23} \, 
\pa^2 D    {\bf W}_{-\frac{1}{3}}^{(\frac{7}{2})}
+ e_{24} \, {\bf T} \pa D  {\bf W}_{-\frac{1}{3}}^{(\frac{7}{2})}
+ e_{25} \, {\bf T} {\bf T} D  {\bf W}_{-\frac{1}{3}}^{(\frac{7}{2})} +
 e_{26} \, [D, \overline{D}] {\bf T} D  {\bf W}_{-\frac{1}{3}}^{(\frac{7}{2})}
\right. \nonu \\
&& + 
\left. 
e_{27} \, D {\bf T}  [D, \overline{D}] {\bf W}_{-\frac{1}{3}}^{(\frac{7}{2})}
+ e_{28} \, D {\bf T} \pa  {\bf W}_{-\frac{1}{3}}^{(\frac{7}{2})}
 +  
 e_{29} \, \pa {\bf T} D  {\bf W}_{-\frac{1}{3}}^{(\frac{7}{2})}
+ e_{30} \, \pa D {\bf T} {\bf W}_{-\frac{1}{3}}^{(\frac{7}{2})}
+ e_{31} \, {\bf T} D {\bf T}  {\bf W}_{-\frac{1}{3}}^{(\frac{7}{2})}
\right](Z_2)
\nonu \\
&& +   \frac{1}{z_{12}}
\left[  e_{32} \, \pa^2 D    {\bf W}_{-\frac{1}{3}}^{(\frac{7}{2})}
+ e_{33} \, {\bf T} \pa D  {\bf W}_{-\frac{1}{3}}^{(\frac{7}{2})}
+ e_{34} \, {\bf T} {\bf T} D  {\bf W}_{-\frac{1}{3}}^{(\frac{7}{2})}
 + e_{35} \, {\bf T} D {\bf T}  {\bf W}_{-\frac{1}{3}}^{(\frac{7}{2})} 
 + e_{36} \, [D, \overline{D}] {\bf T} D  {\bf W}_{-\frac{1}{3}}^{(\frac{7}{2})}
\right. \nonu \\
&& + \left.
e_{37} \, D {\bf T}  [D, \overline{D}] {\bf W}_{-\frac{1}{3}}^{(\frac{7}{2})}
+ e_{38} \, D {\bf T} \pa  {\bf W}_{-\frac{1}{3}}^{(\frac{7}{2})}
+ 
 e_{39} \, \pa D {\bf T} {\bf W}_{-\frac{1}{3}}^{(\frac{7}{2})}
+ e_{40} \, \pa {\bf T} D  {\bf W}_{-\frac{1}{3}}^{(\frac{7}{2})}
\right](Z_2)
\nonu \\
 && +  
\frac{\theta_{12}}{z_{12}} \left[ 
e_{41} \, {\bf T} D {\bf T} D {\bf W}_{-\frac{1}{3}}^{(\frac{7}{2})}
+ e_{42} \, D {\bf T} \pa D {\bf W}_{-\frac{1}{3}}^{(\frac{7}{2})}
+ e_{43} \, \pa D {\bf T} D {\bf W}_{-\frac{1}{3}}^{(\frac{7}{2})}
\right](Z_2) 
\nonu \\
&& +  
\frac{ \bar{\theta}_{12}}{z_{12}} 
 \left[ e_{44} \,
\overline{D} {\bf T} D {\bf T}  {\bf W}_{-\frac{1}{3}}^{(\frac{7}{2})}
+  e_{45}\, \pa^2 [D, \overline{D}] {\bf W}_{-\frac{1}{3}}^{(\frac{7}{2})}
+  e_{46} \, {\bf T} \pa [D, \overline{D}] {\bf W}_{-\frac{1}{3}}^{(\frac{7}{2})}
+ e_{47} \, 
{\bf T} {\bf T}  [D, \overline{D}] {\bf W}_{-\frac{1}{3}}^{(\frac{7}{2})}
\right. \nonu \\
&& + e_{48} \, {\bf T} {\bf T} {\bf T}  {\bf W}_{-\frac{1}{3}}^{(\frac{7}{2})}
+ e_{49} \, {\bf T} {\bf T}  \pa {\bf W}_{-\frac{1}{3}}^{(\frac{7}{2})}
 + e_{50} \, {\bf T} \overline{D} {\bf T} D   {\bf W}_{-\frac{1}{3}}^{(\frac{7}{2})}
+ e_{51} \, 
{\bf T}  [D, \overline{D}]  {\bf T}  {\bf W}_{-\frac{1}{3}}^{(\frac{7}{2})}
+ e_{52} \, {\bf T} D {\bf T} \overline{D}  {\bf W}_{-\frac{1}{3}}^{(\frac{7}{2})}
\nonu \\
& & + e_{53} \, {\bf T} \pa^2  {\bf W}_{-\frac{1}{3}}^{(\frac{7}{2})}
+  e_{54} \, \overline{D} {\bf T} \pa D   {\bf W}_{-\frac{1}{3}}^{(\frac{7}{2})}
+ e_{55} \, \pa \overline{D} {\bf T}  D   {\bf W}_{-\frac{1}{3}}^{(\frac{7}{2})}
 +   e_{56} \,
[D, \overline{D}]  {\bf T}  [D, \overline{D}]  
{\bf W}_{-\frac{1}{3}}^{(\frac{7}{2})} 
\nonu \\
&& + e_{57} \, [D, \overline{D}]  {\bf T} \pa  {\bf W}_{-\frac{1}{3}}^{(\frac{7}{2})}
+  e_{58} \, \pa [D, \overline{D}]  {\bf T} {\bf W}_{-\frac{1}{3}}^{(\frac{7}{2})}
 + e_{59} \, D  {\bf T} \pa  \overline{D}  {\bf W}_{-\frac{1}{3}}^{(\frac{7}{2})}
+ e_{60} \, \pa D  {\bf T}   \overline{D}  {\bf W}_{-\frac{1}{3}}^{(\frac{7}{2})}
\nonu \\
&&
+ \left.
e_{61} \, \pa {\bf T}  [D, \overline{D}]  {\bf W}_{-\frac{1}{3}}^{(\frac{7}{2})}
 +  e_{62} \, \pa {\bf T} {\bf T}  {\bf W}_{-\frac{1}{3}}^{(\frac{7}{2})}
+ e_{63} \, \pa {\bf T} \pa  {\bf W}_{-\frac{1}{3}}^{(\frac{7}{2})}
+ e_{64} \, \pa^2 {\bf T}  {\bf W}_{-\frac{1}{3}}^{(\frac{7}{2})}
+  e_{65} \, \pa^3  {\bf W}_{-\frac{1}{3}}^{(\frac{7}{2})}
\right](Z_2) 
\nonu \\
& & +\frac{\theta_{12} \, \bar{\theta}_{12}}{z_{12}} 
\left[ e_{66} \,
\pa^3 D  {\bf W}_{-\frac{1}{3}}^{(\frac{7}{2})}
+ e_{67} \,{\bf T} \pa^2 D  {\bf W}_{-\frac{1}{3}}^{(\frac{7}{2})}
+ e_{68} \, {\bf T} {\bf T} \pa D  {\bf W}_{-\frac{1}{3}}^{(\frac{7}{2})} 
 + e_{69} \, {\bf T} {\bf T} {\bf T} D  {\bf W}_{-\frac{1}{3}}^{(\frac{7}{2})} 
\right. \nonu \\
&& + e_{70} \, 
{\bf T}  [D, \overline{D}] {\bf T} D {\bf W}_{-\frac{1}{3}}^{(\frac{7}{2})} 
+ e_{71} \, 
{\bf T} D {\bf T}  [D, \overline{D}] {\bf W}_{-\frac{1}{3}}^{(\frac{7}{2})} 
 + e_{72} \, {\bf T} D {\bf T} \pa  {\bf W}_{-\frac{1}{3}}^{(\frac{7}{2})} 
+ e_{73} \, \overline{D} {\bf T} D {\bf T} D  {\bf W}_{-\frac{1}{3}}^{(\frac{7}{2})} 
\nonu \\
&& + e_{74} \,
[D, \overline{D}]  {\bf T} \pa D {\bf W}_{-\frac{1}{3}}^{(\frac{7}{2})} 
 +  e_{75}\,  \pa 
[D, \overline{D}]  {\bf T} D  {\bf W}_{-\frac{1}{3}}^{(\frac{7}{2})}
+ e_{76} \, D {\bf T} \pa  [D, \overline{D}]   {\bf W}_{-\frac{1}{3}}^{(\frac{7}{2})}
+ e_{77} \, D {\bf T} \pa^2  {\bf W}_{-\frac{1}{3}}^{(\frac{7}{2})}
\nonu \\
& & + e_{78} \, 
\pa D {\bf T}  [D, \overline{D}]   {\bf W}_{-\frac{1}{3}}^{(\frac{7}{2})}
+ e_{79}\, \pa D {\bf T} \pa  {\bf W}_{-\frac{1}{3}}^{(\frac{7}{2})}
+ e_{80} \, \pa {\bf T} \pa D  {\bf W}_{-\frac{1}{3}}^{(\frac{7}{2})}
 + e_{81} \, \pa {\bf T} {\bf T} D {\bf W}_{-\frac{1}{3}}^{(\frac{7}{2})}
\nonu \\
&& + e_{82} \, \pa^2 {\bf T} D {\bf W}_{-\frac{1}{3}}^{(\frac{7}{2})}
+ e_{83} \, \pa^2 D {\bf T} {\bf W}_{-\frac{1}{3}}^{(\frac{7}{2})}
+ e_{84} \, \pa {\bf T} D {\bf T}  {\bf W}_{-\frac{1}{3}}^{(\frac{7}{2})}
+ e_{85} \, \pa D {\bf T} {\bf T}  {\bf W}_{-\frac{1}{3}}^{(\frac{7}{2})}
+  e_{86} \,
[D, \overline{D}] {\bf T} D {\bf T} {\bf W}_{-\frac{1}{3}}^{(\frac{7}{2})}
\nonu \\
& & +  \left. \left.
e_{87} \,
{\bf T} {\bf T} D {\bf T} {\bf W}_{-\frac{1}{3}}^{(\frac{7}{2})}
  \right](Z_2) \right) +\cdots,
\label{33OPE}
\eea
}
where the coefficients are given by
{\small
\bea
e_2& = & -\frac{1}{2}, \qquad
e_3 =  -\frac{3}{11},\qquad
e_4 =  \frac{3 }{44},\qquad
e_5 =  0,\qquad
e_6 =  \frac{25 }{44},\qquad
e_7 =  -\frac{(7 c+3) }{22 (c+9)},\nonu \\
e_8& = & -\frac{30 }{11 (c+9)},\qquad
e_9 =  \frac{10 }{(c+9)},\qquad
e_{10} =  -\frac{3}{22},\qquad
e_{11} =  0,\qquad
e_{12} =  0,\qquad
\nonu \\
e_{13} & = &  \frac{30 }{11 (c+9)},\nonu \\
e_{14}& = & \frac{3 (45 c^4+1326 c^3-6567 c^2-24340 c+82596) }
{44 (c+9) (27 c-46) (3 c^2+90 c-265)},
\nonu \\
e_{15} & = &  -\frac{9 (39 c^3-984 c^2+18923 c-22118) }
{44 (c+9) (27 c-46) (3 c^2+90 c-265)},\nonu \\
e_{16}& = & -\frac{36 (31 c-35) }{(27 c-46) (3 c^2+90 c-265)},
\qquad
e_{17} =  \frac{3 (15 c^3-6552 c^2-121229 c+352266) }
{44 (c+9) (27 c-46) (3 c^2+90 c-265)},\nonu \\
e_{18}& = & -\frac{9 (687 c^3+7144 c^2-25501 c-6790) }
{22 (c+9) (27 c-46) (3 c^2+90 c-265)},\qquad
e_{19} =  -\frac{(51 c^2+614 c-1725) }{4 (c+9) (3 c^2+90 c-265)},\nonu \\
e_{20}& = & \frac{3 (45 c^3-2904 c^2+4281 c+5998) }{2 (c+9) (27 c-46) (3 c^2+90 c-265)},\qquad
e_{21} =  -\frac{3 (3 c^3-1544 c^2-35529 c+108050) }{4 (c+9) (27 c-46) (3 c^2+90 c-265)},\nonu \\
e_{22}& = & \frac{(630 c^4+16881 c^3+18150 c^2-385343 c+529542) }
{44 (c+9) (27 c-46) (3 c^2+90 c-265)},\nonu \\
e_{23}& = & -\frac{(765 c^4+11139 c^3-276591 c^2+897037 c-685470) }
{88 (c+9) (27 c-46) (3 c^2+90 c-265)},\nonu \\
e_{24}& = & -\frac{3 (1569 c^3+44040 c^2-314899 c+453110) }
{44 (c+9) (27 c-46) (3 c^2+90 c-265)},\qquad
e_{25} =  \frac{18 (31 c-35) }{(27 c-46) (3 c^2+90 c-265)},\nonu \\
e_{26}& = & \frac{(8235 c^3+41664 c^2-187409 c+40170) }
{44 (c+9) (27 c-46) (3 c^2+90 c-265)},\qquad
e_{27} =  \frac{3 (279 c^3+1368 c^2-15255 c+30428) }{11 (c+9) (27 c-46) (3 c^2+90 c-265)},\nonu \\
e_{28}& = & \frac{(19485 c^3+630744 c^2-2712543 c+2420134) }{44 (c+9) (27 c-46) (3 c^2+90 c-265)},
\nonu \\
e_{29}  & = &  
-\frac{3 (861 c^3+21992 c^2-45423 c-59690) }{11 (c+9) (27 c-46) (3 c^2+90 c-265)},\nonu \\
e_{30}& = & \frac{(2313 c^3+48408 c^2-334171 c+468230) }{4 (c+9) (27 c-46) (3 c^2+90 c-265)},\qquad
e_{31} =  \frac{36 (31 c-35) }{(27 c-46) (3 c^2+90 c-265)},\nonu \\
e_{32}& = & -\frac{3 
}{44 d_2(c)}
(3 c-1)(3c-2) (45 c^4+1302 c^3+1407 c^2-42794 c+75240),
\nonu \\
e_{33}& = & \frac{9 
}{22 d_2(c)}
(3c-2)(9 c^4-528 c^3-8139 c^2+103598 c-178320),
\nonu \\
e_{34}& = & \frac{216 
}{11 d_2(c) }
(3c-2)(42 c^3+614 c^2-1651 c+1385),
\nonu \\
e_{35}& = & \frac{36 
}{d_2(c)}
(3c-2)(9 c^3-372 c^2+3333 c-4030),
\nonu \\
e_{36}& = & \frac{3 
}{22 d_2(c)}
(3c-2)(2025 c^4+27756 c^3-38283 c^2-269878 c+353640),
\nonu \\
e_{37}& = & \frac{9 
}{11 d_2(c)}
(3c-2)(27 c^4-1575 c^3+507 c^2+16059 c-10778),
\nonu \\
e_{38}& = & -\frac{9 
}{22 d_2(c) }
(3c-2)(45 c^4-4242 c^3-36687 c^2+154420 c-87036),
\nonu \\
e_{39}& = & \frac{9 
}{2  d_2(c)}
(3c-2)(9 c^4-884 c^3-7131 c^2+26506 c-11080),
\nonu \\
e_{40}& = & -\frac{18 
}{11 d_2(c)}
(3c-2)(9 c^4-390 c^3+4957 c^2+17064 c-78580), 
\nonu \\
e_{41}& = & \frac{90 }{11 (c+9) (3 c+4)},\qquad
e_{42} =  \frac{15 (3 c+1) }{11 (c+9) (3 c+4)},\qquad
e_{43} =  \frac{30 (2 c+5) }{11 (c+9) (3 c+4)},\nonu \\
e_{44}& = & \frac{18 
}{d_2(c)}
(3c-2)(9 c^3-372 c^2+3333 c-4030),
\nonu \\
e_{45}& = & \frac{3} 
{88 d_2(c)}
(3c-2)(81 c^5+1395 c^4-32979 c^3-48651 c^2+299050 c+73344),
\nonu \\
e_{46}& = & -\frac{9} 
{44 d_2(c)}
(3c-2)(63 c^4-1662 c^3+22347 c^2+56468 c-133396),
\nonu \\
e_{47} & = & -\frac{54} 
{11 d_2(c)}
(3c-2)(42 c^3+614 c^2-1651 c+1385),
\qquad
e_{48} =  0, \nonu \\
e_{49}& = & -\frac{18} 
{11 d_2(c)}
(3c-2)(1149 c^3+11258 c^2-4612 c-9705),
\nonu \\
e_{50}& = & \frac{216 }
{11 d_2(c)}
(3c-2)(42 c^3+614 c^2-1651 c+1385),
\nonu \\
e_{51}& = & -\frac{9} 
{d_2(c)}
(3c-2)(9 c^3-372 c^2+3333 c-4030),
\nonu \\
e_{52}& = & -\frac{18 }
{d_2(c)}
(3c-2)(9 c^3-372 c^2+3333 c-4030),
\nonu \\
e_{53}& = & \frac{3} 
{44 d_2(c)}
(3c-2)(9 c^4-9006 c^3-182739 c^2+130544 c+972012),
\nonu \\
e_{54}& = & -\frac{9 }
{11 d_2(c)}
(3c-2)(513 c^4+6177 c^3-9947 c^2-56493 c+37790),
\nonu \\
e_{55}& = & -\frac{3} 
{11 d_2(c)}
(3c-2)(2079 c^4+22392 c^3-52749 c^2-48622 c-217560),
\nonu \\
e_{56}  & = &  -\frac{3 
(81 c^2+87 c-698) }{88 (3 c+4) (3 c^2+90 c-265)},\nonu \\
e_{57}& = & -\frac{1}
{88 d_2(c)}
(3c-2)(51111 c^4+651690 c^3-1627485 c^2-2872900 c+5335644),
\nonu \\
e_{58}& = & -\frac{1}
{2 d_2(c)}
(3c-2)(1053 c^4+10062 c^3-57927 c^2+44332 c+54420),
\nonu \\
e_{59}& = & \frac{3} 
{2 d_2(c)}
(3c-2)(81 c^4-5274 c^3-4251 c^2+42996 c-2812),
\nonu \\
e_{60}& = & \frac{3} 
{d_2(c)}
(3c-2)(27 c^4-1470 c^3+5655 c^2-11100 c+14308),
\nonu \\
e_{61}& = & -\frac{9} 
{88 d_2(c)}
(3c-2)(135 c^4-4386 c^3+34019 c^2-26800 c+204292),
\nonu \\
e_{62}& = & -\frac{9 
}{d_2(c)}
(3c-2)(381 c^3+3052 c^2+3457 c-9070),
\nonu \\
e_{63}& = & \frac{3} 
{88 d_2(c)}
(3c-2)(189 c^4+13818 c^3+799921 c^2-762464 c-4549204),
\nonu \\
e_{64}& = & -\frac{3} 
{2 d_2(c)}
(3c-2)(9 c^4-1818 c^3-33539 c^2+65388 c+99740),
\nonu \\
e_{65}& = & \frac{1}
{264 d_2(c)}
(3c-2)(2511 c^5+31365 c^4-242229 c^3-252861 c^2+1795390 c-786096),
\nonu \\
e_{66}& = & -\frac{5} 
{66 d_2(c)}
(243 c^6-513 c^5-109449 c^4+391569 c^3-21342 c^2-1057292 c+1013112),
\nonu \\
e_{67}& = & -\frac{15} 
{11 d_2(c)}
(189 c^5+3501 c^4-46959 c^3+36513 c^2+251732 c-301892),
\nonu \\
e_{68}& = & \frac{90} 
{d_2(c)}
(36 c^4-21 c^3-575 c^2+4846 c-6616),
\nonu \\
e_{69}& = & \frac{1080} 
{11 d_2(c)}
(252 c^3+141 c^2-3017 c+3694),
\nonu \\
e_{70}& = & \frac{15} 
{11 d_2(c)}
(6075 c^4-11664 c^3+56673 c^2-338112 c+406316), \nonu \\
e_{71}& = & \frac{45} 
{11 d_2(c)}
(108 c^4-1809 c^3-23475 c^2+85134 c-71288),
\nonu \\
e_{72}& = & \frac{15} 
{11 d_2(c)}
(3888 c^4+37935 c^3-39735 c^2-289890 c+522872),
\nonu \\
e_{73}& = & \frac{90} 
{11 d_2(c)}
(3456 c^4-6345 c^3-29889 c^2+71576 c-55076),
\nonu \\
e_{74}& = & \frac{5} 
{22 d_2(c)}
(4617 c^5-4077 c^4-41121 c^3-114363 c^2+822908 c-1024116),
\nonu \\
e_{75}& = & \frac{5} 
{11 d_2(c)}
(1944 c^5+18630 c^4-20313 c^3-82179 c^2-351932 c+491628),
\nonu \\
e_{76}& = & \frac{15} 
{22 d_2(c)}
(567 c^5-2214 c^4-31635 c^3+132786 c^2-138528 c+108592),
\nonu \\
e_{77}& = & \frac{5} 
{22 d_2(c)}
(5265 c^5+134136 c^4-548721 c^3+254544 c^2+735192 c-751784),
\nonu \\
e_{78}& = & \frac{15} 
{22 d_2(c)}
(648 c^5+5409 c^4+444 c^3-84405 c^2+295772 c-236180),
\nonu \\
e_{79}& = & \frac{5} 
{22 d_2(c)}
(12636 c^5+303129 c^4-1117764 c^3-365481 c^2+2621708 c-954148),
\nonu \\
e_{80}& = & -\frac{15} 
{22 d_2(c)}
(1539 c^5+37377 c^4-54291 c^3-201209 c^2-136724 c+509396),
\nonu \\
e_{81}& = & \frac{45} 
{11 d_2(c)}
(1179 c^4+11010 c^3+7403 c^2-144728 c+206180),
\nonu \\
e_{82}& = & -\frac{15} 
{11 d_2(c)}
(783 c^5+17685 c^4-41751 c^3-60493 c^2+107556 c-108844),
\nonu \\
e_{83}& = & \frac{5} 
{d_2(c)}
(405 c^5+3150 c^4-57864 c^3+110343 c^2+142176 c-200156),
\nonu \\
e_{84}& = & \frac{15} 
{d_2(c)}
(459 c^4+6075 c^3-444 c^2-39272 c+2648),
\nonu \\
e_{85}& = & \frac{30} 
{d_2(c)}
(216 c^4+1917 c^3+1653 c^2+14448 c-27964), 
\nonu \\
e_{86} & = &  -\frac{5}{d_2(c)}
(3c+4)(27c-46)(81 c^2-183 c-44),
\nonu \\
e_{87}& = & -\frac{360}{d_2(c)} (252 c^3+141 c^2-3017 c+3694),
\nonu \\
d_2(c) & \equiv & (c+9) (3 c-2) (3 c+4) (27 c-46) (3 c^2+90 c-265).
\label{ecoeff}
\eea }
In the large $c$ limit,
all the nonlinear terms corresponding to the coefficients 
Appendix (\ref{ecoeff}) disappear.
One can also analyze the $\frac{1}{c}, \cdots, \frac{1}{c^3}$-terms. 
We introduce $d_2(c)$ in Appendix 
(\ref{ecoeff}) which appears in the denominators 
of the coefficients.

Now the OPE 
between the third higher spin ${\cal N}=2$ multiplet 
and the fourth higher spin ${\cal N}=2$ multiplet 
in (\ref{manyW}) can be described as
{\small
\bea
&& {\bf W}_{\frac{1}{3}}^{(\frac{7}{2})}(Z_1) \, 
{\bf W}_{-\frac{1}{3}}^{(\frac{7}{2})}(Z_2) = 
\frac{\theta_{12} \, \bar{\theta}_{12}}{z_{12}^8} \, g_{379} +
\frac{1}{z_{12}^7} \, \frac{2c}{7}+
\frac{\theta_{12} \, \bar{\theta}_{12}}{z_{12}^7} \, g_{380} \, {\bf T}(Z_2)
+ \frac{1}{z_{12}^6} \, g_{381} \, {\bf T}(Z_2)
\nonu \\
&& +\frac{\theta_{12}}{z_{12}^6} \, g_{382} \, D {\bf T}(Z_2)
\nonu \\
&&+ \frac{\bar{\theta}_{12}}{z_{12}^6} \, g_{383} \, \overline{D} {\bf T}(Z_2)
+ \frac{\theta_{12} \, \bar{\theta}_{12}}{z_{12}^6} \, 
\left[ g_{384} \, [ D, \overline{D} ] {\bf T}
+ g_{385} \, {\bf T} {\bf T} + g_{386} \, \pa {\bf T} \right]
(Z_2)
\nonu \\
&&+ \frac{1}{z_{12}^5} \, \left[ g_{387} \,
[D, \overline{D} ] {\bf T} + g_{388} \, {\bf T} 
{\bf T} + g_{399} \, \pa {\bf T} \right](Z_2) 
 + \frac{\theta_{12}}{z_{12}^5} \, \left[ 
g_{389} \, \pa D {\bf T} + g_{390} \, {\bf T} D {\bf T}
\right](Z_2)
\nonu \\
&&+ \frac{\bar{\theta}_{12}}{z_{12}^5} \, \left[ 
g_{391}\, \pa \overline{D} {\bf T} 
+ g_{392} \, {\bf T} \overline{D} {\bf T}
\right](Z_2)
\nonu \\
&& +  \frac{\theta_{12} \, \bar{\theta}_{12}}{z_{12}^5} \, \left[  
g_{393} \, \pa [ D,  \overline{D} ] {\bf T} + 
g_{394} \, {\bf T} [D, \overline{D}] {\bf T}
+ g_{395} \, {\bf T} {\bf T} {\bf T}
+ g_{396} \, \overline{D} {\bf T} D {\bf T}
+   g_{397} \, \pa {\bf T} {\bf T} 
\right. \nonu \\
&& \left. + 
g_{398} \, \pa^2 {\bf T}
\right](Z_2)
\nonu \\
&&+ \frac{1}{z_{12}^4} \, \left[ g_{400} \, \pa [D, \overline{D}] {\bf T}
+ g_{401} \, {\bf T} [D, \overline{D}] {\bf T} + 
g_{402} \, {\bf T} {\bf T} {\bf T}
+ g_{403} \, \overline{D} {\bf T} D {\bf T} 
 +  g_{404} \, \pa {\bf T} {\bf T}
+ g_{428} \, \pa^2 {\bf T} \right](Z_2)
\nonu \\
&&+ \frac{\theta_{12}}{z_{12}^4} \, \left[ 
g_{405} \, \pa^2 D {\bf T} + g_{406} \, {\bf T} {\bf T} D {\bf T}
+ g_{407} \, [D, \overline{D}] {\bf T} D {\bf T}
+ g_{408} \, \pa D {\bf T} {\bf T}
+   g_{409} \, \pa {\bf T} D {\bf T}
\right](Z_2)
\nonu \\
&&+ 
 \frac{\bar{\theta}_{12}}{z_{12}^4} \, \left[ 
g_{410} \, \pa^2 \overline{D} {\bf T} + 
g_{411} \, {\bf T} {\bf T} \overline{D} {\bf T}
+ g_{412} \, \overline{D} {\bf T} [D, \overline{D}] {\bf T} 
+ g_{413} \, \pa \overline{D} {\bf T} {\bf T}
+   g_{414} \, \pa {\bf T} \overline{D} {\bf T}
\right](Z_2)
\nonu \\
&&+ \frac{\theta_{12} \, \bar{\theta}_{12}}{z_{12}^4}
\left[ g_{415} \, \pa^2 [ D, \overline{D} ] {\bf T}
+ g_{416} \, {\bf T} {\bf T} [ D, \overline{D} ] {\bf T}
+ g_{417} {\bf T} {\bf T} {\bf T} {\bf T} 
+ g_{418} \, {\bf T} \overline{D} {\bf T}
D {\bf T} + g_{419} \,
\pa \overline{D} {\bf T} D {\bf T}
\right. \nonu \\
& & +  g_{420} \, [ D, \overline{D} ] {\bf T}  [ D, \overline{D} ] {\bf T}
+ g_{421}\, \pa  [ D, \overline{D} ] {\bf T} {\bf T}
+  g_{422} \, \pa D {\bf T}  \overline{D} {\bf T}
\nonu \\
&&+ \left.  g_{423} \, \pa {\bf T} [ D, \overline{D} ] {\bf T}
+ g_{424} \, \pa {\bf T} {\bf T} {\bf T} + 
g_{425} \, \pa {\bf T} \pa {\bf T}
+ g_{426} \, \pa^2 {\bf T} {\bf T} + 
g_{427} \, \pa^3 {\bf T}
\right](Z_2)
\nonu \\
&&+  \frac{1}{z_{12}^3}
\left[ g_1 \, \pa^2 [ D, \overline{D} ] {\bf T}
+ g_2 \, {\bf T} {\bf T} [ D, \overline{D} ] {\bf T}
+ g_3 \, {\bf T} {\bf T} {\bf T} {\bf T} +
g_4 \, {\bf T} \overline{D} {\bf T}
D {\bf T} + g_5 \,
\pa \overline{D} {\bf T} D {\bf T}
\right. \nonu \\
&& +  g_6 \, [ D, \overline{D} ] {\bf T}  [ D, \overline{D} ] {\bf T}
+ g_7 \, \pa  [ D, \overline{D} ] {\bf T} {\bf T}
+  g_8 \, \pa D {\bf T}  \overline{D} {\bf T}
\nonu \\
&&+ \left.  g_9 \, \pa {\bf T} [ D, \overline{D} ] {\bf T}
+ g_{10} \, \pa {\bf T} {\bf T} {\bf T} 
+ g_{11} \, \pa {\bf T} \pa {\bf T}
+ g_{12} \, \pa^2 {\bf T} {\bf T} 
+ g_{57} \, \pa^3 {\bf T}
\right](Z_2)
\nonu \\
&&+ \frac{\theta_{12}}{z_{12}^3}
\, \left[ 
g_{13} \, \pa^3 D {\bf T} + 
g_{14} \, {\bf T} {\bf T} {\bf T} D {\bf T} 
+ g_{15} \, {\bf T}  [ D, \overline{D} ] {\bf T} D {\bf T}
+ g_{16} \, \pa [ D, \overline{D} ] {\bf T} D {\bf T}
+   g_{17} \, \pa D {\bf T}  [ D, \overline{D} ] {\bf T}
\right. \nonu \\
&& + \left.
g_{18} \, \pa D {\bf T} {\bf T} {\bf T}
+ g_{19} \, \pa D {\bf T}  \pa {\bf T}
+   g_{20} \, \pa^2 D {\bf T} {\bf T}
+ g_{21}\,  \pa {\bf T} {\bf T} D {\bf T}
+g_{22} \, \pa^2 {\bf T} D {\bf T}
\right](Z_2)
\nonu \\
&&+ \frac{\bar{\theta}_{12}}{z_{12}^3}
\, \left[ g_{23} \,
\pa^3 \overline{D} {\bf T} + 
g_{24} \, {\bf T} {\bf T} {\bf T} \overline{D} {\bf T} 
+ g_{25} \, {\bf T}  \overline{D} {\bf T} [ D, \overline{D} ] {\bf T} 
+ g_{26} \, \pa \overline{D} {\bf T}  [ D, \overline{D} ] {\bf T}
 +   g_{27} \, \pa \overline{D} {\bf T}  {\bf T} {\bf T}
\right. \nonu \\
&& +  \left. g_{28}\,  \pa \overline{D} {\bf T} \pa {\bf T}
+ g_{29} \, \pa^2 \overline{D} {\bf T}   {\bf T}
 +   g_{30} \, [ D, \overline{D} ] {\bf T} \overline{D} {\bf T}
+  g_{31} \,\pa {\bf T} {\bf T} \overline{D} {\bf T}
+ g_{32} \, \pa^2 {\bf T} \overline{D} {\bf T}
\right](Z_2)
\nonu \\
&&+ \frac{\theta_{12} \, \bar{\theta}_{12}}{z_{12}^3}
\left[ g_{33} \, \pa^3 [ D, \overline{D} ] {\bf T} + 
g_{34} \, {\bf T}  {\bf T} {\bf T} 
[ D, \overline{D} ] {\bf T} +
g_{35} \, {\bf T}  {\bf T} {\bf T}   {\bf T}  {\bf T} 
+  g_{36} \, {\bf T}  {\bf T}   \overline{D} {\bf T} D {\bf T}
\right. \nonu \\
&& +  
g_{37} \, {\bf T} [ D, \overline{D} ] {\bf T} [ D, \overline{D} ] {\bf T}
+  
g_{38} \, \overline{D} {\bf T} [ D, \overline{D} ] {\bf T} D {\bf T}
+  g_{39} \, \pa \overline{D} {\bf T} \pa  D {\bf T}
+  g_{40} \,  \pa \overline{D} {\bf T} {\bf T}  D {\bf T}
\nonu \\
&& +  g_{41} \, \pa^2 \overline{D} {\bf T} D {\bf T} +
 g_{42} \, \pa  [ D, \overline{D} ] {\bf T}  [ D, \overline{D} ] {\bf T}
+ g_{43} \, \pa [ D, \overline{D} ] {\bf T} {\bf T} {\bf T} 
 + g_{44} \, \pa [ D, \overline{D} ] {\bf T} \pa {\bf T} 
\nonu \\
&& + g_{45} \,
\pa^2 [ D, \overline{D} ] {\bf T}  {\bf T} 
+  g_{46} \, \pa D {\bf T} {\bf T} \overline{D} {\bf T}
+  g_{47} \, \pa^2 D {\bf T}  \overline{D} {\bf T}
+  g_{48} \, \pa {\bf T}  {\bf T} [ D, \overline{D} ] {\bf T}
+ 
 g_{49} \, \pa {\bf T} {\bf T} {\bf T} {\bf T} 
\nonu \\
&&
+  g_{50} \, \pa {\bf T} \overline{D} {\bf T} D {\bf T}
+  g_{51} \, \pa {\bf T} \pa {\bf T} {\bf T}
\nonu \\
&& +   \left. g_{52} \, \pa^2 {\bf T} [ D, \overline{D} ] {\bf T}
+   g_{53} \, \pa^2 {\bf T} {\bf T} {\bf T}
+  g_{54} \, \pa^2 {\bf T} \pa {\bf T}
+  g_{55} \, \pa^3 {\bf T}  {\bf T}
+  g_{56} \, \pa^4 {\bf T}
\right](Z_2)
\nonu \\
&&+  \frac{1}{z_{12}^2}
\left[ g_{58}\, \pa^3 [ D, \overline{D} ] {\bf T} 
+ g_{59} \, {\bf T}  {\bf T} {\bf T} 
[ D, \overline{D} ] {\bf T} 
+ g_{60} \, {\bf T}  {\bf T} {\bf T}   {\bf T}  {\bf T} 
+  g_{61} \, {\bf T}  {\bf T}   \overline{D} {\bf T} D {\bf T}
\right. 
\nonu \\
&&  +  g_{62} \,
{\bf T} [ D, \overline{D} ] {\bf T} [ D, \overline{D} ] {\bf T}
 \nonu \\
&& +  g_{63} \, \overline{D} {\bf T} [ D, \overline{D} ] {\bf T} D {\bf T}
+  g_{64} \, \pa \overline{D} {\bf T} \pa  D {\bf T}
+   g_{65} \, \pa \overline{D} {\bf T} {\bf T}  D {\bf T}
+  g_{66} \, \pa^2 \overline{D} {\bf T} D {\bf T} 
\nonu \\
&& +
 g_{67} \, \pa [ D, \overline{D} ] {\bf T} [ D, \overline{D}] {\bf T} 
+  g_{68} \, \pa [ D, \overline{D} ] {\bf T} {\bf T} {\bf T} 
+  g_{69} \, \pa [ D, \overline{D} ] {\bf T} \pa {\bf T}
+ g_{70} \, \pa^2 [ D, \overline{D} ] {\bf T}  {\bf T} 
\nonu \\
&& +  g_{71} \, \pa D {\bf T} {\bf T} \overline{D} {\bf T}
+  g_{72} \, \pa^2 D {\bf T}  \overline{D} {\bf T}
+   g_{73} \, \pa {\bf T}  {\bf T} [ D, \overline{D} ] {\bf T}
+  g_{74} \,
\pa {\bf T} {\bf T} {\bf T} {\bf T} 
+  g_{75} \, \pa {\bf T} \overline{D} {\bf T} D {\bf T}
\nonu \\
&& +
\left.    g_{76} \, \pa {\bf T} \pa {\bf T} {\bf T}
+    g_{77} \, \pa^2 {\bf T} [ D, \overline{D} ] {\bf T}
+   g_{78} \, \pa^2 {\bf T} {\bf T} {\bf T}
+  g_{79} \, \pa^2 {\bf T} \pa {\bf T}
+  g_{80} \, \pa^3 {\bf T}  {\bf T}
+  g_{170} \, \pa^4 {\bf T}
\right](Z_2)
\nonu \\
&&+ \frac{\theta_{12}}{z_{12}^2}
\left[ g_{81} \,
\pa^4 D {\bf T}
+ g_{82} \, {\bf T} {\bf T} {\bf T} {\bf T} D {\bf T}
+ g_{83} \, {\bf T} {\bf T}  [ D, \overline{D} ] {\bf T} D {\bf T}
 +  g_{84} \, [ D, \overline{D} ] {\bf T} [ D, \overline{D} ] {\bf T}
D {\bf T}
\right. 
\nonu \\
&&+ 
 g_{85} \, \pa [ D, \overline{D} ] {\bf T} 
\pa D {\bf T}
+ g_{86} \, \pa [ D, \overline{D} ] {\bf T} 
{\bf T} D {\bf T}
+ g_{87} \, \pa^2 [ D, \overline{D} ] {\bf T} 
 D {\bf T}
+ 
g_{88} \, \pa D {\bf T} {\bf T} [ D, \overline{D} ] {\bf T} 
\nonu \\
&& 
+ g_{89} \, \pa D {\bf T} {\bf T} {\bf T} {\bf T}
+ 
  g_{90} \, \pa D {\bf T} \overline{D} {\bf T} D {\bf T}
+ g_{91} \, \pa D {\bf T} \pa {\bf T} {\bf T}
+ g_{92} \, \pa^2 D {\bf T}  [ D, \overline{D} ] {\bf T} 
+  g_{93} \, \pa^2 D {\bf T} {\bf T} {\bf T}
\nonu \\
&&+ 
g_{94} \, \pa^2 D {\bf T} \pa {\bf T} + 
g_{95} \, \pa^3 D {\bf T} {\bf T} 
+  g_{96} \, \pa {\bf T} {\bf T} {\bf T} D {\bf T} 
+ g_{97}\, \pa {\bf T} [ D, \overline{D} ] {\bf T} D {\bf T}
\nonu \\
&&+ \left. 
g_{98} \, \pa {\bf T} \pa {\bf T} D {\bf T}
+  g_{99}\, \pa^2 {\bf T} \pa D {\bf T}
+ g_{100}\,  \pa^2 {\bf T} {\bf T} D {\bf T}
+ g_{101} \, \pa^3 {\bf T} D {\bf T}
\right](Z_2)
\nonu \\
&&+ \frac{\bar{\theta}_{12}}{z_{12}^2}
\left[ 
g_{102} \, \pa^4 \overline{D} {\bf T}
+ g_{103} \, {\bf T} {\bf T} {\bf T} {\bf T} \overline{D} {\bf T}
+ g_{104} \, {\bf T} {\bf T}  \overline{D} {\bf T} [ D, \overline{D} ] {\bf T}
+  g_{105} \, \overline{D} {\bf T}
[ D, \overline{D} ] {\bf T} [ D, \overline{D} ] {\bf T}
\right. \nonu \\
&& + g_{106}
 \, \pa \overline{D} {\bf T} \pa [ D, \overline{D} ] {\bf T} 
+ g_{107} \, \pa  \overline{D} {\bf T} {\bf T} [ D, \overline{D} ] {\bf T} 
+   g_{108} \, \pa  \overline{D}  {\bf T} 
  {\bf T} {\bf T} {\bf T}
+ g_{109} \,
\pa \overline{D} {\bf T}  \overline{D}  {\bf T} D {\bf T} 
\nonu \\
&& + g_{110} \, \pa  \overline{D} {\bf T} \pa {\bf T}  {\bf T} 
+ g_{111} \, \pa^2 \overline{D} {\bf T} [ D, \overline{D} ] {\bf T}
+ g_{112} \,
 \pa^2 \overline{D} {\bf T}  {\bf T}  {\bf T}
+  g_{113} \, \pa^2 \overline{D} {\bf T}  \pa  {\bf T}
\nonu \\
&& +  g_{114} \, \pa^3 \overline{D} {\bf T} {\bf T} +
g_{115} \, \pa [ D, \overline{D} ] {\bf T} {\bf T} \overline{D} {\bf T} 
 +  g_{116} \,
\pa^2 [ D, \overline{D} ] {\bf T}  \overline{D} {\bf T} 
+ g_{117} \, \pa {\bf T} {\bf T} {\bf T} \overline{D} {\bf T} 
\nonu \\
&& \left. 
+ g_{118} \, \pa {\bf T}  \overline{D} {\bf T} [ D, \overline{D} ] {\bf T}
+  g_{119} \,
\pa {\bf T} \pa {\bf T} \overline{D} {\bf T}
+ g_{120}\,  \pa^2 {\bf T} \pa \overline{D} {\bf T}
+ g_{121}\, \pa^2 {\bf T} {\bf T} \overline{D} {\bf T}
+ g_{122} \, \pa^3 {\bf T} \overline{D} {\bf T}
\right](Z_2)
\nonu \\
&&+ \frac{\theta_{12} \, \bar{\theta}_{12}}{z_{12}^2}
\left[ g_{123} \, \pa^4  [ D, \overline{D} ] {\bf T}
+ g_{124}\, {\bf T} {\bf T} {\bf T} {\bf T} [ D, \overline{D} ]  {\bf T}
+ g_{125} \, {\bf T} {\bf T} {\bf T} {\bf T} {\bf T}  {\bf T}
+  g_{126} \, {\bf T} {\bf T} {\bf T} \overline{D} {\bf T} D {\bf T}
\right. \nonu \\
&& + g_{127} \,
 {\bf T} {\bf T}  [ D, \overline{D} ] {\bf T}  [ D, \overline{D} ] {\bf T}
+ g_{128} \, {\bf T} \overline{D} {\bf T}  [ D, \overline{D} ] {\bf T} 
D {\bf T}
+  g_{129} \, \pa \overline{D} {\bf T} {\bf T} {\bf T} D {\bf T}
\nonu \\
&& 
+ g_{130}\, \pa \overline{D} {\bf T}  [ D, \overline{D} ] {\bf T}  D {\bf T}
\nonu \\
&& 
+ g_{131}\, \pa \overline{D} {\bf T}  \pa  D {\bf T} {\bf T}
+  g_{132}\, \pa \overline{D} {\bf T} \pa {\bf T} D {\bf T}
+  g_{133} \, \pa^2 \overline{D} {\bf T} \pa  D {\bf T}
+  g_{134} \, \pa^2 \overline{D} {\bf T} {\bf T}  D {\bf T}
\nonu \\
&& + g_{135} \, \pa^3 \overline{D} {\bf T} D {\bf T}
+ g_{136} \, [ D, \overline{D} ] {\bf T}  [ D, \overline{D} ] {\bf T}  
[ D, \overline{D} ] {\bf T}
+g_{137} \, \pa [ D, \overline{D} ] {\bf T} \pa [ D, \overline{D} ] {\bf T}
\nonu \\
&& +  g_{138} \, 
\pa [ D, \overline{D} ] {\bf T} {\bf T}  [ D, \overline{D} ] {\bf T}
+ g_{139} \, \pa [ D, \overline{D} ] {\bf T} {\bf T} {\bf T} {\bf T}
+ g_{140} \, \pa [ D, \overline{D} ] {\bf T} \overline{D} {\bf T}   D {\bf T}
\nonu \\
&& 
+g_{141}\,  \pa [ D, \overline{D} ] {\bf T} \pa {\bf T} {\bf T}
\nonu \\
&& + g_{142} \,
\pa^2 [ D, \overline{D} ] {\bf T}  [ D, \overline{D} ] {\bf T} 
+ g_{143} \, \pa^2 [ D, \overline{D} ] {\bf T}   {\bf T} {\bf T}
+ g_{144} \, \pa^2 [ D, \overline{D} ] {\bf T}   \pa {\bf T} 
+  g_{145}\, \pa^3 [ D, \overline{D} ] {\bf T}   {\bf T} 
\nonu \\
& &
+   g_{146} \, \pa D {\bf T} {\bf T} {\bf T} \overline{D} {\bf T}
+  g_{147} \, \pa D {\bf T}  \overline{D} {\bf T} [ D, \overline{D} ] {\bf T}
+  g_{148} \, \pa D {\bf T} \pa {\bf T}  \overline{D} {\bf T}
+   g_{149} \, \pa^2 D {\bf T}  \pa \overline{D} {\bf T}
\nonu \\
&& +  g_{150} \, \pa^2 D {\bf T}  {\bf T} \overline{D} {\bf T}
+ g_{151} \, \pa^3 D {\bf T}   \overline{D} {\bf T}
+ g_{152} \, \pa {\bf T} {\bf T} {\bf T}  [ D, \overline{D} ] {\bf T}
+ g_{153} \, \pa {\bf T} {\bf T}   {\bf T} {\bf T} {\bf T}
\nonu \\
&& 
+ g_{154}\, \pa {\bf T} {\bf T}   \overline{D} {\bf T} D {\bf T}
\nonu \\
&&+  g_{155} \,
\pa {\bf T}  [ D, \overline{D} ] {\bf T}   [ D, \overline{D} ] {\bf T} 
+ g_{156} \, \pa {\bf T}  \pa {\bf T}   [ D, \overline{D} ] {\bf T} 
+ g_{157} \, \pa {\bf T}  \pa {\bf T}   {\bf T}   {\bf T} 
\nonu \\
&&+  g_{158} \, \pa {\bf T}  \pa {\bf T}  \pa {\bf T}
+ g_{159}\, \pa^2 {\bf T} \pa  [ D, \overline{D} ] {\bf T} 
+ g_{160} \, \pa^2 {\bf T} {\bf T}  [ D, \overline{D} ] {\bf T} 
+ g_{161} \, \pa^2 {\bf T} {\bf T}  {\bf T}   {\bf T} 
\nonu \\
&&+  g_{162} \, \pa^2 {\bf T} \overline{D} {\bf T} D {\bf T}
+  g_{163} \, \pa^2 {\bf T}  \pa {\bf T} {\bf T}
+ g_{164} \, \pa^2 {\bf T} \pa^2 {\bf T}
+ g_{165} \, \pa^3 {\bf T}   [ D, \overline{D} ] {\bf T} 
\nonu \\
&&+ \left.  g_{166} \, \pa^3 {\bf T}  {\bf T} {\bf T}
+  g_{167}\, \pa^3 {\bf T}  \pa {\bf T}
+ g_{168} \, \pa^4 {\bf T} {\bf T} + 
g_{169} \, \pa^5 {\bf T}
\right](Z_2)
\nonu \\
&&+ \frac{1}{z_{12}}
\left[ g_{171} \, \pa^4  [ D, \overline{D} ] {\bf T}
+  g_{172} \, {\bf T} {\bf T} {\bf T} {\bf T} [ D, \overline{D} ]  {\bf T}
+  g_{173} \, {\bf T} {\bf T} {\bf T} {\bf T} {\bf T}  {\bf T}
+  g_{174} \, {\bf T} {\bf T} {\bf T} \overline{D} {\bf T} D {\bf T}
\right. \nonu \\
&&
+ g_{175} \,   {\bf T} {\bf T}  [ D, \overline{D} ] {\bf T}  [ D, \overline{D} ] {\bf T}
+ g_{176} \,  {\bf T} \overline{D} {\bf T}  [ D, \overline{D} ] {\bf T} 
D {\bf T}
+  g_{177} \, \pa \overline{D} {\bf T} {\bf T} {\bf T} D {\bf T}
\nonu \\
&& 
+ g_{178} \,  \pa \overline{D} {\bf T}  [ D, \overline{D} ] {\bf T}  D {\bf T}
+ g_{179} \,  \pa \overline{D} {\bf T}  \pa  D {\bf T} {\bf T}
+  g_{180} \,  \pa \overline{D} {\bf T} \pa {\bf T} D {\bf T}
+  g_{181} \, \pa^2 \overline{D} {\bf T} \pa  D {\bf T}
\nonu \\
&& + g_{182} \,  \pa^2 \overline{D} {\bf T} {\bf T}  D {\bf T}
+  g_{183} \, \pa^3 \overline{D} {\bf T} D {\bf T}
+ g_{184} \,  [ D, \overline{D} ] {\bf T}  [ D, \overline{D} ] {\bf T}  
[ D, \overline{D} ] {\bf T}
\nonu \\
&&+ g_{185} \, \pa [ D, \overline{D} ] {\bf T} \pa [ D, \overline{D} ] {\bf T}
+ g_{186} \, \pa [ D, \overline{D} ] {\bf T} {\bf T}  [ D, \overline{D} ] {\bf T}
+ g_{187} \, \pa [ D, \overline{D} ] {\bf T} {\bf T} {\bf T} {\bf T}
\nonu \\
&&+  g_{188} \,\pa [ D, \overline{D} ] {\bf T} \overline{D} {\bf T}   D {\bf T}
+  g_{189} \, \pa [ D, \overline{D} ] {\bf T} \pa {\bf T} {\bf T}
+  g_{190} \,\pa^2 [ D, \overline{D} ] {\bf T}  [ D, \overline{D} ] {\bf T} 
\nonu \\
&&+ g_{191} \, \pa^2 [ D, \overline{D} ] {\bf T}   {\bf T} {\bf T}
+  g_{192} \,\pa^2 [ D, \overline{D} ] {\bf T}   \pa {\bf T} 
+ g_{193} \,  \pa^3 [ D, \overline{D} ] {\bf T}   {\bf T}
+   g_{194} \, \pa D {\bf T} {\bf T} {\bf T} \overline{D} {\bf T}
\nonu \\
&&
+ g_{195} \, \pa D {\bf T}  \overline{D} {\bf T} [ D, \overline{D} ] {\bf T}
+ g_{196} \,  \pa D {\bf T} \pa {\bf T}  \overline{D} {\bf T}
 +   g_{197} \, \pa^2 D {\bf T}  \pa \overline{D} {\bf T}
+  g_{198} \, \pa^2 D {\bf T}  {\bf T} \overline{D} {\bf T}
\nonu \\
&&+  g_{199} \,\pa^3 D {\bf T}   \overline{D} {\bf T}
+ g_{200} \, \pa {\bf T} {\bf T} {\bf T}  [ D, \overline{D} ] {\bf T}
+ g_{201} \, \pa {\bf T} {\bf T}   {\bf T} {\bf T} {\bf T}
+  g_{202} \, \pa {\bf T} {\bf T}   \overline{D} {\bf T} D {\bf T}
\nonu \\
&&+ g_{203} \,  \pa {\bf T}  [ D, \overline{D} ] {\bf T}   [ D, \overline{D} ] {\bf T} 
+ g_{204} \, \pa {\bf T}  \pa {\bf T}   [ D, \overline{D} ] {\bf T} 
+ g_{205} \, \pa {\bf T}  \pa {\bf T}   {\bf T}   {\bf T} 
\nonu \\
&&+ g_{206} \,  \pa {\bf T}  \pa {\bf T}  \pa {\bf T}
+ g_{207} \, \pa^2 {\bf T} \pa  [ D, \overline{D} ] {\bf T} 
+   g_{208} \,\pa^2 {\bf T} {\bf T}  [ D, \overline{D} ] {\bf T} 
+  g_{209} \, \pa^2 {\bf T} {\bf T}  {\bf T}   {\bf T} 
\nonu \\
&&+  g_{210} \, \pa^2 {\bf T} \overline{D} {\bf T} D {\bf T}
+  g_{211} \, \pa^2 {\bf T}  \pa {\bf T} {\bf T}
+ g_{212} \, \pa^2 {\bf T} \pa^2 {\bf T}
+  g_{213} \, \pa^3 {\bf T}   [ D, \overline{D} ] {\bf T} 
\nonu \\
&&+ \left.   g_{214} \,\pa^3 {\bf T}  {\bf T} {\bf T}
+  g_{215} \, \pa^3 {\bf T}  \pa {\bf T}
+ g_{216} \, \pa^4 {\bf T} {\bf T} + g_{378} \, \pa^5 {\bf T}
\right](Z_2)
\nonu \\
&&+ \frac{\theta_{12}}{z_{12}} \left[ 
 g_{217} \, \pa^5 D {\bf T}
+  g_{218} \,{\bf T}  {\bf T} {\bf T} {\bf T}  {\bf T} D {\bf T}
+ g_{219} \, {\bf T}  {\bf T} {\bf T}   [ D, \overline{D} ] {\bf T} D {\bf T} 
\right. \nonu \\
& &
+  g_{220} \, {\bf T}   [ D, \overline{D} ] {\bf T}
  [ D, \overline{D} ] {\bf T}  D {\bf T}
+ g_{221} \, \pa \overline{D} {\bf T}  \pa D {\bf T}  D {\bf T}
+  g_{222} \, \pa [ D, \overline{D} ] {\bf T}  {\bf T}  {\bf T} D {\bf T}
\nonu \\
&&+  g_{223} \,\pa [ D, \overline{D} ] {\bf T} [ D, \overline{D} ] {\bf T}
D {\bf T}
+  g_{224} \, \pa [ D, \overline{D} ] {\bf T} \pa D {\bf T} {\bf T}
+  g_{225} \, \pa [ D, \overline{D} ] {\bf T}  \pa {\bf T} D {\bf T}
\nonu \\
&&+ g_{226} \,  \pa^2 [ D, \overline{D} ] {\bf T} \pa D {\bf T}
+  g_{227} \, \pa^2 [ D, \overline{D} ] {\bf T} {\bf T} D {\bf T}
+  g_{228} \, \pa^3 [ D, \overline{D} ] {\bf T}  D {\bf T}
\nonu \\
&&+  g_{229} \, \pa D {\bf T}  {\bf T} {\bf T} [ D, \overline{D} ] {\bf T} 
+ g_{230} \,  \pa D {\bf T}  {\bf T} {\bf T} {\bf T} {\bf T}
+  g_{231} \,  \pa D {\bf T}  {\bf T}  \overline{D} {\bf T} D {\bf T}
\nonu \\
&& +  g_{232} \, \pa D {\bf T}  [ D, \overline{D} ] {\bf T}  [ D, \overline{D} ] {\bf T}
\nonu \\
&& +  g_{233} \, \pa D {\bf T} \pa {\bf T}  [ D, \overline{D} ] {\bf T}
+  g_{234} \, \pa D {\bf T} \pa {\bf T}   {\bf T} {\bf T}
+  g_{235} \, \pa D {\bf T} \pa {\bf T} \pa {\bf T}
+  g_{236} \,  \pa^2 D {\bf T}  \pa [ D, \overline{D} ] {\bf T} 
\nonu \\
&& +  g_{237} \, \pa^2 D {\bf T}  {\bf T} [ D, \overline{D} ] {\bf T} 
+ g_{238} \,  \pa^2 D {\bf T}  {\bf T} {\bf T} {\bf T}
+ g_{239} \,  \pa^2 D {\bf T}  \overline{D} {\bf T} D {\bf T}
+ g_{240} \,  \pa^2 D {\bf T} \pa  {\bf T} {\bf T}
\nonu \\
&&
+  g_{241} \, \pa^2 D {\bf T} \pa^2  {\bf T}
+  g_{242} \,\pa^3 D {\bf T}   [ D, \overline{D} ] {\bf T} 
+  g_{243} \, \pa^3 D {\bf T}  {\bf T} {\bf T}
+  g_{244} \, \pa^3 D {\bf T}  \pa {\bf T}
+  g_{245} \, \pa^4 D {\bf T}  {\bf T}
\nonu \\
&&+ g_{246} \,  \pa {\bf T}  {\bf T} {\bf T} {\bf T} D {\bf T}
+ g_{247} \,  \pa {\bf T} {\bf T} [ D, \overline{D} ] {\bf T} D {\bf T}
+  g_{248} \, \pa {\bf T}  \pa {\bf T}  {\bf T} D {\bf T}
\nonu \\
&&+  g_{249} \, \pa^2 {\bf T}  {\bf T} {\bf T} D {\bf T}
+ g_{250} \,  \pa^2 {\bf T}   [ D, \overline{D} ] {\bf T}  D {\bf T}
+ g_{251} \,  \pa^2 {\bf T}  \pa D {\bf T} {\bf T} 
\nonu \\
&&+  \left. g_{252} \, \pa^2 {\bf T}  \pa  {\bf T} D {\bf T} 
+  g_{253} \, \pa^3 {\bf T}  \pa D {\bf T}
+ g_{254} \, \pa^3 {\bf T}  {\bf T} D {\bf T}
+  g_{255} \, \pa^4 {\bf T}  D {\bf T} \right](Z_2)
\nonu \\
&&+ \frac{\bar{\theta}_{12}}{z_{12}} \left[ 
 g_{256} \, \pa^5 \overline{D} {\bf T}
+ g_{257} \,  {\bf T}  {\bf T} {\bf T} {\bf T}  {\bf T} \overline{D} {\bf T}
+ g_{258} \,  {\bf T}  {\bf T} {\bf T}  \overline{D} {\bf T}   
[ D, \overline{D} ] {\bf T}
+  g_{259} \, {\bf T}  \overline{D} {\bf T}   [ D, \overline{D} ] {\bf T}
  [ D, \overline{D} ] {\bf T}
\right. \nonu \\ 
&& 
+ g_{260} \, \pa \overline{D} {\bf T}   {\bf T}   {\bf T}  [ D, \overline{D} ] {\bf T}
+  g_{261} \,  \pa  \overline{D} {\bf T}  {\bf T}  {\bf T}  {\bf T}  {\bf T} 
+  g_{262} \, \pa  \overline{D} {\bf T}  {\bf T} \overline{D} {\bf T}   D {\bf T}
\nonu \\
&& +  g_{263} \, \pa \overline{D} {\bf T} [ D, \overline{D} ] {\bf T}  
[ D, \overline{D} ] {\bf T} 
+  g_{264} \, \pa \overline{D} {\bf T}   \pa [ D, \overline{D} ] {\bf T}   {\bf T}
+ g_{265} \, \pa \overline{D} {\bf T}  \pa D {\bf T} \overline{D} {\bf T} 
\nonu \\
&& + g_{266} \, \pa \overline{D} {\bf T}  \pa {\bf T} [ D, \overline{D} ] {\bf T}
+  g_{267} \, \pa \overline{D} {\bf T}  \pa {\bf T}  {\bf T}  {\bf T}
+ g_{268} \, \pa \overline{D} {\bf T}  \pa {\bf T}  \pa {\bf T}
+ g_{269} \, \pa^2 \overline{D} {\bf T} \pa   [ D, \overline{D} ] {\bf T}
\nonu \\
&& +  g_{270} \,\pa^2 \overline{D} {\bf T}  {\bf T}   [ D, \overline{D} ] {\bf T}
+ g_{271} \, \pa^2 \overline{D} {\bf T}  {\bf T} {\bf T} {\bf T}
+  g_{272} \, \pa^2 \overline{D} {\bf T}  \overline{D} {\bf T} D {\bf T} 
+  g_{273} \,\pa^2 \overline{D} {\bf T}  \pa {\bf T}  {\bf T} 
\nonu \\
&&+  g_{274} \, \pa^2 \overline{D} {\bf T}  \pa^2 {\bf T}
+  g_{275} \, \pa^3 \overline{D} {\bf T}  [ D, \overline{D} ] {\bf T}
+ g_{276} \,  \pa^3 \overline{D} {\bf T}   {\bf T} {\bf T}
+  g_{277} \, \pa^3 \overline{D} {\bf T} \pa  {\bf T}
\nonu \\
&&+  g_{278} \,\pa^4 \overline{D} {\bf T}    {\bf T} 
+  g_{279} \, \pa   [ D, \overline{D} ] {\bf T}  {\bf T} {\bf T} \overline{D} {\bf T}
+ g_{280} \, \pa   [ D, \overline{D} ] {\bf T}  \overline{D} {\bf T}    
[ D, \overline{D} ] {\bf T}
 +  
g_{281} \, \pa [ D, \overline{D} ] {\bf T}  \pa {\bf T} \overline{D}  {\bf T}
\nonu \\
&&+ g_{282} \, \pa^2 [ D, \overline{D} ] {\bf T}  \pa  \overline{D} {\bf T}
+  g_{283} \, \pa^2 [ D, \overline{D} ] {\bf T}  {\bf T} \overline{D} {\bf T}
+  g_{284} \, \pa^3 [ D, \overline{D} ] {\bf T}  \overline{D} {\bf T}
+  g_{285} \, \pa {\bf T}  {\bf T} {\bf T} {\bf T}  \overline{D} {\bf T}
\nonu \\
&& +  g_{286} \, \pa {\bf T}   {\bf T}  \overline{D} {\bf T} [ D, \overline{D} ] {\bf T}
+  g_{287} \, \pa {\bf T}  \pa  {\bf T} {\bf T} \overline{D} {\bf T} 
+  g_{288} \, \pa^2 {\bf T}    {\bf T}  {\bf T} \overline{D} {\bf T} 
+   g_{289} \,\pa^2 {\bf T}   \overline{D} {\bf T} [ D, \overline{D} ] {\bf T}
\nonu \\
&& \left. + g_{290} \, \pa^2 {\bf T}  \pa \overline{D} {\bf T}  {\bf T}
+  g_{291} \, \pa^2 {\bf T}  \pa {\bf T} \overline{D} {\bf T}
+  g_{292} \, \pa^3 {\bf T}  \pa  \overline{D} {\bf T}
+  g_{293} \,\pa^3 {\bf T}  {\bf T}  \overline{D} {\bf T}
+  g_{294} \, \pa^4 {\bf T} \overline{D} {\bf T}
\right](Z_2)
\nonu \\
&&+ \frac{\theta_{12} \, \bar{\theta}_{12}}{z_{12}} \,
\left[  g_{295} \, \pa^5 [ D, \overline{D} ] {\bf T}
+ g_{296} \, {\bf T} {\bf T}  {\bf T} {\bf T}   {\bf T} [ D, \overline{D} ] {\bf T}
+  g_{297} \, {\bf T} {\bf T}  {\bf T} {\bf T}   {\bf T} {\bf T} {\bf T}
\right. \nonu \\
&&+  g_{298} \, {\bf T} {\bf T}  {\bf T} {\bf T}  \overline{D} {\bf T} D {\bf T}
+  g_{299} \, {\bf T} {\bf T}  {\bf T}  [ D, \overline{D} ] {\bf T}  
[ D, \overline{D} ] {\bf T}
+ g_{300} \,  {\bf T} {\bf T}  \overline{D} {\bf T} [ D, \overline{D} ] {\bf T}  
D {\bf T}
\nonu \\
&&+  g_{301} \, {\bf T} 
[ D, \overline{D} ] {\bf T}  [ D, \overline{D} ] {\bf T}  [ D, \overline{D} ] {\bf T} 
+ g_{302} \,  \overline{D} {\bf T} [ D, \overline{D} ] {\bf T} [ D, \overline{D} ] {\bf T}
D {\bf T} 
+ 
 g_{303} \,  \pa \overline{D} {\bf T}   {\bf T} {\bf T}  {\bf T} D {\bf T}
\nonu \\
&&+  g_{304} \,  \pa \overline{D} {\bf T}  {\bf T}  [ D, \overline{D} ] {\bf T} 
 D {\bf T}
+ g_{305} \,  \pa \overline{D} {\bf T}  \pa [ D, \overline{D} ] {\bf T} 
D {\bf T}
+  g_{306} \, \pa \overline{D} {\bf T} \pa D {\bf T} [ D, \overline{D} ] {\bf T}
\nonu \\
&& + g_{307} \,  \pa \overline{D} {\bf T} \pa D {\bf T} {\bf T} {\bf T} 
\nonu \\
&&
+ g_{308} \,  \pa \overline{D} {\bf T} \pa D {\bf T} \pa {\bf T}
+ g_{309} \, \pa \overline{D} {\bf T} \pa  {\bf T} {\bf T} D {\bf T}
+ g_{310} \,   \pa^2 \overline{D} {\bf T}   \pa^2 D {\bf T} 
+  g_{311} \, \pa^2 \overline{D} {\bf T}  {\bf T} {\bf T}  D {\bf T}
\nonu \\
&&
+ g_{312} \,  \pa^2 \overline{D} {\bf T}  [ D, \overline{D} ] {\bf T} D {\bf T}
+  g_{313} \,  \pa^2 \overline{D} {\bf T} \pa  D {\bf T} {\bf T}
+ g_{314} \, \pa^2 \overline{D} {\bf T} \pa {\bf T} D {\bf T} 
+  g_{315} \, \pa^3 \overline{D} {\bf T} \pa  D {\bf T}
\nonu \\
&&+  g_{316} \, \pa^3 \overline{D} {\bf T}  {\bf T} D {\bf T}
+ g_{317} \, \pa^4 \overline{D} {\bf T}   D {\bf T}
+ g_{318} \, \pa  [ D, \overline{D} ] {\bf T} {\bf T}  {\bf T}
 [ D, \overline{D} ] {\bf T}
+ g_{319} \, \pa  [ D, \overline{D} ] {\bf T}  {\bf T}  {\bf T} {\bf T}   {\bf T}
\nonu \\
&&
+ g_{320} \,\pa  [ D, \overline{D} ] {\bf T}  {\bf T}  \overline{D} {\bf T}   D {\bf T}
 + 
 g_{321} \, \pa  [ D, \overline{D} ] {\bf T}  [ D, \overline{D} ] {\bf T}  [ D, \overline{D} ] {\bf T} 
+  g_{322} \,
\pa  [ D, \overline{D} ] {\bf T} \pa  [ D, \overline{D} ] {\bf T} {\bf T}
\nonu \\
&& +  
g_{323} \,\pa  [ D, \overline{D} ] {\bf T} \pa  D {\bf T} \overline{D} {\bf T} 
+ g_{324} \, \pa  [ D, \overline{D} ] {\bf T} \pa {\bf T}  [ D, \overline{D} ] {\bf T}
+ g_{325} \, 
\pa  [ D, \overline{D} ] {\bf T} \pa {\bf T}  {\bf T} {\bf T}
\nonu \\
&& + g_{326} \, \pa  [ D, \overline{D} ] {\bf T} \pa {\bf T} \pa {\bf T}
+  g_{327} \, \pa^2  [ D, \overline{D} ] {\bf T} \pa  [ D, \overline{D} ] {\bf T} 
+  g_{328} \,
 \pa^2  [ D, \overline{D} ] {\bf T} {\bf T}  [ D, \overline{D} ] {\bf T}
\nonu \\
&&
+ g_{329} \, \pa^2  [ D, \overline{D} ] {\bf T}  {\bf T} {\bf T}
+ g_{330} \,  \pa^2  [ D, \overline{D} ] {\bf T} \overline{D} {\bf T} D {\bf T}
+ g_{331} \,
\pa^2  [ D, \overline{D} ] {\bf T}  \pa {\bf T} {\bf T}
+ g_{332} \, \pa^2  [ D, \overline{D} ] {\bf T} \pa^2 {\bf T}
\nonu \\
&& 
+ g_{333} \, \pa^3  [ D, \overline{D} ] {\bf T} [ D, \overline{D} ] {\bf T}
+ g_{334} \,
\pa^3  [ D, \overline{D} ] {\bf T}  {\bf T} {\bf T}
+ g_{335} \,\pa^3  [ D, \overline{D} ] {\bf T}  \pa {\bf T}
+ g_{336} \,\pa^4  [ D, \overline{D} ] {\bf T}  {\bf T}
\nonu \\
&&+ g_{337} \,
\pa D {\bf T}  {\bf T}  {\bf T} {\bf T} \overline{D} {\bf T} 
+ g_{338} \, \pa D {\bf T}  {\bf T}  \overline{D} {\bf T}  [ D, \overline{D} ] {\bf T}
+ g_{339} \, \pa D {\bf T} \pa  {\bf T}   {\bf T} \overline{D} {\bf T} 
+ g_{340} \,
\pa^2 D {\bf T}   {\bf T} {\bf T} \overline{D} {\bf T} 
\nonu \\
&&
+ g_{341} \,\pa^2 D {\bf T} \overline{D} {\bf T}  [ D, \overline{D} ] {\bf T}
+ g_{342} \,\pa^2 D {\bf T} \pa \overline{D} {\bf T}  {\bf T}
+ g_{343} \,
\pa^2 D {\bf T} \pa {\bf T} \overline{D} {\bf T} 
+ g_{344} \,\pa^3 D {\bf T} \pa \overline{D} {\bf T}
\nonu \\
&& + g_{345} \, \pa^3 D {\bf T} {\bf T} \overline{D} {\bf T}
+ g_{346} \, \pa^4 D {\bf T}  \overline{D} {\bf T} 
+ g_{347} \, \pa {\bf T} {\bf T} {\bf T} {\bf T}  [ D, \overline{D} ] {\bf T}
+ g_{348} \, \pa {\bf T}  {\bf T} {\bf T} {\bf T} {\bf T} {\bf T}
\nonu \\
&&+ g_{349} \,
\pa {\bf T}  {\bf T} {\bf T}  \overline{D} {\bf T}  D {\bf T}
+ g_{350} \, \pa {\bf T}  {\bf T}  [ D, \overline{D} ] {\bf T}  [ D, \overline{D} ] {\bf T} 
+ g_{351} \, \pa {\bf T}   \overline{D} {\bf T}  [ D, \overline{D} ] {\bf T}
D {\bf T}
\nonu \\
&&+ g_{352} \,
\pa {\bf T} \pa {\bf T}  {\bf T}  [ D, \overline{D} ] {\bf T}  
+ g_{353} \,\pa {\bf T} \pa {\bf T}  {\bf T}  {\bf T} {\bf T} 
+ g_{354} \,\pa {\bf T} \pa {\bf T}   \overline{D} {\bf T}   D {\bf T}
+ g_{355} \, \pa {\bf T} \pa {\bf T}  \pa {\bf T} {\bf T}
\nonu \\
&& +  g_{356} \, \pa^2 {\bf T}   {\bf T}  {\bf T}  [ D, \overline{D} ] {\bf T}  
+ g_{357} \, \pa^2 {\bf T}  {\bf T}  {\bf T}  {\bf T} {\bf T}
+  g_{358} \, \pa^2 {\bf T}  {\bf T}   \overline{D} {\bf T}   D {\bf T}
+  g_{359} \, \pa^2 {\bf T} \pa  \overline{D} {\bf T}  D {\bf T}
\nonu \\
&&
+  g_{360} \, \pa^2 {\bf T}  [ D, \overline{D} ] {\bf T}   [ D, \overline{D} ] {\bf T} 
+  g_{361} \, \pa^2 {\bf T}  \pa [ D, \overline{D} ] {\bf T} {\bf T}
+  g_{362} \,\pa^2 {\bf T} \pa  D {\bf T}  \overline{D} {\bf T}  
+  g_{363} \,\pa^2 {\bf T}  \pa  {\bf T}   [ D, \overline{D} ] {\bf T}
\nonu \\
&&+ g_{364} \,
 \pa^2 {\bf T}  \pa  {\bf T}  {\bf T} {\bf T}
+  g_{365} \, \pa^2 {\bf T}  \pa  {\bf T}  \pa  {\bf T}  
+ g_{366} \, \pa^2 {\bf T}  \pa^2 {\bf T} {\bf T}
+  g_{367} \,\pa^3 {\bf T} \pa  [ D, \overline{D} ] {\bf T}
\nonu \\
&&  + g_{368} \,\pa^3 {\bf T} {\bf T}  [ D, \overline{D} ] {\bf T}
\nonu \\
&& 
+ g_{369} \, \pa^3 {\bf T} {\bf T} {\bf T} {\bf T}
+  g_{370} \, \pa^3 {\bf T}  \overline{D} {\bf T} D {\bf T}
+  g_{371} \, \pa^3 {\bf T} \pa {\bf T} {\bf T}
+  g_{372} \, \pa^3 {\bf T} \pa^2 {\bf T}
\nonu \\
&& \left. + 
 g_{373} \,  \pa^4 {\bf T} [ D, \overline{D} ] {\bf T}
+ g_{374} \,  \pa^4 {\bf T}  {\bf T} {\bf T}
+  g_{375} \, \pa^4 {\bf T}  \pa {\bf T}
+    g_{376} \, \pa^5 {\bf T} \pa {\bf T}
+ g_{377} \,  \pa^6 {\bf T}
\right](Z_2),
\label{34OPE}
\eea
}
where the various coefficients are given by
{\small
\bea
g_1 & = & -\frac{(63 c^5-255 c^4+177 c^3+1835 c^2-2304 c-36)}{42 (c-1) (c+1) (c+6) (2 c-3) (5 c-9)},\nonu \\
g_2  & = & 
\frac{2 (1113 c^3-230 c^2+59 c-402)}{21 (c-1) (c+1) (c+6) (2 c-3) (5 c-9)},\nonu \\
g_3& = & \frac{4 (796 c^2-309 c+248)}{21 (c-1) (c+1) (c+6) (2 c-3) (5 c-9)},
\nonu \\
g_4  & = &  
\frac{4 (29 c^3-1720 c^2+1945 c+86)}{7 (c-1) (c+1) (c+6) (2 c-3) (5 c-9)},\nonu \\
g_5& = & \frac{2 (24 c^4-1401 c^3+1692 c^2-473 c+18)}{21 (c-1) (c+1) (c+6) (2 c-3) (5 c-9)},\nonu \\
g_6 & = &  \frac{(2331 c^4+162 c^3-15457 c^2+15540 c-36)}{126 (c-1) (c+1) (c+6) (2 c-3) (5 c-9)},\nonu \\
g_7& = & -\frac{(105 c^4+274 c^3+5065 c^2-6996 c-1068)}{21 (c-1) (c+1) (c+6) (2 c-3) (5 c-9)},\nonu \\
g_8  & = &  \frac{2 (39 c^4-2358 c^3+2451 c^2+1204 c-36)}{21 (c-1) (c+1) (c+6) (2 c-3) (5 c-9)},\nonu \\
g_9& = & -\frac{(3 c+5) (7 c+18)}{21 (c-1) (c+6) (2 c-3)},
\qquad
g_{10} =  -\frac{2 (31 c+134)}{7 (c-1) (c+6) (2 c-3)},\nonu \\
g_{11}& = & -\frac{(121 c^4-492 c^3-1435 c^2+5674 c-4328)}{14 (c-1) (c+1) (c+6) (2 c-3) (5 c-9)},\nonu \\
g_{12}  & = &  -\frac{(179 c^4+1544 c^3+555 c^2-9834 c+9016)}{21 (c-1) (c+1) (c+6) (2 c-3) (5 c-9)},\nonu \\
g_{13}& = & -\frac{40 (3 c^5-9 c^4-23 c^3+145 c^2-72 c+36)}{63 (c-1) (c+1) (c+6) (2 c-3) (5 c-9)},\nonu \\
g_{14}  & = &  
\frac{80 (157 c^2+57 c+98)}{21 (c-1) (c+1) (c+6) (2 c-3) (5 c-9)},\nonu \\
g_{15}& = & \frac{80 (57 c^3-91 c^2+104 c-12)}{21 (c-1) (c+1) (c+6) (2 c-3) (5 c-9)},\qquad
g_{16} =  \frac{20 (57 c^3-91 c^2+104 c-12)}{21 (c+1) (c+6) (2 c-3) (5 c-9)},\nonu \\
g_{17}& = & \frac{40 (111 c^4-218 c^3-497 c^2+468 c+108)}{63 (c-1) (c+1) (c+6) (2 c-3) (5 c-9)},\nonu \\
g_{18}  & = &  \frac{80 (47 c^3-291 c^2-68 c-60)}{21 (c-1) (c+1) (c+6) (2 c-3) (5 c-9)},\nonu \\
g_{19}& = & -\frac{40 (11 c^3-26 c^2+45 c-50)}{21 (c-1) (c+1) (2 c-3) (5 c-9)},\qquad
g_{20} =  -\frac{80 (3 c^4-9 c^3+148 c^2-128 c+240)}{21 (c-1) (c+1) (c+6) (2 c-3) (5 c-9)},\nonu \\
g_{21}& = & \frac{40 (43 c^3+82 c^2-167 c-74)}{7 (c-1) (c+1) (c+6) (2 c-3) (5 c-9)},\qquad
g_{22} =  -\frac{20 (14 c^3-173 c^2+271 c+62)}{21 (c+1) (c+6) (2 c-3) (5 c-9)},\nonu \\
g_{23}& = & \frac{11 (12 c^4-72 c^3+223 c^2-263 c-150)}{63 (c-1) (c+6) (2 c-3) (5 c-9)},\qquad
g_{24} =  \frac{176 (c^2-54 c-22)}{21 (c-1) (c+1) (c+6) (2 c-3) (5 c-9)},\nonu \\
g_{25}& = & \frac{22 (3 c^3-180 c^2+163 c-6)}{21 (c-1) (c+1) (c+6) (2 c-3) (5 c-9)},\nonu \\
g_{26} & = &  
-\frac{22 (222 c^4-205 c^3-374 c^2+351 c-54)}{63 (c-1) (c+1) (c+6) (2 c-3) (5 c-9)},\nonu \\
g_{27}& = & -\frac{44 (109 c^3+88 c^2-59 c-258)}{21 (c-1) (c+1) (c+6) (2 c-3) (5 c-9)},\qquad
g_{28} =  -\frac{22 (2 c^3+15 c^2+12 c-89)}{21 (c-1) (c+1) (2 c-3) (5 c-9)},\nonu \\
g_{29}& = & \frac{11 (6 c^4-147 c^3+634 c^2+745 c-1098)}{21 (c-1) (c+1) (c+6) (2 c-3) (5 c-9)},\nonu \\
g_{30}  & = &  
-\frac{22 (57 c^4-142 c^3+33 c^2+44 c-12)}{21 (c-1) (c+1) (c+6) (2 c-3) (5 c-9)},
\nonu \\
g_{31}& = & -\frac{22 (111 c^3+12 c^2-345 c-158)}{7 (c-1) (c+1) (c+6) (2 c-3) (5 c-9)},\nonu \\
g_{32}  & = &  
-\frac{22 (4 c^4-227 c^3+468 c^2+65 c-370)}{21 (c-1) (c+1) (c+6) (2 c-3) (5 c-9)},\nonu \\
g_{33}& = & \frac{1
}{567 d(c)}
(c+18)(4c-9)(7c-15)(9 c^7-3246 c^6-91551 c^5+242852 c^4 \nonu \\
& + & 399148 c^3-1203616 c^2+641568 c-180864),
\nonu \\
g_{34}& = & \frac{8
}{189 d(c) }
 (c+18)(4c-9)(7c-15) (26112 c^4+4477 c^3-35377 c^2-36338 c+19776),
\nonu \\
g_{35}& = & \frac{4 
}{63 d(c) }
(c+18)(4c-9)(7c-15) (24896 c^3+40675 c^2-153502 c+48816),
\nonu \\
g_{36}& = & \frac{8 
}{21 d(c)}
(c+18)(4c-9)(7c-15)(10823 c^4-19437 c^3-135958 c^2+266168 c-92576),
\nonu \\
g_{37}& = & \frac{1
}{378 d(c)}
(c+18)(4c-9)(7c-15)(72927 c^5-94098 c^4 \nonu \\
& + & 365507 c^3-2601712 c^2+4749420 c-2325744), \nonu \\
g_{38}& = & \frac{2 
}{63 d(c)}
(c+18)(4c-9)(7c-15) \nonu \\
& \times & 
(45249 c^5-147298 c^4-67143 c^3+578128 c^2-465852 c+7536), \nonu \\
g_{39}& = & -\frac{4 
}{189 d(c) }
(c+18)(4c-9)(7c-15) (7542 c^6+46245 c^5-140660 c^4\nonu \\
& - & 183039 c^3+363272 c^2-28284 c+67824),
\nonu \\
g_{40}& = & -\frac{2(c+18) 
}{63 d(c)}
(4c-9)(7c-15) \nonu \\
& \times & (8691 c^5-75270 c^4 -365597 c^3+1751072 c^2-1445108 c+370512),
\nonu \\
g_{41}& = & \frac{1
}{63 d(c)}
(c+18)(4c-9)(7c-15)(-5661 c^6-12339 c^5+287493 c^4 \nonu \\
& - & 1117141 c^3+1957332 c^2-1307252 c+340368),
\nonu \\
g_{42}& = & \frac{1
}{378 d(c)}
(c+18)(4c-9)(7c-15)(-333 c^6+114951 c^5-560311 c^4 \nonu \\
& + & 645245 c^3+20508 c^2-133596 c+158256), \nonu \\
g_{43}& = & \frac{1
}{63 d(c)}
(c+18)(4c-9)(7c-15)\nonu \\
& \times & 
(-3459 c^5+33178 c^4-33663 c^3-1194712 c^2+2002836 c+103920), \nonu \\
g_{44}& = & \frac{1
}{126 d(c)}
(c+18)(4c-9)(7c-15)(-6585 c^6-63517 c^5-443 c^4 \nonu \\
& + & 1004305 c^3-504420 c^2-2488748 c+1479408), \nonu \\
g_{45}& = & \frac{1
}{126 d(c)}
(c+18)(4c-9)(7c-15)(-1971 c^6+3729 c^5+214035 c^4 \nonu \\
& - & 625425 c^3-571004 c^2+2710156 c-1838640), \nonu \\
g_{46}& = & \frac{2 
}{63 d(c)}
(c+18)(4c-9)(7c-15)\nonu \\
& \times & (6891 c^5+177570 c^4-571981 c^3-796064 c^2+1673228 c+718416),
\nonu \\
g_{47}& = & \frac{1
}{63 d(c)}
(c+18)(4c-9)(7c-15)(5652 c^6-26781 c^5-111158 c^4 \nonu \\
& + & 728295 c^3-1513232 c^2+644156 c+808848),
\nonu \\
g_{48}& = & -\frac{2 (5109 c^3+5750 c^2+17467 c+9654)}{63 (c-1) (c+1) (c+6) (2 c-3) (5 c-9)},\nonu \\
g_{49}  & = &  
-\frac{8 (6203 c^2+6438 c+5614)}{63 (c-1) (c+1) (c+6) (2 c-3) (5 c-9)},\nonu \\
g_{50}& = & -\frac{2 (2047 c^3-11780 c^2-1105 c-1622)}{21 (c-1) (c+1) (c+6) (2 c-3) (5 c-9)},\nonu \\
g_{51}& = & \frac{(c+18)
}{42 d(c)}
(4c-9)(7c-15) \nonu \\
& \times & 
(-16327 c^5-128478 c^4+335397 c^3+708008 c^2-1608956 c+338416), \nonu \\
g_{52}& = & \frac{1
}{189 d(c)}
(c+18)(4c-9)(7c-15)(-9885 c^6-64879 c^5+243945 c^4 \nonu \\
& + & 19251 c^3-701636 c^2+1310300 c-1277616),
\nonu \\
g_{53}& = & -\frac{(15013 c^4-38838 c^3+47701 c^2-119852 c+148196)}{63 (c-2) (c-1) (c+1) (c+6) (2 c-3) (5 c-9)},\nonu \\
g_{54}& = & \frac{(803 c^4+947 c^3-7125 c^2+19437 c-1982)}{63 (c-1) (c+1) (c+6) (2 c-3) (5 c-9)},\nonu \\
g_{55}& = & \frac{1
}{378 d(c)}
(c+18)(4c-9)(7c-15)(2391 c^6+11061 c^5+99553 c^4 \nonu \\
& - & 59901 c^3-5463220 c^2+10206588 c-2254512),
\nonu \\
g_{56}& = & \frac{1
}{378 d(c)}
(c+18)(4c-9)(7c-15)(471 c^7-936 c^6-85715 c^5+223232 c^4 \nonu \\
& + & 357536 c^3-1118032 c^2+382720 c-28896),
\nonu \\
g_{57}& = & \frac{(15 c^5-177 c^4-341 c^3-10307 c^2+13410 c+1920)}{126 (c-1) (c+1) (c+6) (2 c-3) (5 c-9)},\nonu \\
g_{58}& = & \frac{1
}{189 d(c)}
(c+18)(4c-9)(7c-15)(-63 c^7-378 c^6+3957 c^5-2884 c^4 \nonu \\
& - & 9716 c^3+44192 c^2-34656 c-1152),
\nonu \\
g_{59}& = & \frac{8 
}{63 d(c)}
(c+18)(4c-9)(7c-15)(861 c^4+8206 c^3-1621 c^2+5986 c+1488),
\nonu \\
g_{60}& = & \frac{4 
}{21 d(c)}
(c+18)(4c-9)(7c-15)(848 c^3+8095 c^2-5686 c+6768),
\nonu \\
g_{61}& = & \frac{8 
}{7 d(c)}
(c+18)(4c-9)(7c-15)(29 c^4-1551 c^3-11014 c^2+16064 c-4928),
\nonu \\
g_{62}& = & \frac{1
}{126 d(c)}
(c+18)(4c-9)(7c-15) \nonu \\
& \times & (2331 c^5+22446 c^4-5449 c^3-180016 c^2+300540 c-98352),
\nonu \\
g_{63}& = & \frac{2 
}{21 d(c)}
(c+18)(4c-9)(7c-15)(57 c^5-3574 c^4+2061 c^3+12544 c^2-15276 c+48),
\nonu \\
g_{64}& = & -\frac{4 
}{63 d(c)}
(c+18)(4c-9)(7c-15) \nonu \\
& \times & (6 c^6-315 c^5-2960 c^4-10347 c^3+8216 c^2+5268 c+432),
\nonu \\
g_{65}& = & \frac{2 
}{21 d(c)}
(c+18)(4c-9)(7c-15) \nonu \\
& \times & (117 c^5-5490 c^4-78019 c^3+173344 c^2-96076 c-4176),
\nonu \\
g_{66}& = & \frac{(c+18)
}{21  d(c)}
(4c-9)(7c-15) \nonu \\
& \times & (27 c^6-1407 c^5-10671 c^4+123947 c^3-208284 c^2+102604 c+3984),
\nonu \\
g_{67}& = & \frac{1
}{126 d(c)}
(c+18)(4c-9)(7c-15)(2331 c^6+23643 c^5-80503 c^4 \nonu \\
& - & 136735 c^3+563964 c^2-419148 c+1008),
\nonu \\
g_{68}& = & \frac{1
}{21 d(c)}
(c+18)(4c-9)(7c-15) \nonu \\
& \times & (1113 c^5+11074 c^4-38259 c^3-60376 c^2+90948 c-19920),
\nonu \\
g_{69}& = & \frac{1
}{42 d(c)}
(c+18)(4c-9)(7c-15)(-105 c^6-1321 c^5-5699 c^4 \nonu \\
& - & 22475 c^3+178140 c^2-
209564 c+45744), \nonu \\
g_{70}& = & \frac{1
}{42 d(c)}
(c+18)(4c-9)(7c-15) \nonu \\
& \times & (-63 c^6-363 c^5+3255 c^4-15645 c^3-32492 c^2+83548 c-44400),
\nonu \\
g_{71}& = & -\frac{2 
}{21 d(c)}
(c+18)(4c-9)(7c-15) \nonu \\
& \times & (57 c^5-3090 c^4-17687 c^3+191552 c^2-178844 c-8208),
\nonu \\
g_{72}& = & \frac{(c+18)
}{21 d(c)}
(4c-9)(7c-15) \nonu \\
& \times & (36 c^6-1953 c^5-12134 c^4+102075 c^3-261296 c^2+169868 c+8784),
\nonu \\
g_{73}& = & \frac{2 (1113 c^3-230 c^2+59 c-402)
}{21 (c-1) (c+1) (c+6) (2 c-3) (5 c-9)},\nonu \\
g_{74}  & = &  \frac{8 (796 c^2-309 c+248)
}{21 (c-1) (c+1) (c+6) (2 c-3) (5 c-9)},\nonu \\
g_{75}& = & \frac{2 (29 c^3-1720 c^2+1945 c+86)
}{7 (c-1) (c+1) (c+6) (2 c-3) (5 c-9)},\nonu \\
g_{76}& = & \frac{1
}{14 d(c)}
(c+18)(4c-9)(7c-15)(-331 c^5-3534 c^4+6441 c^3+47624 c^2-95468 c+35248),
\nonu \\
g_{77}& = & \frac{1
}{63 d(c)}
(c+18)(4c-9)(7c-15) \nonu \\
& \times & (-105 c^6-1357 c^5-4155 c^4-1767 c^3+37132 c^2+17300 c-90768),
\nonu \\
g_{78}& = & -\frac{(289 c^4+1266 c^3+1153 c^2-11996 c+14228)
}{21 (c-2) (c-1) (c+1) (c+6) (2 c-3) (5 c-9)},\nonu \\
g_{79}& = & -\frac{(121 c^4-521 c^3+285 c^2+3729 c-4414)
}{21 (c-1) (c+1) (c+6) (2 c-3) (5 c-9)},\nonu \\
g_{80}& = & \frac{1
}{126 d(c)}
(c+18)(4c-9)(7c-15)(-237 c^6-1527 c^5+13069 c^4 \nonu \\
& - & 51273 c^3-10180 c^2-422196 c
+459024),
\nonu \\
g_{81}& = & -\frac{5 
}{378 d(c)}
(c+18)(4c-9)(7c-15)(30 c^7-40 c^6-8007 c^5+16870 c^4
\nonu \\
& + & 57971 c^3-154880 c^2+73036 c+336),
\nonu \\
g_{82}& = & -\frac{40 
}{21 d(c)}
(c+18)(4c-9)(7c-15)(632 c^3-533 c^2-3790 c+48),
\nonu \\
g_{83}& = & -\frac{20 
}{21 d(c)}
(c+18)(4c-9)(7c-15)(1077 c^4-627 c^3-6128 c^2+5356 c-3024),
\nonu \\
g_{84}& = & -\frac{5 
}{63 d(c)}
(c+18)(4c-9)(7c-15) \nonu \\
& \times & (2673 c^5-2826 c^4-24367 c^3+51008 c^2-23964 c-3024),
\nonu \\
g_{85}& = & \frac{10 
}{63 d(c)}
(c+18)(4c-9)(7c-15) \nonu \\
& \times & (225 c^6+1510 c^5-4129 c^4-27426 c^3+78172 c^2-38904 c-10080),
\nonu \\
g_{86}& = & \frac{10 
}{21 d(c)}
(c+12)(c+18)(4c-9)(7c-15)(231 c^4-1002 c^3+1319 c^2-292 c-116),
\nonu \\
g_{87}& = & \frac{5 
}{21 d(c)}
(3 c-8) (c+18)(4c-9)(7c-15)(23 c^5+42 c^4-885 c^3+1732 c^2-836 c-96),
\nonu \\
g_{88}& = & \frac{20 
}{63 d(c)}
(c+18)(4c-9)(7c-15)(393 c^5+9236 c^4-22673 c^3+11096 c^2-23940 c+7632),
\nonu \\
g_{89}& = & \frac{40 
}{63 d(c)}
(c+18)(4c-9)(7c-15)(553 c^4+12765 c^3-21576 c^2-18980 c-9744),
\nonu \\
g_{90}& = & -\frac{40 
}{21 d(c)}
(c+18)(4c-9)(7c-15)(16 c^5-1065 c^4+9163 c^3-20196 c^2+10588 c+1296),
\nonu 
\\
g_{91}& = & \frac{40 
}{21 d(c)}
(c+18)(4c-9)(7c-15) (91 c^5+401 c^4-3342 c^3-2152 c^2+10576 c+3840),
\nonu \\
g_{92}& = & \frac{10 
}{63 d(c)}
(c+18)(4c-9)(7c-15) \nonu \\
& \times & (165 c^6-259 c^5-644 c^4+3064 c^3+3064 c^2-21792 c+4608),
\nonu \\
g_{93}& = & \frac{40 
}{21 d(c)}
(c+18)(4c-9)(7c-15)(36 c^5-73 c^4+725 c^3+2550 c^2-10480 c-1680),
\nonu \\
g_{94}& = & -\frac{10 
}{21 d(c)}
(c+18)(4c-9)(7c-15) \nonu \\
& \times & (15 c^6-34 c^5-522 c^4+295 c^3+17664 c^2-38292 c+13584),
\nonu \\
g_{95}& = & -\frac{10 
}{63 d(c)}
(c+18)(4c-9)(7c-15) \nonu \\
& \times & (18 c^6+151 c^5+3384 c^4-4921 c^3-27816 c^2+77244 c-50160),
\nonu \\
g_{96}& = & \frac{20 
}{7 d(c)}
(c+18)(4c-9)(7c-15)(339 c^4+2011 c^3-3852 c^2-20 c-4528),
\nonu \\
g_{97}& = & \frac{10 
}{21 d(c)}
(c+18)(4c-9)(7c-15)(285 c^5-1042 c^4-651 c^3+9816 c^2-10428 c-5136),
\nonu \\
g_{98}& = & \frac{5 
}{7 d(c)}
(c+18)(4c-9)(7c-15)(103 c^5+498 c^4-4353 c^3+11528 c^2-10868 c-304),
\nonu \\
g_{99}& = & -\frac{10 
}{63 d(c)}
(c+18)(4c-9)(7c-15) \nonu \\
& \times & (50 c^6+267 c^5+928 c^4+20651 c^3-81944 c^2+59372 c+12624),
\nonu \\
g_{100}& = & \frac{10 
}{21 d(c)}
(c+18)(4c-9)(7c-15)(145 c^5+3142 c^4+3225 c^3-36640 c^2+22436 c+19440),
\nonu \\
g_{101}& = & -\frac{5  
}{63 d(c)}
(c+18)(4c-9)(7c-15) \nonu \\
& \times & (45 c^6-758 c^5-11243 c^4+56292 c^3-72460 c^2+10752 c+13632),
\nonu \\
g_{102}& = & \frac{11  
}{3024 d(c) }
(c+18)(4c-9)(7c-15) (120 c^7+580 c^6-11035 c^5+29766 c^4 \nonu \\
& - & 106759 c^3+125104 c^2+7396 c-40272),
\nonu \\
g_{103}& = & \frac{220  
}{7 d(c)}
(c+18)(4c-9)(7c-15)(64 c^3+213 c^2-402 c+208),
\nonu \\
g_{104}& = & \frac{88  
}{21 d(c)}
(c+18)(4c-9)(7c-15)(327 c^4+392 c^3-3042 c^2+3952 c-1224),
\nonu \\
g_{105}& = & \frac{11  
}{126 d(c)}
(c+18)(4c-9)(7c-15)(2673 c^5-2478 c^4-21523 c^3+36848 c^2-2796 c-11664),
\nonu \\
g_{106}& = & -\frac{11  
}{63 d(c) }
(c+12)(c+18)(4c-9)(7c-15)(225 c^5-846 c^4+1617 c^3-1476 c^2+244 c+96),
\nonu \\
g_{107}& = & -\frac{22  
}{63 d(c) }
(c+18)(4c-9)(7c-15)(51 c^5-5314 c^4+6231 c^3-6656 c^2+14940 c-11952),
\nonu \\
g_{108}& = & -\frac{176  
}{63 d(c)}
(c+18)(4c-9)(7c-15)(19 c^4-1931 c^3-1924 c^2+6584 c-4728),
\nonu \\
g_{109}& = & -\frac{22  
}{21 d(c))}
(c+18)(4c-9)(7c-15)(37 c^5-2402 c^4+15797 c^3-24392 c^2+4100 c+7920),
\nonu \\
g_{110}& = & -\frac{22  
}{21 d(c)}
(c+18)(4c-9)(7c-15)(221 c^5+2114 c^4+1953 c^3-22088 c^2+21332 c-5712),
\nonu \\
g_{111}& = & -\frac{11  
}{63 d(c)}
(c+18)(4c-9)(7c-15) \nonu \\
& \times & (165 c^6+913 c^5-2619 c^4+5505 c^3+1760 c^2-20148 c+18864),
\nonu \\
g_{112}& = & -\frac{22  
}{21 d(c)}
(c+18)(4c-9)(7c-15)(81 c^5+560 c^4+2669 c^3+10636 c^2-30492 c+24576),
\nonu \\
g_{113}& = & \frac{11  
}{42 d(c)}
(c+12)(c+18)(4c-9)(7c-15)(109 c^4+1062 c^3-4223 c^2+4900 c-2508),
\nonu \\
g_{114}& = & \frac{11  
}{63 d(c)}
(c+18)(4c-9)(7c-15) \nonu \\
& \times & (6 c^6+25 c^5-1380 c^4+17101 c^3+4888 c^2-65164 c+52464),
\nonu \\
g_{115}& = & \frac{11  
}{21 d(c)}
(c+12)(c+18)(4c-9)(7c-15)(3 c^4-178 c^3-349 c^2+1388 c-644),
\nonu \\
g_{116}& = & -\frac{11  
}{42 d(c)}
(c+18)(4c-9)(7c-15) \nonu \\
& \times & (69 c^6-50 c^5-3279 c^4+8780 c^3-4044 c^2-5600 c+4224),
\nonu \\
g_{117}& = & \frac{88  
}{7 d(c)}
(c+18)(4c-9)(7c-15)(8 c^4-397 c^3-553 c^2+1978 c+344),
\nonu \\
g_{118}& = & \frac{44  
}{21 d(c) }
(c+18)(4c-9)(7c-15)(15 c^5-791 c^4+1448 c^3+1082 c^2-1332 c-1392),
\nonu \\
g_{119}& = & -\frac{11  
}{14 d(c)}
(c+18)(4c-9)(7c-15)(133 c^5+394 c^4-9279 c^3+20024 c^2-2412 c-12880),
\nonu \\
g_{120}& = & -\frac{11  
}{63 d(c)}
(c+18)(4c-9)(7c-15) \nonu \\
& \times & (10 c^6-233 c^5+780 c^4+22963 c^3-69768 c^2+33356 c+34512),
\nonu \\
g_{121}& = & -\frac{11  
}{21 d(c)}
(c+18)(4c-9)(7c-15)(201 c^5+3038 c^4-3943 c^3-12992 c^2-1244 c+34800),
\nonu \\
g_{122}& = & -\frac{11  
}{126 d(c)}
(c+18)(4c-9)(7c-15)(15 c^6-724 c^5-8077 c^4
\nonu \\
& + & 42482 c^3-20772 c^2-75624 c+60000),
\nonu \\
g_{123}& = & \frac{1}{27216 d(c)}   (c+18) (4c-9)(
270 c^8+33930 c^7-3440805 c^6+10604631 c^5+45769551 c^4  \nonu \\
& - &  215694881 c^3+168965116 c^2+73731228 c-114855600),\nonu \\
g_{124}& = & \frac{1 
}{189 d(c)}
(c+18) (4c-9)(1534128 c^4-420433 c^3+2987923 c^2-6448618 c+7860240),
\nonu \\
g_{125}& = & \frac{16  
}{189 d(c)}
(c+18) (4c-9)(141040 c^3+135667 c^2-658666 c+449864),
\nonu \\
g_{126}& = & \frac{8  
}{63 d(c)}
(c+18) (4c-9)(243776 c^4-963011 c^3-2649879 c^2+7047194 c-4592960),
\nonu \\
g_{127}& = & \frac{(c+18)
}{189 d(c)}
(4c-9)(262953 c^5-103656 c^4 \nonu \\
& + & 1139981 c^3-11673926 c^2+24972848 c-13112400),
\nonu \\
g_{128}& = & \frac{8  
}{63 d(c)}
(c+18) (4c-9)(82419 c^5-472936 c^4 \nonu \\
& + & 592296 c^3+532459 c^2-1667198 c+691120),
\nonu \\
g_{129}& = & \frac{4  
}{63 d(c)}
(c+18) (4c-9)(294069 c^5-256143 c^4 \nonu \\
& - & 5037032 c^3+9147382 c^2-5511476 c+349600),
\nonu \\
g_{130}& = & \frac{1 
}{189 d(c)}
(c+18) (4c-9)(1251639 c^6-5154105 c^5-3627207 c^4 \nonu \\
& + & 45878577 c^3-78136316 c^2+41281732 c-271440),
\nonu \\
g_{131}& = & \frac{1 
}{189 d(c)}
(c+18) (4c-9)(-216603 c^6-882087 c^5+1561151 c^4 \nonu \\
& + & 4247859 c^3+45396484 c^2-91851364 c+58661840),
\nonu \\
g_{132}& = & \frac{1 
}{189 d(c) }
(c+18) (4c-9)(-174429 c^6+1839363 c^5-1101231 c^4 \nonu \\
& - & 21457239 c^3+38219868 c^2+7048628 c-26386000),
\nonu \\
g_{133}& = & \frac{1}{
1134 d(c)}
 (c+18)(4c-9) 
(-475245 c^7+469812 c^6+7397670 c^5-17815704 c^4 \nonu \\
&+ & 47920239 c^3
-   243327056 c^2+338526844 c-169470960 ),\nonu \\
g_{134}& = & -\frac{2  
}{189 d(c)}
(c+18)(4c-9) (74313 c^6+110397 c^5-3093819 c^4\nonu \\
& + & 5586579 c^3-14669262 c^2+9934912 c-4791440),
\nonu \\
g_{135}& = & \frac{1}{
1134 d(c)}
 (c+18)(4c-9)
(-190152 c^7+832755 c^6+9239727 c^5-67190247 c^4
  \nonu \\
& + &  
158826669 c^3-175083092 c^2+57252900 c+4653680 )
,\nonu \\
g_{136}& = & \frac{1
}{13608 d(c)}
(c+18)(4c-9)(54675 c^6+6848379 c^5-43308495 c^4\nonu \\
& + & 113526145 c^3-170739620 c^2+157450836 c-63193680),
\nonu \\
g_{137}& = & -\frac{1
}{2268 d(c)}
(3 c-8) (c+18)(4c-9) (585 c^6-894591 c^5+4652247 c^4\nonu \\
& - & 9340773 c^3+13034992 c^2-12846812 c+4791120),
\nonu \\
g_{138}& = & \frac{1
}{756 d(c)}
(c+18)(4c-9)(1023309 c^6-3506835 c^5+9746335 c^4 \nonu \\
& - & 68041505 c^3+209164956 c^2-265302660 c+110873040),
\nonu \\
g_{139}& = & \frac{2 
}{189 d(c)}
(c+18)(4c-9)(366429 c^5-678193 c^4 \nonu \\
& - & 174152 c^3-12317548 c^2+25111224 c-15778320), 
\nonu \\
g_{140}& = & \frac{1
}{126 d(c)}
(c+18)(4c-9)(7c-15)(92769 c^5-441878 c^4 \nonu \\
& + & 850917 c^3-1068352 c^2+703668 c+18096),
\nonu \\
g_{141}& = & \frac{1
}{756 d(c)}
(c+18)(4c-9)(-422505 c^6+2359047 c^5-24584763 c^4 \nonu \\
& - & 19229307 c^3+369783732 c^2-559725484 c+162900080),
\nonu \\
g_{142}& = & \frac{1}{
4536 d(c)}
(c+18)(4c-9)(-4185 c^7-526401 c^6+3653373 c^5+10280229 c^4
 \nonu \\
& - &  86800524 c^3+114801940 c^2+39099088 c-115366080
),\nonu \\
g_{143}& = & \frac{1
}{378 d(c)}
(c+18)(4c-9)(-42579 c^6+38943 c^5+5072241 c^4 \nonu \\
& - & 17465121 c^3-16467840 c^2+86619556 c-75520880),
\nonu \\
g_{144}& = & \frac{1
}{504 d(c)}
(c+18)(4c-9)(7c-15)(-7911 c^6-1080 c^5+228897 c^4 \nonu \\
& - & 815234 c^3+1501700 c^2-3250904 c+3327072),
\nonu \\
g_{145}& = & \frac{1}{
1134 d(c)}
(c+18)(4c-9)(-27657 c^7+84678 c^6+2719818 c^5-14454542 c^4
 \nonu \\
& + &  4424159 c^3+99156580 c^2-181929196 c+101753840 ),\nonu \\
g_{146}& = & -\frac{4 
}{63 d(c)}
(c+18)(4c-9) (300504 c^5-1530069 c^4 \nonu \\
& - & 2079871 c^3+22480376 c^2-29629180 c+10548320),
\nonu \\
g_{147}& = & -\frac{2 
}{189 d(c)}
(c+18)(4c-9) (625455 c^6-3210129 c^5-300153 c^4
\nonu \\
& + & 27655623 c^3-53627648 c^2+32145412 c-271440),
\nonu \\
g_{148}& = & \frac{1
}{189 d(c)}
(c+18)(4c-9)(162459 c^6+2933967 c^5-32595351 c^4
\nonu \\
& + & 75850941 c^3-16839780 c^2-67186556 c+17845360),
\nonu \\
g_{149}& = & \frac{1}{
1134 d(c)}
 (c+18)(4c-9)(475110 c^7-3251841 c^6+9988227 c^5+14280429 c^4
 \nonu \\
&- & 
145988349 c^3+167392540 c^2+72391324 c-185757360),\nonu \\
g_{150}& = & \frac{4 
}{189 d(c)}
(c+18)(4c-9)
(23769 c^6-254280 c^5+1408404 c^4 \nonu \\
& + & 2464431 c^3-37531800 c^2+69153716 c-42505600),
\nonu \\
g_{151}& = & \frac{1}{
1134 d(c)}
(c+18)(4c-9)(189990 c^7-432285 c^6-11776965 c^5+53366073 c^4
 \nonu \\
& - &  47183853 c^3-146220332 c^2+325480812 c-203762800),\nonu \\
g_{152}& = & \frac{2 
}{63 d(c)}
(c+18)(4c-9)(7c-15)(69837 c^4+80102 c^3-273197 c^2-8398 c-9264), \nonu \\
g_{153}& = & \frac{5 
}{63 d(c)}
(c+18)(4c-9)(7c-15)(59920 c^3+99239 c^2-369542 c+118512), \nonu \\
g_{154}& = & \frac{4 
}{21 d(c)}
(c+18)(4c-9)(7c-15)(22483 c^4-80817 c^3-234518 c^2+603208 c-194656), \nonu \\
g_{155}& = & \frac{1
}{1512  d(c)}
(c+18)(4c-9)(7c-15)(292707 c^5+774462 c^4 \nonu \\
& - & 3643553 c^3-1591472 c^2+8949660 c-2864304), \nonu \\
g_{156}& = & \frac{1
}{1512 d(c)}
(c+18)(4c-9)
(-562455 c^6-5398647 c^5+25184739 c^4 \nonu \\
& - & 35895549 c^3+74930604 c^2-108934132 c+14813840), \nonu \\
g_{157}& = & \frac{1
}{21 d(c)}
(c+18)(4c-9)(-105463 c^5-819822 c^4 \nonu \\
& + & 3550777 c^3+35728 c^2-5642220 c
-104560), \nonu \\
g_{158}& = & \frac{1
}{504 d(c) }
(c+18)(4c-9)(7c-15)(-45581 c^5-75866 c^4 \nonu \\
& + & 1615031 c^3-1802776 c^2-4285428 c
+5960400), \nonu \\
g_{159}& = & \frac{1}{
756 d(c)}
 (c+18)(4c-9)(-138495 c^7-520346 c^6+606288 c^5+26463330 c^4
 \nonu \\
& - &  
90266845 c^3+99376456 c^2-47040788 c+30331600 ),\nonu \\
g_{160}& = & \frac{1
}{756 d(c)}
(c+18)(4c-9)(-373575 c^6-2267971 c^5+6491907 c^4 \nonu \\
& - & 31017 c^3-6577068 c^2+55992044 c-116103920), \nonu \\
g_{161}& = & -\frac{2 
}{189 d(c)}
(c+18)(4c-9)(199091 c^5+1646137 c^4 \nonu \\
& - & 6103252 c^3+10943272 c^2-24441808 c+33035440), \nonu \\
g_{162}& = & \frac{1
}{378 d(c)}
(c+18)(4c-9)(-178347 c^6-21231 c^5+15272175 c^4 \nonu \\
& - & 59486805 c^3+84471732 c^2-76464484 c+65345360), \nonu \\
g_{163}& = & \frac{1
}{252 d(c)}
(c+18)(4c-9)(7c-15)(-75337 c^5-554838 c^4 \nonu \\
& + & 561551 c^3+3297104 c^2-2733092 c-2489328), \nonu \\
g_{164}& = & \frac{1}{
2268 d(c)}
 (c+18)(4c-9) (39990 c^7+290601 c^6-1768743 c^5+7591803 c^4
 \nonu \\
& - &  19564683 c^3+28507012 c^2-53076860 c+66726960), \nonu \\
g_{165}& = & \frac{1}{
4536  d(c)}
(c+18)(4c-9)(-554085 c^7-769497 c^6+23913393 c^5-60404875 c^4
 \nonu \\
& - &  
77152300 c^3+539972532 c^2-857056848 c+457760640),\nonu \\
g_{166}& = & \frac{
}{378 d(c)}
(c+18)(4c-9)(-183939 c^6-145125 c^5+9357785 c^4 \nonu \\
& - & 32632385 c^3+15566564 c^2-1629780 c+8217840), \nonu \\
g_{167}& = & \frac{1}{
4536 d(c)}
 (c+18)(4c-9) (119835 c^7+215001 c^6-683421 c^5-59477757 c^4
\nonu \\
& + &  73321842 c^3+452224012 c^2-823406552 c+236772000
),\nonu \\
g_{168}& = & \frac{1}{
2268 d(c)}
 (c+18)(4c-9)(23886 c^7-138501 c^6-1098825 c^5+8561277 c^4
 \nonu \\
& - &  70074753 c^3+208979872 c^2-197776524 c+2931248),\nonu \\
g_{169}& = & \frac{1}{
45360 d(c)}
 (c+18)(4c-9)(79170 c^8-50580 c^7-9407661 c^6+63719587 c^5-153782941 c^4
 \nonu \\
& + & 
87595321 c^3+185469740 c^2-191366716 c-5370960),\nonu \\
g_{170}& = & \frac{1
}{126 d(c)}
(c+18)(4c-9)(7c-15)(3 c^7-48 c^6-815 c^5+1796 c^4 \nonu \\
& - & 20932 c^3+62864 c^2-50480 c
+1632), \nonu \\
g_{171}& = & \frac{1}{
4536 d(c)}
 (c+18)(4c-9)(-1890 c^8+90 c^7+293955 c^6-1336737 c^5+3168903 c^4
 \nonu \\
& - &  2937673 c^3-2876452 c^2+3876924 c-682800),\nonu \\
g_{172}& = & -\frac{2 
}{63 d(c)}
(c+18)(4c-9)(131376 c^4-202711 c^3-302219 c^2+294074 c-101520),
\nonu \\
g_{173}& = & -\frac{128  
}{63 d(c)}
(c+18)(4c-9)(20 c-13) (98 c^2-129 c-194),  \nonu \\
g_{174}& = & -\frac{16  
}{21 d(c)}
(c+18)(4c-9)(1652 c^4-98177 c^3+263487 c^2-101002 c-48320), \nonu \\
g_{175}& = & -\frac{2  
}{63 d(c)}
(c+18)(4c-9)(45801 c^5-47772 c^4-440083 c^3+910558 c^2-363304 c-152400), \nonu \\
g_{176}& = & -\frac{16  
}{21 d(c)}
(c+18)(4c-9)(588 c^5-36292 c^4+161817 c^3-243347 c^2+127114 c-6560),
\nonu \\
g_{177}& = & -\frac{8  
}{21 d(c)}
(c+18)(4c-9)(273 c^5-14301 c^4-6314 c^3+108034 c^2-27092 c-60800), \nonu \\
g_{178}& = & -\frac{2  
}{63 d(c)}
(c+18)(4c-9)(1953 c^6-119295 c^5+579231 c^4 \nonu \\
& - & 1030281 c^3+803548 c^2-255716 c+720), \nonu \\
g_{179}& = & \frac{2(c+18)  
}{63  d(c)}
(4c-9)(861 c^6-29391 c^5-1133657 c^4 \nonu \\
& + & 4368507 c^3-4375708 c^2+974908 c+362320), 
\nonu \\
g_{180}& = & \frac{2(c+18)  
}{63 d(c)}
(4c-9)(1323 c^6-72861 c^5-243063 c^4 \nonu \\
& + & 1554153 c^3-1902276 c^2+898804 c-203600), \nonu \\
g_{181}& = & \frac{ 1
}{189 d(c)}
(c+18)(4c-9)(315 c^7-17724 c^6-18570 c^5+3771528 c^4 \nonu \\
& - & 12915513 c^3+12105392 c^2-2283748 c-1069680), \nonu \\
g_{182}& = & \frac{4 
}{63 d(c)}
(c+18)(4c-9)(441 c^6-23181 c^5-168783 c^4 \nonu \\
& + & 698013 c^3-631674 c^2+484624 c-328880), \nonu \\
g_{183}& = & \frac{1
}{189 d(c)}
(c+18)(4c-9)(504 c^7-27885 c^6-122169 c^5+2445849 c^4 \nonu \\
& - & 7192203 c^3+7285004 c^2-3313020 c+793840), \nonu \\
g_{184}& = & \frac{1
}{2268 d(c)}
(c+18)(4c-9)(-382725 c^6+175347 c^5+9121545 c^4 \nonu \\
& - & 29407735 c^3+33144380 c^2-11888652 c-530640), \nonu \\
g_{185}& = & \frac{1
}{378 d(c)}
(3 c-8) (c+18)(4c-9) (4095 c^6+42183 c^5-88431 c^4 \nonu \\
& - & 298851 c^3+690304 c^2-294404 c-42960), \nonu \\
g_{186}& = & \frac{1
}{126 d(c)}
(c+18)(4c-9)(16317 c^6+108045 c^5+496175 c^4 \nonu \\
& - & 5061985 c^3+10644348 c^2-8247300 c+1632720),
\nonu \\
g_{187}& = & \frac{4 
}{63 d(c)}
(c+18)(4c-9)(6027 c^5+39571 c^4+160094 c^3-724244 c^2+223632 c+122640), \nonu \\
g_{188}& = & \frac{1
}{21 d(c)}
 (c+18)(4c-9)(7c-15)(57 c^5-3574 c^4+2061 c^3+12544 c^2-15276 c+48), \nonu \\
g_{189}& = & \frac{1
}{126 d(c)}
 (c+18)(4c-9)(46935 c^6+344391 c^5-1960539 c^4 \nonu \\
& - & 3602811 c^3+19759956 c^2-16297772 c+918640),
\nonu \\
g_{190}& = & \frac{1 
}{756 d(c)}
(c+18)(4c-9)(29295 c^7+2007 c^6-1069131 c^5+3951597 c^4 \nonu \\
& - & 3487692 c^3-4812940 c^2+7455824 c-911040), \nonu \\
g_{191}& = & \frac{(c+18)
}{63 d(c)}
(4c-9)(6993 c^6-1341 c^5-224067 c^4 \nonu \\
& + & 1719387 c^3-4142880 c^2+2882708 c-40240),
\nonu \\
g_{192}& = & \frac{1 
}{84 d(c)}
(c+18)(4c-9)(7c-15)(-63 c^6-360 c^5+2841 c^4 \nonu \\
& - & 1042 c^3-46940 c^2+31208 c+26976), \nonu \\
g_{193}& = & \frac{1
}{189 d(c)}
(c+18)(4c-9)(-441 c^7-1386 c^6+28674 c^5-904366 c^4 \nonu \\
& + & 3121207 c^3-2904460 c^2-116108 c+945520), \nonu \\
g_{194}& = & -\frac{8 
}{21 d(c)}
(c+18)(4c-9)(882 c^5-48177 c^4-167813 c^3+941008 c^2-853460 c+160960),
\nonu \\
g_{195}& = & -\frac{4 
}{63 d(c)}
(c+18)(4c-9)(1575 c^6-98457 c^5+391671 c^4 \nonu \\
& - & 429801 c^3-40864 c^2+216356 c-720), \nonu \\
g_{196}& = & -\frac{2(c+18)
}{63 d(c)}
(4c-9)(693 c^6-46311 c^5+336423 c^4 \nonu \\
& + & 387867 c^3-3181020 c^2+2718748 c-112880), \nonu \\
g_{197}& = & \frac{1
}{189 d(c)}
(c+18)(4c-9)(630 c^7-30873 c^6-379989 c^5+4992837 c^4 \nonu \\
& - & 14666997 c^3+15196220 c^2-4653508 c-1112880),  \nonu \\
g_{198}& = & -\frac{8
}{63 d(c)}
 (c+18)(4c-9)(63 c^6-4890 c^5+67728 c^4 \nonu \\
& - & 912513 c^3+2593440 c^2-2163148 c+543200), \nonu \\
g_{199}& = & \frac{1
}{189 d(c)}
(c+18)(4c-9)(630 c^7-35085 c^6-162885 c^5+1152489 c^4 \nonu \\
& - & 567069 c^3-3787276 c^2+5366316 c-2322800), \nonu \\
g_{200}& = & \frac{4 
}{21 d(c)}
(c+18)(4c-9)(7c-15)(861 c^4+8206 c^3-1621 c^2+5986 c+1488), 
\nonu \\
g_{201}& = & \frac{10 
}{21 d(c)}
(c+18)(4c-9)(7c-15)(848 c^3+8095 c^2-5686 c+6768), \nonu \\
g_{202}& = & \frac{8
}{7 d(c)}
 (c+18)(4c-9)(7c-15)(29 c^4-1551 c^3-11014 c^2+16064 c-4928), \nonu \\
g_{203}& = & \frac{
}{252  d(c)}
(c+18)(4c-9)(7c-15)(2331 c^5+22446 c^4 \nonu \\
& - & 5449 c^3-180016 c^2+300540 c-98352), \nonu \\
g_{204}& = & \frac{1
}{252 d(c)}
(c+18)(4c-9) (56385 c^6-11631 c^5-2686773 c^4 \nonu \\
& + & 11981883 c^3-18743028 c^2+6918124 c+5071120),
\nonu \\
g_{205}& = & \frac{2 
}{7 d(c)}
(c+18)(4c-9)
(5971 c^5+25134 c^4-229189 c^3+390104 c^2-140100 c-33680), \nonu \\
g_{206}& = & \frac{1
}{84 d(c)}
(c+18)(4c-9)(7c-15)(-373 c^5-2218 c^4 \nonu \\
& + & 23023 c^3+53032 c^2-256404 c+221520), \nonu \\
g_{207}& = & \frac{1
}{126 d(c)}
(c+18)(4c-9)(-735 c^7-7378 c^6-31896 c^5-510630 c^4 \nonu \\
& + & 3822475 c^3-7366072 c^2+4269836 c+146000), \nonu \\
g_{208}& = & \frac{1
}{126 d(c)}
(c+18)(4c-9)(27825 c^6+365797 c^5-1149429 c^4 \nonu \\
& - & 730641 c^3+2394996 c^2+1405132 c-1870960),
\nonu \\
g_{209}& = & \frac{4 
}{63 d(c)}
(c+18)(4c-9)(9947 c^5+129379 c^4-458014 c^3+295024 c^2-328216 c+644080),
\nonu \\
g_{210}& = & \frac{
}{63 d(c)}
(c+18)(4c-9)(1029 c^6-37503 c^5-1446465 c^4 \nonu \\
& + & 8032635 c^3-13411404 c^2+6475228 c+780880),
\nonu \\
g_{211}& = & \frac{1
}{42 d(c)}
(c+18)(4c-9)(7c-15)(-641 c^5-7494 c^4-7817 c^3+4432 c^2+18044 c-62064), \nonu \\
g_{212}& = & \frac{1
}{378 d(c)}
(c+18)(4c-9)(-2730 c^7+1233 c^6+57681 c^5-2034861 c^4 \nonu \\
& + & 10003101 c^3-16987804 c^2+9551780 c-125520), \nonu \\
g_{213}& = & \frac{1
}{756 d(c)}
(c+18)(4c-9)(-2205 c^7+8799 c^6+198729 c^5+816925 c^4 \nonu \\
& - & 8304140 c^3+20430036 c^2-23033424 c+10561920), \nonu \\
g_{214}& = & \frac{1
}{63 d(c)}
(c+18)(4c-9)(-1407 c^6-1185 c^5+89605 c^4
\nonu \\
& + & 893675 c^3-1742348 c^2-2855460 c+3459120),
\nonu \\
g_{215}& = & \frac{1
}{756 d(c)}
(c+18)(4c-9)(-7245 c^7+47313 c^6+1512987 c^5-5063061 c^4 \nonu \\
& - & 17619294 c^3+79973996 c^2-86415256 c+26704800), \nonu \\
g_{216}& = & \frac{1
}{378 d(c)}
(c+18)(4c-9)(-882 c^7+11187 c^6+37095 c^5-1265499 c^4\nonu \\
& + & 5601471 c^3-11901904 c^2+14259828 c-6980816), \nonu \\
g_{217}& = & \frac{1
}{189 d(c)}
(c+18)(4c-9)(-90 c^8-90 c^7+19455 c^6-104869 c^5+107515 c^4 \nonu \\
& + & 308109 c^3-634042 c^2+291364 c-62640), \nonu \\
g_{218}& = & -\frac{80
}{21 d(c)}
 (c+18)(4c-9) (6096 c^3-2809 c^2-17966 c+13808), \nonu \\
g_{219}& = & -\frac{80
}{63 d(c)}
 (c+18)(4c-9)(13257 c^4-27821 c^3-7135 c^2+53020 c-38640), \nonu \\
g_{220}& = & -\frac{10 
}{63 d(c)}
(c+18)(4c-9)(19197 c^5-72906 c^4+75965 c^3-31032 c^2+80596 c-80400), \nonu \\
g_{221}& = & \frac{40 
}{63 d(c)}
(c+18)(4c-9)(228 c^6-14949 c^5+123220 c^4 \nonu \\
& - & 301133 c^3+252994 c^2-27140 c-34320), \nonu \\
g_{222}& = & -\frac{20 
}{21 d(c)}
(c+18)(4c-9)(3819 c^5-13562 c^4+15695 c^3+35476 c^2-110548 c+69520), \nonu \\
g_{223}& = & -\frac{10 
}{63 d(c)}
(c+18)(4c-9)(9477 c^6-37995 c^5+5173 c^4 \nonu \\
& + & 153891 c^3-205310 c^2+44324 c+31440), \nonu \\
g_{224}& = & \frac{20 
}{63 d(c)}
(c+18)(4c-9)(1395 c^6+27692 c^5-110585 c^4 \nonu \\
& - & 21390 c^3+272348 c^2+24864 c-191040), \nonu \\
g_{225}& = & \frac{10 
}{21 d(c)}
(c+18)(4c-9)(1011 c^6-6685 c^5+12515 c^4\nonu \\
& - & 28659 c^3+106830 c^2-148052 c+43760), \nonu \\
g_{226}& = & \frac{10 
}{63 d(c)}
(c+18)(4c-9) (477 c^7-1746 c^6-6298 c^5+71104 c^4\nonu \\
& - & 246755 c^3+368286 c^2-199708 c+11280), \nonu \\
g_{227}& = & \frac{10
}{21 d(c)}
 (c+18)(4c-9) \nonu \\
& \times & (489 c^6-1364 c^5-18887 c^4+67270 c^3-49488 c^2-40760 c+45600),
\nonu \\
g_{228}& = & \frac{10 
}{189 d(c)}
(c+18)(4c-9) (486 c^7-1419 c^6-33744 c^5+231697 c^4 \nonu \\
& - & 508856 c^3+402488 c^2-29272 c-62880), \nonu \\
g_{229}& = & -\frac{40 
}{63 d(c)}
(c+18)(4c-9)(7761 c^5-58352 c^4+23253 c^3+121752 c^2-45140 c+10320),
\nonu \\
g_{230}& = & -\frac{80 
}{63 d(c)}
(c+18)(4c-9)(4882 c^4-68019 c^3+38490 c^2+151372 c-78720),
\nonu \\
g_{231}& = & -\frac{40 
}{21 d(c)}
(c+18)(4c-9)(343 c^5-22056 c^4+159035 c^3-268298 c^2+52988 c+67120),
\nonu \\
g_{232}& = & -\frac{10 
}{189 d(c)}
(c+18)(4c-9) (18225 c^6-46848 c^5-332323 c^4 \nonu \\
& + & 1476154 c^3-1946332 c^2+756984 c+90720),
\nonu \\
g_{233}& = & \frac{20 
}{63 d(c)}
(c+12)(c+18)(4c-9)(1731 c^5-8585 c^4+20519 c^3-33363 c^2+20866 c+10220),
\nonu \\
g_{234}& = & \frac{40 
}{21 d(c)}
(c+18)(4c-9)(2099 c^5+32122 c^4-110529 c^3-19982 c^2+199604 c+2320),
\nonu \\
g_{235}& = & \frac{10 
}{21 d(c)}
(c+18)(4c-9)(763 c^6-3594 c^5-19489 c^4 \nonu \\
& + & 134932 c^3-213840 c^2+30880 c+96640),
\nonu \\
g_{236}& = & \frac{10 
}{63 d(c)}
(3 c-8) (c+18)(4c-9)(195 c^6-493 c^5+8747 c^4 \nonu \\
& - & 40059 c^3+68806 c^2-49588 c+11280), \nonu \\
g_{237}& = & \frac{10 
}{63 d(c)}
(c+18)(4c-9)(1866 c^6+1757 c^5+23458 c^4 \nonu \\
& + & 119349 c^3-499894 c^2+398660 c-53040), \nonu \\
g_{238}& = & \frac{40 
}{21 d(c)}
(c+18)(4c-9)(443 c^5+724 c^4+23935 c^3+17768 c^2-173660 c+134000),
\nonu \\
g_{239}& = & -\frac{20 
}{21 d(c)}
(c+18)(4c-9) \nonu \\
& \times & (111 c^6-7548 c^5+72827 c^4-223666 c^3+301112 c^2-178096 c+29760), \nonu \\
g_{240}& = & \frac{10 
}{21 d(c)}
(c+18)(4c-9)(984 c^6-3659 c^5+46656 c^4 \nonu \\
& - & 35279 c^3-402294 c^2+687268 c-122800), \nonu \\
g_{241}& = & -\frac{10 
}{63 d(c)}
(c+18)(4c-9) (120 c^7-1287 c^6+4254 c^5+28319 c^4 \nonu \\
& - & 64680 c^3-4418 c^2-116352 c+274080), \nonu \\
g_{242}& = & \frac{10 
}{189 d(c)}
(c+18)(4c-9) (918 c^7+1836 c^6-50245 c^5+143604 c^4\nonu \\
& + & 121361 c^3-765994 c^2+565460 c+129360), \nonu \\
g_{243}& = & \frac{20 
}{63 d(c)}
(c+18)(4c-9)(408 c^6+1021 c^5-30454 c^4 \nonu \\
& + & 192145 c^3-250296 c^2-168204 c+370160), \nonu \\
g_{244}& = & -\frac{10 
}{63 d(c)}
(c+18)(4c-9) (78 c^7-108 c^6+2243 c^5-16880 c^4 \nonu \\
& + & 54353 c^3-189246 c^2+399540 c-301680), \nonu \\
g_{245}& = & -\frac{5 
}{189 d(c)}
(c+18)(4c-9) (150 c^7+385 c^6-23625 c^5+143885 c^4 \nonu \\
& - & 769841 c^3+2257442 c^2-2765932 c+902160), \nonu \\
g_{246}& = & -\frac{80 
}{21 d(c)}
(c+18)(4c-9) (4195 c^4-1047 c^3-40025 c^2+55008 c+3760), \nonu \\
g_{247}& = & -\frac{40 
}{21 d(c)}
(c+18)(4c-9)(3609 c^5-1335 c^4-49737 c^3+104901 c^2-72164 c+32000),
\nonu \\
g_{248}& = & \frac{10 
}{7  d(c)}
(c+18)(4c-9) (2701 c^5-5894 c^4-31363 c^3+99732 c^2-70492 c-5200), \nonu \\
g_{249}& = & \frac{20 
}{21 d(c)}
(c+18)(4c-9) (2319 c^5+9138 c^4-31589 c^3+13120 c^2-113284 c+214160), \nonu \\
g_{250}& = & \frac{20 
}{63 d(c)}
(c+18)(4c-9) (1116 c^6-13153 c^5+44010 c^4 \nonu \\
& - & 9795 c^3-142538 c^2+126284 c+42480), \nonu \\
g_{251}& = & \frac{20 
}{63 d(c)}
(c+18)(4c-9) (1091 c^6+10161 c^5-27479 c^4 \nonu \\
& - & 377765 c^3+1175300 c^2-618356 c-441840), \nonu \\
g_{252}& = & \frac{20 
(50 c^3-153 c^2+315 c-54)
}{21 (c-1) (c+1) (c+6) (2 c-3) (5 c-9)},\nonu \\
g_{253}& = & -\frac{10 
}{189 d(c)}
(c+18)(4c-9) (285 c^7-3522 c^6-35938 c^5+269676 c^4 \nonu \\
& - & 556887 c^3+595950 c^2-471660 c+85200), \nonu \\
g_{254}& = & \frac{10 
}{63 d(c)}
(c+18)(4c-9)(627 c^6+5374 c^5+44127 c^4 \nonu \\
& - & 383192 c^3+475524 c^2+433392 c-689280), \nonu \\
g_{255}& = & -\frac{5 
}{63 d(c)}
(c+18)(4c-9) (66 c^7-2211 c^6-5180 c^5+96103 c^4 \nonu \\
& - & 224456 c^3+58658 c^2+239816 c-146176), \nonu \\
g_{256}& = & \frac{11 
}{3780 d(c)}
(c+18)(4c-9) (180 c^8-180 c^7-21765 c^6+157788 c^5-528517 c^4\nonu \\
& + & 460462 c^3+821708 c^2-1216416 c+198720), \nonu \\
g_{257}& = & -\frac{176 
}{21 d(c)}
(c+18)(4c-9)(80 c^3-4329 c^2+4962 c-2608), \nonu \\
g_{258}& = & -\frac{176 
}{63 d(c)}
(c+18)(4c-9)(171 c^4-9956 c^3+25436 c^2-24506 c+6360), \nonu \\
g_{259}& = & -\frac{22 
}{63 d(c)}
(c+18)(4c-9)(243 c^5-15138 c^4+61983 c^3-85808 c^2+35980 c+1680), \nonu \\
g_{260}& = & \frac{88 
}{63 d(c)}
(c+18)(4c-9)(4479 c^5+452 c^4-43237 c^3+109792 c^2-112436 c+38400), \nonu \\
g_{261}& = & \frac{88 
}{63 d(c)}
(c+18)(4c-9)(6604 c^4+11431 c^3-38011 c^2+40346 c-21360), \nonu \\
g_{262}& = & \frac{176 
}{21 d(c)}
(c+18)(4c-9)(28 c^5-1577 c^4-2573 c^3+14678 c^2-12966 c+1880), \nonu \\
g_{263}& = & \frac{11(c+18)  
}{189 d(c)}
(4c-9)(18225 c^6-27165 c^5-237457 c^4 \nonu \\
& + & 752397 c^3-644252 c^2-43068 c+174960), \nonu \\
g_{264}& = & -\frac{88  }{63 d(c)} 
(c+18)(4c-9)(45 c^6-4887 c^5+22329 c^4-12699 c^3-41686 c^2+49518 c-7320),
\nonu \\ 
g_{265}& = & \frac{44  
}{63 d(c)}
(c+18)(4c-9)(249 c^6-16059 c^5+101543 c^4 \nonu \\
& - & 253253 c^3+227044 c^2+22196 c-88080), \nonu \\
g_{266}& = & \frac{22
 }{63 d(c)}
(c+12)(c+18)(4c-9)(177 c^5-457 c^4+4119 c^3-13923 c^2+11424 c-260), \nonu \\
g_{267}& = & -\frac{176 
}{21 d(c)}
 (c+18)(4c-9)(23 c^5-4919 c^4+6774 c^3+24481 c^2-45744 c+27280), \nonu \\
g_{268}& = & -\frac{11  
}{21 d(c)}
(c+18)(4c-9)(931 c^6-335 c^5-18139 c^4 \nonu \\
& + & 38639 c^3+88420 c^2-402836 c+375440),
\nonu \\
g_{269}& = & -\frac{11  
}{126 d(c)}
(c+18)(4c-9) (1170 c^7+2985 c^6-28259 c^5+141419 c^4 \nonu \\
& - & 560379 c^3+883164 c^2-433140 c-42480), \nonu \\
g_{270}& = & -\frac{22
}{63 d(c)}
  (c+18)(4c-9)(222 c^6-4807 c^5+46014 c^4
\nonu \\
& - & 161769 c^3+440212 c^2-576072 c+258720), \nonu \\
g_{271}& = & -\frac{88  
}{63 d(c))}
(c+18)(4c-9)(165 c^5-3333 c^4+25253 c^3+31402 c^2-80902 c+36600), \nonu \\
g_{272}& = & -\frac{88  
}{21 d(c)}
(c+18)(4c-9)(33 c^6-2232 c^5+19960 c^4 \nonu \\
& - & 62545 c^3+70802 c^2-12248 c-15360),
\nonu \\
g_{273}& = & -\frac{22 
}{21 d(c)}
(c+18)(4c-9)(576 c^6+2159 c^5+18482 c^4 \nonu \\
& - & 70077 c^3-28984 c^2+184884 c-155280), \nonu \\
g_{274}& = & -\frac{11 
}{126 d(c)}
(c+18)(4c-9) (30 c^7-2199 c^6-14551 c^5+52307 c^4 \nonu \\
& - & 75747 c^3+228476 c^2-336076 c+55920), \nonu \\
g_{275}& = & -\frac{11 
}{189 d(c)}
(c+18)(4c-9)(918 c^7-21 c^6-33223 c^5+150105 c^4 \nonu \\
& - & 171483 c^3-412484 c^2+973548 c-526320), \nonu \\
g_{276}& = & -\frac{44   
}{63 d(c)}
(c+18)(4c-9)(225 c^6+236 c^5+1897 c^4 \nonu \\
& + & 30008 c^3-247068 c^2+388012 c-208240),
\nonu \\
g_{277}& = & \frac{11(c+18) 
}{63 d(c)}
(4c-9)(6 c^7+609 c^6+887 c^5+31567 c^4 \nonu \\
& - & 189813 c^3+449308 c^2-659124 c+413840),
\nonu \\
g_{278}& = & \frac{11
}{756 d(c)}
 (c+18)(4c-9) (120 c^7-1760 c^6-75337 c^5+455562 c^4 \nonu \\
& + & 57343 c^3-3932888 c^2+6392420 c-3266640), \nonu \\
g_{279}& = & \frac{44 
}{21 d(c)}
(c+18)(4c-9)(2319 c^5-3956 c^4-23059 c^3+62404 c^2-53148 c+14800),
\nonu \\
g_{280}& = & \frac{11 
}{63 d(c)}
(c+18)(4c-9)(9477 c^6-36411 c^5-18185 c^4
\nonu \\
& + & 216495 c^3-225252 c^2-37004 c+86640), \nonu \\
g_{281}& = & \frac{11 
}{21 d(c)}
(c+18)(4c-9)(213 c^6-11855 c^5+52943 c^4 \nonu \\
& - & 98693 c^3+140940 c^2-169548 c+91120), \nonu \\
g_{282}& = & -\frac{11 
}{63 d(c)}
(c+18)(4c-9)(477 c^7-1230 c^6-7150 c^5+32158 c^4
\nonu \\
& - & 21915 c^3-45632 c^2+34732 c+24720),
\nonu \\
g_{283}& = & \frac{22 
}{21 d(c)}
(c+18)(4c-9)(3 c^6-202 c^5+503 c^4-11528 c^3+57580 c^2-79976 c+28320),
\nonu \\
g_{284}& = & -\frac{22}{189 d(c)} 
(c+18)(4c-9) \nonu \\
& \times & (243 c^7-795 c^6-10749 c^5+49307 c^4-45450 c^3-58048 c^2+110932 c-45840),
\nonu \\
g_{285}& = & \frac{176 
}{21 d(c)}
(c+18)(4c-9)(3389 c^4+2916 c^3-47276 c^2+61726 c-23240),
\nonu \\
g_{286}& = & \frac{44 
}{21 d(c)}
(c+18)(4c-9)(4617 c^5-5848 c^4-53777 c^3+154292 c^2-139344 c+29440), \nonu \\
\nonu \\
g_{287}& = & \frac{22 
}{7 d(c)}
(c+18)(4c-9)(213 c^5-11306 c^4+26581 c^3+21664 c^2-61892 c+4880),
\nonu \\
g_{288}& = & \frac{44 
}{21 d(c)}
(c+18)(4c-9) (169 c^5-9148 c^4+5303 c^3+65592 c^2-69716 c-27760), \nonu \\
g_{289}& = & \frac{11 
}{63 d(c)}
(c+18)(4c-9) (657 c^6-37867 c^5+147795 c^4 \nonu \\
&- & 115425 c^3-124292 c^2+46412 c+130320), \nonu \\
g_{290}& = & -\frac{44 
}{63 d(c)}
(c+18)(4c-9) (703 c^6+8125 c^5+13077 c^4 \nonu \\
& - & 177397 c^3+257304 c^2+5908 c-109200), \nonu \\
g_{291}& = & -\frac{11 (133 c^3-888 c^2+745 c+270)}{21 (c-1) (c+1) (c+6) (2 c-3) (5 c-9)},\nonu \\
g_{292}& = & -\frac{11(c+12) 
}{189 d(c)}
(c+18)(4c-9) (75 c^6-3450 c^5+24846 c^4 \nonu \\
& - & 61394 c^3+51555 c^2+29628 c-52380),
\nonu \\
g_{293}& = & -\frac{22 
}{63 d(c)}
(c+18)(4c-9) (471 c^6-94 c^5-14965 c^4 \nonu \\
& + & 58500 c^3-168796 c^2+430464 c-392640),
\nonu \\
g_{294}& = & -\frac{11 
}{63 d(c)}
(c+18)(4c-9) (12 c^7-717 c^6-440 c^5+20163 c^4\nonu \\
& - & 12750 c^3-166852 c^2+334948 c-174544), 
\nonu \\
g_{295}& = & -\frac{11}{
6804 d(c)}
  (2700 c^9+19890 c^8+982473 c^7-2869278 c^6-38080623 c^5+225907902 c^4
 \nonu \\
& - &  435221980 c^3+281645120 c^2+49821936 c-82434240),\nonu \\
g_{296}& = & -\frac{440
}{189 d(c)}
  (75072 c^5-22400 c^4-949713 c^3+1594868 c^2-1242708 c+182304),
\nonu \\
g_{297}& = & -\frac{1760
}{189 d(c)}
  (17920 c^4+18322 c^3-367777 c^2+584066 c-272880),
\nonu \\
g_{298}& = & -\frac{880 
}{63 d(c)}
(50048 c^5-87950 c^4-1615395 c^3+6639000 c^2-7876228 c+3215520), \nonu \\
g_{299}& = & -\frac{440 
}{567 d(c)}
(78516 c^6-128676 c^5-272257 c^4 \nonu \\
& - & 2046118 c^3+10495212 c^2-14684472 c+5922720), \nonu \\
g_{300}& = & -\frac{880 
}{63 d(c)}
(34896 c^6-109310 c^5-720956 c^4 \nonu \\
& + & 4384211 c^3-8814282 c^2+7933696 c-2580480), 
\nonu \\
g_{301}& = & -\frac{55}{
1701 d(c)
}
 (218700 c^7-654759 c^6+1552032 c^5-24397223 c^4
 \nonu \\
& + & 
119501698 c^3-247092684 c^2+236806776 c-87726240),\nonu \\
g_{302}& = & -\frac{110}{
189 d(c)
}
 (145800 c^7-659337 c^6-1084556 c^5+11396927 c^4 \nonu \\
& - &  24532882 c^3+20544604 c^2-4304856 c-1317600),
\nonu \\
g_{303}& = & \frac{1760 
}{189 d(c)}
(9636 c^6-30138 c^5-798891 c^4+3871226 c^3-6086812 c^2+1337064 c+2168640), \nonu \\
g_{304}& = & \frac{220}{
189 d(c)}
 (26352 c^7-142275 c^6-2066646 c^5+13702985 c^4
 \nonu \\
& - &  
26614412 c^3+11039360 c^2+14381736 c-9223200),\nonu \\
g_{305}& = & \frac{220}{
189 d(c)}
 (11664 c^8+114705 c^7-1146123 c^6+2204193 c^5+2296855 c^4
 \nonu \\
& - &  10585714 c^3+7644412 c^2+2585208 c-2570400),\nonu \\
g_{306}& = & \frac{440}{
567 d(c))}
 (22464 c^8+256896 c^7-1522029 c^6-2278006 c^5+29552061 c^4
 \nonu \\
& - &  79406486 c^3+98880444 c^2-51900984 c+4263840),\nonu \\
g_{307}& = & \frac{880 
}{189 d(c)}
(10674 c^7+133065 c^6-193007 c^5-1867691 c^4  \nonu \\
&  + &  921587 c^3+11754034 c^2-12898912 c+2498400),
\nonu \\
g_{308}& = & -\frac{440 
}{189 d(c)}
(c+18)(4c-9) (252 c^6+831 c^5-7178 c^4+1761 c^3+6158 c^2-54924 c+60240),
\nonu \\
g_{309}& = & \frac{220}{
189 d(c)}
 (66600 c^7+864357 c^6-5671110 c^5-14850519 c^4 \nonu \\
& + &  113717484 c^3-194772256 c^2+124433864 c-14677920)
,\nonu \\
g_{310}& = & -\frac{55}{
567 d(c)}
 (6480 c^9+26406 c^8-702807 c^7+7542303 c^6-43143369 c^5
+137338479 c^4  \nonu \\
& - &  270293228 c^3+311793112 c^2-110789736 c-26598240),
\nonu \\
g_{311}& = & \frac{440}{
189 d(c)}
 (11736 c^7+99789 c^6-800994 c^5+3414513 c^4
 \nonu \\
& - &  12036999 c^3+21724034 c^2-3972424 c-10565280)
,\nonu \\
g_{312}& = & \frac{110}{
189 d(c)}
 (16920 c^8+121245 c^7-1925880 c^6+7768803 c^5-11118178 c^4
 \nonu \\
& - &  8040462 c^3+43138372 c^2-42244320 c+10670400),\nonu \\
g_{313}& = & -\frac{110}{
567 d(c)}
 (9576 c^8+321936 c^7+983931 c^6-8031738 c^5+35068203 c^4
 \nonu \\
& - &  247077040 c^3+685259288 c^2-626900136 c+177547680),\nonu \\
g_{314}& = & -\frac{110}{
189 d(c)}
 (2592 c^8+39573 c^7+244185 c^6-588723 c^5-25366917 c^4
 \nonu \\
& + &  121262562 c^3-208758820 c^2+154418408 c-33521760),\nonu \\
g_{315}& = & -\frac{110}{
1701 d(c)}
 (6912 c^9+70704 c^8-913830 c^7+2463801 c^6
+6707100 c^5-76191867 c^4  \nonu \\
& + &  303245488 c^3-562399784 c^2+397912296 c-78153120),\nonu \\
g_{316}& = & -\frac{220}{
567 d(c)}
 (2304 c^8+47226 c^7-327027 c^6-2111052 c^5+15011133 c^4
 \nonu \\
& - &  38563088 c^3+48602704 c^2-4415880 c-16247520),\nonu \\
g_{317}& = & -\frac{55}{
2268 d(c)}
 (5760 c^9+68664 c^8-1149888 c^7-2753637 c^6+54526944 c^5
-170726221 c^4  \nonu \\
& + &  188145570 c^3-13709740 c^2-121599512 c+53350560),
\nonu \\
g_{318}& = & \frac{110}{
189 d(c)}
 (9324 c^7-78771 c^6+141400 c^5+6246773 c^4
 \nonu \\
& - &  39604982 c^3+95449044 c^2-104485128 c+40249440),\nonu \\
g_{319}& = & \frac{440 
}{189 d(c)}
(6888 c^6+20750 c^5-294623 c^4 \nonu \\
& + & 3691028 c^3-12970656 c^2+15344808 c-7447680), \nonu \\
g_{320}& = & \frac{220 
}{63 d(c)}
(c+18)(4c-9)(1533 c^5-10840 c^4+27621 c^3-34274 c^2+9820 c+5680),
\nonu \\
g_{321}& = & -\frac{55}{
567 d(c)}
 (182250 c^7-1340337 c^6+2778044 c^5-177025 c^4
 \nonu \\
& - & 
280080 c^3-17509728 c^2+32278536 c-15564960),\nonu \\
g_{322}& = & \frac{110}{
567 d(c)}
 (7020 c^8+83745 c^7-512802 c^6+521697 c^5+665592 c^4
\nonu \\
& - &  901952 c^3-6257576 c^2+22610976 c-17712000),
\nonu \\
g_{323}& = & -\frac{220}{
189 d(c)}
 (11664 c^8+102477 c^7-1295826 c^6+3360573 c^5+316212 c^4
 \nonu \\
& - & 
10271768 c^3+8582392 c^2+1976736 c-2453760),\nonu \\
g_{324}& = & \frac{110}{
189 d(c)}
 (9324 c^8+146139 c^7-202511 c^6-4250207 c^5+10925659 c^4
 \nonu \\
& + & 
13144068 c^3-70183216 c^2+76275624 c-23410080),\nonu \\
g_{325}& = & \frac{110}{
189 d(c)}
 (53532 c^7+767685 c^6-2193444 c^5-10896171 c^4
 \nonu \\
& + & 
9742374 c^3+98626204 c^2-180480680 c+88161120),\nonu \\
g_{326}& = & -\frac{55}{
63 d(c)}
(840 c^8+9860 c^7+48679 c^6+1019934 c^5-6951781 c^4
 \nonu \\
& + & 
3898456 c^3+38360160 c^2-74863128 c+39506400),\nonu \\
g_{327}& = & \frac{55}{
189 d(c)}
 (4464 c^8-7341 c^7-558519 c^6+4947737 c^5-18320529 c^4 \nonu \\
& + &  34654036 c^3-33672936 c^2+14996088 c-2008800),
\nonu \\
g_{328}& = & \frac{55}{
567 d(c)}
 (16740 c^8+28035 c^7-2868195 c^6+14771073 c^5-28031037 c^4
 \nonu \\
& + &  34901672 c^3-69359032 c^2+93286344 c-42487200),\nonu \\
g_{329}& = & 
\frac{220 
}{189 d(c)}
(3996 c^7+12855 c^6-689520 c^5+2504119 c^4  \nonu \\
& + &  1801300 c^3-18685216 c^2+23304536 c-8197920), \nonu \\
g_{330}& = & \frac{220}{
63 d(c)}
 (1800 c^8+231 c^7-226962 c^6+1842287 c^5-6541848 c^4
 \nonu \\
& + &  11783392 c^3-10109556 c^2+2626296 c+704160),\nonu \\
g_{331}& = & -\frac{55 
}{63 d(c)}
(c+18)(4c-9)(189 c^6-63 c^5-12315 c^4 \nonu \\
& + & 52871 c^3-23558 c^2-57564 c+54160),
\nonu \\
g_{332}& = & -\frac{55}{
189 d(c)}
 (756 c^9+7755 c^8-65685 c^7+47744 c^6-848063 c^5+9440203 c^4
 \nonu \\
& - &  28907678 c^3+30662324 c^2+209304 c-12972960),\nonu \\
g_{333}& = & \frac{55}{
1701 d(c)}
 (18756 c^8+659607 c^7-2818998 c^6-4991441 c^5+43101944 c^4
 \nonu \\
& - &  94625792 c^3+123669768 c^2-115595424 c+58164480),\nonu \\
g_{334}& = & -\frac{110}{
567 d(c)}
(504 c^8-3432 c^7-456099 c^6-159398 c^5+14022357 c^4
 \nonu \\
& - & 
33026844 c^3-1201720 c^2+58928632 c-47512800),\nonu \\
g_{335}& = & -\frac{55}{
567 d(c)}
 (1008 c^9+13908 c^8-75549 c^7+165818 c^6-152501 c^5-7415864 c^4
 \nonu \\
&+ &  
45003360 c^3-112579544 c^2+138976224 c-66389760),\nonu \\
g_{336}& = & -\frac{55}{
6804 d(c)}
 (2160 c^9+13860 c^8-586974 c^7-1430121 c^6+29833068 c^5-16564081 c^4
 \nonu \\
& - &  
467615860 c^3+1447219200 c^2-1517889672 c+551279520),
\nonu \\
g_{337}& = & -\frac{1760 
}{189 d(c)}
(7584 c^6+256845 c^5-824244 c^4 \nonu \\
& - & 5117351 c^3+22590322 c^2-26351376 c+11050560), \nonu \\
g_{338}& = & -\frac{220}{
189 d(c)}
 (20520 c^7+669645 c^6-3174690 c^5-4390427 c^4
 \nonu \\
& + &  43436732 c^3-82034432 c^2+61874472 c-13093920),
\nonu \\
g_{339}& = & -\frac{220}{
189 d(c)}
 (67608 c^7+602169 c^6-8845074 c^5+17484441 c^4
 \nonu \\
& + &  
81590556 c^3-320183248 c^2+332866088 c-94770720),\nonu \\
g_{340}& = & -\frac{
440}{189 d(c)}
 (12240 c^7+18663 c^6-390342 c^5+2451513 c^4
 \nonu \\
& - & 
22030368 c^3+75267800 c^2-94835896 c+47020320),\nonu \\
g_{341}& = & -\frac{
110}{189 d(c)}
 (16920 c^8-7179 c^7-1144119 c^6+6051443 c^5-11704793 c^4
 \nonu \\
& + &  
7385432 c^3-413664 c^2+9106200 c-10095840),\nonu \\
g_{342}& = & \frac{220
}{567 d(c)}
 (4032 c^8-59421 c^7-467307 c^6+8919069 c^5-50115093 c^4
 \nonu \\
& + & 
131158472 c^3-146937904 c^2-5764488 c+69927840),\nonu \\
g_{343}& = & \frac{110
}{189 d(c)}
 (1944 c^8+13365 c^7-789843 c^6+533307 c^5+30900147 c^4
 \nonu \\
& - &  164552136 c^3+337871824 c^2-282108968 c+74034720),\nonu \\
g_{344}& = & \frac{
220}{1701 d(c)}
 (3456 c^9+45972 c^8-347322 c^7+643431 c^6+6279756 c^5
-50971293 c^4 \nonu \\
& + &  132957668 c^3-122144272 c^2-19673256 c+63171360),
\nonu \\
g_{345}& = & \frac{
440}{567 d(c)}
 (612 c^8-1212 c^7+171213 c^6+126246 c^5-7296891 c^4
 \nonu \\
& + & 
24212776 c^3-26589932 c^2+697008 c+8484480),\nonu \\
g_{346}& = & \frac{
55
}{567 d(c)}
 (1440 c^9+7422 c^8-577035 c^7+1083753 c^6
+13870939 c^5-61672407 c^4  \nonu \\
& + & 
67529760 c^3+56787992 c^2-142427144 c+63113760),\nonu \\
g_{347}& = & \frac{880 
}{189 d(c)}
(c+18)(4c-9)(4305 c^4+1609 c^3-913 c^2-10322 c+15120), \nonu \\
g_{348}& = & \frac{440 
}{63 d(c)}
(c+18)(4c-9)(5936 c^3+5849 c^2-27890 c+19024), \nonu \\
g_{349}& = & \frac{880 
}{21 d(c)}
(c+18)(4c-9) (1148 c^4-6356 c^3-9777 c^2+34004 c-21440), \nonu \\
g_{350}& = & \frac{55 
}{189 d(c)}
(c+18)(4c-9)\nonu \\
& \times & (6993 c^5+10086 c^4-42523 c^3-138836 c^2+425780 c-221520),
\nonu \\
g_{351}& = & \frac{440
}{63 d(c)}
(c+18)(4c-9) (777 c^5-5747 c^4+9969 c^3+515 c^2-12784 c+3920), \nonu \\
g_{352}& = & \frac{
55}{189 d(c)}
 (139068 c^7+1823397 c^6-6898464 c^5-17405571 c^4
 \nonu \\
& + &  96004698 c^3-144865388 c^2+119329720 c-61120800),\nonu \\
g_{353}& = & \frac{440 
}{63 d(c)}
(22972 c^6+311824 c^5-1170463 c^4 \nonu \\
& - & 3006342 c^3+15376220 c^2-17709176 c+5450400),
\nonu \\
g_{354}& = & \frac{110 
}{21 d(c)}
(7064 c^7+9079 c^6-1085304 c^5+4623711 c^4  \nonu \\
&- &  2173070 c^3-17521628 c^2+30388328 c-14116320),
\nonu \\
g_{355}& = & -\frac{55 (c+18)(4c-9)
}{63 d(c)}
(2359 c^5+11914 c^4-82357 c^3+20228 c^2+232572 c-193200),
\nonu \\
g_{356}& = & \frac{
110}{189 d(c)}
 (42612 c^7+636959 c^6-3010552 c^5+1866583 c^4
 \nonu \\
& - & 
892930 c^3+19975644 c^2-33472216 c+12201120),\nonu \\
g_{357}& = & \frac{440 
}{189 d(c)}
(20920 c^6+347936 c^5-1426883 c^4 \nonu \\
& - & 349602 c^3+4745732 c^2-2378728 c-2568960), \nonu \\
g_{358}& = & \frac{
220
}{189 d(c)}
 (31908 c^7+441945 c^6-5567610 c^5+13472361 c^4
 \nonu \\
& + & 
7973136 c^3-54016184 c^2+38621704 c+6444000),\nonu \\
g_{359}& = & -\frac{
440}{189 d(c)}
 (480 c^8+6009 c^7-178962 c^6+244423 c^5+3951962 c^4
 \nonu \\
& - &  
16359516 c^3+23025464 c^2-11163840 c+449280),\nonu \\
g_{360}& = & \frac{
55
}{567 d(c)}
 (27972 c^8+368259 c^7-1520038 c^6-5892689 c^5+36035712 c^4
 \nonu \\
& - & 
64733828 c^3+61012632 c^2-51558480 c+31700160),\nonu \\
g_{361}& = & -\frac{
110}{189 d(c))}
 (1680 c^8+18604 c^7-148821 c^6+2494884 c^5-9644493 c^4
 \nonu \\
& + & 
4505594 c^3+22334308 c^2-16509656 c-11950560),\nonu \\
g_{362}& = & \frac{
220}{189 d(c)}
 (1140 c^8+24789 c^7+137025 c^6-3014987 c^5+10097159 c^4
 \nonu \\
& - & 
6796098 c^3-10457476 c^2+3581688 c+12048480),\nonu \\
g_{363}& = & -\frac{110 
}{189 d(c)}
(c+18)(4c-9)(525 c^6+3799 c^5-17013 c^4 \nonu \\
& + & 35349 c^3-72648 c^2+31940 c+84720), \nonu \\
g_{364}& = & -\frac{110 
}{63 d(c)}
(c+18)(4c-9)(2149 c^5+15422 c^4-45335 c^3+26084 c^2-72140 c+185840),
\nonu \\
g_{365}& = & -\frac{
55}{63 d(c)}
 (3388 c^8+44693 c^7-303824 c^6-542215 c^5+6922578 c^4
 \nonu \\
& - & 
14228704 c^3-790240 c^2+30490624 c-24491520),\nonu \\
g_{366}& = & -\frac{
110}{567 d(c)}
 (8004 c^8+169263 c^7+395325 c^6-3165477 c^5-12906741 c^4
 \nonu \\
& + & 
70703614 c^3-85379492 c^2-3641736 c+40802400),\nonu \\
g_{367}& = & -\frac{
55}{1701 d(c)}
 (7560 c^9+124812 c^8-315657 c^7-2082843 c^6-2104887 c^5
+54569691 c^4  \nonu \\
& - &  142143196 c^3+197622056 c^2-255923016 c+183716640),
\nonu \\
g_{368}& = & -\frac{
55}{567 d(c)}
 (6300 c^8+101037 c^7+517305 c^6-4463881 c^5-15442625 c^4
 \nonu \\
& + & 
152773328 c^3-419060760 c^2+542904936 c-279776160),\nonu \\
g_{369}& = & -\frac{
220 }{567 d(c)}
 (6132 c^7+98787 c^6+240098 c^5-3258161 c^4 \nonu \\
& - &  11117866 c^3+59336412 c^2-44413752 c-2743200),\nonu \\
g_{370}& = & -\frac{220 
}{189 d(c)}
(c+18)(4c-9)(105 c^6-1731 c^5+3031 c^4 \nonu \\
& + & 6573 c^3-27982 c^2+69168 c-72480),
\nonu \\
g_{371}& = & -\frac{
55}{567 d(c)}
 (24228 c^8+338607 c^7-2454189 c^6-8206011 c^5+67040301 c^4
 \nonu \\
& + & 
11941624 c^3-471344152 c^2+633509752 c-222351840),\nonu \\
g_{372}& = & \frac{
55}{567 d(c)}
 (420 c^9+6339 c^8-24317 c^7-43908 c^6-3206283 c^5+
13399955 c^4  \nonu \\
& + & 
2587018 c^3-63802028 c^2+66375544 c-4753440), \nonu \\
g_{373}& = & -\frac{
55 }{3402 d(c)}
 (7560 c^9+74718 c^8-1829385 c^7-751944 c^6+62307729 c^5-196513344 c^4
 \nonu \\
& + & 
77768554 c^3+516936580 c^2-821150832 c+363021696),
\nonu \\
g_{374}& = & -\frac{
110}{567 d(c)}
    (2304 c^8+16596 c^7-526905 c^6+265554 c^5+18027105 c^4
 \nonu \\
& - &  73760894 c^3+123819628 c^2-105616440 c+39566880),\nonu \\
g_{375}& = & \frac{
55}{378 d(c)}
 (168 c^9+1458 c^8-35047 c^7-279404 c^6-4601 c^5+7937884 c^4
 \nonu \\
& - &  9851418 c^3-42288452 c^2+100462416 c-57763584),\nonu \\
g_{376}& = & \frac{
11 }{2268 d(c)}
 (1680 c^9+23940 c^8-271626 c^7+1785817 c^6+27001532 c^5
-241942423 c^4  \nonu \\
& + & 
626558564 c^3-567683360 c^2+19348296 c+185436000),
\nonu \\
g_{377}& = & \frac{
11}{20412 d(c)}
(2160 c^{10}+21060 c^9-521748 c^8-4464708 c^7+46409487 c^6-
85550322 c^5 \nonu \\
& - & 227188973 c^4   +  
1081582664 c^3-1401210248 c^2+356920248 c+275888160),\nonu \\
g_{378}& = & \frac{(c+18)(4c-9)}{7560 d(c)} 
(210 c^8-2340 c^7-17013 c^6+357851 c^5-2719733 c^4\nonu \\
&+ & 4848833 c^3+1172620 c^2-4953308 c+847920),
\nonu \\
g_{379}& = & -\frac{c}{21},\qquad g_{380} =  \frac{62}{21},
\qquad
g_{381} =  \frac{2}{7},\qquad
g_{382} =  -\frac{20}{7},\qquad
g_{383} =  \frac{22}{7}, \nonu \\
g_{384}  & = &   \frac{5 (3 c+25)}{126 (c-1)}, \nonu \\
g_{385}& = & \frac{10}{3 (c-1)}, \qquad
g_{386} =  \frac{125}{42},\qquad
g_{387} =  -\frac{(21 c-1)}{21 (c-1)}, \qquad
g_{388} =  -\frac{20}{7 (c-1)},
\nonu \\
g_{389}  & = &  -\frac{40 (c-2)}{21 (c-1)},
\nonu \\
g_{390}& = & -\frac{40}{7 (c-1)},\qquad
g_{391} =  \frac{44}{21},\qquad
g_{392} =  0, \qquad
g_{393} =  
\frac{2 (3 c^3-505 c^2+1656 c-564)}{63 (c-1) (c+6) (2 c-3)},\nonu \\
g_{394}& = & -\frac{4 (327 c^2-272 c+1260)}{63 (c-1) (c+6) (2 c-3)},
\qquad
g_{395} =  -\frac{8 (467 c+358)}{63 (c-1) (c+6) (2 c-3)},
\nonu \\
g_{396} & = &  \frac{4 (377 c^2-981 c+94)}{21 (c-1) (c+6) (2 c-3)},
\qquad
g_{397} =  \frac{100}{21 (c-1)}, 
\nonu \\
g_{398} & = &  
\frac{4 (47 c^3-24 c^2-144 c+376)}{63 (c-1) (c+6) (2 c-3)},\nonu \\
g_{399}& = & \frac{1}{7},\qquad
g_{400} =  -\frac{(21 c^3+95 c^2-288 c+12)}{21 (c-1) (c+6) (2 c-3)},
\qquad
g_{401} =  -\frac{2 (3 c+5) (7 c+18)}{21 (c-1) (c+6) (2 c-3)},
\nonu \\
g_{402} & = &  -\frac{4 (31 c+134)}{21 (c-1) (c+6) (2 c-3)},
\qquad
g_{403} =  \frac{2 (c^2-63 c+2)}{7 (c-1) (c+6) (2 c-3)},
\qquad
g_{404} =  -\frac{20}{7 (c-1)},\nonu \\
g_{405}& = & -\frac{10 (c^3-7 c^2+24 c-12)}{7 (c-1) (c+6) (2 c-3)},
\qquad
g_{406} =  \frac{40 (9 c+2)}{7 (c-1) (c+6) (2 c-3)},\nonu \\
g_{407}& = & \frac{10 (5 c-6) (9 c+2)}{21 (c-1) (c+6) (2 c-3)},
\qquad
g_{408} =  -\frac{20 (7 c^2+124 c-132)}{21 (c-1) (c+6) (2 c-3)},
\nonu \\
g_{409}& = & -\frac{10 (5 c^2-34 c+32)}{7 (c-1) (c+6) (2 c-3)},
\qquad
g_{410} =  \frac{11 (6 c^3-3 c^2+85 c+42)}{42 (c-1) (c+6) (2 c-3)},\nonu \\
g_{411}& = & -\frac{44 (11 c+14)}{7 (c-1) (c+6) (2 c-3)},
\qquad
g_{412} =  -\frac{11 (45 c^2-41 c+6)}{21 (c-1) (c+6) (2 c-3)},
\nonu \\
g_{413}& = & \frac{22 (c^2-97 c+6)}{21 (c-1) (c+6) (2 c-3)},
\qquad
g_{414} =  -\frac{11 (c^2-53 c+62)}{7 (c-1) (c+6) (2 c-3)},
\nonu \\
g_{415}& = & \frac{(27 c^5+645 c^4-7467 c^3+14615 c^2+8064 c+9036)}{252 (c-1) (c+1) (c+6) (2 c-3) (5 c-9)},\nonu \\
g_{416}& = & -\frac{(7077 c^3+2290 c^2+36671 c+29622)}{63 (c-1) (c+1) (c+6) (2 c-3) (5 c-9)},\nonu \\
g_{417}  & = &  
-\frac{2 (10084 c^2+10239 c+9032)}{63 (c-1) (c+1) (c+6) (2 c-3) (5 c-9)},\nonu \\
g_{418}& = & -\frac{2 (3941 c^3-15760 c^2-5735 c-9706)}{21 (c-1) (c+1) (c+6) (2 c-3) (5 c-9)},
\nonu \\
g_{419} & = &  
-\frac{(15096 c^4-44349 c^3-7572 c^2+23683 c-4518)}{63 (c-1) (c+1) (c+6) (2 c-3) (5 c-9)},\nonu \\
g_{420}& = & -\frac{(999 c^4+27978 c^3-80293 c^2+91140 c+9036)}{756 (c-1) (c+1) (c+6) (2 c-3) (5 c-9)},\nonu \\
g_{421}& = & -\frac{(6555 c^4-10426 c^3+8075 c^2-29436 c-125508)}{126 (c-1) (c+1) (c+6) (2 c-3) (5 c-9)},\nonu \\
g_{422}& = & \frac{(15069 c^4-58578 c^3+23841 c^2+112124 c-9036)}{63 (c-1) (c+1) (c+6) (2 c-3) (5 c-9)},\nonu \\
g_{423}& = & -\frac{(2631 c^2+2539 c-1170)}{126 (c-1) (c+6) (2 c-3)},
\qquad
g_{424} =  -\frac{(2501 c+1954)}{21 (c-1) (c+6) (2 c-3)},\nonu \\
g_{425}& = & \frac{(1309 c^4+4692 c^3-30535 c^2+30466 c-6632)}{84 (c-1) (c+1) (c+6) (2 c-3) (5 c-9)},\nonu \\
g_{426}& = & \frac{(2591 c^4+1856 c^3+5895 c^2-22506 c+89224)}{126 (c-1) (c+1) (c+6) (2 c-3) (5 c-9)},
\label{gcoeff} 
\\
g_{427}& = & \frac{(3765 c^5-18027 c^4+4169 c^3+26543 c^2+55350 c+196800)}{756 (c-1) (c+1) (c+6) (2 c-3) (5 c-9)},
\nonu \\
g_{428} & = &  \frac{2 (c^3+3 c^2+18 c+8)}{21 (c-1) (c+6) (2 c-3)}.
\nonu
\eea}
We introduce the $c$-dependent coefficient $d(c)$ in Appendix 
(\ref{gcoeff})
as follows:
{\small
\bea
d(c) \equiv  (c-2) (c-1) (c+1) (c+6) (c+12) (c+18) (2 c-3) (4 c-9) (5 c-9) (7 c-15).
\label{ddefinition}
\eea}
One can also analyze the $\frac{1}{c}, \cdots, \frac{1}{c^6}$-terms
in the large $c$ limit. 

Finally, the OPE 
between the fourth higher spin ${\cal N}=2$ multiplet and itself 
in (\ref{manyW}) can be summarized as
{\small
\bea
&& {\bf W}_{-\frac{1}{3}}^{(\frac{7}{2})}(Z_1) \, 
{\bf W}_{-\frac{1}{3}}^{(\frac{7}{2})}(Z_2) = 
C_{(\frac{7}{2}) \, (\frac{7}{2})}^{(\frac{7}{2}) \, -} \, \left(
\frac{\theta_{12}}{z_{12}^4} \, {\bf W}_{\frac{1}{3}}^{(\frac{7}{2})}(Z_2)
+\frac{\theta_{12} \, \bar{\theta}_{12}}{z_{12}^4} \,
f_2 \, \overline{D} {\bf W}_{\frac{1}{3}}^{(\frac{7}{2})}(Z_2) +
+\frac{1}{z_{12}^3} \,
f_3 \, \overline{D} {\bf W}_{\frac{1}{3}}^{(\frac{7}{2})}(Z_2) \right.
\nonu \\
&& +
\frac{ \theta_{12}}{z_{12}^3} \, \left[
f_4 \, [D, \overline{D}]  {\bf W}_{\frac{1}{3}}^{(\frac{7}{2})}
+ f_5 \, {\bf T}  {\bf W}_{\frac{1}{3}}^{(\frac{7}{2})} 
+  f_6 \, \pa {\bf W}_{\frac{1}{3}}^{(\frac{7}{2})}
\right](Z_2) 
\nonu \\
&&+
\frac{\theta_{12} \, \bar{\theta}_{12}}{z_{12}^3} 
\, 
\left[ 
f_7 \, \pa \overline{D}  {\bf W}_{\frac{1}{3}}^{(\frac{7}{2})}
+ f_8 \, {\bf T}  \overline{D}  {\bf W}_{\frac{1}{3}}^{(\frac{7}{2})}
+ f_9\, \overline{D} {\bf T}    {\bf W}_{\frac{1}{3}}^{(\frac{7}{2})}
\right](Z_2)
\nonu \\
&& + \frac{1}{z_{12}^2} \, 
\left[ 
f_{10} \, \pa \overline{D}  {\bf W}_{\frac{1}{3}}^{(\frac{7}{2})}
+ f_{11} \, {\bf T}  \overline{D}  {\bf W}_{\frac{1}{3}}^{(\frac{7}{2})}
+ f_{12} \, \overline{D} {\bf T}    {\bf W}_{\frac{1}{3}}^{(\frac{7}{2})}
\right]
(Z_2)
+  \frac{\bar{\theta}_{12}}{z_{12}^2} \, f_{13}\,
\overline{D}  {\bf T} \overline{D}
 {\bf W}_{\frac{1}{3}}^{(\frac{7}{2})}(Z_2)
\nonu \\
&& + \frac{\theta_{12}}{z_{12}^2} \, \left[ 
f_{14} \, \pa [D, \overline{D}]  {\bf W}_{\frac{1}{3}}^{(\frac{7}{2})}
+ f_{15} \, {\bf T} [D, \overline{D}]  {\bf W}_{\frac{1}{3}}^{(\frac{7}{2})}
+ f_{16} \, {\bf T} {\bf T}  {\bf W}_{\frac{1}{3}}^{(\frac{7}{2})}
 +  f_{17} \,
{\bf T} \pa {\bf W}_{\frac{1}{3}}^{(\frac{7}{2})}
+ f_{18} \, D {\bf T} \overline{D}  {\bf W}_{\frac{1}{3}}^{(\frac{7}{2})}
\right. \nonu \\
&& \left.
+  f_{19} \, [D, \overline{D}] {\bf T}  {\bf W}_{\frac{1}{3}}^{(\frac{7}{2})}
+  f_{20} \,
\overline{D} {\bf T} D  {\bf W}_{\frac{1}{3}}^{(\frac{7}{2})}
+ f_{21} \, \pa {\bf T}  {\bf W}_{\frac{1}{3}}^{(\frac{7}{2})}
+ f_{22} \, \pa^2  {\bf W}_{\frac{1}{3}}^{(\frac{7}{2})}
\right](Z_2)
\nonu \\
&& +   \frac{\theta_{12} \, \bar{\theta}_{12}}{z_{12}^2}
\left[ f_{23} \,
\pa^2 \overline{D}    {\bf W}_{\frac{1}{3}}^{(\frac{7}{2})}
+ f_{24} \, {\bf T} \pa  \overline{D}  {\bf W}_{\frac{1}{3}}^{(\frac{7}{2})}
+ f_{25} \, {\bf T} {\bf T}  \overline{D}  {\bf W}_{\frac{1}{3}}^{(\frac{7}{2})}
+ f_{26} \,
 [D, \overline{D}] {\bf T}  \overline{D}  {\bf W}_{\frac{1}{3}}^{(\frac{7}{2})}
\right. \nonu \\
&& \left. + f_{27} \,
\overline{D} {\bf T}  [D, \overline{D}] {\bf W}_{\frac{1}{3}}^{(\frac{7}{2})}
+ f_{28} \, 
\overline{D} {\bf T} \pa  {\bf W}_{\frac{1}{3}}^{(\frac{7}{2})}
+  
f_{29} \,
 \pa {\bf T} \overline{D}  {\bf W}_{\frac{1}{3}}^{(\frac{7}{2})}
+ f_{30} \, \pa \overline{D} {\bf T} {\bf W}_{\frac{1}{3}}^{(\frac{7}{2})}
+ f_{31} \, {\bf T} \overline{D} {\bf T}  {\bf W}_{\frac{1}{3}}^{(\frac{7}{2})}
\right](Z_2)
\nonu \\
&& +   \frac{1}{z_{12}}
\left[  f_{32}\, \pa^2  \overline{D}    {\bf W}_{\frac{1}{3}}^{(\frac{7}{2})}
+ f_{33} \, {\bf T} \pa  \overline{D}  {\bf W}_{\frac{1}{3}}^{(\frac{7}{2})}
+ f_{34} \, {\bf T} {\bf T}  \overline{D}  {\bf W}_{\frac{1}{3}}^{(\frac{7}{2})}
+ f_{35} \, {\bf T} \overline{D} {\bf T}  {\bf W}_{\frac{1}{3}}^{(\frac{7}{2})}
+
 f_{36} \, 
[D, \overline{D}] {\bf T}  \overline{D}  {\bf W}_{\frac{1}{3}}^{(\frac{7}{2})}
\right. \nonu \\
&& \left. + f_{37} \, 
\overline{D} {\bf T}  [D, \overline{D}] {\bf W}_{\frac{1}{3}}^{(\frac{7}{2})}
+ f_{38}\, \overline{D} {\bf T} \pa  {\bf W}_{\frac{1}{3}}^{(\frac{7}{2})}
+ 
 f_{39} \, \pa \overline{D} {\bf T}   {\bf W}_{\frac{1}{3}}^{(\frac{7}{2})}
+ f_{40} \, \pa   {\bf T} \overline{D} {\bf W}_{\frac{1}{3}}^{(\frac{7}{2})}
\right](Z_2)
\nonu \\
 && +  
\frac{\bar{\theta}_{12}}{z_{12}} \left[ 
f_{41} \, {\bf T} 
\overline{D} {\bf T} \overline{D} {\bf W}_{\frac{1}{3}}^{(\frac{7}{2})}
+ f_{42} \, 
\overline{D} {\bf T} \pa \overline{D} {\bf W}_{\frac{1}{3}}^{(\frac{7}{2})}
+ f_{43} \, 
\pa  \overline{D} {\bf T} \overline{D} {\bf W}_{\frac{1}{3}}^{(\frac{7}{2})}
\right](Z_2) 
\nonu \\
&& +  
\frac{ \theta_{12}}{z_{12}} 
 \left[ 
 f_{44} \, \overline{D} {\bf T} D {\bf T} {\bf W}_{\frac{1}{3}}^{(\frac{7}{2})}
+  f_{45} \, \pa^2 [D, \overline{D}] {\bf W}_{\frac{1}{3}}^{(\frac{7}{2})}
+  f_{46} \, {\bf T} \pa [D, \overline{D}] {\bf W}_{\frac{1}{3}}^{(\frac{7}{2})}
+ f_{47} \, {\bf T} {\bf T}  [D, \overline{D}] {\bf W}_{\frac{1}{3}}^{(\frac{7}{2})}
\right. \nonu \\
&& + f_{48} \, {\bf T} {\bf T} {\bf T}  {\bf W}_{\frac{1}{3}}^{(\frac{7}{2})}
+ f_{49} \, {\bf T} {\bf T}  \pa {\bf W}_{\frac{1}{3}}^{(\frac{7}{2})}
+ f_{50} \, {\bf T} D {\bf T} \overline{D}   {\bf W}_{\frac{1}{3}}^{(\frac{7}{2})}
+ f_{51} \, {\bf T}  [D, \overline{D}]  {\bf T}  {\bf W}_{\frac{1}{3}}^{(\frac{7}{2})}
+ f_{52} \, {\bf T} \overline{D} {\bf T} D  {\bf W}_{\frac{1}{3}}^{(\frac{7}{2})}
\nonu \\
&&+ f_{53} \, {\bf T} \pa^2  {\bf W}_{\frac{1}{3}}^{(\frac{7}{2})}
+  f_{54} \, D {\bf T} \pa \overline{D}   {\bf W}_{\frac{1}{3}}^{(\frac{7}{2})}
+ f_{55} \, \pa D {\bf T}  \overline{D}   {\bf W}_{\frac{1}{3}}^{(\frac{7}{2})}
+   f_{56}\,
 [D, \overline{D}]  {\bf T}  [D, \overline{D}]  {\bf W}_{\frac{1}{3}}^{(\frac{7}{2})}
\nonu \\
&& 
+ f_{57} \, [D, \overline{D}]  {\bf T} \pa  {\bf W}_{\frac{1}{3}}^{(\frac{7}{2})}
+  f_{58} \, \pa [D, \overline{D}]  {\bf T} {\bf W}_{\frac{1}{3}}^{(\frac{7}{2})}
+ f_{59} \, \overline{D}  {\bf T} \pa  D  {\bf W}_{\frac{1}{3}}^{(\frac{7}{2})}
+ f_{60} \, \pa  \overline{D}  {\bf T}   D  {\bf W}_{\frac{1}{3}}^{(\frac{7}{2})}
\nonu \\
&&
\left. 
+ f_{61} \, \pa {\bf T}  [D, \overline{D}]  {\bf W}_{\frac{1}{3}}^{(\frac{7}{2})}
+  f_{62} \, \pa {\bf T} {\bf T}  {\bf W}_{\frac{1}{3}}^{(\frac{7}{2})}
+ f_{63} \, \pa {\bf T} \pa  {\bf W}_{\frac{1}{3}}^{(\frac{7}{2})}
+ f_{64} \, \pa^2 {\bf T}  {\bf W}_{\frac{1}{3}}^{(\frac{7}{2})}
+ f_{65} \, \pa^3  {\bf W}_{\frac{1}{3}}^{(\frac{7}{2})}
\right](Z_2) 
\nonu \\
&&+\frac{\theta_{12} \, \bar{\theta}_{12}}{z_{12}} 
\left[ 
f_{66} \, \pa^3 \overline{D}  {\bf W}_{\frac{1}{3}}^{(\frac{7}{2})}
+ f_{67} \, {\bf T} \pa^2  \overline{D}  {\bf W}_{\frac{1}{3}}^{(\frac{7}{2})}
+ f_{68} \, {\bf T} {\bf T} \pa  \overline{D}  {\bf W}_{\frac{1}{3}}^{(\frac{7}{2})}
+ f_{69} \, {\bf T} 
{\bf T} {\bf T}  \overline{D}  {\bf W}_{\frac{1}{3}}^{(\frac{7}{2})}
\right. \nonu \\
&& 
+ f_{70} \, {\bf T}  [D, \overline{D}] {\bf T} \overline{D} {\bf W}_{\frac{1}{3}}^{(\frac{7}{2})}
+ f_{71} \, {\bf T} \overline{D} {\bf T}  [D, \overline{D}] {\bf W}_{\frac{1}{3}}^{(\frac{7}{2})}
+ f_{72} \, {\bf T} \overline{D} {\bf T} \pa  {\bf W}_{\frac{1}{3}}^{(\frac{7}{2})}
+  f_{73} \, 
\overline{D} {\bf T}  D {\bf T} \overline{D}  {\bf W}_{\frac{1}{3}}^{(\frac{7}{2})}
\nonu \\
&&
+ f_{74} \,
[D, \overline{D}]  {\bf T} \pa \overline{D} {\bf W}_{\frac{1}{3}}^{(\frac{7}{2})}
+  f_{75} \,
\pa 
[D, \overline{D}]  {\bf T} \overline{D}  {\bf W}_{\frac{1}{3}}^{(\frac{7}{2})}
+ f_{76} \,
\overline{D} {\bf T} \pa  [D, \overline{D}]   {\bf W}_{\frac{1}{3}}^{(\frac{7}{2})}
+ f_{77} \,
\overline{D} {\bf T} \pa^2  {\bf W}_{\frac{1}{3}}^{(\frac{7}{2})}
\nonu \\
&&+ f_{78} \,
\pa \overline{D} {\bf T}  [D, \overline{D}]   {\bf W}_{\frac{1}{3}}^{(\frac{7}{2})}
+ f_{79} \, \pa \overline{D} {\bf T} \pa  {\bf W}_{\frac{1}{3}}^{(\frac{7}{2})}
+ f_{80} \, \pa {\bf T} \pa  \overline{D}  {\bf W}_{\frac{1}{3}}^{(\frac{7}{2})}
+  f_{81} \, \pa {\bf T} {\bf T}  \overline{D} {\bf W}_{\frac{1}{3}}^{(\frac{7}{2})}
\nonu \\
&&
+  f_{82} \, \pa^2 {\bf T}  \overline{D} {\bf W}_{\frac{1}{3}}^{(\frac{7}{2})}
+  f_{83} \, \pa^2  \overline{D} {\bf T} {\bf W}_{\frac{1}{3}}^{(\frac{7}{2})}
+ f_{84} \, \pa {\bf T}  \overline{D} {\bf T}  {\bf W}_{\frac{1}{3}}^{(\frac{7}{2})}
+ f_{85} \, \pa \overline{D} {\bf T} {\bf T}  {\bf W}_{\frac{1}{3}}^{(\frac{7}{2})}
\nonu \\
&& \left. \left. +  f_{86} \,
\overline{D} {\bf T} [D, \overline{D}] {\bf T} {\bf W}_{\frac{1}{3}}^{(\frac{7}{2})}
+  
 f_{87} \,
{\bf T} {\bf T} \overline{D} {\bf T} {\bf W}_{\frac{1}{3}}^{(\frac{7}{2})}
  \right](Z_2) \right) +\cdots,
\label{44OPE} 
\eea}
where the coefficients in Appendix (\ref{44OPE}) 
 can be written in terms of previous ones 
in Appendix (\ref{ecoeff})
{\small
\bea
f_2 & = & - e_2, \qquad
f_{3}   =   e_3, \qquad f_4 =- e_4, \qquad f_5 =0, \qquad
f_6 = e_6, \qquad f_7 = - e_7, \nonu \\
f_8  & = & e_8, \qquad f_9 = e_{9},
\qquad f_{10} =e_{10}, \qquad
f_{11} = 0, \qquad f_{12} = 0, \qquad
f_{13} = -e_{13}, \nonu \\
f_{14} & = & -e_{14},
\qquad
f_{15} = e_{15}, \qquad
f_{16} = e_{16}, \qquad
f_{17} =- e_{17}, \qquad
f_{18} = - e_{18}, \qquad
f_{19} = e_{19},
\nonu \\
f_{20} & = & - e_{20}, \qquad
f_{21}   =   -e_{21}, \qquad f_{22} = e_{22}, \qquad f_{23} = - e_{23}, \qquad
f_{24} = e_{24}, \qquad f_{25} = - e_{25}, \nonu \\
f_{26}  & = & -e_{26}, \qquad f_{27} = -e_{27},
\qquad f_{28} =e_{28}, \qquad
f_{29} = e_{29}, \qquad f_{30} = e_{30}, \qquad
f_{31} = -e_{31}, \nonu \\
f_{32} & = & e_{32},
\qquad
f_{33} = -e_{33}, \qquad
f_{34} = e_{34}, \qquad
f_{35} = e_{35}, \qquad
f_{36} =  e_{36}, \qquad
f_{37} = e_{37},
\nonu \\
f_{38} & = & - e_{38}, \qquad
f_{39}   =   -e_{39}, \qquad f_{40} =- e_{40}, 
\qquad f_{41} =  e_{41}, \qquad
f_{42} = -e_{42}, \qquad f_{43} = - e_{43}, \nonu \\
f_{44}  & = & -e_{44}, \qquad f_{45} = -e_{45},
\qquad f_{46} =e_{46}, \qquad
f_{47} = -e_{47}, \qquad f_{48} = 0, \qquad
f_{49} = e_{49}, \nonu \\
f_{50} & = & e_{50},
\qquad
f_{51} = -e_{51}, \qquad
f_{52} = e_{52}, \qquad
f_{53} =- e_{53}, \qquad
f_{54} = - e_{54}, \qquad
f_{55} = -e_{55},
\nonu \\
f_{56} & = & - e_{56}, \qquad
f_{57}   =   e_{57}, \qquad  
f_{59}   =   - e_{59}, \qquad
f_{60} = -e_{60}, \qquad f_{61} =  e_{61}, \nonu \\
f_{62}  & = & e_{62}, \qquad f_{63} = -e_{63},
\qquad f_{64} = -e_{64}, \qquad
f_{65} = e_{65}, \qquad 
f_{66} = -e_{66}, \qquad
f_{67} = e_{67}, \nonu \\
f_{68} & = & -e_{68},
\qquad
f_{69} = e_{69}, \qquad
f_{70} = e_{70}, \qquad
f_{71} = e_{71}, \qquad
f_{72} = - e_{72}, \qquad
f_{73} = e_{73},
\nonu  \\
f_{74} & = & -e_{74},
\qquad
f_{76}  =   -e_{76}, \qquad
f_{77} = e_{77}, \qquad
f_{78} = - e_{78}, \qquad
f_{79} = e_{79},
\qquad
f_{80}   =   e_{80},
\nonu \\
f_{81} & = &  - e_{81}, 
\qquad
f_{82}  =   e_{82}, 
\qquad
f_{84}  =   - e_{84}, \qquad
f_{85} = -e_{85},
\qquad
f_{86}   =   e_{86}, \qquad
f_{87} = e_{87},
\nonu \\
f_{58} & = & -\frac{(1053 c^4+10224 c^3-64623 c^2+104326 c-18120) }
{2 (c+9) (3 c+4) (27 c-46) (3 c^2+90 c-265)},
\nonu \\
f_{75} & = &  -\frac{10 (972 c^5-6237 c^4+18396 c^3+93411 c^2-498058 c+493656) }
{11 (c+9) (3 c-2) (3 c+4) (27 c-46) (3 c^2+90 c-265)}, 
\nonu \\
f_{83}  &  = & 
\frac{5 (810 c^5+12861 c^4-132981 c^3+207708 c^2+319344 c-392216) }
{2 (c+9) (3 c-2) (3 c+4) (27 c-46) (3 c^2+90 c-265)}. 
\label{fcoeff}
\eea
}
In this case, the last three coefficients in Appendix (\ref{fcoeff})
are different from the ones in Appendix (\ref{ecoeff}). 
One can also analyze the $\frac{1}{c}, \cdots, \frac{1}{c^3}$-terms in
the large $c$ limit. 

\section{The component OPEs  in the OPE  
 ${\bf W}_{  \pm \frac{2}{3}}^{(2)}(Z_1) 
\, {\bf W}_{  \pm \frac{2}{3}}^{(2)}(Z_2)$ }

Although the complete OPEs between the higher spin currents 
in ${\cal N}=2$ superspace are determined in Appendix $G$, 
sometimes one should reexpress them in terms of its component results.
In this Appendix and Appendices $I$ and $J$, we will present them 
\footnote{ Although the ${\cal N}=2$ results in ${\cal N}=2$ superspace in section $7$ provides the component results automatically using the command $\tt N2OPEToComponents$ in \cite{KT}, 
for convenience, we would like to present them explicitly.}.

\subsection{The component OPEs  in the OPE 
${\bf W}_{  \frac{2}{3}}^{(2)}(Z_1) 
\, {\bf W}_{  \frac{2}{3}}^{(2)}(Z_2)$ }

The ten component OPEs corresponding to (\ref{oneone})
can be summarized by
{\small
\bea
 \frac{1}{C_{(2)(2)}^{(\frac{7}{2})+}} \, 
W_{  \frac{2}{3}}^{(2)}(z) \, W_{  \frac{2}{3}}^{(2)}(w) & = & + \cdots,
\nonu \\
\frac{1}{C_{(2)(2)}^{(\frac{7}{2})+}} \, 
W_{  \frac{2}{3}}^{(2)}(z) \, W_{  \frac{5}{3}}^{(\frac{5}{2})}(w)  & = &  + \cdots,
\nonu \\
\frac{1}{C_{(2)(2)}^{(\frac{7}{2})+}} \, 
W_{  \frac{2}{3}}^{(2)}(z) \, W_{  -\frac{1}{3}}^{(\frac{5}{2})}(w) &= &
- \frac{1}{(z-w)} \,  W_{  \frac{1}{3}}^{(\frac{7}{2})}(w) + \cdots,
\nonu \\
\frac{1}{C_{(2)(2)}^{(\frac{7}{2})+}} \,
W_{  \frac{2}{3}}^{(2)}(z) \, W_{  \frac{2}{3}}^{(3)}(w) &= &
\frac{1}{(z-w)} \,  \frac{1}{2} \, W_{  \frac{4}{3}}^{(4)}(w) + \cdots,
\nonu \\
\frac{1}{C_{(2)(2)}^{(\frac{7}{2})+}} \, 
W_{  \frac{5}{3}}^{(\frac{5}{2})}(z) \, W_{  \frac{5}{3}}^{(\frac{5}{2})}(w) &= & +
\cdots,
\nonu \\
\frac{1}{C_{(2)(2)}^{(\frac{7}{2})+}} \,
W_{  \frac{5}{3}}^{(\frac{5}{2})}(z) \, W_{  -\frac{1}{3}}^{(\frac{5}{2})}(w) &= &
-\frac{1}{(z-w)} \,  \frac{1}{2} \, W_{  \frac{4}{3}}^{(4)}(w) + \cdots,
\nonu \\
\frac{1}{C_{(2)(2)}^{(\frac{7}{2})+}} \, 
W_{  \frac{5}{3}}^{(\frac{5}{2})}(z) \, W_{  \frac{2}{3}}^{(3)}(w) &= & + \cdots,
\nonu \\
\frac{1}{C_{(2)(2)}^{(\frac{7}{2})+}} \,
W_{  -\frac{1}{3}}^{(\frac{5}{2})}(z) \, W_{  -\frac{1}{3}}^{(\frac{5}{2})}(w) &= &
-\frac{1}{(z-w)} \,   W_{  -\frac{2}{3}}^{(4)}(w) + \cdots,
\nonu \\
\frac{1}{C_{(2)(2)}^{(\frac{7}{2})+}} \,
W_{  -\frac{1}{3}}^{(\frac{5}{2})}(z) \, W_{  \frac{2}{3}}^{(3)}(w) &= &
\frac{1}{(z-w)^2} \, \frac{1}{2} \, 
 W_{  \frac{1}{3}}^{(\frac{7}{2})}(w) +
\frac{1}{(z-w)} \, \left[ \frac{1}{2}   W_{  \frac{1}{3}}^{(\frac{9}{2})}+
+ \frac{1}{4} \, \pa  W_{  \frac{1}{3}}^{(\frac{7}{2})}\right](w) + \cdots, 
\nonu \\
\frac{1}{C_{(2)(2)}^{(\frac{7}{2})+}} \,
W_{  \frac{2}{3}}^{(3)}(z) \, W_{  \frac{2}{3}}^{(3)}(w) &= &
-\frac{1}{(z-w)^2} \, \frac{1}{2}  W_{  \frac{4}{3}}^{(4)}(w) -
\frac{1}{(z-w)} \, \frac{1}{4} \, \pa  W_{  \frac{4}{3}}^{(4)}(w)+ \cdots.
\label{h1operesult}
\eea
}
Intentionally, we put the overall structure constant in the left hand side.
For convenience, we also presented the trivial OPEs
which can be checked by $U(1)$ charge counting.
For example, the first equation of Appendix (\ref{h1operesult})
implies that the left hand side contains the $U(1)$ charge 
$\frac{4}{3}$. Then 
the right hand side of this OPE should preserve
this $U(1)$ charge. The possible spin contents 
in the right hand side are given by $1, 2$ or $3$.
Then there are no composite fields 
having an $U(1)$ charge $\frac{4}{3}$
for given spins. 
Note that $W_{\frac{1}{3}}^{(\frac{9}{2})}(w)$ is not a primary field.
The coefficient $-\frac{1}{4} = -\frac{1}{2} \times \frac{1}{2}$ 
appearing in the last OPE in 
Appendix (\ref{h1operesult}) is an expected expression from the spin counting 
of the left hand side and right hand side.

\subsection{The component OPEs  in the OPE 
${\bf W}_{  \frac{2}{3}}^{(2)}(Z_1)
\, {\bf W}_{  -\frac{2}{3}}^{(2)}(Z_2)$ }

The sixteen component OPEs corresponding to (\ref{onetwo})
can be written as 
{\small
\bea
&& W_{  \frac{2}{3}}^{(2)}(z) \, W_{  -\frac{2}{3}}^{(2)}(w) = 
\frac{1}{(z-w)^4} \, \frac{c}{2} 
+\frac{1}{(z-w)^3} \, J(w) 
\nonu \\
&& +  \frac{1}{(z-w)^2} \left[
-\frac{2
}{(c-1)}
 \, J J +
\frac{2(3 c-1)}{3 (c-1)} T +\frac{1}{2} \, \pa J 
 \right](w) \nonu \\
&&  +  
\frac{1}{(z-w)} \, \frac{1}{(c-1) (c+6) (2 c-3)} \, \left[ 
-(c^2-36 c+8) 
 \, G^{-} G^{+}
+\frac{1}{3} \, (3 c+8) (8 c+9) 
\,  J T
\right. \nonu \\
&& 
- 2 (c+6)(2c -3)  \, 
\pa J J
 -   \frac{1}{3} \, (32 c+127) 
 \, 
 J J J
+\frac{1}{6} \, (c+2) (2 c^2+2 c+5) 
 \, \pa^2 J
 \nonu \\
&& \left. +   \frac{1}{6} \, (12 c^3+53 c^2-234 c+60)
 \,
 \pa T 
\right](w)
 + \cdots,
\nonu \\
&& W_{  \frac{2}{3}}^{(2)}(z) \, W_{  \frac{1}{3}}^{(\frac{5}{2})}(w) = 
\frac{1}{(z-w)^3} \, \frac{7}{2} G^{+}(w) +
\frac{1}{(z-w)^2} \, \left[ 
\frac{7 
}{2 (c-1)} \,  J G^{+}
+\frac{7 (c-2) 
}{6 (c-1)} \, \pa G^{+}
\right](w)
\nonu \\
& & + \frac{1}{(z-w)} \,  \frac{1}{(c-1) (c+6) (2 c-3)} \, \left[ 
\frac{7}{6} \, (27 c^2-32 c+6) 
\, T G^{+}
+\frac{7}{6} \, (c^2+64 c-42) 
 \,
 G^{+} \pa J
\right. \nonu \\
&&  \left. +   \frac{7}{4} (3 c^2-22 c+20) 
 \, \pa J G^{+}
-\frac{7}{2} \, (12 c+7) 
 J J G^{+}
+  \frac{7}{12} \, (c^3-17 c^2+54 c-18) 
 \, \pa^2 G^{+} 
\right](w) + \cdots,
\nonu \\
&& W_{  \frac{2}{3}}^{(2)}(z) \, W_{  -\frac{5}{3}}^{(\frac{5}{2})}(w) = 
-\frac{1}{(z-w)^3} \, \frac{5}{2} \, G^{-}(w) 
+\frac{1}{(z-w)^2} \left[ 
-\frac{15}{2 (c-1)}\, J G^{-}
-\frac{5 (c+2) 
}{6 (c-1)} \, \pa G^{-} 
\right](w) \nonu \\
& & +  
\frac{1}{(z-w)} \, \frac{1}{(c-1) (c+6) (2 c-3)} \, \left[ 
-\frac{5}{6} \, (27 c^2-38 c-30) 
 \, G^{-} T
-\frac{5}{4} \, (5 c^2+56 c-104)
  \, \pa J G^{-}
\right. \nonu \\
&&  \left. -   \frac{5}{6} \, (7 c^2-10 c+78) 
 \, \pa G^{-}  J
+\frac{5}{2} \, (4 c-41) 
 \, J J G^{-}
 -   \frac{5}{24} \, (2 c^3-11 c^2+34 c-138) 
 \, \pa^2 G^{-}
\right](w) + \cdots,
\nonu \\
&& W_{  \frac{2}{3}}^{(2)}(z) \, W_{  -\frac{2}{3}}^{(3)}(w) = 
-\frac{1}{(z-w)^5} \, \frac{c}{6} +\frac{1}{(z-w)^4} \, \frac{8}{3} \, J(w)
\nonu \\
&& +  \frac{1}{(z-w)^3} \, \left[ 
\frac{37}{6 (c-1)} \, J J 
-\frac{(3 c+34)}{9 (c-1)} T 
-\frac{1}{6}\,  \pa J
\right](w) \nonu \\
&&  +  \frac{1}{(z-w)^2}  \, \frac{1}{(c-1) (c+6) (2 c-3)} \, \left[ 
-\frac{1}{3}\, (131 c^2-306 c+208) 
 \, G^{-} G^{+}
+\frac{1}{9} \, (204 c^2-469 c-18) 
 \, J T 
\right. \nonu \\
&& +   \frac{41}{12} \, (c+6)(2c-3) \, \pa J J
-\frac{2}{9} \, (101 c-239) 
\, J J J
 -  
\frac{1}{36} \, (18 c^3-641 c^2+1962 c-1824) 
\, \pa T
\nonu \\
&& \left. -\frac{1}{18} \, (2 c^3+138 c^2-333 c+226) \, 
\pa^2 J
\right](w)\nonu \\
&& +  \frac{1}{(z-w)} \, \frac{1}{ (c-1) (c+1) (c+6) (2 c-3) (5 c-9)} \,
\left[ 
-\frac{1}{18} (c+1)(5c-9)(12 c^2+973 c-1224) 
 \, \pa J T
\right. \nonu \\
&&  -   \frac{35}{3} \, (40 c^3-131 c^2+23 c-26) 
 \, J G^{-} G^{+}
+  \frac{70}{9} \, (30 c^3-23 c^2+59 c+57) 
  \, J J T
\nonu \\
&& -  
\frac{35}{54} (264 c^3-827 c^2+759 c+90) 
 \,
 T T 
 +  \frac{1}{18} \, (1305 c^4-6759 c^3+6510 c^2+13562 c-8712) 
\,
\pa G^{+}  G^{-} 
\nonu \\
&& -  \frac{1}{18} \, (1305 c^4-1089 c^3-2520 c^2-6458 c+1368) 
 \, 
\pa G^{-}  G^{+}
\nonu \\
&& +  \frac{1}{18} \, (990 c^4+3268 c^3-18755 c^2+5331 c+3264) 
 \,  \pa T J
\nonu \\
&& +  \frac{1}{36} \, (410 c^4-1423 c^3+7011 c^2-12444 c+17212) 
 \, \pa^2 J J
\nonu \\
&& +  \frac{1}{24} \, (40 c^4+288 c^3-4321 c^2+10429 c-8102)
  \, \pa J \pa J
\nonu \\
&& -  \frac{1}{6} \, (c+1)(5c-9) (124 c-361)
\,
\pa J J J 
 -  \frac{175}{18} \, (2 c-1) (16 c+5) 
\, J J J J 
\nonu \\
&& -  \frac{1}{36} \, (30 c^5+171 c^4-3759 c^3+7607 c^2+1125 c-2682) 
 \,
\pa^2 T
\nonu \\
&& +  \left. 
\frac{1}{108} \, (30 c^5+3366 c^4+2069 c^3-35110 c^2+14727 c+2370) 
 \, \pa^3 J
\right](w) + \cdots,
\nonu \\
&& W_{  \frac{5}{3}}^{(\frac{5}{2})}(z) \, W_{  -\frac{2}{3}}^{(2)}(w) = 
-\frac{1}{(z-w)^3} \, \frac{5}{2} \, G^{+}(w)
+ 
\frac{1}{(z-w)^2} \,  \left[ 
-\frac{15 }{2 (c-1)} \, J G^{+}
-\frac{5 (2 c-5) }{6 (c-1)} \, 
\pa G^{+}
\right](w) 
\nonu \\
&& +  \frac{1}{(z-w)} \, \frac{1}{(c-1) (c+6) (2 c-3)} \, \left[ - 
\frac{5}{6} \, (27 c^2-38 c-30)  \,
 T G^{+}
-\frac{5}{4} \, (7 c^2-2 c-4)  \,
\pa J G^{+}
\right. \nonu \\
&&
\left. -  
\frac{5}{6} \, (11 c^2+91 c-240) \,
\pa G^{+} J
+\frac{5}{2} \, (4 c-41) 
  \, J J G^{+}
-   \frac{5}{4} \, (c^3-2 c^2+12 c-30) 
 \, \pa^2 G^{+}
 \right](w) +
\cdots, 
\nonu \\
&& W_{  \frac{5}{3}}^{(\frac{5}{2})}(z) \, W_{  \frac{1}{3}}^{(\frac{5}{2})}(w) = 
-\frac{1}{(z-w)}  \, 
\frac{35
}{6 (c-1)} \, \pa G^{+} G^{+}(w)
   + \cdots,
\nonu \\
&& W_{  \frac{5}{3}}^{(\frac{5}{2})}(z) \, W_{  -\frac{5}{3}}^{(\frac{5}{2})}(w) = 
\frac{1}{(z-w)^5} \, \frac{5c}{6}+ \frac{1}{(z-w)^4} \,
\frac{25}{6} \, J(w) \nonu \\
&& +   \frac{1}{(z-w)^3} \,
\left[ 
\frac{25}{6 (c-1)} \,  J J
+\frac{25 (3 c-5) }{18(c-1)} \, T 
+\frac{25}{12}  \, \pa J
\right](w) \nonu \\
&& + \frac{1}{(z-w)^2}\, \frac{1}{(c-1) (c+6) (2 c-3)} \, \left[ 
\frac{25}{9} \, (15 c^2+11 c-45) \,
J T
-\frac{25}{6} \, (c-5) (7 c-10) \,
G^{-} G^{+}
\right.
\nonu \\
&& +  \frac{25 }{6} \, (c+6) (2c-3) \,
\pa J J
-\frac{25}{18} \, (20 c-23) \,
J J J
+\frac{25}{18} \, (c^3+9 c-16) \,
\pa^2 J
\nonu \\
& & +  \left. \frac{25}{18} \, (3 c-8) (c^2+9 c-15 ) 
 \,
\pa T
\right](w)
\nonu \\
&& + \frac{1}{(z-w)} \, \frac{1}{(c-1) (c+1) (c+6) (2 c-3) (5 c-9)} \, \left[
\frac{25}{18} \,  (c+1)(5c-9) (15 c^2+11 c-45) \,
\pa J T
\right. \nonu \\
&& +  
\frac{25}{18} \, (3 c-8) (24 c^3-168 c^2+5 c+225)  \,
\pa G^{+} G^{-}
 -  
\frac{25}{3} \, (68 c^3-316 c^2+109 c+185) 
 \,
J G^{-} G^{+}
\nonu \\
&& +  
\frac{25}{18} \, (3 c-8) (25 c^3+133 c^2-96 c-120) 
 \,
\pa T J
 + 
\frac{25}{54} \, (3 c-8) (81 c^3-86 c^2+12 c-45) 
 \,
T T 
\nonu \\
&& +  
\frac{25}{18} \,(132 c^3+269 c^2+100 c-345) 
 \,
J J T
 -  
\frac{25}{18} \, (33 c^4-63 c^3-258 c^2-20 c+450) 
 \,
\pa G^{-} G^{+}
\nonu \\
& & +  
\frac{25}{36} \, (23 c^4-157 c^3+708 c^2-618 c+34) 
\,
\pa^2 J J
 +  
\frac{25}{24} \, (7 c^4+132 c^3-508 c^2+373 c+82) 
 \,
\pa J \pa J 
\nonu \\
&& -  \frac{25}{12} \, (c+1) (5c-9)(20 c-23) 
 \,
\pa J J J 
 -  
\frac{175}{18} \, (2 c-1) (16 c+5) 
 \,
J J J J 
\nonu \\
&& +  
\frac{25}{36} \, (3 c-8) (3 c^4-6 c^3+46 c^2-18 c-45) 
 \,
\pa^2 T
\nonu \\
&& +  \left. \frac{25}{216} 
\, (15 c^5-162 c^4-290 c^3+2356 c^2-867 c-1488) 
 \,
\pa^3 J
 \right](w) + \cdots,
\nonu \\
&& W_{  \frac{5}{3}}^{(\frac{5}{2})}(z) \, W_{  -\frac{2}{3}}^{(3)}(w) = 
\frac{1}{(z-w)^4} \, \frac{95}{12} \, G^{+}(w) + 
\frac{1}{(z-w)^3} \,
\left[ 
\frac{95 }{6 (c-1)} \, J G^{+}
+\frac{5 (21 c-59) }{36 (c-1)} \, 
\pa G^{+}
\right](w)
\nonu \\
&& +  \frac{1}{(z-w)^2} \, \frac{1}{(c-1) (c+6) (2 c-3)} \,
\left[ 
\frac{5}{36} \, (591 c^2-1244 c+510) \,
 T G^{+}
\right. \nonu \\
&& +\frac{5}{24} \, (167 c^2-114 c-196)  \,
\pa J G^{+}
-  \frac{5}{36} \, (3 c^2-1276 c+1986)  \,
\pa G^{+} J
-\frac{5}{12} \, (212 c-353)  \,
J J G^{+}
\nonu \\
& & +  \left.
\frac{5}{72} \,
(23 c^3-406 c^2+1395 c-1296)  \,
\pa^2 G^{+}
\right](w)
\nonu \\
&& +  \frac{1}{(z-w)} \, \frac{1}{(c-1) (c+1) (c+6) (2 c-3) (5 c-9)} \, 
\left[ 
\frac{175}{9} (16 c+5) \, (3 c^2-7 c+6)
 \, J T G^{+}
\right. \nonu \\
& & +  \frac{5}{72} \, (c+6)(825 c^3-2776 c^2+2543 c-16)
 \, 
\pa G^{+} \pa J 
- 
\frac{5}{6} \, (220 c^3-58 c^2-601 c+1217) 
 \, \pa J J G^{+}
\nonu \\
& &-  \frac{5}{36}  \, (2340 c^3-17009 c^2+2852 c+6801) 
 \,
\pa G^{+} J J 
\nonu \\
&& +  \frac{5}{36} (1485 c^4-3614 c^3+2137 c^2-1274 c+3810) 
 \,
\pa T G^{+}
\nonu \\
&  & +  \frac{5}{108} \, (3195 c^4-18612 c^3+11171 c^2+24948 c-32670) 
 \,
\pa G^{+} T
\nonu \\
& &
+  \frac{5}{72} \, (875 c^4-2542 c^3+3825 c^2-6074 c+5164) 
 \,
\pa^2 J G^{+}
\nonu \\
& & -  \frac{5}{36} \, (95 c^4+2047 c^3-13570 c^2+8539 c-579) 
 \,
\pa^2 G^{+} J
-  \frac{350}{9} \, (2 c-1) (16 c+5) 
 \,
J J J G^{+}
\nonu \\
& & +  \left.
\frac{5}{216} (75 c^5-723 c^4-1539 c^3+9977 c^2-4176 c-2574) 
 \,
\pa^3 G^{+} 
\right](w)+ \cdots,
\nonu \\
&& W_{  -\frac{1}{3}}^{(\frac{5}{2})}(z) \, W_{  -\frac{2}{3}}^{(2)}(w) = 
\frac{1}{(z-w)^3} \, \frac{7}{2} G^{-}(w) +
\frac{1}{(z-w)^2} \left[ 
\frac{7 }{2 (c-1)} \, J G^{-}
+\frac{7 (2 c-1) }{6 (c-1)} \,
\pa G^{-}
\right](w) +
\nonu \\
& & + \frac{1}{(z-w)} \, \frac{1}{(c-1) (c+6) (2 c-3)} \, \left[ 
\frac{7}{6} \, (27 c^2-32 c+6)\,
G^{-} T
+\frac{7}{6} \, (5 c^2-37 c-12)  \,
\pa G^{-} J
\right. \nonu \\
& & +  \left. \frac{7}{4} \, (c^2+40 c-56)  \,
\pa J G^{-}
-\frac{7}{2} \, (12 c+7) \,
J J G^{-}
+  
\frac{7}{24} \, (6 c^3+3 c^2+4 c+42)  \,
\pa^2 G^{-}
\right](w) +  
\cdots,
\nonu \\
&& W_{  -\frac{1}{3}}^{(\frac{5}{2})}(z) \, W_{  \frac{1}{3}}^{(\frac{5}{2})}(w) 
= \frac{1}{(z-w)^5} \, \frac{7c}{6} -\frac{1}{(z-w)^4} \, \frac{7}{6}
\, J(w) \nonu \\
& & + 
 \frac{1}{(z-w)^3} \, \left[ 
-\frac{49 }{6 (c-1)} \, J J
+\frac{7 (15 c-1) }{18 (c-1)} \,
T
-\frac{7}{12} \, \pa J 
\right](w) +\nonu \\
& & + \frac{1}{(z-w)^2} \, \frac{1}{(c-1) (c+6) (2 c-3)}  \, \left[ 
-\frac{7}{9} \, (15 c^2-127 c+63)  \,
J T
+\frac{7}{6} (c-1)(43 c-2)  \,
G^{-} G^{+}
\right. \nonu \\
& &
-  \frac{49 }{6} \, (c+6) (2 c-3) \,
\pa J J
+\frac{7}{18} \, (44 c-191)  \,
J J J
-    
\frac{7}{18} \, (c-1)(c^2-17 c-8)  \,
\pa^2 J 
\nonu \\
&& \left.
+\frac{7}{18} \, (15 c^3+2 c^2-72 c+6)  \,
\pa T 
\right](w)
\nonu \\
& & + \frac{1}{(z-w)} \, \frac{1}{ (c-1) (c+1) (c+6) (2 c-3) (5 c-9)} 
\, \left[ 
-\frac{7}{18} \, (c+1)(5c-9)(15 c^2-127 c+63)  \,
\pa J T
\right.
\nonu \\
& &
+  \frac{7}{3} \, (20 c^3-616 c^2+523 c+59) \,
J G^{-} G^{+}
-  \frac{7}{18} \, (1140 c^3-577 c^2-44 c+573) 
 \,
J J T
\nonu \\
& &
+  \frac{7}{18} \, (450 c^4-1410 c^3+129 c^2+853 c-36) 
 \,
\pa G^{-} G^{+}
\nonu \\
&& -  \frac{7}{18} \, (195 c^4+219 c^3-720 c^2+338 c-18) 
 \,
\pa G^{+} G^{-}
\nonu \\
& &
-  \frac{35}{18} (15 c^4-127 c^3-232 c^2+492 c-78) 
 \,
 \pa T J
+  \frac{7}{54} \, (1215 c^4-1092 c^3-4900 c^2+6171 c-36) 
\, 
T T 
\nonu \\
& &
-  \frac{7}{36} \, (125 c^4+935 c^3-1164 c^2-3048 c+4426) 
 \,
\pa^2 J J
\nonu \\
&& -  \frac{7}{24} \, (85 c^4-198 c^3-628 c^2+1813 c-1142) 
 \,
\pa J \pa J
\nonu \\
& &
+  \frac{7}{12} \, (c+1)(5c-9)(44 c-191) \,
\pa J J J
+  \frac{175}{18} \, (2 c-1) (16 c+5)  \,
J J J J 
\nonu \\
& &
+  \frac{7}{36} \, (45 c^5-156 c^4+48 c^3+728 c^2-657 c-36) 
 \,
\pa^2 T
\nonu \\
& &
-  \left. 
\frac{7}{216} \, (15 c^5-432 c^4+1594 c^3+2818 c^2-5259 c+564) 
 \,
\pa^3 J 
\right](w) +
\cdots,
\nonu \\
&& 
W_{  -\frac{1}{3}}^{(\frac{5}{2})}(z) \, W_{  -\frac{5}{3}}^{(\frac{5}{2})}(w) = -
\frac{1}{(z-w)} \, 
\frac{35 }{6 (c-1)} \,
\pa G^{-} G^{-}(w)
+ \cdots,
\nonu \\
&& W_{  -\frac{1}{3}}^{(\frac{5}{2})}(z) \, W_{  -\frac{2}{3}}^{(3)}(w) = 
\frac{1}{(z-w)^4} \, \frac{77}{12} \, G^{-}(w)
+ \frac{1}{(z-w)^3} \, \left[ 
\frac{119 }{6 (c-1)} \,
J G^{-}
+\frac{7 (9 c+25) }{36 (c-1)} \,
\pa G^{-}
\right](w) 
\nonu \\
& & + \frac{1}{(z-w)^2} \, \frac{1}{(c-1) (c+6) (2 c-3)} \, \left[ 
\frac{7}{36} \, (399 c^2-874 c+282) \,
G^{-} T
\right. \nonu \\
&& +\frac{7}{36} \, (147 c^2-170 c+318) \,
\pa G^{-} J
 \nonu \\
&& -  \frac{7}{24} \, (15 c^2-516 c+784) \,
\pa J G^{-}
 -\frac{7}{12} \, (124 c-361) \,
J J G^{-}
\nonu \\
&& + 
\left.
\frac{7}{144} \, (14 c^3-305 c^2+800 c-1086) 
\,
\pa^2 G^{-}
\right](w)
\nonu \\
& & +  \frac{1}{(z-w)} \, \frac{1}{(c-1) (c+1) (c+6) (2 c-3) (5 c-9)}\, 
\left[ 
\frac{175}{9} \, (16 c+5) (3 c^2-7 c+6) 
 \,
J G^{-} T
\right.
\nonu \\
& &
-  \frac{7}{72} \, (c+6)(45 c^3+274 c^2-1997 c+2174) 
 \,
\pa G^{-} \pa J
-  \frac{7}{6} \, 420 c^3-1786 c^2+483 c+1589) 
 \,
\pa J J G^{-}
\nonu \\
& &
-  \frac{7}{36} \, (60 c^3+6667 c^2-1876 c+2517) 
 \,
\pa G^{-} J J
\nonu \\
&& 
 +  \frac{7}{36} \, (945 c^4-1636 c^3-4015 c^2+6388 c-978) 
 \,
\pa T G^{-}
\nonu \\
& &
+  \frac{7}{108} \, (1935 c^4-6018 c^3+10475 c^2+234 c-594) 
 \,
\pa G^{-} T
\nonu \\
& &
+  \frac{7}{36} \, (275 c^4-975 c^3+6513 c^2-3394 c+2043) 
 \,
\pa^2 G^{-} J
\nonu \\
& &
-  \frac{7}{72} \, (95 c^4-3196 c^3+4929 c^2+7996 c-13424) 
 \,
\pa^2 J G^{-}
-  \frac{350}{9} (2 c-1) (16 c+5)  \,
J J J G^{-}
\nonu \\
& & +  \left. \frac{7}{216} \, (c+1)
(15 c^4-852 c^3+4131 c^2-10370 c+6402) 
\,
\pa^3 G^{-}
\right](w) 
+ \cdots,
\nonu \\
&& W_{  \frac{2}{3}}^{(3)}(z) \, W_{  -\frac{2}{3}}^{(2)}(w) = 
-\frac{1}{(z-w)^5} \, \frac{c}{6} +\frac{1}{(z-w)^4} \, \frac{8}{3} \, J(w)
\nonu \\
& & +  
\frac{1}{(z-w)^3} \, \left[ \frac{37 }{6 (c-1)} \,
J J 
-\frac{(3 c+34) }{9 (c-1)} \,
T
+\frac{17}{6} \, \pa J 
\right](w) \nonu \\
& & + 
\frac{1}{(z-w)^2} \, \frac{1}{ (c-1) (c+6) (2 c-3)}\, \left[ 
-\frac{1}{3 }\, (131 c^2-306 c+208) \,
G^{-} G^{+}
+\frac{1}{9 } \, (204 c^2-469 c-18) \,
J T
\right.
\nonu \\
& &
+  \frac{107 }{12} \, (c+6)(2c-3) \,
\pa J J
-\frac{2}{9} \, (101 c-239)  \,
J J J
\nonu \\
& & - 
\left.
\frac{1}{36 } \, (6 c^3-551 c^2+2718 c-3120) 
\,
\pa T
+ 
\frac{1}{18} \, (52 c^3+51 c^2-396 c+260) 
\,
\pa^2 J 
\right](w) \nonu \\
& & +  \frac{1}{(z-w)} \, \frac{1}{ (c-1) (c+1) (c+6) (2 c-3) (5 c-9)}
\, \left[
\frac{35}{18} \, (c+1)(5c-9)(12 c^2+c-36) \,
\pa J T
\right.
\nonu \\
& & -  \frac{35}{18} \, (75 c^4-159 c^3-58 c-72) 
\,
\pa G^{-} G^{+}
 -  \frac{35}{3} \, (40 c^3-131 c^2+23 c-26) 
\,
J G^{+} G^{-}
\nonu \\
& & +  \frac{70}{9} \, (30 c^3-23 c^2+59 c+57) 
\,
J J T
 -  \frac{35}{54} \, (264 c^3-827 c^2+759 c+90) 
 \,
T T
\nonu \\
& & +  \frac{35}{24} \, (c-2) (20 c^3+118 c^2-159 c-37) 
 \,
\pa J \pa J
 +  \frac{35}{18} \, (75 c^4-321 c^3+258 c^2+514 c-360) 
\,
\pa G^{+} G^{-}
\nonu \\
& & +  \frac{35}{18} \, (30 c^4-34 c^3-253 c^2+231 c-240) 
\,
\pa T J
 +  \frac{35}{36} \, (40 c^4+64 c^3-207 c^2-381 c+950) 
\,
\pa^2 J J
\nonu \\
& & -  \frac{35}{6} \, (c+1)(5c-9)(8 c-17) \,
\pa J J J 
-  \frac{175}{18} \, (2 c-1) (16 c+5) \,
J J J J
\nonu \\
& & -  \frac{35}{36} \, (12 c^4-57 c^3+70 c^2+9 c+90) 
\,
\pa^2 T 
\nonu \\
& & +  \left.
\frac{35}{216} \, (30 c^5-54 c^4-235 c^3+782 c^2-177 c+210) 
 \,
\pa^3 J
 \right](w) 
+ \cdots,
\nonu \\
&& W_{  \frac{2}{3}}^{(3)}(z) \, W_{  \frac{1}{3}}^{(\frac{5}{2})}(w) = 
\frac{1}{(z-w)^4} \, \frac{77}{12} \, G^{+}(w)+
\frac{1}{(z-w)^3} \, \left[ 
\frac{119 }{6 (c-1)} \,
J G^{+}
+\frac{7 (12 c-29) }{18 (c-1)} \,
\pa G^{+}
\right](w) \nonu \\
& & +  \frac{1}{(z-w)^2} \, \frac{1}{(c-1) (c+6) (2 c-3)} \, \left[  
\frac{7}{36} \, (399 c^2-874 c+282) \,
T G^{+}
\right.
\nonu \\
&& +\frac{7}{36} \, (57 c^2+1088 c-2154) \,
G^{+} \pa J 
\nonu \\
& &  +  \frac{7}{24} \, (151 c^2+96 c-440) \,
\pa J G^{+}
-\frac{7}{12} \, (124 c-361) \,
J J G^{+}
\nonu \\
&& \left. +   \frac{7}{72} \, (37 c^3-179 c^2+738 c-1026)  \,
\pa^2 G^{+} 
\right](w) 
\nonu \\
& & +  \frac{1}{(z-w)} \, \frac{1}{ (c-1) (c+1) (c+6) (2 c-3) (5 c-9)} 
\, \left[ 
\frac{175}{9} \, (16 c+5) (3 c^2-7 c+6) 
\,
J T G^{+}
\right.
\nonu \\
& & +  \frac{35}{36} \, (c+6)(75 c^3-254 c^2+187 c+76) 
\,
\pa G^{+}  \pa J
 -  \frac{35}{6} \, (40 c^3-103 c^2-31 c+332) 
\,
\pa J J G^{+}
\nonu \\
& & -  \frac{35}{18} \, (180 c^3-1357 c^2+286 c+723) 
\,
\pa G^{+} J J
+  \frac{35}{18} \, (105 c^4-433 c^3+533 c^2+35 c-156)
\,
\pa T G^{+}
\nonu \\
& & +  \frac{35}{54} \, (405 c^4-1188 c^3-653 c^2+1998 c-702) 
\,
\pa G^{+} T
\nonu \\
&&  +  \frac{35}{18} \, (5 c^4-71 c^3+755 c^2-662 c+267) 
\,
\pa^2 G^{+} J
\nonu \\
& & +  \frac{35}{36} \, (115 c^4-95 c^3-207 c^2-439 c+878) 
\,
\pa^2 J G^{+}
-  \frac{350}{9} \, (2 c-1) (16 c+5) 
\,
J J J  G^{+}
\nonu \\
& & +  \left.
\frac{35}{108} \, (15 c^5-24 c^4-117 c^3+406 c^2-198 c+198) 
\,
\pa^3 G^{+}
\right](w) +
 \cdots,
\nonu \\
&& W_{  \frac{2}{3}}^{(3)}(z) \, W_{  -\frac{5}{3}}^{(\frac{5}{2})}(w) = 
\frac{1}{(z-w)^4} \, \frac{95}{12} \, G^{-}(w)+
\frac{1}{(z-w)^3} \, \left[ 
\frac{95 }{6 (c-1)}\,
J G^{-}
+\frac{5 (18 c+1) }{18 (c-1)}\,
\pa G^{-}
\right](w) \nonu \\
& & +  \frac{1}{(z-w)^2}\,
\frac{1}{(c-1) (c+6) (2 c-3)}
\,
\left[ 
\frac{5}{36} \, (591 c^2-1244 c+510) \,
G^{-} T
\right. \nonu \\
&& +\frac{5}{36} \, (231 c^2-250 c-66) \,
\pa G^{-}  J
 \nonu \\
& &   -  \frac{5}{24} \, (15 c^2-798 c+1172) \,
\pa J G^{-}
-\frac{5}{12} \, (212 c-353) \,
J J G^{-}
 \nonu \\
&& \left. +  
\frac{5}{144} \, (106 c^3-163 c^2+52 c-174) \,
\pa^2 G^{-} 
\right](w)
\nonu \\
& & +   
\frac{1}{(z-w)} \, \frac{1}{(c-1) (c+1) (c+6) (2 c-3) (5 c-9)} \, \left[ 
\frac{175}{9} \, (16 c+5) (3 c^2-7 c+6) 
\,
J G^{-} T
\right.
\nonu \\
& & -  \frac{175}{6} \, (3 c-8) (8 c^2-3 c-7) 
\,
\pa J J G^{-}
 +  \frac{175}{18} \, (3 c-8) (7 c^3-5 c^2-13 c+15) 
\,
\pa T G^{-}
\nonu \\
& & +  \frac{175}{36} \, (c+6)(9 c^3-32 c^2+37 c-10) 
\,
\pa G^{-} \pa J
 -  \frac{175}{18} \, (12 c^3+131 c^2-62 c+39) 
\,
\pa G^{-} J J 
\nonu \\
& & +  \frac{175}{54} \, (81 c^4-102 c^3-65 c^2+36 c+270)
\, 
\pa G^{-} T
 +  \frac{175}{36} \, (14 c^4-42 c^3+261 c^2-139 c+72) 
\,
\pa^2 G^{-} J
\nonu \\
& & -  \frac{175}{36} \, (7 c^4-77 c^3+93 c^2+179 c-262) 
\,
\pa^2 J G^{-}
 -  \frac{350}{9} \, (2 c-1) (16 c+5) 
\,
J J J G^{-}
\nonu \\
& & +  \left.
\frac{175}{108} \, (c+1)(3 c^4-21 c^3+54 c^2-121 c+60) 
\,
\pa^3 G^{-}
\right](w)
+ \cdots,
\nonu \\
&& W_{  \frac{2}{3}}^{(3)}(z) \, W_{  -\frac{2}{3}}^{(3)}(w) = 
\frac{1}{(z-w)^6} \, \frac{5c}{2} +\frac{1}{(z-w)^5} \, 3 J(w)
\nonu \\
& & + 
\frac{1}{(z-w)^4} \, \left[ 
-\frac{3 }{(c-1)}\, J J
+\frac{2(22 c-19)}{3 (c-1)}\,
 T
+\frac{3}{2} \,\pa J
\right](w) \nonu \\
& & +  \frac{1}{(z-w)^3} \,  \frac{1}{(c-1) (c+6) (2 c-3)} \, \left[ 
\frac{1}{6} \, (277 c^2+1368 c-2544) \,
G^{-} G^{+}
\right. \nonu \\
&& +\frac{1}{6} \, (304 c^2+1351 c-2448) \,
J T
\nonu \\
& &
-  3 \, (c+6)(2c-3)
\pa J J
-\frac{1}{6} \, (32 c+127) \,
J J J
\nonu \\
& & +  \left.
\frac{1}{12} 
\, (176 c^3+363 c^2-3636 c+3912) \,
\pa T 
+\frac{1}{36 } \, (24 c^3+361 c^2+1044 c-2328) 
\,
\pa^2 J
\right](w) \nonu \\
& & +  \frac{1}{(z-w)^2}  \, \frac{1}{(c-1) (c+6) (2 c-3)}\, \left[ 
\frac{1}{36} \, (2301 c^2-3596 c+2448) 
\,
\pa G^{-} G^{+}
\right. \nonu \\
& & +  \frac{1}{36} \, (639 c^2-11804 c+17712) 
\,
\pa G^{+} G^{-}
 +  \frac{1}{12} \, (304 c^2+2471 c-4828) 
\,
\pa T J
\nonu \\
& & 
+\frac{1}{12} \, (304 c^2+1351 c-2448)
\,
\pa J T
 +  \frac{35}{18} \, (48 c^2-97 c+30) 
\,
T T
+\frac{1}{18} \, (87 c^2-1376 c+2262) 
\,
\pa^2 J J
\nonu \\
& & -  \frac{1}{24} \, (304 c^2-1047 c+974) 
\,
\pa J \pa J
-\frac{70}{3} \, (8 c-17) \,
J G^{-} G^{+}
-  \frac{35}{3} \, (8 c-17) \,
J J T
\nonu \\
&& -\frac{1}{4} \, (32 c+127) \,
\pa J J J 
\nonu \\
& & +  
\left.
\frac{1}{36} \, (158 c^3-870 c^2+2331 c-2466) \,
\pa^2 T 
+\frac{1}{144} \, (12 c^3-289 c^2-3736 c+7220) 
\,
\pa^3 J
\right](w) \nonu \\
& & +  \frac{1}{(z-w)} \, \frac{1}{ (c-1) (c+1) (c+6) (2 c-3) (5 c-9)} \, 
\left[ 
-\frac{350}{9} \, (2 c-1) (16 c+5) 
\,
J J J T
\right.
\nonu \\
& &
+  
\frac{175}{9} (16 c+5) (3 c^2-7 c+6)
\,
J T T
+  
\frac{35}{27} \, (45 c^4-60 c^3+676 c^2+627 c-594) 
\,
\pa G^{-} \pa G^{+}
\nonu \\
& &
+  
\frac{35}{36} \, (145 c^4-489 c^3+1698 c^2-822 c+366) 
\,
\pa^2 G^{-} G^{+}
 \nonu \\
&& +  
\frac{35}{18} \, (240 c^4-917 c^3+591 c^2+448 c-420) 
\,
\pa T T
\nonu \\
& & +  
\frac{35}{12} \, (40 c^3-437 c^2+56 c+203) 
\,
\pa T J J
 +  
\frac{35}{18} \, (480 c^3-3107 c^2+641 c+1368) 
\,
\pa G^{+} J G^{-}
\nonu \\
& & +  
\frac{35}{36} \, (65 c^4-59 c^3-1645 c^2+1907 c-972) 
\,
\pa^2 G^{+} G^{-}
 -  \frac{35}{3} (c+1) (5c-9)(8 c-17) \,
\pa J J T
\nonu \\
& & +  \frac{35}{18} \, (30 c^3-503 c^2+149 c+132) 
\,
\pa^2 J J J
 +  \frac{35}{36} \, (15 c^4-77 c^3-570 c^2+201 c+19)
\, 
\pa^3 J J
\nonu \\
& & -  \frac{35}{3} (c+1)(5c-9) (8 c-17) \,
\pa J G^{-} G^{+}
 -  \frac{35}{72} \, (c+1)(5c-9)(18 c^2-25 c-18)
\,
\pa^2 J \pa J 
\nonu \\
& & -  
\frac{350}{3} (2 c-1) (16 c+5) \,
J J G^{-} G^{+}
 -  
\frac{35}{12} \, (40 c^3-131 c^2+23 c-26) 
\,
\pa J \pa J J
\nonu \\
& & +  
\frac{175}{9} \, (16 c+5) (3 c^2-7 c+6) 
\,
G^{-} T G^{+}
 - 
\frac{455}{18} \, (131 c^2-53 c+36) 
\,
\pa G^{-} J G^{+}
\nonu \\
& & +  
\frac{35}{18} \, (20 c^4-148 c^3+825 c^2-581 c+186) 
\,
\pa^2 T J
 \nonu \\
&& +  
\frac{35}{34} \, (60 c^4+552 c^3-2415 c^2+1529 c+696) 
\,
\pa T \pa J
\nonu \\
& & +  
\frac{35}{54} \, (60 c^4+387 c^3-422 c^2-1455 c+2154) 
\,
\pa^2 J T
 \nonu \\
&& - 
\frac{35}{432} \, (66 c^4-162 c^3+131 c^2-153 c+588) 
\,
\pa^4 J 
\nonu \\
& & +  
\left.
\frac{35}{648} \, (90 c^5-657 c^4+264 c^3-71 c^2-4362 c+2880) 
\,
\pa^3 T 
\right](w) 
+\cdots.
\label{h2ope}
\eea
}
In the large $c$ limit,
all the nonlinear terms in Appendix (\ref{h2ope}) disappear.
One can also analyze the subleading
$\frac{1}{c}, \cdots, \frac{1}{c^3}$-terms. 
The OPE in the expression of $W_q^{(h)}(z) \, W_{-q}^{(h)}(w)$
contains the first four singular terms similar to the one in (\ref{twospin2ope}).

\subsection{The component OPEs  in the OPE  ${\bf W}_{  -\frac{2}{3}}^{(2)}(Z_1)
\, {\bf W}_{  -\frac{2}{3}}^{(2)}(Z_2)$ }

The ten component OPEs corresponding to (\ref{twotwo})
can be summarized by
{\small
\bea
\frac{1}{C_{(2)(2)}^{(\frac{7}{2})-}} \,
W_{  -\frac{2}{3}}^{(2)}(z) \, W_{  -\frac{2}{3}}^{(2)}(w) &= & + \cdots,
\nonu \\
\frac{1}{C_{(2)(2)}^{(\frac{7}{2})-}} \,
W_{  -\frac{2}{3}}^{(2)}(z) \, W_{  \frac{1}{3}}^{(\frac{5}{2})}(w) &= & 
-\frac{1}{(z-w)} \, W_{  -\frac{1}{3}}^{(\frac{7}{2})}(w) +
\cdots,
\nonu \\
\frac{1}{C_{(2)(2)}^{(\frac{7}{2})-}} \,
W_{  -\frac{2}{3}}^{(2)}(z) \, W_{  -\frac{5}{3}}^{(\frac{5}{2})}(w) &= & + \cdots,
\nonu \\
\frac{1}{C_{(2)(2)}^{(\frac{7}{2})-}} \,
W_{  -\frac{2}{3}}^{(2)}(z) \, W_{  -\frac{2}{3}}^{(3)}(w) &= &
-\frac{1}{(z-w)} \, \frac{1}{2} \, W_{  -\frac{4}{3}}^{(4)}(w) +
\cdots,
\nonu \\
\frac{1}{C_{(2)(2)}^{(\frac{7}{2})-}} \,
W_{  \frac{1}{3}}^{(\frac{5}{2})}(z) \, W_{  \frac{1}{3}}^{(\frac{5}{2})}(w) &= &
-\frac{1}{(z-w)} \,  W_{  \frac{2}{3}}^{(4)}(w) +\cdots,
\nonu \\
\frac{1}{C_{(2)(2)}^{(\frac{7}{2})-}} \,
W_{  \frac{1}{3}}^{(\frac{5}{2})}(z) \, W_{  -\frac{5}{3}}^{(\frac{5}{2})}(w) &= &
-\frac{1}{(z-w)} \, \frac{1}{2}\,  W_{  -\frac{4}{3}}^{(4)}(w) + \cdots,
\nonu \\
\frac{1}{C_{(2)(2)}^{(\frac{7}{2})-}} \,
W_{  \frac{1}{3}}^{(\frac{5}{2})}(z) \, W_{  -\frac{2}{3}}^{(3)}(w) &= &
-\frac{1}{(z-w)^2} \, \frac{1}{2}\, W_{  -\frac{1}{3}}^{(\frac{7}{2})}(w)+
\frac{1}{(z-w)} \, \left[  -\frac{1}{2} \,  W_{  -\frac{1}{3}}^{(\frac{9}{2})}
-\frac{1}{4} \, \pa  W_{  -\frac{1}{3}}^{(\frac{7}{2})}
\right](w) + \cdots,
\nonu \\
\frac{1}{C_{(2)(2)}^{(\frac{7}{2})-}} \,
W_{  -\frac{5}{3}}^{(\frac{5}{2})}(z) \, W_{  -\frac{5}{3}}^{(\frac{5}{2})}(w) &= &
+ \cdots,
\nonu \\
\frac{1}{C_{(2)(2)}^{(\frac{7}{2})-}} \,
W_{  -\frac{5}{3}}^{(\frac{5}{2})}(z) \, W_{  -\frac{2}{3}}^{(3)}(w) &= &
+ \cdots,
\nonu \\
\frac{1}{C_{(2)(2)}^{(\frac{7}{2})-}} \,
W_{  -\frac{2}{3}}^{(3)}(z) \, W_{  -\frac{2}{3}}^{(3)}(w) &= &
-\frac{1}{(z-w)^2} \, \frac{1}{2} \, 
  W_{  -\frac{4}{3}}^{(4)}(w) -\frac{1}{(z-w)} \,
\frac{1}{4} \pa  W_{  -\frac{4}{3}}^{(4)}(w)
+ \cdots.
\label{h3ope}
\eea
}
As observed in section $7$, these OPEs in Appendix (\ref{h3ope}) 
are very similar to the previous ones
in Appendix $(H.1)$.

\section{The component OPEs  in the OPE 
${\bf W}_{  \pm \frac{2}{3}}^{(2)}(Z_1) 
\, {\bf W}_{  \pm \frac{1}{3}}^{(\frac{7}{2})}(Z_2)$ }

As in previous Appendix $H$, we continue to describe the component 
OPEs corresponding to the ${\cal N}=2$ version in Appendix $G$.

\subsection{The component OPEs in the OPE  
${\bf W}_{  \frac{2}{3}}^{(2)}(Z_1)
\, {\bf W}_{  \frac{1}{3}}^{(\frac{7}{2})}(Z_2)$ }

The $16$ component OPEs corresponding to Appendix (\ref{13OPE})
can be summarized by
{\small
\bea
&& 
\frac{1}{C_{(2)(\frac{7}{2})}^{(4)+}} \,
W_{  \frac{2}{3}}^{(2)}(z) \, W_{  \frac{1}{3}}^{(\frac{7}{2})}(w) = 
-\frac{1}{(z-w)} \, \frac{1}{4} \,W_{1}^{(\frac{9}{2})}(w) + \cdots,
\nonu \\
& & \frac{1}{C_{(2)(\frac{7}{2})}^{(4)+}} \,
 W_{  \frac{2}{3}}^{(2)}(z) \, W_{  \frac{4}{3}}^{(4)}(w) =  +\cdots,
\nonu \\
&& 
\frac{1}{C_{(2)(\frac{7}{2})}^{(4)+}} \,
W_{  \frac{2}{3}}^{(2)}(z) \, W_{  -\frac{2}{3}}^{(4)}(w) = 
-\frac{1}{(z-w)^2} \, W_{0}^{(4)}(w) +\frac{1}{(z-w)}
\left[ 
-\frac{1}{6} \, W_{0}^{(5)}  -\frac{1}{4} \, \pa   W_{0}^{(4)}
\right](w) + \cdots,
\nonu \\
&& 
\frac{1}{C_{(2)(\frac{7}{2})}^{(4)+}} \,
W_{  \frac{2}{3}}^{(2)}(z) \, W_{  \frac{1}{3}}^{(\frac{9}{2})}(w) = 
\frac{1}{(z-w)^2} \, \frac{17}{24} \, W_{1}^{(\frac{9}{2})}(w)
\nonu \\
& & +   \frac{1}{(z-w)} \, \frac{1}{(c+9)} \, \left[ 
\frac{20  }{3 } \,
G^{+}  W_{0}^{(4)}
-\frac{5  }{3 } \,
J  W_{1}^{(\frac{9}{2})}
+\frac{ (11 c+219) }{72 } \,
\pa W_{1}^{(\frac{9}{2})}
\right](w) + \cdots,
\nonu \\
&& \frac{1}{C_{(2)(\frac{7}{2})}^{(4)+}} \,
W_{  \frac{5}{3}}^{(\frac{5}{2})}(z) \, W_{  \frac{1}{3}}^{(\frac{7}{2})}(w) = + \cdots,
\nonu \\
&& \frac{1}{C_{(2)(\frac{7}{2})}^{(4)+}} \,
W_{  \frac{5}{3}}^{(\frac{5}{2})}(z) \, W_{  \frac{4}{3}}^{(4)}(w) =  +\cdots,
\nonu \\
&& 
\frac{1}{C_{(2)(\frac{7}{2})}^{(4)+}} \,
W_{  \frac{5}{3}}^{(\frac{5}{2})}(z) \, W_{  -\frac{2}{3}}^{(4)}(w) = 
-\frac{1}{(z-w)^2} \, \frac{5}{12} \, W_{1}^{(\frac{9}{2})}(w) 
\nonu \\
&& +  
\frac{1}{(z-w)} \,  \frac{1}{(c+9)} \,\left[ 
\frac{20  }{3 } \,
G^{+}  W_{0}^{(4)} 
-\frac{5  }{3 } \,
J  W_{1}^{(\frac{9}{2})}
-\frac{5 (c-3) }{36 } \,
\pa  W_{1}^{(\frac{9}{2})}
\right](w) + \cdots,
\nonu \\
&& 
\frac{1}{C_{(2)(\frac{7}{2})}^{(4)+}} \,
W_{  \frac{5}{3}}^{(\frac{5}{2})}(z) \, W_{  \frac{1}{3}}^{(\frac{9}{2})}(w) = 
-\frac{1}{(z-w)} \, 
\frac{25  }{3 (c+9)} \,
G^{+}  W_{1}^{(\frac{9}{2})}(w)
+\cdots,
\nonu \\
&& 
\frac{1}{C_{(2)(\frac{7}{2})}^{(4)+}} \,
W_{  -\frac{1}{3}}^{(\frac{5}{2})}(z) \, W_{  \frac{1}{3}}^{(\frac{7}{2})}(w) = 
\frac{1}{(z-w)^2} \, W_{0}^{(4)}(w) +
\frac{1}{(z-w)} \, \left[ 
- \frac{1}{12} \,
W_{0}^{(5)}
+\frac{3}{8}  
\pa W_{0}^{(4)}
\right](w) + \cdots,
\nonu \\
&& 
\frac{1}{C_{(2)(\frac{7}{2})}^{(4)+}} \,
W_{  -\frac{1}{3}}^{(\frac{5}{2})}(z) \, W_{  \frac{4}{3}}^{(4)}(w) = 
-\frac{1}{(z-w)^2}\, \frac{5}{6} \,  W_{1}^{(\frac{9}{2})}(w) 
\nonu \\
& & +   \frac{1}{(z-w)} \,  \frac{1}{(c+9)} \, \left[ 
\frac{5  }{3 }\,
J  W_{1}^{(\frac{9}{2})}
-\frac{20  }{3 } \,
G^{+}  W_{0}^{(4)}
-\frac{5  (c+15) }{18 } \,
\pa W_{1}^{(\frac{9}{2})}
\right](w) 
+ \cdots,
\nonu \\
&& 
\frac{1}{C_{(2)(\frac{7}{2})}^{(4)+}} \,
W_{  -\frac{1}{3}}^{(\frac{5}{2})}(z) \, W_{  -\frac{2}{3}}^{(4)}(w) = 
-\frac{1}{(z-w)^2} \,  W_{-1}^{(\frac{9}{2})}(w)
-\frac{1}{(z-w)} \, \frac{1}{3}  \, \pa  W_{-1}^{(\frac{9}{2})}(w)
+ \cdots,
\nonu \\
&& 
\frac{1}{C_{(2)(\frac{7}{2})}^{(4)+}} \,
W_{  -\frac{1}{3}}^{(\frac{5}{2})}(z) \, W_{  \frac{1}{3}}^{(\frac{9}{2})}(w) = 
\frac{1}{(z-w)^3} \,  W_{0}^{(4)}(w) +
\frac{1}{(z-w)^2} \, \left[ 
\frac{13}{48}\,  \pa   W_{0}^{(4)}
+\frac{19}{24} \,   W_{0}^{(5)}  
\right](w) \nonu \\
& & + \frac{1}{(z-w)} \,  \frac{1}{(c+9)} \,\left[ 
-\frac{5  }{3 } \,
G^{-}   W_{1}^{(\frac{9}{2})}
-\frac{20  }{3 } \,
G^{+}  W_{-1}^{(\frac{9}{2})}
-\frac{5}{3 } \,
J  W_{0}^{(5)}
\right. \nonu \\
& &  \left. +   \frac{20}{3 }\, 
T W_{0}^{(4)}
-\frac{10 }{3 } 
\pa J  W_{0}^{(4)}
+\frac{5 }{6 } \, 
J \pa  W_{0}^{(4)}
+\frac{ (17 c+273)}{72 } \, 
\pa  W_{0}^{(5)}
+   \frac{ (7 c-57) }{144 }\,
\pa^2  W_{0}^{(4)}
\right](w)+\cdots,
\nonu \\
&& 
\frac{1}{C_{(2)(\frac{7}{2})}^{(4)+}} \,
W_{  \frac{2}{3}}^{(3)}(z) \, W_{  \frac{1}{3}}^{(\frac{7}{2})}(w) = 
-\frac{1}{(z-w)^2} \, \frac{7}{24} \,  W_{1}^{(\frac{9}{2})}(w) 
\nonu \\
& & +  
\frac{1}{(z-w)} \,  \frac{1}{(c+9)} \, \left[ 
\frac{20  }{3 } \,
G^{+}  W_{0}^{(4)}
-\frac{5  }{3 } \,
J  W_{1}^{(\frac{9}{2})}
-\frac{5  (c-3) }{36 }
\pa  W_{1}^{(\frac{9}{2})}
\right](w) +\cdots,
\nonu \\
&& 
\frac{1}{C_{(2)(\frac{7}{2})}^{(4)+}} \,
W_{  \frac{2}{3}}^{(3)}(z) \, W_{  \frac{4}{3}}^{(4)}(w) = 
-\frac{1}{(z-w)} \, 
\frac{25  }{3 (c+9)}
G^{+} W_{1}^{(\frac{9}{2})}(w)
 + \cdots,
\nonu \\
&& 
\frac{1}{C_{(2)(\frac{7}{2})}^{(4)+}} \,
W_{  \frac{2}{3}}^{(3)}(z) \, W_{  -\frac{2}{3}}^{(4)}(w) =
\frac{1}{(z-w)^3} \, W_{0}^{(4)}(w) + 
\frac{1}{(z-w)^2} \, \left[ 
-\frac{1}{3} \,  W_{0}^{(5)} 
+\frac{1}{3}  \pa W_{0}^{(4)}
\right](w)
\nonu \\
& & + \frac{1}{(z-w)} \,  \frac{1}{(c+9)} \,\left[ 
-\frac{5 }{3 }\,
G^{-}  W_{1}^{(\frac{9}{2})}
-\frac{20  }{3 } \,
G^{+}  W_{-1}^{(\frac{9}{2})}
-\frac{5 }{3 } \, 
J  W_{0}^{(5)}
\right. \nonu \\
& &
\left.
+  \frac{20  }{3 } \,
T W_{0}^{(4)}
-\frac{10 }{3 } \, 
\pa J  W_{0}^{(4)}
+\frac{5 }{6 } \, 
J \pa W_{0}^{(4)} 
-   \frac{5  (c-3) }{36 } \,
\pa  W_{0}^{(5)}
+\frac{5  (c-3) }{72 } \,
\pa^2 W_{0}^{(4)} 
\right](w)+ \cdots,
\nonu \\
&& 
\frac{1}{C_{(2)(\frac{7}{2})}^{(4)+}} \,
W_{  \frac{2}{3}}^{(3)}(z) \, W_{  \frac{1}{3}}^{(\frac{9}{2})}(w) = 
-\frac{1}{(z-w)^3} \, \frac{9}{8}\,
 W_{1}^{(\frac{9}{2})}(w)-\frac{1}{(z-w)^2} \, \frac{17}{48} \, 
\pa  W_{1}^{(\frac{9}{2})}(w) +\nonu \\
& & + \frac{1}{(z-w)} \,  \frac{1}{(c+9)} \,\left[ 
-\frac{25  }{3 } \,
T  W_{1}^{(\frac{9}{2})}
+\frac{25 }{3 } \, 
G^{+}   W_{0}^{(5)}
+\frac{10 }{3 } \, 
\pa G^{+}  W_{0}^{(4)} 
\right. \nonu \\
& &
\left. +  \frac{10 }{3 } \, 
\pa J  W_{1}^{(\frac{9}{2})}
-\frac{5  }{6 }\,
J \pa  W_{1}^{(\frac{9}{2})}
-\frac{5  }{6 } \,
G^{+} \pa  W_{0}^{(4)} 
-   \frac{5  (c-3) }{72 }\,
\pa^2  W_{1}^{(\frac{9}{2})}
\right](w)
+ \cdots.
\label{i1ope}
\eea
}
In this case, the overall structure constant appearing in the 
above OPEs is written in the left hand side for simplicity.
In the large $c$ limit,
all the nonlinear terms in Appendix (\ref{i1ope}) disappear.
One can also analyze the $\frac{1}{c}$-term. 
Note that the higher spin current 
$W_0^{(5)}(w)$ is not a primary field.
For the trivial OPEs in Appendix (\ref{i1ope}),
one can check them by taking the $U(1)$ charges for given spins.

\subsection{The component OPEs  in the OPE  
${\bf W}_{  \frac{2}{3}}^{(2)}(Z_1) 
\, {\bf W}_{  -\frac{1}{3}}^{(\frac{7}{2})}(Z_2)$ }

The $4$ component OPEs corresponding to Appendix (\ref{14OPE})
can be summarized by
{\small
\bea
&& 
\frac{1}{C_{(2)(\frac{7}{2})}^{(2)+}} \,
W_{  \frac{2}{3}}^{(2)}(z) \, W_{  -\frac{1}{3}}^{(\frac{7}{2})}(w) = 
-\frac{1}{(z-w)^3} \, \frac{3}{7} \, W_{  \frac{1}{3}}^{(\frac{5}{2})}(w)
\nonu \\
& & +   \frac{1}{(z-w)^2}\,  \frac{1}{(c+6)} \,
\left[ 
\frac{9 }{5 }\, 
G^{+ } W_{  -\frac{2}{3}}^{(2)}
-\frac{27 }{35 } \,
J  W_{  \frac{1}{3}}^{(\frac{5}{2})}
-\frac{3 (c-3) }{35 }\,
\pa  W_{  \frac{1}{3}}^{(\frac{5}{2})}
\right](w) 
\nonu \\
& & +  \frac{1}{(z-w)} \, 
\frac{1}{ (c+6) (3 c+2) (9 c-11) (3 c^2+54 c-169)} \, \left[ 
\right. 
\nonu \\
&& \frac{54}{35} \, (132 c^3+1258 c^2-995 c+2279) 
 \,
J J  W_{  \frac{1}{3}}^{(\frac{5}{2})}
\nonu \\
& &
+  \frac{18}{5} \, (63 c^3+237 c^2+3075 c-1499) 
 \,
 J G^{+}  W_{  -\frac{2}{3}}^{(2)}
 \nonu \\
&& -  \frac{15}{7} \, (81 c^4+810 c^3-1836 c^2-3428 c+2217) 
 \,
 T   W_{  \frac{1}{3}}^{(\frac{5}{2})}
\nonu \\
& & -  \frac{9}{35} \, (54 c^4-3591 c^3+6033 c^2+19605 c-1073) 
 \,
G^{+}   W_{  -\frac{2}{3}}^{(3)}
\nonu \\
& & +  \frac{9}{5} \, (36 c^4+167 c^3-3893 c^2+4967 c+2391)  
 \,
 \pa G^{+}  W_{  -\frac{2}{3}}^{(2)}
\nonu \\
& & -  \frac{9}{70} \, (261 c^4+3852 c^3+8902 c^2-41898 c-75809) 
 \,
\pa J  W_{  \frac{1}{3}}^{(\frac{5}{2})}
\nonu \\
& & +  \frac{9}{70} \, (180 c^4+8589 c^3+1245 c^2-53167 c+4597)  
 \,
G^{+} \pa W_{  -\frac{2}{3}}^{(2)}
\nonu \\
& & -  \frac{9}{35} \, (45 c^4+678 c^3-8628 c^2-8020 c+29841)  
 \,
 J  \pa  W_{  \frac{1}{3}}^{(\frac{5}{2})}
\nonu \\
& & -  
\left. 
\frac{3}{70} \, (27 c^5-747 c^4-15963 c^3+26637 c^2+60860 c-38166) 
  \,
\pa^2  W_{  \frac{1}{3}}^{(\frac{5}{2})}
\right](w) +
\cdots,
\nonu \\
&& 
\frac{1}{C_{(2)(\frac{7}{2})}^{(2)+}} \,
W_{  \frac{2}{3}}^{(2)}(z) \, W_{  \frac{2}{3}}^{(4)}(w) = 
-\frac{1}{(z-w)^2} \, 
\frac{33  }{7 (c+6)}\,
G^{+}  W_{  \frac{1}{3}}^{(\frac{5}{2})}(w)
\nonu \\
&& +  \frac{1}{(z-w)} \,  \frac{1}{(c+6) (3c+2)} \, \left[ 
\right.
\nonu \\
&& \left. -\frac{33 (5 c+8) }{35 } \, 
\pa G^{+}  W_{  \frac{1}{3}}^{(\frac{5}{2})}
-\frac{33 (3 c-4)  }{35} \,
G^{+} \pa  W_{  \frac{1}{3}}^{(\frac{5}{2})}
-   \frac{396  }{35} \,
J G^{+}  W_{  \frac{1}{3}}^{(\frac{5}{2})}
\right](w)
 + \cdots,
\nonu \\
&& 
\frac{1}{C_{(2)(\frac{7}{2})}^{(2)+}} \,
W_{  \frac{2}{3}}^{(2)}(z) \, W_{  -\frac{4}{3}}^{(4)}(w) = 
-\frac{1}{(z-w)^4} \,  W_{  -\frac{2}{3}}^{(2)}(w) + \frac{1}{(z-w)^3} 
\, \left[ 
-\frac{12}{35}   \, W_{  -\frac{2}{3}}^{(3)}
-\frac{1}{35}  \pa  W_{  -\frac{2}{3}}^{(2)} 
\right](w) \nonu \\
& & +  \frac{1}{(z-w)^2} \, \frac{1}{ 
(c+6) (9 c-11) (3 c^2+54 c-169)} \, \left[ 
\frac{9}{5} \, (c+1) (6 c^2+27 c+271)  
\,
J  W_{  -\frac{2}{3}}^{(3)}
\right. \nonu \\
& & -  (9c-11)(4 c-3) (3 c+25)  
 \,
T  W_{  -\frac{2}{3}}^{(2)} 
 +  234 \, (c+1)(c+6)
\,  
J J W_{  -\frac{2}{3}}^{(2)} 
\nonu \\
& & +  \frac{18}{7} \, 30 c^3+214 c^2-467 c-1751)  
 \,
G^{-}  W_{  \frac{1}{3}}^{(\frac{5}{2})}
-  6 \, (9 c^3-84 c^2-300 c+893) 
 \, 
G^{+}  W_{  -\frac{5}{3}}^{(\frac{5}{2})}
\nonu \\
& & -  3 \, (6 c^3+203 c^2+738 c-3859) 
\,
\pa J   W_{  -\frac{2}{3}}^{(2)} 
+  \frac{3}{5} \, (12 c^3+621 c^2+2606 c-9003)  
 \,
J \pa W_{  -\frac{2}{3}}^{(2)}
\nonu \\
& & -  
\frac{6}{35} \, (9 c^4+264 c^3+2328 c^2-1750 c-14823) 
\,
\pa  W_{  -\frac{2}{3}}^{(3)}
\nonu \\
& & -   \left.
\frac{1}{70} \, (9 c^4-2991 c^3-13737 c^2+58135 c-19128)  
 \,
\pa^2   W_{  -\frac{2}{3}}^{(2)}
\right](w) \nonu \\
& & + \frac{1}{(z-w)} \, \frac{1}{ (c+6) (3 c+2) (9 c-11) 
(2 c^2+9 c-40) (3 c^2+54 c-169)}\, 
\nonu \\
&& \times \left[ 
-
\frac{6}{245} \,
(54 c^7-243 c^6-14235 c^5-199026 c^4 \right. \nonu \\
&& +1010729 c^3+2743737 c^2
-16277128 c+16901952)  \,
\pa^2  W_{  -\frac{2}{3}}^{(3)}
\nonu \\
& & -  
\frac{18}{245} 
(324 c^6+2160 c^5+54399 c^4-61158 c^3-1714933 c^2+8852748 c-12723040
)  
 \,
J \pa W_{  -\frac{2}{3}}^{(3)}
\nonu \\
& & 
+  \frac{36}{7} \, (c+6) (24 c^3-1413 c^2+7051 c-312)  
\,
J J J  W_{  -\frac{2}{3}}^{(2)}
\nonu \\
& & + 
\frac{18}{245} \, 
(2202 c^5-4103 c^4-39374 c^3+1597111 c^2-2139828 c-3847008)
 \,
J J \pa  W_{  -\frac{2}{3}}^{(2)}
\nonu \\
& & +  
\frac{36}{49} 
\, (1656 c^5+22749 c^4-56517 c^3-247578 c^2+957196 c-1544736) 
   \,
J G^{-}  W_{  \frac{1}{3}}^{(\frac{5}{2})}
\nonu \\
& & 
+  
\frac{3}{245} \,
(468 c^6+51000 c^5+341403 c^4-609536 c^3-6943491 c^2+14475796 c+806880)  
 \,
J \pa^2 W_{  -\frac{2}{3}}^{(2)}
\nonu \\
& & +  
\frac{18}{49} \,
(288 c^6-768 c^5-46948 c^4+225537 c^3+295633 c^2-2939824 c+3711872)  
 \,
G^{-} \pa  W_{  \frac{1}{3}}^{(\frac{5}{2})}
\nonu \\
& &
+ 
\frac{6}{49} \, 
(1080 c^6+27720 c^5+97212 c^4-967463 c^3-267499 c^2+7778204 c-10494432)   \,
\pa G^{-} W_{  \frac{1}{3}}^{(\frac{5}{2})}
\nonu \\
& &
-  
\frac{4}{245} \, 
(1458 c^6-160065 c^5-1269567 c^4+5688544 c^3+15358599 c^2-59269229 c+27494760)  
\nonu \\
&&  \times  
T \pa  W_{  -\frac{2}{3}}^{(2)}
\nonu \\
& & -  
\frac{2}{7} \,
(1134 c^6+19116 c^5+3177 c^4-268730 c^3+2187 c^2+442416 c+415104)  
 \,
\pa T W_{  -\frac{2}{3}}^{(2)}
\nonu \\
& & -  
\frac{36}{245} \,
(396 c^6+10050 c^5+57494 c^4-96296 c^3-519387 c^2-1345359 c+4766392)  
 \,
\pa J W_{  -\frac{2}{3}}^{(3)}
\nonu \\
& & +  
\frac{18}{7} \,
(438 c^5+7309 c^4+14842 c^3-162049 c^2+113692 c+151712) 
\,
\pa J J W_{  -\frac{2}{3}}^{(2)}
\nonu \\
& & +  
\frac{3}{245} \, 
(864 c^6-24330 c^5-155879 c^4+999856 c^3-94513 c^2+4617474 c-18643472) 
 \,
\pa J \pa  W_{  -\frac{2}{3}}^{(2)}
\nonu \\
& & -  
\frac{12}{245} \, (c+6)(9c-11)
(648 c^4+4374 c^3-4193 c^2-109749 c+199240)  
 \,
 T W_{  -\frac{2}{3}}^{(3)} 
\nonu \\
& & 
+  \frac{12}{7} 
\, (216 c^5+3429 c^4-8106 c^3+91185 c^2-415616 c+218880)  
 \,
J T  W_{  -\frac{2}{3}}^{(2)}
\nonu \\
& &  -  \frac{36}{7} \,
(90 c^5+48 c^4-567 c^3+60931 c^2-268420 c+214624)  
 \,
J G^{+} W_{  -\frac{5}{3}}^{(\frac{5}{2})} 
\nonu \\
& & -  
\frac{12}{7} \,
(999 c^5+19143 c^4-99057 c^3-200733 c^2+1202212 c-874656) 
\, 
G^{-} G^{+}  W_{  -\frac{2}{3}}^{(2)}
\nonu \\
& & -  
\frac{6}{7} \, 
(162 c^6-1371 c^5-30618 c^4-10062 c^3+679299 c^2-1107376 c+101248) 
 \,
\pa G^{+}  W_{  -\frac{5}{3}}^{(\frac{5}{2})} 
\nonu \\
& & -  
\frac{1}{7} \,
(324 c^6+10134 c^5+88809 c^4-326400 c^3-1590463 c^2+4706064 c-684672)  
 \,
\pa^2 J  W_{  -\frac{2}{3}}^{(2)}
\nonu \\
& & -  
\frac{6}{35} \, (3 c-8) (90 c^5+2013 c^4+36738 c^3-6574 c^2-579885 c+493004) 
\,
G^{+} \pa  W_{  -\frac{5}{3}}^{(\frac{5}{2})} 
\nonu \\
& & +  
\frac{108}{245} \,
(624 c^5+13424 c^4+48617 c^3-441123 c^2+492804 c-724256)  \,
J J W_{  -\frac{2}{3}}^{(3)}
\nonu \\
& & -  
\frac{1}{1470} \, 
(162 c^7-59949 c^6+134535 c^5+5339307 c^4 \nonu \\
&& \left. - 16284313 c^3-52881834 c^2+212816996 c-151818144) 
\pa^3  W_{  -\frac{2}{3}}^{(2)}
\right](w) +\cdots,
\nonu \\
&& \frac{1}{C_{(2)(\frac{7}{2})}^{(2)+}} \,
W_{  \frac{2}{3}}^{(2)}(z) \, W_{  -\frac{1}{3}}^{(\frac{9}{2})}(w) = 
\frac{1}{(z-w)^4} \,  \frac{13}{14} \, W_{  \frac{1}{3}}^{(\frac{5}{2})}(w) 
\nonu \\
&& + \frac{1}{(z-w)^3} \, \frac{1}{(c+6)} \, \left[
\frac{31}{5} \, G^{+} W_{  -\frac{2}{3}}^{(2)} 
-\frac{93}{35} \, J  W_{  \frac{1}{3}}^{(\frac{5}{2})}
+\frac{(c+192)}{70} \pa W_{  \frac{1}{3}}^{(\frac{5}{2})}
\right](w) \nonu \\
&& + \frac{1}{(z-w)^2} \,\frac{1}{(c+6) (3 c+2) (9 c-11) (3 c^2+54 c-169)} \, \left[ \right. 
\nonu \\
&& -\frac{9}{5} (528 c^3+6262 c^2-15995 c+17871) 
 \, J J W_{  \frac{1}{3}}^{(\frac{5}{2})}
 \nonu \\
&& -\frac{9}{5} (3 c^3-5143 c^2+50495 c-36759)  
\, J G^{+}  W_{  -\frac{2}{3}}^{(2)} 
 \nonu \\
&& -\frac{1}{14} (4131 c^4+20718 c^3+164556 c^2-850228 c+456003)  
T W_{  \frac{1}{3}}^{(\frac{5}{2})}
\nonu \\
&& -\frac{3}{70} (8154 c^4+52551 c^3-423801 c^2+770211 c+61409)  
G^{+}  W_{  -\frac{2}{3}}^{(3)} 
\nonu \\
&& +\frac{1}{10} (1647 c^4+57780 c^3+20028 c^2-438746 c+224359)  
\pa G^{+}  W_{  -\frac{2}{3}}^{(2)} 
\nonu \\
&& +\frac{3}{140} (3501 c^4+32310 c^3+7924 c^2+389472 c-2570643) 
\pa J W_{  \frac{1}{3}}^{(\frac{5}{2})}
\nonu \\
&& +\frac{1}{140} (3510 c^4-288567 c^3-1653351 c^2+6330489 c-2329237)  
G^{+} \pa  W_{  -\frac{2}{3}}^{(2)} 
\nonu \\
&& -\frac{3}{70} (1656 c^4+20229 c^3-184065 c^2+1031677 c-1187885)  
J \pa  W_{  \frac{1}{3}}^{(\frac{5}{2})}
\nonu \\
&& \left. +\frac{1}{140}
(27 c^5-10521 c^4-80097 c^3-676605 c^2+3011162 c-1639782) 
\pa^2  W_{  \frac{1}{3}}^{(\frac{5}{2})}
\right](w) \nonu \\
&& + \frac{1}{(z-w)} \, \frac{1}{
(c+6) (3 c+2) (6 c-5) (9 c-11) (2 c^2+9 c-40) (3 c^2+54 c-169)} \, 
\nonu \\
&& \times \left[ 
\frac{1}{2940} 
(972 c^8-355968 c^7-9455877 c^6-11669739 c^5+104947449 c^4
\right.
\nonu \\
&& +660636269 c^3-2683044612 c^2+2930835770 c-1299814800) 
\, \pa^3  W_{  \frac{1}{3}}^{(\frac{5}{2})}
\nonu \\
&& 
+\frac{396}{7} 
(288 c^5+282 c^4-8208 c^3+31325 c^2-92433 c+70440) 
\, J J J  W_{  \frac{1}{3}}^{(\frac{5}{2})}
\nonu \\
&&
+\frac{132}{49} 
(5508 c^6+19170 c^5-231174 c^4+22842 c^3+2275051 c^2-4668767 c+2971960) 
 \, J T W_{  \frac{1}{3}}^{(\frac{5}{2})}
\nonu \\
&&
+\frac{132}{49} 
(12636 c^6+63801 c^5-637848 c^4+651405 c^3+1738728 c^2-2363090 c+457200) 
\nonu \\
&& \times  G^{-} G^{+} W_{  \frac{1}{3}}^{(\frac{5}{2})}
\nonu \\
&&
+\frac{1}{980} 
(50544 c^7-15976548 c^6-35411724 c^5+375345195 c^4\nonu \\
&& -211554213 c^3-755063573 c^2+169169907 c+262216840) 
\, G^{+} \pa^2  W_{  -\frac{2}{3}}^{(2)} 
\nonu \\
&&
+\frac{3}{35} 
(11124 c^6+699264 c^5-2134113 c^4-9013287 c^3+76849931 c^2-128290435 c+54481400) 
\nonu \\
&& \times \pa G^{+} J  W_{  -\frac{2}{3}}^{(2)} 
\nonu \\
&&
+\frac{1}{980} 
(89424 c^7+793044 c^6+39963276 c^5+16855641 c^4
\nonu \\
&&
-309096607 c^3-1011629119 c^2+3369520433 c-1963795240) 
\, \pa G^{+} \pa  W_{  -\frac{2}{3}}^{(2)} 
\nonu \\
&&
+\frac{1}{35}
(17172 c^7+332064 c^6+3929988 c^5-17644845 c^4
\nonu \\
&& -61522118 c^3+359006763 c^2-470122960 c+173809600) 
\, \pa^2 G^{+} W_{  -\frac{2}{3}}^{(2)} 
\nonu \\
&&
+\frac{3}{980} 
(4212 c^7-96300 c^6-2349771 c^5+42056368 c^4
\nonu \\
&& -259031186 c^3+40739182 c^2+1557580395 c-1349009400) 
\, \pa J \pa  W_{  \frac{1}{3}}^{(\frac{5}{2})}
\nonu \\
&&
+\frac{3}{35} (5184 c^6+742824 c^5+267627 c^4-47562127 c^3+150357871 c^2-112833015 c+11915800) 
\nonu \\
&& \times  \pa J  G^{+} W_{  -\frac{2}{3}}^{(2)} 
\nonu \\
&&
+\frac{1}{980}
(294516 c^7+10366164 c^6+67363461 c^5-393530124 c^4\nonu \\
&& -713824950 c^3+3665027882 c^2-2110555905 c-112805400) 
\, \pa^2 J  W_{  \frac{1}{3}}^{(\frac{5}{2})}
\nonu \\
&&
-\frac{176}{7} (9c-11)
(243 c^5-891 c^4-3948 c^3+28517 c^2-47169 c+14120) 
\, T G^{+} W_{  -\frac{2}{3}}^{(2)} 
\nonu \\
&&
-\frac{396}{7} 
(720 c^5-1920 c^4-19891 c^3+159458 c^2-289411 c+167480) 
\, J J G^{+} W_{  -\frac{2}{3}}^{(2)} 
\nonu \\
&&
+\frac{1}{98}
(4860 c^7-114156 c^6-5195385 c^5+35319918 c^4 \nonu \\
&& -57149944 c^3-9492552 c^2+5296559 c+87151080) 
\, T \pa  W_{  \frac{1}{3}}^{(\frac{5}{2})}
\nonu \\
&&
+\frac{1}{98}
(175932 c^7+1565676 c^6-2193705 c^5-11977530 c^4
\nonu \\
&&
-85711576 c^3+417373800 c^2-461057505 c+123498600) 
\, \pa T W_{  \frac{1}{3}}^{(\frac{5}{2})}
\nonu \\
&&
-\frac{9}{245} 
(41184 c^6+834024 c^5-5042768 c^4
\nonu \\
&&
-10987322 c^3+175445691 c^2-422961195 c+269190200) 
\, J J \pa W_{  \frac{1}{3}}^{(\frac{5}{2})}
\nonu \\
&&
+\frac{396}{245} 
(2376 c^6+43416 c^5-154857 c^4-137112 c^3+1486276 c^2-2517689 c+1637320) 
\nonu \\
&& \times J G^{+}  W_{  -\frac{2}{3}}^{(3)}
\nonu \\
&&
-\frac{3}{245} 
(62316 c^6+781416 c^5+21115533 c^4 \nonu \\
&& +42457581 c^3-472604633 c^2+707153499 c-363267320) 
\, J G^{+} \pa  W_{  -\frac{2}{3}}^{(2)}
\nonu \\
&&
-\frac{72}{245} 
(26631 c^6+517386 c^5+826228 c^4-14619433 c^3+19570384 c^2+16405915 c-23010800) 
\nonu \\
&& \times \pa J J W_{  \frac{1}{3}}^{(\frac{5}{2})}
\nonu \\
&&
-\frac{3}{490} 
(23220 c^7-192276 c^6-3405531 c^5+33286542 c^4
\nonu \\
&& +91434148 c^3-878231224 c^2+1606271605 c-894785800) 
\, J \pa^2  W_{  \frac{1}{3}}^{(\frac{5}{2})}
\nonu \\
&&
+\frac{3}{ 490} 
(114696 c^7+2601396 c^6+6301530 c^5-92272437 c^4
\nonu \\
&& +140771499 c^3+130747355 c^2-331422531 c+95701880) 
\, G^{+} \pa  W_{  -\frac{2}{3}}^{(3)}
\nonu \\
&&
+\frac{3}{490} 
(228744 c^7+2956212 c^6-19184502 c^5-100604277 c^4 \nonu \\
&& 
\left.
+684144571 c^3-1212716261 c^2+690576381 c-48505480) 
\, \pa G^{+}  W_{  -\frac{2}{3}}^{(3)}
\right](w)+
\cdots.
\label{i2ope}
\eea
}
Because the ${\cal N}=2$ version in Appendix $G$ is known, 
the remaining $12$ OPEs can be presented similarly. However, 
due to the space of the paper, we do not present them in this paper 
completely.  
In the large $c$ limit,
all the nonlinear terms in Appendix (\ref{i2ope}) disappear.
One can also analyze the $\frac{1}{c}, \cdots, \frac{1}{c^3}$-terms. 

\subsection{The component OPEs  in the OPE $
{\bf W}_{  -\frac{2}{3}}^{(2)}(Z_1) 
\, {\bf W}_{  \frac{1}{3}}^{(\frac{7}{2})}(Z_2)$ }

The $4$ component OPEs corresponding to Appendix (\ref{23OPE})
can be summarized by
{\small 
\bea
&& 
\frac{1}{C_{(2)(\frac{7}{2})}^{(2)-}} \,
W_{  -\frac{2}{3}}^{(2)}(z) \, W_{  \frac{1}{3}}^{(\frac{7}{2})}(w) = 
-\frac{1}{(z-w)^3} \, \frac{3}{7} \, W_{  -\frac{1}{3}}^{(\frac{5}{2})}(w)
\nonu \\
&&+ \frac{1}{(z-w)^2} \, \frac{1}{(c+6)}\, \left[ 
\frac{27}{35} \,J  W_{  -\frac{1}{3}}^{(\frac{5}{2})}
-\frac{9}{5} \,
G^{-} W_{  \frac{2}{3}}^{(2)}
-\frac{3}{35} \, (c-3) \, 
\pa W_{  -\frac{1}{3}}^{(\frac{5}{2})}
\right](w) \nonu \\
&& + \frac{1}{(z-w)} \, 
\frac{1}{(c+6) (3 c+2) (9 c-11) (3 c^2+54 c-169)}\, 
\left[ 
\right. \nonu \\
&& \frac{54}{35} (132 c^3+1258 c^2-995 c+2279) 
\,
J J  W_{  -\frac{1}{3}}^{(\frac{5}{2})}
 \nonu \\
 && +\frac{18}{5} (63 c^3+237 c^2+3075 c-1499) 
J G^{-} W_{  \frac{2}{3}}^{(2)}
\nonu \\
&& -\frac{9}{35} (54 c^4-3591 c^3+6033 c^2+19605 c-1073)
G^{-} W_{  \frac{2}{3}}^{(3)}
 \nonu \\
&&
-\frac{15}{7} (81 c^4+810 c^3-1836 c^2-3428 c+2217)
 T  W_{  -\frac{1}{3}}^{(\frac{5}{2})}
\nonu \\
&& +\frac{9}{70} (261 c^4+3852 c^3+8902 c^2-41898 c-75809)
\pa J   W_{  -\frac{1}{3}}^{(\frac{5}{2})}
\nonu \\
&&
-\frac{9}{5} (36 c^4+167 c^3-3893 c^2+4967 c+2391) 
\pa G^{-} W_{  \frac{2}{3}}^{(2)}
\nonu \\
&&  
+\frac{9}{35} (45 c^4+678 c^3-8628 c^2-8020 c+29841)
J  \pa   W_{  -\frac{1}{3}}^{(\frac{5}{2})}
\nonu \\
&&
-\frac{9}{70} (180 c^4+8589 c^3+1245 c^2-53167 c+4597)
G^{-} \pa  W_{  \frac{2}{3}}^{(2)}
\nonu \\
&& \left. -\frac{3}{70} 
(27 c^5-747 c^4-15963 c^3+26637 c^2+60860 c-38166) 
\pa^2  W_{  -\frac{1}{3}}^{(\frac{5}{2})}
\right](w) + \cdots,
\nonu \\
&& 
\frac{1}{C_{(2)(\frac{7}{2})}^{(2)-}} \,
W_{  -\frac{2}{3}}^{(2)}(z) \, W_{  \frac{4}{3}}^{(4)}(w) =
-\frac{1}{(z-w)^4} \, W_{  \frac{2}{3}}^{(2)}(w) +
\frac{1}{(z-w)^3} \,\left[ 
\frac{12}{35} \,   W_{  \frac{2}{3}}^{(3)}
-\frac{1}{35} \pa   W_{  \frac{2}{3}}^{(2)}
\right](w) \nonu \\
&& + \frac{1}{(z-w)^2} \, 
\frac{1}{(c+6) (9 c-11) (3 c^2+54 c-169)}
\,\left[ 
\frac{9}{5} (c+1) (6 c^2+27 c+271) 
J  W_{  \frac{2}{3}}^{(3)}
\right. \nonu \\
&& - 2 (2 c-3) (3 c+25)(9c-11) 
T  W_{  \frac{2}{3}}^{(2)}
+ 234 (c+1) (c+6) 
J J  W_{  \frac{2}{3}}^{(2)}
\nonu \\
&& +6 (9 c^3-84 c^2-300 c+893) 
G^{-} W_{  \frac{5}{3}}^{(\frac{5}{2})} 
-\frac{18}{7} 
(30 c^3+214 c^2-467 c-1751) 
G^{+}  W_{  -\frac{1}{3}}^{(\frac{5}{2})} 
\nonu \\
&&
+ 3 (6 c^3+203 c^2+738 c-3859) 
\pa J  W_{  \frac{2}{3}}^{(2)}
-\frac{3}{5} 
(12 c^3+621 c^2+2606 c-9003) 
J \pa W_{  \frac{2}{3}}^{(2)}
\nonu \\
&&  +\frac{6}{35} 
(9 c^4+264 c^3+2328 c^2-1750 c-14823) 
\pa  W_{  \frac{2}{3}}^{(3)}
\nonu \\
&& \left.
-\frac{1}{70}
(9 c^4-2991 c^3-13737 c^2+58135 c-19128) 
\pa^2 W_{  \frac{2}{3}}^{(2)}
\right](w) 
\nonu \\
&& + \frac{1}{(z-w)} \, 
\frac{1}{ (c+6) (3 c+2) (9 c-11) (2 c^2+9 c-40) (3 c^2+54 c-169)}
 \nonu \\
&& \times \left[ 
-\frac{18}{245} (324 c^6+2160 c^5+54399 c^4-61158 c^3-1714933 c^2+8852748 c-12723040)
J \pa   W_{  \frac{2}{3}}^{(3)}
\right. \nonu \\
&&
-\frac{108}{245} (624 c^5+13424 c^4+48617 c^3-441123 c^2+492804 c-724256) 
J J  W_{  \frac{2}{3}}^{(3)}
\nonu \\
&&
+\frac{18}{245} (2202 c^5-4103 c^4-39374 c^3+1597111 c^2-2139828 c-3847008) 
J J \pa  W_{  \frac{2}{3}}^{(2)}
\nonu \\
&&
-\frac{12}{7} (216 c^5+3429 c^4-8106 c^3+91185 c^2-415616 c+218880)
J T   W_{  \frac{2}{3}}^{(2)}
\nonu \\
&&
+\frac{36}{49} (1656 c^5+22749 c^4-56517 c^3-247578 c^2+957196 c-1544736) 
J G^{+}  W_{  -\frac{1}{3}}^{(\frac{5}{2})}  
\nonu \\
&&
+\frac{6 (3 c-8)}{35} (90 c^5+2013 c^4+36738 c^3-6574 c^2-579885 c+493004) 
G^{-} \pa W_{  \frac{5}{3}}^{(\frac{5}{2})} 
\nonu \\
&&
+\frac{12}{7} (999 c^5+19143 c^4-99057 c^3-200733 c^2+1202212 c-874656) 
G^{-} G^{+}  W_{  \frac{2}{3}}^{(2)}
\nonu \\
&&
+\frac{6}{7} (162 c^6-1371 c^5-30618 c^4-10062 c^3+679299 c^2-1107376 c+101248)
\pa G^{-}  W_{  \frac{5}{3}}^{(\frac{5}{2})}
\nonu \\
&&
+\frac{12}{245} (c+6)(9c-11) (648 c^4+4374 c^3-4193 c^2-109749 c+199240)
T  W_{  \frac{2}{3}}^{(3)}  
\nonu \\
&&
+\frac{2}{245} (1458 c^6-160065 c^5-1269567 c^4+5688544 c^3+15358599 c^2-59269229 c+27494760)
\nonu \\
&& \times T \pa   W_{  \frac{2}{3}}^{(2)}
\nonu \\
&&
+\frac{1}{7} (1134 c^6+25110 c^5+118035 c^4-863072 c^3-1202211 c^2+7655688 c-4832832)
\pa T  W_{  \frac{2}{3}}^{(2)}
\nonu \\
&&
-\frac{36}{245} (396 c^6+10050 c^5+57494 c^4-96296 c^3-519387 c^2-1345359 c+4766392)
\pa J  W_{  \frac{2}{3}}^{(3)}  
\nonu \\
&&
+\frac{18}{7} (438 c^5+7309 c^4+14842 c^3-162049 c^2+113692 c+151712)
\pa J J   W_{  \frac{2}{3}}^{(2)}
\nonu \\
&&
+\frac{1}{7} (324 c^6+10134 c^5+88809 c^4-326400 c^3-1590463 c^2+4706064 c-684672)
\pa^2 J   W_{  \frac{2}{3}}^{(2)}
\nonu \\
&&
-\frac{36}{7} (c+6)(24 c^3-1413 c^2+7051 c-312) 
J J J  W_{  \frac{2}{3}}^{(2)}
\nonu \\
&&
-\frac{36}{7} (90 c^5+48 c^4-567 c^3+60931 c^2-268420 c+214624)
J G^{-}  W_{  \frac{5}{3}}^{(\frac{5}{2})} 
\nonu \\
&&
-\frac{18}{49} (288 c^6-768 c^5-46948 c^4+225537 c^3+295633 c^2-2939824 c+3711872)
G^{+} \pa  W_{  -\frac{1}{3}}^{(\frac{5}{2})}
\nonu \\
&&
-\frac{6}{49} (1080 c^6+27720 c^5+97212 c^4-967463 c^3-267499 c^2+7778204 c-10494432)
\pa G^{+}  W_{  -\frac{1}{3}}^{(\frac{5}{2})}
\nonu \\
&&
+\frac{6}{245} (54 c^7-243 c^6-14235 c^5-199026 c^4+1010729 c^3+2743737 c^2-16277128 c+16901952)
\nonu \\
&& \times \pa^2  W_{  \frac{2}{3}}^{(3)}  
\nonu \\
&&
-\frac{3}{245} (468 c^6+51000 c^5+341403 c^4-609536 c^3-6943491 c^2+14475796 c+806880)
J \pa^2 W_{  \frac{2}{3}}^{(2)}  
\nonu \\
&&
-\frac{3}{245} 
(864 c^6-24330 c^5-155879 c^4+999856 c^3-94513 c^2+4617474 c-18643472)
\pa J \pa   W_{  \frac{2}{3}}^{(2)}  
\nonu \\
&&
 -\frac{1}{1470} 
(162 c^7-59949 c^6+134535 c^5+5339307 c^4 \nonu \\
&& \left. -16284313 c^3-52881834 c^2+212816996 c-151818144)
\pa^3  W_{  \frac{2}{3}}^{(2)}
\right](w)
+ \cdots,
\nonu \\
&& 
\frac{1}{C_{(2)(\frac{7}{2})}^{(2)-}} \,
W_{  -\frac{2}{3}}^{(2)}(z) \, W_{  -\frac{2}{3}}^{(4)}(w) = \frac{1}{(z-w)^2} \,
\frac{33}{7 (c+6)} \,
G^{-}   W_{  -\frac{1}{3}}^{(\frac{5}{2})}
\nonu \\
&& +\frac{1}{(z-w)} \, \frac{1}{(c+6) (3 c+2)} \, \left[ 
\frac{33}{35} (5 c+8) 
\pa G^{-}  W_{  -\frac{1}{3}}^{(\frac{5}{2})}
+\frac{33}{35} (3 c-4) \, 
G^{-} \pa W_{  -\frac{1}{3}}^{(\frac{5}{2})}
-\frac{396}{35} 
J G^{-}  W_{  -\frac{1}{3}}^{(\frac{5}{2})}
\right](w)
\nonu \\
&& + \cdots,
\nonu \\
&& 
\frac{1}{C_{(2)(\frac{7}{2})}^{(2)-}} \,
W_{  -\frac{2}{3}}^{(2)}(z) \, W_{  \frac{1}{3}}^{(\frac{9}{2})}(w) =
-\frac{1}{(z-w)^4} \, \frac{13}{14} \, 
W_{  -\frac{1}{3}}^{(\frac{5}{2})}(w)
\nonu \\
&& +\frac{1}{(z-w)^3} \, \frac{1}{(c+6)} \, \left[ 
\frac{31}{5} 
G^{-}  W_{  \frac{2}{3}}^{(2)} 
-\frac{93}{35} 
J  W_{  -\frac{1}{3}}^{(\frac{5}{2})}
-\frac{1}{70} (c+192) 
\pa  W_{  -\frac{1}{3}}^{(\frac{5}{2})}
\right](w)\nonu \\
&& +\frac{1}{(z-w)^2} \, 
\frac{1}{(c+6) (3 c+2) (9 c-11) (3 c^2+54 c-169)}
\,
\left[ 
\right. \nonu \\
&& \frac{9}{5} (528 c^3+6262 c^2-15995 c+17871) 
 J J  W_{  -\frac{1}{3}}^{(\frac{5}{2})}
\nonu \\
&&
+\frac{9}{5} (3 c^3-5143 c^2+50495 c-36759) 
J G^{-}  W_{  \frac{2}{3}}^{(2)} 
\nonu \\
&&
-\frac{3}{70} (8154 c^4+52551 c^3-423801 c^2+770211 c+61409) 
G^{-}  W_{  \frac{2}{3}}^{(3)} 
\nonu \\
&&
+\frac{1}{28} (4131 c^4+20718 c^3+164556 c^2-850228 c+456003) 
T  W_{  -\frac{1}{3}}^{(\frac{5}{2})}
\nonu \\
&&
+\frac{1}{10} (1647 c^4+57780 c^3+20028 c^2-438746 c+224359)
\pa G^{-}  W_{  \frac{2}{3}}^{(2)}  
\nonu \\
&&
+\frac{3}{140} (3501 c^4+32310 c^3+7924 c^2+389472 c-2570643)
 \pa J W_{  -\frac{1}{3}}^{(\frac{5}{2})}
\nonu \\
&&
+\frac{1}{140} (3510 c^4-288567 c^3-1653351 c^2+6330489 c-2329237) 
G^{-} \pa  W_{  \frac{2}{3}}^{(2)} 
\nonu \\
&&
-\frac{3}{170} (1656 c^4+20229 c^3-184065 c^2+1031677 c-1187885)
J \pa   W_{  -\frac{1}{3}}^{(\frac{5}{2})}
\nonu \\
&&
\left. -\frac{1}{140}
(27 c^5-10521 c^4-80097 c^3-676605 c^2+3011162 c-1639782)
\pa^2   W_{  -\frac{1}{3}}^{(\frac{5}{2})}
\right](w) 
\nonu \\
&& +\frac{1}{(z-w)} \, 
\frac{1}{ (c+6) (3 c+2) (6 c-5) (9 c-11) (2 c^2+9 c-40) (3 c^2+54 c-169)}
\nonu \\
&& \times \left[ 
\frac{9}{245} 
(41184 c^6+834024 c^5-5042768 c^4
\right. \nonu \\
&&
-10987322 c^3+175445691 c^2-422961195 c+269190200)
  J J \pa  W_{  -\frac{1}{3}}^{(\frac{5}{2})}
\nonu \\
&&
+\frac{396}{7} (288 c^5+282 c^4-8208 c^3+31325 c^2-92433 c+70440)
J J J   W_{  -\frac{1}{3}}^{(\frac{5}{2})}
\nonu \\
&&
+\frac{3}{245} 
(62316 c^6+781416 c^5+21115533 c^4 \nonu \\
&& +42457581 c^3-472604633 c^2+707153499 c-363267320)
J G^{-} \pa   W_{  \frac{2}{3}}^{(2)} 
\nonu \\
&& 
+\frac{264}{49} (5508 c^6+19170 c^5-231174 c^4+22842 c^3+2275051 c^2-4668767 c+2971960)
J T  W_{  -\frac{1}{3}}^{(\frac{5}{2})} 
\nonu \\
&&
+\frac{3}{980} 
(114696 c^7+2601396 c^6+6301530 c^5-92272437 c^4 \nonu \\
&& +140771499 c^3+130747355 c^2-331422531 c+95701880)
G^{-} \pa  W_{\frac{2}{3}}^{(3)} 
\nonu \\
&&
+\frac{132}{49} 
(12636 c^6+63801 c^5-637848 c^4+651405 c^3+1738728 c^2-2363090 c+457200)
G^{-} G^{+}  W_{  -\frac{1}{3}}^{(\frac{5}{2})}
\nonu \\ 
&& 
+\frac{1}{980}
(50544 c^7-15976548 c^6-35411724 c^5+375345195 c^4
\nonu \\
&& -211554213 c^3-755063573 c^2+169169907 c+262216840)
G^{-} \pa^2   W_{  \frac{2}{3}}^{(2)} 
\nonu \\
&&
-\frac{3}{490} 
(228744 c^7+2956212 c^6-19184502 c^5-100604277 c^4
\nonu \\
&& +684144571 c^3-1212716261 c^2+690576381 c-48505480)
\pa G^{-}  W_{\frac{2}{3}}^{(3)} 
\nonu \\
&&
+\frac{1}{980}
(89424 c^7+793044 c^6+39963276 c^5+16855641 c^4 
\nonu \\
&&-309096607 c^3-1011629119 c^2+3369520433 c-1963795240)
\pa G^{-} \pa   W_{  \frac{2}{3}}^{(2)} 
\nonu \\
&&
+\frac{1}{35}
(17172 c^7+572634 c^6+2753868 c^5-20475255 c^4 \nonu \\
&& -28513208 c^3+277803883 c^2-399069670 c+156724400)
\pa^2 G^{-}   W_{  \frac{2}{3}}^{(2)} 
\nonu \\
&&
+\frac{1}{196}
(4860 c^7-114156 c^6-5195385 c^5+35319918 c^4 \nonu \\
&& -57149944 c^3-9492552 c^2+5296559 c+87151080)
T \pa   W_{  -\frac{1}{3}}^{(\frac{5}{2})}
\nonu \\
&&
+\frac{1}{196}
(175932 c^7+4901580 c^6+14649759 c^5-180369402 c^4
\nonu \\
&&
+86259344 c^3+876397992 c^2-1084913265 c+244199400)
\pa T   W_{  -\frac{1}{3}}^{(\frac{5}{2})}
\nonu \\
&&
+\frac{3}{980} 
(4212 c^7-96300 c^6-2349771 c^5+42056368 c^4
\nonu \\
&& -259031186 c^3+40739182 c^2+1557580395 c-1349009400)
\pa J \pa  W_{  -\frac{1}{3}}^{(\frac{5}{2})}
\nonu \\
&&
+\frac{72}{245} 
(26631 c^6+517386 c^5+826228 c^4-14619433 c^3+19570384 c^2+16405915 c-23010800)
\nonu \\
&& \times \pa J J   W_{  -\frac{1}{3}}^{(\frac{5}{2})}
\nonu \\
&&
+\frac{1}{980}
(294516 c^7+10366164 c^6+67363461 c^5-393530124 c^4
\nonu \\
&& -713824950 c^3+3665027882 c^2-2110555905 c
-112805400)
\pa^2 J   W_{  -\frac{1}{3}}^{(\frac{5}{2})}
\nonu \\
&&
-\frac{44}{7} (9c-11) (243 c^5-891 c^4-3948 c^3+28517 c^2-47169 c+14120)
G^{-} T  W_{  \frac{2}{3}}^{(2)} 
\nonu \\
&&
-\frac{396}{7} 
(720 c^5-1920 c^4-19891 c^3+159458 c^2-289411 c+167480)
J J G^{-}   W_{  \frac{2}{3}}^{(2)} 
\nonu \\
&&
-\frac{3}{35} 
(11124 c^6+699264 c^5-2134113 c^4-9013287 c^3+76849931 c^2-128290435 c+54481400)
\nonu \\
&& \times \pa G^{-} J   W_{  \frac{2}{3}}^{(2)}
\nonu \\
&&
-\frac{3}{35} 
(5184 c^6+742824 c^5+267627 c^4-47562127 c^3+150357871 c^2-112833015 c+11915800)
\nonu \\
&& \times \pa J G^{-}   W_{  \frac{2}{3}}^{(2)}
\nonu \\
&&
+\frac{ 396}{245} 
(2376 c^6+43416 c^5-154857 c^4-137112 c^3+1486276 c^2-2517689 c+1637320)
\nonu \\
&& \times J G^{-}  W_{  \frac{2}{3}}^{(3)} 
\nonu \\
&&
-\frac{3}{490} 
(23220 c^7-192276 c^6-3405531 c^5+33286542 c^4
\nonu \\
&& +91434148 c^3-878231224 c^2+1606271605 c-894785800)
J  \pa^2   W_{  -\frac{1}{3}}^{(\frac{5}{2})}
\nonu \\
&&
-\frac{1}{2940}
(972 c^8-355968 c^7-9455877 c^6-11669739 c^5+104947449 c^4
\nonu \\
&& \left. +660636269 c^3-2683044612 c^2+2930835770 c-1299814800)
\pa^3   W_{  -\frac{1}{3}}^{(\frac{5}{2})}
\right](w) + \cdots.
\label{i3ope}
\eea
}
The remaining twelve OPEs are abbreviated in this paper. 
In the large $c$ limit,
all the nonlinear terms in Appendix (\ref{i3ope}) disappear.
One can also analyze the $\frac{1}{c}, \cdots, \frac{1}{c^3}$-terms. 

\subsection{The component OPEs   in the OPE  
${\bf W}_{  -\frac{2}{3}}^{(2)}(Z_1) 
\, {\bf W}_{  -\frac{1}{3}}^{(\frac{7}{2})}(Z_2)$ }

The $16$ component OPEs corresponding to Appendix (\ref{24OPE})
can be summarized by
{\small
\bea
&& 
\frac{1}{C_{(2)(\frac{7}{2})}^{(4)-}} \,
W_{  -\frac{2}{3}}^{(2)}(z) \, W_{  -\frac{1}{3}}^{(\frac{7}{2})}(w) = 
-\frac{1}{(z-w)} \, \frac{1}{4} \, W_{  -1}^{(\frac{9}{2})}(w)+\cdots,
\nonu \\
&& 
\frac{1}{C_{(2)(\frac{7}{2})}^{(4)-}} \,
W_{  -\frac{2}{3}}^{(2)}(z) \, W_{  \frac{2}{3}}^{(4)}(w) = 
-\frac{1}{(z-w)^2} \,  W_{0}^{(4)}(w) +
\frac{1}{(z-w)} \, \left[ 
\frac{1}{6}   W_{0}^{(5)}
-\frac{1}{4}  \pa   W_{0}^{(4)}
\right](w) +  \cdots,
\nonu \\
& & \frac{1}{C_{(2)(\frac{7}{2})}^{(4)-}} \,
W_{  -\frac{2}{3}}^{(2)}(z) \, W_{  -\frac{4}{3}}^{(4)}(w) = + \cdots,
\nonu \\
&& 
\frac{1}{C_{(2)(\frac{7}{2})}^{(4)-}} \,
W_{  -\frac{2}{3}}^{(2)}(z) \, W_{  -\frac{1}{3}}^{(\frac{9}{2})}(w) = 
-\frac{1}{(z-w)^2} \, \frac{17}{24} \, W_{  -1}^{(\frac{9}{2})}(w)
\nonu \\
& & + 
\frac{1}{(z-w)} \, \frac{1}{(c+9)} \, \left[ 
\frac{20 }{3 } \,
G^{-}  W_{0}^{(4)}
-\frac{5  }{3 } \,
J W_{  -1}^{(\frac{9}{2})}
-\frac{ (11 c+219) }{72 } \,
\pa  W_{  -1}^{(\frac{9}{2})}
\right](w) + \cdots,
\nonu \\
&& 
\frac{1}{C_{(2)(\frac{7}{2})}^{(4)-}} \,
W_{  \frac{1}{3}}^{(\frac{5}{2})}(z) \, W_{  -\frac{1}{3}}^{(\frac{7}{2})}(w) = 
\frac{1}{(z-w)^2} \,  W_{0}^{(4)}(w) +\frac{1}{(z-w)} \, 
\left[ 
\frac{3}{8}  \pa  W_{0}^{(4)}
+\frac{1}{12}    W_{0}^{(5)}
\right](w) + \cdots, 
\nonu \\
&& 
\frac{1}{C_{(2)(\frac{7}{2})}^{(4)-}} \,
W_{  \frac{1}{3}}^{(\frac{5}{2})}(z) \, W_{  \frac{2}{3}}^{(4)}(w) = 
-\frac{1}{(z-w)^2} \, W_{  1}^{(\frac{9}{2})}(w) 
-\frac{1}{(z-w)} \, \frac{1}{3} \, \pa   W_{  1}^{(\frac{9}{2})}(w)+\cdots,
\nonu \\
&& 
\frac{1}{C_{(2)(\frac{7}{2})}^{(4)-}} \,
W_{  \frac{1}{3}}^{(\frac{5}{2})}(z) \, W_{  -\frac{4}{3}}^{(4)}(w) = 
-\frac{1}{(z-w)^2} \, \frac{5}{6}   W_{  -1}^{(\frac{9}{2})}(w) 
\nonu \\
& & +   
 \frac{1}{(z-w)} \, \frac{1}{(c+9)} \, \left[ 
\frac{20 }{3 } \,
G^{-}  W_{0}^{(4)}
-\frac{5  }{3 } \,
J  W_{  -1}^{(\frac{9}{2})}
-\frac{5  (c+15) }{18 } \,
\pa  W_{  -1}^{(\frac{9}{2})}
\right](w) 
+ \cdots,
\nonu \\
&& 
\frac{1}{C_{(2)(\frac{7}{2})}^{(4)-}} \,
W_{  \frac{1}{3}}^{(\frac{5}{2})}(z) \, W_{  -\frac{1}{3}}^{(\frac{9}{2})}(w) = 
-\frac{1}{(z-w)^3} \,  W_{0}^{(4)}(w) +
\frac{1}{(z-w)^2}\, \left[ 
\frac{19}{24}  
 W_{0}^{(5)}
-\frac{13}{48} \pa  W_{0}^{(4)} 
\right](w) \nonu \\
&& +  
\frac{1}{(z-w)} \, \frac{1}{(c+9)} \, \left[ 
-\frac{20}{3 }\, 
G^{-}  W_{  1}^{(\frac{9}{2})}
-\frac{5  }{3 } \,
G^{+} W_{  -1}^{(\frac{9}{2})}
+\frac{10 }{3 }\, 
T W_{0}^{(4)} 
\right.
\nonu \\
& & \left. +  \frac{5  }{3 } \,
J W_{0}^{(5)} 
-\frac{10 }{3 }\, 
\pa J W_{0}^{(4)}
+\frac{5  }{6 }\,
J \pa  W_{0}^{(4)}
-  
\frac{ (17 c+273) }{144 } \,
\pa W_{0}^{(5)}
-\frac{ (7 c-57)}{144 }\,
\pa^2  W_{0}^{(4)}
\right](w) +
\cdots,
\nonu \\
&& \frac{1}{C_{(2)(\frac{7}{2})}^{(4)-}} \,
W_{  -\frac{5}{3}}^{(\frac{5}{2})}(z) \, W_{  -\frac{1}{3}}^{(\frac{7}{2})}(w) = + \cdots,
\nonu \\
&& 
\frac{1}{C_{(2)(\frac{7}{2})}^{(4)-}} \,
W_{  -\frac{5}{3}}^{(\frac{5}{2})}(z) \, W_{  \frac{2}{3}}^{(4)}(w) = 
-\frac{1}{(z-w)^2} \, \frac{5}{12} \, 
W_{  -1}^{(\frac{9}{2})}(w) \nonu \\
& & +  \frac{1}{(z-w)} \, \frac{1}{(c+9)} \, \left[ 
\frac{5}{3 }\, 
J  W_{  -1}^{(\frac{9}{2})}
-\frac{20  }{3 }\,
G^{-}  W_{0}^{(4)}
-\frac{5  (c-3)}{36 } \, 
\pa W_{  -1}^{(\frac{9}{2})}
\right](w) + 
\cdots,
\nonu \\
&& \frac{1}{C_{(2)(\frac{7}{2})}^{(4)-}} \,
W_{  -\frac{5}{3}}^{(\frac{5}{2})}(z) \, W_{  -\frac{4}{3}}^{(4)}(w) = +\cdots,
\nonu \\
&& 
\frac{1}{C_{(2)(\frac{7}{2})}^{(4)-}} \,
W_{  -\frac{5}{3}}^{(\frac{5}{2})}(z) \, W_{  -\frac{1}{3}}^{(\frac{9}{2})}(w) = 
-\frac{1}{(z-w)} \,
\frac{25  }{3 (c+9)} \,
G^{-} W_{  -1}^{(\frac{9}{2})}(w)
+ \cdots,
\nonu \\
&& 
\frac{1}{C_{(2)(\frac{7}{2})}^{(4)-}} \,
W_{  -\frac{2}{3}}^{(3)}(z) \, W_{  -\frac{1}{3}}^{(\frac{7}{2})}(w) = 
\frac{1}{(z-w)^2} \, \frac{7}{24} \,  W_{  -1}^{(\frac{9}{2})}(w)
\nonu \\
&& +  
\frac{1}{(z-w)} \, \frac{1}{(c+9)} \, \left[ 
\frac{20 }{3 }\, 
G^{-}  W_{0}^{(4)}
-\frac{5  }{3 } \,
J  W_{  -1}^{(\frac{9}{2})}
+\frac{5  (c-3) }{36 } \,
\pa  W_{  -1}^{(\frac{9}{2})}
\right](w) + \cdots,
\nonu \\
&& 
\frac{1}{C_{(2)(\frac{7}{2})}^{(4)-}} \,
W_{  -\frac{2}{3}}^{(3)}(z) \, W_{  \frac{2}{3}}^{(4)}(w) = 
-\frac{1}{(z-w)^3}  W_{0}^{(4)}(w) +\frac{1}{(z-w)^2} \,
\left[- 
\frac{1}{3}   W_{0}^{(5)}
-\frac{1}{3}  \pa W_{0}^{(4)}
\right](w) \nonu \\
& & +   \frac{1}{(z-w)} \, \frac{1}{(c+9)} \, \left[ 
-\frac{20 }{3 } \,
G^{-}  W_{  1}^{(\frac{9}{2})}
-\frac{5 }{3 } \, 
G^{+}  W_{  -1}^{(\frac{9}{2})}
 -  \frac{20 }{3 } \,
T  W_{0}^{(4)}
-\frac{5  }{6 } \,
J  W_{0}^{(5)}
-\frac{10 }{3 } \,
\pa J  W_{0}^{(4)}
\right.
\nonu \\
& &
+  \left. 
\frac{5 }{6 } \, 
J \pa W_{0}^{(4)}
-\frac{5  (c-3) }{36 } \,
\pa  W_{0}^{(5)}
-\frac{5  (c-3) }{72 }
\pa^2 W_{0}^{(4)}
\right](w) 
+ \cdots,
\nonu \\
&& 
\frac{1}{C_{(2)(\frac{7}{2})}^{(4)-}} \,
W_{  -\frac{2}{3}}^{(3)}(z) \, W_{  -\frac{4}{3}}^{(4)}(w) = 
-\frac{1}{(z-w)} \,  
\frac{25 }{3 (c+9)} \,
G^{-}  W_{  -1}^{(\frac{9}{2})}(w)
+ \cdots,
\nonu \\
&& 
\frac{1}{C_{(2)(\frac{7}{2})}^{(4)-}} \,
W_{  -\frac{2}{3}}^{(3)}(z) \, W_{  -\frac{1}{3}}^{(\frac{9}{2})}(w) = 
-\frac{1}{(z-w)^3} \, \frac{9}{8} \,  W_{  -1}^{(\frac{9}{2})}(w)-
\frac{1}{(z-w)^2} \,\frac{17}{48}  \, \pa W_{  -1}^{(\frac{9}{2})}(w)
+ \nonu \\
&& + \frac{1}{(z-w)} \, \frac{1}{(c+9)} \,\left[ 
-\frac{25  }{3 } \,
T W_{  -1}^{(\frac{9}{2})}
+\frac{25 }{3 }\, 
G^{-}  W_{0}^{(5)} 
-\frac{10 }{3 } \, 
\pa G^{-}  W_{0}^{(4)} 
\right.
\nonu \\
&& \left.
-  \frac{10  }{3 } \,
\pa J  W_{  -1}^{(\frac{9}{2})}
+\frac{5 }{6 }\, 
J \pa  W_{  -1}^{(\frac{9}{2})}
+\frac{5 }{6 } \, 
G^{-} \pa  W_{0}^{(4)} 
-  
\frac{5  (c-3) }{72} \,
\pa^2  W_{  -1}^{(\frac{9}{2})}
\right](w) + \cdots.
\label{i4ope}
\eea
}
The structure of the OPE looks like as the one in Appendix (\ref{i1ope}).
In the large $c$ limit,
all the nonlinear terms in Appendix (\ref{i4ope}) disappear.
One can also analyze the $\frac{1}{c}$-term. 

\section{The component OPEs  in the OPE  
${\bf W}_{  \pm \frac{1}{3}}^{(\frac{7}{2})}(Z_1)
\, {\bf W}_{  \pm \frac{1}{3}}^{(\frac{7}{2})}(Z_2)$ }

In this final Appendix, we present the component OPEs 
corresponding to the ${\cal N}=2$ description in Appendix $(G.2)$.

\subsection{The component OPEs  in the OPE 
${\bf W}_{  \frac{1}{3}}^{(\frac{7}{2})}(Z_1) \,
{\bf W}_{  \frac{1}{3}}^{(\frac{7}{2})}(Z_2)$ }

The $4$ component OPEs corresponding to Appendix (\ref{33OPE})
can be summarized by
{\small
\bea
&& 
\frac{1}{C_{(\frac{7}{2})(\frac{7}{2})}^{(\frac{7}{2})+}} \,
W_{  \frac{1}{3}}^{(\frac{7}{2})}(z) \, W_{  \frac{1}{3}}^{(\frac{7}{2})}(w) = 
- \frac{1}{(z-w)^3} \, \frac{3}{11} \,W_{  \frac{2}{3}}^{(4)}(w) + 
\frac{1}{(z-w)^2} \, \frac{3}{22} \, \pa W_{  \frac{2}{3}}^{(4)}(w) 
\nonu \\
& & + \frac{1}{(z-w)} \, 
\frac{1}{ (c+9) (3 c+4) (27 c-46) 
(3 c^2+90 c-265)} \, \left[ 
\right. 
\nonu \\
&& \frac{216}{11} (42 c^3+614 c^2-1651 c+1385) 
 \,
J J W_{  \frac{2}{3}}^{(4)} 
 \nonu \\
&&
+ 36 (9 c^3-372 c^2+3333 c-4030) 
 \,
J G^{+} W_{  -\frac{1}{3}}^{(\frac{7}{2})}
 \nonu \\
&& 
-\frac{3}{11} (2025 c^4+27756 c^3-38283 c^2-269878 c+353640) 
 \,
T W_{  \frac{2}{3}}^{(4)}
\nonu \\
&& -\frac{18}{11} (27 c^4-1575 c^3+507 c^2+16059 c-10778) 
\,
G^{+}   W_{  -\frac{1}{3}}^{(\frac{9}{2})}
\nonu \\
&&
+\frac{9}{2} (9 c^4-884 c^3-7131 c^2+26506 c-11080) 
 \,
\pa G^{+}   W_{  -\frac{1}{3}}^{(\frac{7}{2})}
\nonu \\
&& -\frac{18}{11} (9 c^4-390 c^3+4957 c^2+17064 c-78580) 
 \,
\pa J  W_{  \frac{2}{3}}^{(4)} 
\nonu \\
&&
+\frac{9}{22} (9 c^4-528 c^3-8139 c^2+103598 c-178320) 
 \,
J  \pa  W_{  \frac{2}{3}}^{(4)} 
\nonu \\
&& -\frac{9}{22} (45 c^4-4242 c^3-36687 c^2+154420 c-87036) 
 \,
G^{+} \pa   W_{  -\frac{1}{3}}^{(\frac{7}{2})}
\nonu \\
&&
- \left. 
\frac{3}{44} (3 c-1) (45 c^4+1302 c^3+1407 c^2-42794 c+75240) 
 \,
\pa^2   W_{  \frac{2}{3}}^{(4)} 
\right](w) 
+\cdots,
\nonu \\
&& 
\frac{1}{C_{(\frac{7}{2})(\frac{7}{2})}^{(\frac{7}{2})+}} \,
W_{  \frac{1}{3}}^{(\frac{7}{2})}(z) \, W_{  \frac{4}{3}}^{(4)}(w) =
-\frac{1}{(z-w)^2} \, \frac{30 }{11 (c+9)} \,
G^{+}  W_{  \frac{2}{3}}^{(4)}(w)
\nonu \\
&& + 
\frac{1}{(z-w)} \, \frac{1}{(c+9) (3 c+4)} \, \left[ 
\frac{90 }{11 } \,
J G^{+}   W_{  \frac{2}{3}}^{(4)}
-\frac{15}{11} \, (3 c+7)  \,
G^{+} \pa  W_{  \frac{2}{3}}^{(4)}
-\frac{30}{11} \, (c-1)  \,
\pa G^{+}  W_{  \frac{2}{3}}^{(4)}
\right](w)  \nonu \\
&& + \cdots ,
\nonu \\
&& 
\frac{1}{C_{(\frac{7}{2})(\frac{7}{2})}^{(\frac{7}{2})+}} \,
W_{  \frac{1}{3}}^{(\frac{7}{2})}(z) \, W_{  -\frac{2}{3}}^{(4)}(w) = 
-\frac{1}{(z-w)^4} \,  W_{  -\frac{1}{3}}^{(\frac{7}{2})}(w) 
+ \frac{1}{(z-w)^3} \, \left[ 
-\frac{3}{22} \,
 W_{  -\frac{1}{3}}^{(\frac{9}{2})}
 -\frac{19}{44}  \,
\pa  W_{  -\frac{1}{3}}^{(\frac{7}{2})}
\right](w) \nonu \\
&& + \frac{1}{(z-w)^2} \, 
\frac{1}{(c+9) (27 c-46) (3 c^2+90 c-265)}
\,
\left[ 
-\frac{1}{2} (27c-46) (51 c^2+614 c-1725) 
 \,
T  W_{  -\frac{1}{3}}^{(\frac{7}{2})}
\right. \nonu \\
&& 
+ 36 (c+9)(31 c-35) \,
 J J W_{  -\frac{1}{3}}^{(\frac{7}{2})}
\nonu \\
&&
+\frac{9}{22} (687 c^3+7144 c^2-25501 c-6790)
 \, 
G^{-}  W_{  \frac{2}{3}}^{(4)}
-\frac{3}{2} (45 c^3-2904 c^2+4281 c+5998) 
 \,
G^{+}  W_{  -\frac{4}{3}}^{(4)}
\nonu \\
&&
-\frac{9}{22} (39 c^3-984 c^2+18923 c-22118) 
\,
J W_{  -\frac{1}{3}}^{(\frac{9}{2})}
+\frac{3}{4} (3 c^3-1544 c^2-35529 c+108050) 
 \,
\pa J  W_{  -\frac{1}{3}}^{(\frac{7}{2})}
\nonu \\
&&
-\frac{3}{44} (15 c^3-6552 c^2-121229 c+352266) 
 \,
J \pa  W_{  -\frac{1}{3}}^{(\frac{7}{2})}
\nonu \\
&& +\frac{3}{44} (36 c^4+1695 c^3+15900 c^2-65125 c+27114) 
 \,
 \pa W_{  -\frac{1}{3}}^{(\frac{9}{2})}
\nonu \\
&&
\left. 
-\frac{1}{44} (387 c^4+7818 c^3-9849 c^2-116948 c+200412) 
 \,
\pa^2   W_{  -\frac{1}{3}}^{(\frac{7}{2})}
\right](w)
 \nonu \\
&& + \frac{1}{(z-w)} \, 
\frac{1}{(c+9) (3 c+4) (27 c-46) (3 c^2+90 c-265)}
\,
\left[ 
\right. \nonu \\
&& -\frac{3}{22} (c+9) (27c-46) (81 c^2+87 c-698) 
\,
T W_{  -\frac{1}{3}}^{(\frac{9}{2})}
\nonu \\
&&
-\frac{3}{22} (27 c^5+1233 c^4+17949 c^3-40569 c^2-15268 c-74292) 
 \,
\pa^2 W_{  -\frac{1}{3}}^{(\frac{9}{2})}
\nonu \\
&&
-\frac{9}{11} 27 c^4-567 c^3+15243 c^2-23565 c+22462) 
 \,
J \pa  W_{  -\frac{1}{3}}^{(\frac{9}{2})}
\nonu \\
&&
+\frac{18}{11} (897 c^3+7574 c^2+5294 c-18015) 
\,
J J \pa W_{  -\frac{1}{3}}^{(\frac{7}{2})}
\nonu \\
&&
+\frac{216}{11} (42 c^3+614 c^2-1651 c+1385) 
 \,
J G^{-} W_{  \frac{2}{3}}^{(4)} 
\nonu \\
&&
+\frac{9}{22} (1035 c^4+11826 c^3-28033 c^2-9388 c-102740) 
 \,
G^{-} \pa  W_{  \frac{2}{3}}^{(4)} 
\nonu \\
&&
+ 18 (9 c^3-372 c^2+3333 c-4030) 
\,
G^{-} G^{+}  W_{  -\frac{1}{3}}^{(\frac{7}{2})}
\nonu \\
&&
+\frac{3}{22} (2025 c^4+27756 c^3-38283 c^2-269878 c+353640) 
 \,
\pa G^{-}  W_{  \frac{2}{3}}^{(4)} 
\nonu \\
&&
-\frac{1}{44} (39771 c^4+408798 c^3-2058153 c^2+1525928 c+1647156) 
 \,
T \pa W_{  -\frac{1}{3}}^{(\frac{7}{2})}
\nonu \\
&&
-\frac{5}{2} (405 c^4+5616 c^3-10335 c^2-29978 c+41712) 
 \,
\pa T  W_{  -\frac{1}{3}}^{(\frac{7}{2})}
\nonu \\
&&
-\frac{9}{44} (99 c^4-1206 c^3+71647 c^2+45476 c-381236) 
\,
\pa J W_{  -\frac{1}{3}}^{(\frac{9}{2})}
\nonu \\
&&
+9 (363 c^3+3796 c^2-3209 c-1010) 
\,
\pa J J  W_{  -\frac{1}{3}}^{(\frac{7}{2})}
\nonu \\
&&
+\frac{3}{55} (297 c^4-48630 c^3-901075 c^2+2098520 c+2141068) 
\,
\pa J \pa W_{  -\frac{1}{3}}^{(\frac{7}{2})}
\nonu \\
&&
+18 (9 c^3-372 c^2+3333 c-4030) 
\,
J T  W_{  -\frac{1}{3}}^{(\frac{7}{2})}
\nonu \\
&&
-18 (9 c^3-372 c^2+3333 c-4030) 
\, J G^{+} W_{  -\frac{4}{3}}^{(4)} 
\nonu \\
&&
-3 (27 c^4-1629 c^3+2739 c^2-3939 c+13402) 
 \,
G^{+} \pa  W_{  -\frac{4}{3}}^{(4)} 
\nonu \\
&&
-\frac{3}{2} (81 c^4-5592 c^3-10083 c^2+57318 c-4624) 
\,
\pa G^{+} W_{  -\frac{4}{3}}^{(4)}
\nonu \\
&&
-\frac{3}{4} (9 c^4+984 c^3+45685 c^2-51258 c-232720) 
\,
\pa^2 J W_{  -\frac{1}{3}}^{(\frac{7}{2})}
\nonu \\
&&
+\frac{108}{11} (42 c^3+614 c^2-1651 c+1385) 
\,
J J W_{  -\frac{1}{3}}^{(\frac{9}{2})}
\nonu \\
&&
-\frac{3}{22} (18 c^4-5295 c^3-103578 c^2+220669 c+218526) 
 \,
J \pa^2 W_{  -\frac{1}{3}}^{(\frac{7}{2})}
\nonu \\
&&
\left. 
-\frac{1}{66}
(324 c^5-846 c^4-67125 c^3+228810 c^2-155309 c-27234) 
 \,
\pa^3 W_{  -\frac{1}{3}}^{(\frac{7}{2})}
\right](w)+ \cdots,
\nonu \\
&& 
\frac{1}{C_{(\frac{7}{2})(\frac{7}{2})}^{(\frac{7}{2})+}} \,
W_{  \frac{1}{3}}^{(\frac{7}{2})}(z) \, W_{  \frac{1}{3}}^{(\frac{9}{2})}(w) = 
\frac{1}{(z-w)^4} \, \frac{1}{2} \, W_{  \frac{2}{3}}^{(4)} 
\nonu \\
&& +
\frac{1}{(z-w)^3} \, \frac{1}{(c+9)} \, \left[ 
10  \,
G^{+}  W_{  -\frac{1}{3}}^{(\frac{7}{2})}
-\frac{30 }{11} \,
J  W_{  \frac{2}{3}}^{(4)} 
+\frac{2}{11} \,(c+24)  \,
\pa W_{  \frac{2}{3}}^{(4)} 
\right](w)  \nonu \\
&& + \frac{1}{(z-w)^2} \, 
\frac{1}{(c+9) (27 c-46) (3 c^2+90 c-265)} \, \left[ 
-18 (c+9)(31 c-35) 
\, J J W_{  \frac{2}{3}}^{(4)} 
\right. \nonu \\
&& -36 (c+9) (31 c-35) 
 \,
J G^{+}  W_{  -\frac{1}{3}}^{(\frac{7}{2})}
\nonu \\
&&
+\frac{6}{11} (279 c^3+1368 c^2-15255 c+30428) 
\,
G^{+} W_{  -\frac{1}{3}}^{(\frac{9}{2})}
\nonu \\
&& +\frac{1}{22} (8235 c^3+41664 c^2-187409 c+40170) 
\,
T  W_{  \frac{2}{3}}^{(4)} 
\nonu \\
&&
+\frac{1}{4} (927 c^3+43272 c^2-117629 c+19370) 
 \,
\pa G^{+} W_{  -\frac{1}{3}}^{(\frac{7}{2})}
\nonu \\
&& +\frac{3}{11} (51 c^3-928 c^2+67527 c-181590) 
 \,
\pa J  W_{  \frac{2}{3}}^{(4)} 
\nonu \\
&&
+\frac{1}{44} (16155 c^3+377736 c^2-2257257 c+2943466) 
 \,
G^{+} \pa  W_{  -\frac{1}{3}}^{(\frac{7}{2})}
\nonu \\
&& 
-\frac{3}{44} (1671 c^3+47640 c^2-136901 c+34490) 
 \,
J  \pa W_{  \frac{2}{3}}^{(4)}
\nonu \\
&& \left.
+\frac{1}{88}
(279 c^4+12453 c^3+217491 c^2-1276973 c+1581870) 
\,
\pa^2 W_{  \frac{2}{3}}^{(4)}
\right](w)  \nonu \\
&& + \frac{1}{(z-w)} \, \frac{1}{ (c+9) (3 c-2) (3 c+4) (27 c-46) 
(3 c^2+90 c-265)} \,
\left[ 
\right.
\nonu \\
&& \frac{1}{264}
(1215 c^6+33723 c^5+474705 c^4
 \nonu \\
&& -5268615 c^3+5682720 c^2+18658892 c-21364560) 
\,
\pa^3  W_{  \frac{2}{3}}^{(4)}
 \nonu \\
&&
+\frac{1080}{11} (252 c^3+141 c^2-3017 c+3694) 
\,
J J J  W_{  \frac{2}{3}}^{(4)}
\nonu \\
&&
-\frac{30}{11} (6075 c^4-11664 c^3+56673 c^2-338112 c+406316) 
\,
J T W_{  \frac{2}{3}}^{(4)}
\nonu \\
&&
-\frac{90}{11} (108 c^4-1809 c^3-23475 c^2+85134 c-71288)
\, J G^{+}  W_{  -\frac{1}{3}}^{(\frac{9}{2})}
\nonu \\
&&
+\frac{90}{11} (3456 c^4-6345 c^3-29889 c^2+71576 c-55076) 
\,
G^{-} G^{+}  W_{  \frac{2}{3}}^{(4)}
\nonu \\
&&
+\frac{1}{44}
(37665 c^5+192834 c^4-4184127 c^3+7278324 c^2+7100772 c-9611168) 
 \,
G^{+} \pa^2  W_{  -\frac{1}{3}}^{(\frac{7}{2})}
\nonu \\
&&
+\frac{1}{22} 
(21384 c^5+791397 c^4+367884 c^3-6884313 c^2+812852 c+4057516)  \,
\pa G^{+} \pa  W_{  -\frac{1}{3}}^{(\frac{7}{2})} 
\nonu \\
&&
+\frac{1}{4}
(1863 c^5+15642 c^4-364209 c^3+973728 c^2+907396 c-2389120) 
 \,
\pa^2 G^{+} W_{  -\frac{1}{3}}^{(\frac{7}{2})} 
\nonu \\
&& 
+\frac{3}{44} (567 c^5-180 c^4+367575 c^3-1038530 c^2+86068 c-2907960) 
 \,
\pa J \pa  W_{  \frac{2}{3}}^{(4)}
\nonu \\
&&
+\frac{3}{11} 
(189 c^5+9099 c^4+367767 c^3-801239 c^2-1350036 c+1509340) 
 \,
\pa^2 J  W_{  \frac{2}{3}}^{(4)}
\nonu \\
&&
-5 (3 c+4) (27 c-46) (81 c^2-183 c-44) 
 \,
T G^{+}  W_{  -\frac{1}{3}}^{(\frac{7}{2})} 
\nonu \\
&&
-18 (99 c^4+2487 c^3+1256 c^2-28072 c+35600) 
 \,
J J \pa W_{  \frac{2}{3}}^{(4)}
\nonu \\
&&
- 360 (252 c^3+141 c^2-3017 c+3694) 
 \,
J J G^{+} W_{  -\frac{1}{3}}^{(\frac{7}{2})}
\nonu \\
&&
- 6 (594 c^4+4707 c^3-17979 c^2-95292 c+154940) 
 \,
\pa G^{+} J  W_{  -\frac{1}{3}}^{(\frac{7}{2})}
\nonu \\
&&
-3 (1053 c^4-1791 c^3-17208 c^2+150256 c+17000) 
 \,
\pa J G^{+}  W_{  -\frac{1}{3}}^{(\frac{7}{2})}
\nonu \\
&&
-\frac{3}{11} (17388 c^4+124749 c^3-15033 c^2+942306 c-2281720) 
 \,
J G^{+} \pa W_{  -\frac{1}{3}}^{(\frac{7}{2})}
\nonu \\
&&
-\frac{9}{11} (6381 c^4+49758 c^3-108251 c^2+554592 c-920020) 
 \,
\pa J J  W_{  \frac{2}{3}}^{(4)}
\nonu \\
&&
+\frac{3}{11} 
(2187 c^5+39042 c^4-104463 c^3-321174 c^2+1301856 c-1029808)  \,
G^{+} \pa  W_{  -\frac{1}{3}}^{(\frac{9}{2})}
\nonu \\
&&
+\frac{3}{11} (1782 c^5+927 c^4-264858 c^3+764781 c^2-869644 c+694052) 
\,
\pa G^{+}  W_{  -\frac{1}{3}}^{(\frac{9}{2})}
\nonu \\
&&
-\frac{3}{44} 
(4239 c^5+86526 c^4-127797 c^3-633936 c^2+3066788 c-4363360)  \,
J \pa^2   W_{  \frac{2}{3}}^{(4)} 
\nonu \\
&&
-\frac{1}{44}
(27945 c^5+465156 c^4-1091367 c^3+47394 c^2-6488788 c+9919800)  \,
T \pa  W_{  \frac{2}{3}}^{(4)}
\label{j1ope} \\
&&
\left.
-\frac{1}{44}
(35235 c^5+51786 c^4-1096317 c^3+547344 c^2+8778932 c-10153920) 
\,
\pa T W_{  \frac{2}{3}}^{(4)}
\right](w) + \cdots.
\nonu 
\eea
}
In the large $c$ limit,
all the nonlinear terms in Appendix (\ref{j1ope}) disappear.
One can also analyze the $\frac{1}{c}, \cdots, \frac{1}{c^3}$-terms. 
One can easily observe that from the above four OPEs,
the remaining ten OPEs in the component approach by ${\cal N}=2$ 
supersymmetry can be determined
even if one does not know the ${\cal N}=2$ superspace results in 
section $7$. 
That is, one can have the ${\cal N}=2$ superspace generalization
with fixed coefficients  
from the above four OPEs along the line of \cite{AK1509}. 

\subsection{The component OPEs  in the OPE  
${\bf W}_{  \frac{1}{3}}^{(\frac{7}{2})}(Z_1)
\, {\bf W}_{ - \frac{1}{3}}^{(\frac{7}{2})}(Z_2)$ }

The $4$ component OPEs corresponding to Appendix (\ref{34OPE})
can be summarized by
{\small
\bea
&& W_{  \frac{1}{3}}^{(\frac{7}{2})}(z) \, W_{  -\frac{1}{3}}^{(\frac{7}{2})}(w) = 
\frac{1}{(z-w)^7} \, \frac{2c}{7} + \frac{1}{(z-w)^6} \, \frac{2}{7} \, J(w)
\nonu \\
&& +\frac{1}{(z-w)^5} \, \left[ 
-\frac{20 }{7 (c-1)}\,
J J
+\frac{2(21 c-1) }{21 (c-1)}\,
T
+\frac{1}{7} \, \pa J
\right](w) \nonu \\
&& + \frac{1}{(z-w)^4} \, \frac{1}{(c-1) (c+6) (2 c-3)} \, \left[ 
-\frac{2}{7} (c^2-63 c+2)\,
G^{-} G^{+} 
+\frac{4}{21} (3 c+5) (7 c+18)  \,
 J T
\right. \nonu \\
&&  -\frac{20 }{7} (c+6) (2c-3) \,
\pa J J
-\frac{4}{21} (31 c+134) \,
J J J
+\frac{2}{21} 
(21 c^3+95 c^2-288 c+12)\,
\pa T
\nonu \\
&& \left. +\frac{2}{21} (c^3+3 c^2+18 c+8) 
\,
\pa^2 J 
\right](w) \nonu \\
&&+ \frac{1}{(z-w)^3} \, 
\frac{1}{ (c-1) (c+1) (c+6) (2 c-3) (5 c-9)}
\,
\left[ 
\frac{4}{21} (796 c^2-309 c+248)
\, J J J J 
\right. 
\nonu \\
&&
+\frac{4}{7} (29 c^3-1720 c^2+1945 c+86) 
 \,
J G^{-} G^{+}
-\frac{4}{21} (1113 c^3-230 c^2+59 c-402) 
\,
J J T
\nonu \\
&& 
+\frac{2}{21} (24 c^4-1401 c^3+1692 c^2-473 c+18) 
\, \pa G^{-} G^{+}
\nonu \\
&& +\frac{2}{21} (39 c^4-2358 c^3+2451 c^2+1204 c-36) 
\,
\pa G^{+} G^{-}
\nonu \\
&& 
+\frac{2}{21} (105 c^4+274 c^3+5065 c^2-6996 c-1068) 
 \,
\pa T J
\nonu \\
&& +\frac{2}{63} (2331 c^4+162 c^3-15457 c^2+15540 c-36) 
 \,
T T 
\nonu \\
&&
-\frac{1}{21} (179 c^4+1544 c^3+555 c^2-9834 c+9016) 
\,
\pa^2 J J
\nonu \\
&& -\frac{1}{14}
(121 c^4-492 c^3-1435 c^2+5674 c-4328) 
 \,
\pa J \pa J
\nonu \\
&&
+\frac{2}{21} (c+1)(5c-9)(3 c+5) (7 c+18) 
\,
\pa J T 
-\frac{2}{7} (c+1)(5c-9)(31 c+134) 
 \,
\pa J J J
\nonu \\
&&
+\frac{1}{21}
(63 c^5-255 c^4+177 c^3+1835 c^2-2304 c-36) 
\,
\pa^2 T
\nonu \\
&& \left.
+\frac{1}{126}
(15 c^5-177 c^4-341 c^3-10307 c^2+13410 c+1920) 
\,
\pa^3  J
\right](w) \nonu \\
&& + \frac{1}{(z-w)^2} \, 
\frac{(c+18)(4c-9)(7c-15)}{d(c)} \,
\left[ 
\right. \nonu \\
&& 
-\frac{16}{63} (861 c^4+8206 c^3-1621 c^2+5986 c+1488) 
\,
J J J T
 \nonu \\
&& +\frac{4}{21} (848 c^3+8095 c^2-5686 c+6768) 
\,
J J J J J
\nonu \\
&&
+\frac{8}{7} (29 c^4-1551 c^3-11014 c^2+16064 c-4928) 
 \,
J J G^{-} G^{+}
\nonu \\
&&
+\frac{2}{63} (2331 c^5+22446 c^4-5449 c^3-180016 c^2+300540 c-98352) 
\,
J T T 
\nonu \\
&&
+\frac{2}{21} (57 c^5-3574 c^4+2061 c^3+12544 c^2-15276 c+48) 
\,
G^{-} T G^{+}
\nonu \\
&&
+\frac{2}{21} (117 c^5-5490 c^4-78019 c^3+173344 c^2-96076 c-4176) 
 \,
\pa G^{-} J G^{+}
\nonu \\
&&
+\frac{1}{21} 
(27 c^6-1407 c^5-10671 c^4+123947 c^3-208284 c^2+102604 c+3984) \,
\pa^2 G^{-} G^{+}
\nonu \\
&&
+\frac{2}{63}
(2331 c^6+23643 c^5-80503 c^4-136735 c^3+563964 c^2-419148 c+1008)  \,
\pa T T
\nonu \\
&&
-\frac{2}{21} (1113 c^5+11074 c^4-38259 c^3-60376 c^2+90948 c-19920) 
\,
\pa T J J 
\nonu \\
&&
+\frac{1}{21}
(36 c^6-1953 c^5-12134 c^4+102075 c^3-261296 c^2+169868 c+8784) \,
\pa^2 G^{+} G^{-}
\nonu \\
&&
-\frac{4}{21} (c-2)(c+12) (1113 c^3-230 c^2+59 c-402) 
\,
\pa J J T
\nonu \\
&&
+\frac{8}{21} (c-2)(c+12) (796 c^2-309 c+248) 
\,
\pa J J J J 
\nonu \\
&& 
+\frac{2}{7} (c-2)(c+12) (29 c^3-1720 c^2+1945 c+86) 
\,
\pa J G^{-} G^{+}
\nonu \\
&&
+\frac{1}{126} (3 c^7-48 c^6-815 c^5+1796 c^4-20932 c^3+62864 c^2-50480 c
+1632) 
\,
\pa^4 J
\nonu \\
&&
-\frac{1}{21} (c-2)(c+12) (121 c^4-521 c^3+285 c^2+3729 c-4414) 
 \,
\pa^2 J \pa J
\nonu \\
&&
-\frac{1}{21} (c+12) (289 c^4+1266 c^3+1153 c^2-11996 c+14228) 
 \,
\pa^2 J J J
\nonu \\
&&
-\frac{1}{14} (331 c^5+3534 c^4-6441 c^3-47624 c^2+95468 c-35248) 
 \,
\pa J \pa J J
\nonu \\
&&
-\frac{2}{21} (57 c^5-3090 c^4-17687 c^3+191552 c^2-178844 c-8208) 
\,
\pa G^{+} J G^{-}
\nonu \\
&&
+\frac{1}{21}
(105 c^6+1321 c^5+5699 c^4+22475 c^3-178140 c^2+209564 c-45744) \,
\pa T \pa J
\nonu \\
&&
+\frac{1}{21} 
(63 c^6+363 c^5-3255 c^4+15645 c^3+32492 c^2-83548 c+44400) 
\,
\pa^2 T J
\nonu \\
&&
-\frac{4}{63} 
(6 c^6-315 c^5-2960 c^4-10347 c^3+8216 c^2+5268 c+432) 
\,
\pa G^{-} \pa G^{+}
\nonu \\
&&
+\frac{2}{63}
(105 c^6+1357 c^5+4155 c^4+1767 c^3-37132 c^2-17300 c+90768) 
\,
\pa^2 J T
\nonu \\
&&
-\frac{1}{126}
(237 c^6+1527 c^5-13069 c^4+51273 c^3+10180 c^2+422196 c-459024) \,
\pa^3 J J
\nonu \\
&&\left.
+\frac{2}{189}
(63 c^7+378 c^6-3957 c^5+2884 c^4+9716 c^3-44192 c^2+34656 c+1152) 
\,
\pa^3 T
\right](w) \nonu \\
&& + \frac{1}{(z-w)} \, 
\frac{(c+18)(4c-9)}{d(c)} \,
\left[ 
\frac{2}{63} 
(861 c^6-29391 c^5-1133657 c^4 \right. \nonu \\
&& +4368507 c^3-4375708 c^2+974908 c+362320) 
\,
\pa G^{-} \pa G^{+} J
\nonu \\
&&
+\frac{2}{63} (1323 c^6-72861 c^5-243063 c^4+1554153 c^3-1902276 c^2+898804 c-203600) 
\nonu \\
&& \times
\pa G^{-} \pa J G^{+}
\nonu \\
&&
+\frac{1}{189}
(315 c^7-17724 c^6-18570 c^5+3771528 c^4\nonu \\
&& -12915513 c^3+12105392 c^2-2283748 c-1069680) 
\,
\pa^2 G^{-} \pa G^{+}
\nonu \\
&&
+\frac{4}{63} 
(441 c^6-23181 c^5-168783 c^4+698013 c^3-631674 c^2+484624 c-328880) 
\,
\pa^2 G^{-} J G^{+}
\nonu \\
&&
+\frac{1}{189}
(504 c^7-27885 c^6-122169 c^5+2445849 c^4\nonu \\
&& -7192203 c^3+7285004 c^2-3313020 c+793840) 
\,
\pa^3 G^{-} G^{+}
\nonu \\
&&
+\frac{2}{189}
(3 c-8) (4095 c^6+42183 c^5-88431 c^4-298851 c^3+690304 c^2-294404 c-42960) 
\,
\pa T \pa T
\nonu \\
&&
+\frac{2}{63} 
(16317 c^6+108045 c^5+496175 c^4-5061985 c^3+10644348 c^2-8247300 c+1632720) 
\,
\pa T J T
\nonu \\
&&
-\frac{8}{63} (6027 c^5+39571 c^4+160094 c^3-724244 c^2+223632 c+122640) 
\,
\pa T J J J
\nonu \\
&&
-\frac{2}{21} (7c-15) (57 c^5-3574 c^4+2061 c^3+12544 c^2-15276 c+48) 
\,
\pa T G^{-} G^{+}
\nonu \\
&&
-\frac{1}{63} (46935 c^6+344391 c^5-1960539 c^4-3602811 c^3+19759956 c^2-16297772 c+918640) 
\nonu \\
&& \times 
\pa T \pa J J 
\nonu \\
&&
+\frac{1}{189} (29295 c^7+2007 c^6-1069131 c^5+3951597 c^4-3487692 c^3-4812940 c^2+7455824 c
\nonu \\
&& -911040) \, 
\pa^2 T T 
\nonu \\
&&
-\frac{2}{63} (6993 c^6-1341 c^5-224067 c^4+1719387 c^3-4142880 c^2+2882708 c-40240) 
\,
\pa^2 T J J
\nonu \\
&&
+\frac{1}{189} (630 c^7-30873 c^6-379989 c^5+4992837 c^4
\nonu \\
&& -14666997 c^3+15196220 c^2-4653508 c-1112880) 
\,
\pa^2 G^{+} \pa G^{-}
\nonu \\
&&
+\frac{1}{189} (630 c^7-35085 c^6-162885 c^5+1152489 c^4
\nonu \\
&& -567069 c^3-3787276 c^2+5366316 c-2322800) 
\,
\pa^3 G^{+} G^{-}
\nonu \\
&&
-\frac{8}{21} (7c-15) (861 c^4+8206 c^3-1621 c^2+5986 c+1488) 
\,
\pa J J J T
\nonu \\
&&
+\frac{10}{21} (7c-15) (848 c^3+8095 c^2-5686 c+6768) 
\,
\pa J J J J J 
\nonu \\
&&
+\frac{8}{7} (7c-15) (29 c^4-1551 c^3-11014 c^2+16064 c-4928) 
\,
\pa J J G^{-} G^{+}
\nonu \\
&&
+\frac{1}{63} (7c-15)
(2331 c^5+22446 c^4-5449 c^3-180016 c^2+300540 c-98352) 
\,
\pa J T T
\nonu \\
&&
-\frac{1}{126} 
(56385 c^6-11631 c^5-2686773 c^4+11981883 c^3-18743028 c^2+6918124 c+5071120) 
\nonu \\
&& \times 
\pa J \pa J T
\nonu \\
&&
+\frac{2}{7} (5971 c^5+25134 c^4-229189 c^3+390104 c^2-140100 c-33680) 
\,
\pa J \pa J J J 
\nonu \\
&&
-\frac{1}{63} (27825 c^6+365797 c^5-1149429 c^4-730641 c^3+2394996 c^2+1405132 c-1870960) 
\nonu \\
&& \times
\pa^2 J J T
\nonu \\
&&
+\frac{4}{63} (9947 c^5+129379 c^4-458014 c^3+295024 c^2-328216 c+644080) 
\,
\pa^2 J J J J 
\nonu \\
&&
+\frac{1}{63} (1029 c^6-37503 c^5-1446465 c^4+8032635 c^3-13411404 c^2+6475228 c+780880) 
\nonu \\
&& \times 
\pa^2 J G^{-} G^{+}
\nonu \\
&&
+\frac{1}{7560} (210 c^8-2340 c^7-17013 c^6+357851 c^5-2719733 c^4
\nonu \\
&& +4848833 c^3+1172620 c^2-4953308 c+847920) 
\,
\pa^5 J
\nonu \\
&&
-\frac{1}{42} 
(7c-15) (641 c^5+7494 c^4+7817 c^3-4432 c^2-18044 c+62064) 
\,
\pa^2 J \pa J J
\nonu \\
&&
+\frac{1}{42} (7c-15)
(63 c^6+360 c^5-2841 c^4+1042 c^3+46940 c^2-31208 c-26976) 
\,
\pa^2 T \pa J
\nonu \\
&&
-\frac{1 }{84} (7c-15)
(373 c^5+2218 c^4-23023 c^3-53032 c^2+256404 c-221520) 
\,
\pa J \pa J \pa J
\nonu \\
&&
-\frac{16}{21} (1652 c^4-98177 c^3+263487 c^2-101002 c-48320) 
\,
J J J G^{-} G^{+}
\nonu \\
&&
+\frac{32}{21} (588 c^5-36292 c^4+161817 c^3-243347 c^2+127114 c-6560) 
\,
J G^{-} T G^{+}
\nonu \\
&&
-\frac{8}{21} (273 c^5-14301 c^4-6314 c^3+108034 c^2-27092 c-60800) 
\,
\pa G^{-} J J G^{+}
\nonu \\
&&
-\frac{8}{21} (882 c^5-48177 c^4-167813 c^3+941008 c^2-853460 c+160960) 
\,
\pa G^{+} J J G^{-}
\nonu \\
&&
+\frac{4}{63} (131376 c^4-202711 c^3-302219 c^2+294074 c-101520) 
\,
J J J J T
\nonu \\
&&
-\frac{128}{63} (20 c-13) (98 c^2-129 c-194) 
\,
J J J J J J 
\nonu \\
&&
-\frac{8}{63} (45801 c^5-47772 c^4-440083 c^3
+910558 c^2-363304 c-152400) 
\,
J J T T 
\nonu \\
&&
-\frac{2}{63} 
(1953 c^6-119295 c^5+579231 c^4-1030281 c^3+803548 c^2-255716 c+720) 
\,
\pa G^{-} T G^{+}
\nonu \\
&&
+\frac{8}{63}
(1575 c^6-98457 c^5+391671 c^4-429801 c^3-40864 c^2+216356 c-720) 
\,
\pa G^{+} G^{-} T
\nonu \\
&&
-\frac{2}{63} 
(693 c^6-46311 c^5+336423 c^4+387867 c^3-3181020 c^2+2718748 c-112880) 
\,
\pa G^{+} \pa J G^{-}
\nonu \\
&&
-\frac{8}{63} 
(63 c^6-4890 c^5+67728 c^4-912513 c^3+2593440 c^2-2163148 c+543200) 
\,
\pa^2 G^{+} J G^{-}
\nonu \\
&&
-\frac{1}{63} (1407 c^6+1185 c^5-89605 c^4-893675 c^3+1742348 c^2+2855460 c-3459120) 
\, \pa^3 J J J 
\nonu \\
&&
+\frac{1}{63} (735 c^7+7378 c^6+31896 c^5+510630 c^4-3822475 c^3+7366072 c^2-4269836 c-146000) 
\nonu \\
&& \times
\pa^2 J \pa T
\nonu \\
&&
+\frac{2}{189}
(441 c^7+1386 c^6-28674 c^5+904366 c^4 \nonu \\
&& -3121207 c^3+2904460 c^2+116108 c-945520) 
\,
\pa^3 T J
\nonu \\
&&
-\frac{1}{378}
(2730 c^7-1233 c^6-57681 c^5+2034861 c^4
\nonu \\
&& -10003101 c^3+16987804 c^2-9551780 c+125520) 
\,
\pa^2 J \pa^2 J
\nonu \\
&&
-\frac{1}{378} (882 c^7-11187 c^6-37095 c^5+1265499 c^4
\nonu \\
&&
-5601471 c^3+11901904 c^2-14259828 c+6980816) 
\,
\pa^4 J J
\nonu \\
&&
+\frac{1}{378} 
(2205 c^7-8799 c^6-198729 c^5-816925 c^4
\nonu \\
&&
+8304140 c^3-20430036 c^2+23033424 c-10561920) 
\,
\pa^3 J T 
\nonu \\
&&
-\frac{1}{756}
(7245 c^7-47313 c^6-1512987 c^5+5063061 c^4 \nonu \\
&& +17619294 c^3-79973996 c^2+86415256 c-26704800) 
\,
\pa^3 J \pa J
\nonu \\
&&
+\frac{2}{567}
(382725 c^6-175347 c^5-9121545 c^4+29407735 c^3-33144380 c^2+11888652 c+530640) 
\nonu \\
&& \times
T T T
\nonu \\
&&
+\frac{1}{2268}
(1890 c^8-90 c^7-293955 c^6+1336737 c^5-3168903 c^4
\nonu \\
&& \left. +2937673 c^3+2876452 c^2-3876924 c+682800) 
\,
\pa^4 T
\right](w)+ \cdots,
\nonu \\
&& W_{  \frac{1}{3}}^{(\frac{7}{2})}(z) \, W_{  \frac{2}{3}}^{(4)}(w) = 
\frac{1}{(z-w)^6} \, \frac{22}{7} G^{+}(w) +\frac{1}{(z-w)^5}
\, \frac{22}{21} \, \pa G^{+}(w) \nonu \\
&& + 
\frac{1}{(z-w)^4}  \, \frac{1}{(c-1) (c+6) (2 c-3)} \, \left[ 
\frac{22}{21} (45 c^2-41 c+6) 
\,
T G^{+}
+\frac{11}{7} (c^2-53 c+62) 
\,
\pa J G^{+}
\right. \nonu \\
&& \left.
-\frac{22}{21} (c^2-97 c+6) 
\,
\pa G^{+} J
-\frac{44 (11 c+14) }{7 }\,
J J G^{+} 
+\frac{11}{21} c (c^2-31 c+90) 
\,
\pa^2 G^{+}
\right](w) \nonu \\
&& + \frac{1}{(z-w)^3} \, \frac{1}{(c-1) (c+1) (c+6) (2 c-3) (5 c-9)} \, 
\left[ 
\frac{176}{21} (c^2-54 c-22) 
\,
J J J G^{+}
\right. \nonu \\
&&
-\frac{44}{21} (3 c^3-180 c^2+163 c-6) 
\,
J T G^{+}
+\frac{11}{21} (c+6) (c^3+111 c^2-149 c-83) 
\,
\pa G^{+} \pa J
\nonu \\
&&
-\frac{22}{7} (109 c^3+92 c^2-275 c-346) 
\,
\pa J J G^{+}
-\frac{88}{21} (28 c^3-5 c^2-203 c-60) 
\,
\pa G^{+} J J 
\nonu \\
&&
+\frac{22}{21} (111 c^4-101 c^3-277 c^2+257 c-30) 
\,
\pa T G^{+}
\nonu \\
&& +\frac{22}{63} (231 c^4-745 c^3+115 c^2+333 c-54) 
\,
\pa G^{+} T
\nonu \\
&&
+\frac{11}{21} (7 c^4-353 c^3+603 c^2+557 c-934) 
\,
\pa^2 J G^{+}
\nonu \\
&& -\frac{22}{21} (2 c^4-87 c^3+338 c^2-535 c-258) 
\,
\pa^2 G^{+} J
\nonu \\
&&
\left.
+\frac{11}{63} (c-1) (3 c^4-69 c^3+167 c^2+81 c+18) 
\,
\pa^3 G^{+}
\right](w) 
\nonu \\
&& + \frac{1}{(z-w)^2}  \, \frac{(c+18)(4c-9)(7c-15)}{d(c)} \, 
\left[ 
\frac{11}{756} 
(6 c^7-200 c^6-6117 c^5+17490 c^4 \right. \nonu \\
&& +21165 c^3-78976 c^2+16948 c+26544) 
\,
\pa^4 G^{+}
 \nonu \\
&&
+\frac{220}{7} (64 c^3+213 c^2-402 c+208) 
\,
J J J J G^{+}
\nonu \\
&&
-\frac{178}{21} (327 c^4+392 c^3-3042 c^2+3952 c-1224) 
\,
J J T G^{+}
\nonu \\
&&
+\frac{22}{63} (2673 c^5-2478 c^4-21523 c^3+36848 c^2-2796 c-11664) 
\,
T T G^{+}
\nonu \\
&&
-\frac{22}{21} (c+12) (3 c^4-194 c^3+1395 c^2-2052 c+668) 
\,
\pa T J G^{+}
\nonu \\
&&
-\frac{178}{63} (15 c^5-1441 c^4+276 c^3+2794 c^2+756 c-2880) 
\,
\pa G^{+} J T
\nonu \\
&&
+\frac{176}{63} (22 c^4-2063 c^3-3682 c^2+9812 c-3144) 
\,
\pa G^{+} J J J 
\nonu \\
&&
+\frac{22}{21} (37 c^5-2402 c^4+15797 c^3-24392 c^2+4100 c+7920) 
\,
\pa G^{+} G^{-} G^{+}
\nonu \\
&&
+\frac{11}{63} (4 c^6-47 c^5+3354 c^4+15985 c^3-76224 c^2+57572 c+20976) 
\,
\pa^2 J \pa G^{+}
\nonu \\
&&
+\frac{11}{126} (12 c^6-439 c^5-6574 c^4+19665 c^3+39208 c^2-128076 c+74544)
\,
\pa^3 J G^{+}
\nonu \\
&&
-\frac{88}{7} (6 c^4-309 c^3+619 c^2-174 c-712) 
\,
\pa J J J G^{+}
\nonu \\
&&
-\frac{11}{14} (129 c^5+514 c^4-7443 c^3+17208 c^2-9532 c-3856) 
\,
\pa J \pa J G^{+}
\nonu \\
&&
-\frac{88}{21} (28 c^5+237 c^4+805 c^3-6540 c^2+8852 c-1392) 
\,
\pa G^{+} \pa J J
\nonu \\
&&
-\frac{44}{21} (14 c^5-367 c^4+1179 c^3+11386 c^2-20800 c+8688) 
\,
\pa^2 G^{+} J J 
\nonu \\
&&
+\frac{44}{21} (27 c^5-1432 c^4+4605 c^3-3780 c^2+1308 c-2928) 
\,
\pa J T G^{+}
\nonu \\
&&
-\frac{11}{21} (195 c^5+3218 c^4-1189 c^3-17216 c^2-11924 c+48336) 
\,
\pa^2 J J G^{+}
\nonu \\
&&
+\frac{11}{21} (66 c^6+103 c^5-1720 c^4+1127 c^3+5872 c^2-9716 c+4368)
\,
\pa^2 T G^{+}
\nonu \\
&&
-\frac{11}{42} (c^6+80 c^5-1659 c^4+17354 c^3-53820 c^2+65304 c-22560) 
\pa^2 G^{+} \pa J
\nonu \\
&&
+\frac{44}{63} (3 c-8) (19 c^5+174 c^4+851 c^3-3196 c^2+1212 c+1440) 
\,
\pa T \pa G^{+}
\nonu \\
&&
-\frac{11}{63} (3 c^6-44 c^5-4425 c^4+12714 c^3+22228 c^2-59832 c+35616) 
\,
\pa^3 G^{+} J
\nonu \\
&&
\left.
+\frac{11}{63} 
(117 c^6-3330 c^5+765 c^4+59088 c^3-93340 c^2-4272 c+47232) 
\,
\pa^2 G^{+} T
\right](w) 
\nonu \\
&& +\frac{1}{(z-w)} \, 
\frac{(c+18)(4c-9)}{d(c)}
\, \left[ 
\frac{11}{1890} (15 c^8-415 c^7-15392 c^6+108617 c^5-184721 c^4
\right. \nonu \\ 
&& -157044 c^3+611676 c^2-324976 c-45360) 
\,
\pa^5 G^{+}
 \nonu \\
&&
-\frac{88}{21} (2259 c^5-366 c^4-31289 c^3+84184 c^2-82548 c+21920) 
\,
\pa T J J G^{+}
\nonu \\
&&
+\frac{88}{63} (4617 c^6-10515 c^5-47653 c^4+182143 c^3-173520 c^2-1352 c+44160) 
\,
\pa T T G^{+}
\nonu \\
&&
-\frac{22}{63} 
(417 c^6-40955 c^5+93387 c^4+233663 c^3-773980 c^2+545868 c+11280) 
\,
\pa T \pa G^{+} J
\nonu \\
&&
-\frac{44}{21} 
(3 c^6-230 c^5+3279 c^4+13256 c^3-77944 c^2+97456 c-30720) 
\,
\pa^2 T J G^{+} 
\nonu \\
&&
-\frac{178}{63} (2388 c^5-6935 c^4-38285 c^3+110090 c^2-91108 c+16680) 
\,
\pa G^{+} J J T
\nonu \\
&&
+\frac{44}{63} (6952 c^4+1033 c^3-194383 c^2+256178 c-97680) 
\,
\pa G^{+} J J J J
\nonu \\
&&
+\frac{22}{189} (19683 c^6-117993 c^5+134441 c^4+237549 c^3-428372 c^2-32988 c+174960) 
\,
\pa G^{+} T T
\nonu \\
&&
+\frac{88}{21} (94 c^5-11007 c^4-9458 c^3+132443 c^2-165102 c+55240) 
\,
\pa G^{+} \pa J J J 
\nonu \\
&&
-\frac{22}{63} (333 c^6-15147 c^5+103899 c^4-530989 c^3+1108196 c^2-981932 c+271920) 
\,
\pa^2 G^{+} J T
\nonu \\
&&
+\frac{88}{63} (122 c^5-5251 c^4+29266 c^3-128381 c^2+141234 c-65280) 
\,
\pa^2 G^{+} J J J
\nonu \\
&&
+\frac{22}{21} (127 c^6-8441 c^5+66769 c^4-157519 c^3+111372 c^2+42932 c-57360) 
\,
\pa^2 G^{+} G^{-} G^{+}
\nonu \\
&&
+\frac{11}{126} (5 c^7-1342 c^6+22836 c^5-165546 c^4+658371 c^3-1243584 c^2+980940 c-168240) 
\nonu \\
&& \times 
\pa^2 G^{+} \pa^2 J
\nonu \\
&&
+\frac{176}{21} (3331 c^4+5049 c^3-42859 c^2+50564 c-23560) 
\,
\pa J J J J G^{+}
\nonu \\
&&
-\frac{88}{21} (4539 c^5-2796 c^4-54919 c^3+138884 c^2-132048 c+44000) 
\,
\pa J J T G^{+}
\nonu \\
&&
+\frac{22}{189} (15 c^7-297 c^6-1738 c^5+46753 c^4-212513 c^3+345900 c^2-75960 c-160800) 
\,
\pa^3 J \pa G^{+}
\nonu \\
&&
+\frac{11}{63}
(9 c^7-413 c^6-266 c^5+6807 c^4+62339 c^3-327008 c^2+479288 c-222176) 
\,
\pa^4 J G^{+}
\nonu \\
&&
-\frac{88}{21} (c-2)(c+12)(7c-15)(16 c^3-86 c^2+115 c-25) 
\,
\pa^2 J \pa J G^{+}
\nonu \\
&&
+\frac{44}{63} (c+12) (33 c^5+4613 c^4-16173 c^3+12023 c^2+4524 c-3700) 
\,
\pa G^{+} \pa J T
\nonu \\
&&
-\frac{22}{7} (179 c^5-9302 c^4+17627 c^3+24888 c^2-56652 c+17200) 
\,
\pa J \pa J J G^{+}
\nonu \\
&&
-\frac{176}{21} (80 c^3-4329 c^2+4962 c-2608) 
\,
J J J J J G^{+}
\nonu \\
&&
+\frac{44}{21} 
(93 c^6-5021 c^5+19611 c^4-37351 c^3+72428 c^2-102480 c+56320) 
\,
\pa T \pa J G^{+}
\nonu \\
&&
-\frac{176}{21} (28 c^5-1577 c^4-2573 c^3+14678 c^2-12966 c+1880) 
\,
\pa G^{+} J G^{-} G^{+}
\nonu \\
&&
-\frac{11}{21} (395 c^6-11573 c^5+66909 c^4+162741 c^3-987916 c^2+1181724 c-454960) 
\,
\pa^2 G^{+} \pa J J
\nonu \\
&&
-\frac{44}{21} (125 c^5-6308 c^4+27853 c^3-30678 c^2-19192 c+44320) 
\,
\pa^2 J J J G^{+}
\nonu \\
&&
-\frac{11}{42} 
(931 c^6-1297 c^5+26569 c^4-166259 c^3+481188 c^2-772652 c+508400) 
\,
\pa G^{+} \pa J \pa J
\nonu \\
&&
+\frac{ 352}{63} (171 c^4-9956 c^3+25436 c^2-24506 c+6360) 
\,
J J J T G^{+}
\nonu \\
&&
-\frac{88}{63} (243 c^5-15138 c^4+61983 c^3-85808 c^2+35980 c+1680) 
\,
J T T G^{+}
\nonu \\
&&
-\frac{22}{63} 
(279 c^6-19989 c^5+236741 c^4-716591 c^3+729652 c^2-62572 c-180240) 
\,
\pa G^{-} \pa G^{+} G^{+}
\nonu \\
&&
+\frac{22}{63} 
(237 c^7-734 c^6+9760 c^5-75014 c^4+214507 c^3-247960 c^2+58004 c+57360) 
\,
\pa^2 T \pa G^{+}
\nonu \\
&&
-\frac{22}{63} 
(117 c^6-2401 c^5+17047 c^4+38493 c^3-194740 c^2+245924 c-81520) 
\,
\pa^3 G^{+} J J 
\nonu \\
&&
+\frac{88}{63} 
(135 c^6-6998 c^5+36339 c^4-64014 c^3+42134 c^2-26336 c+31680) 
\,
\pa^2 J T G^{+} 
\nonu \\
&&
-\frac{11}{63} 
(1409 c^6+15869 c^5+32499 c^4-776369 c^3+1943940 c^2-1361428 c-109680) 
\,
\pa^2 J \pa G^{+} J
\nonu \\
&&
-\frac{22}{63} 
(453 c^6+550 c^5-13635 c^4+51660 c^3-302008 c^2+744120 c-553440) 
\,
\pa^3 J J G^{+}
\nonu \\
&&
+\frac{11}{63} 
(405 c^7-12912 c^6+73802 c^5-103624 c^4-191411 c^3+638808 c^2-542108 c+103920) 
\nonu \\
&& \times
\pa^2 G^{+} \pa T
\nonu \\
&&
-\frac{11}{126} 
(9 c^7+452 c^6-21946 c^5+93252 c^4+34389 c^3-395280 c^2+163044 c+163440) 
\,
\pa^3 G^{+} \pa J
\nonu \\
&&
+\frac{11}{189} 
(918 c^7+207 c^6-54177 c^5+284165 c^4-485793 c^3+66668 c^2+546412 c-372720) 
\,
\pa^3 T G^{+}
\nonu \\
&&
+\frac{11}{189} 
(495 c^7-11064 c^6-11362 c^5+359328 c^4-698193 c^3-255568 c^2+1389484 c-744240) 
\nonu \\
&& \times
\pa^3 G^{+} T
\nonu \\
&&
-\frac{11}{378} 
(24 c^7-1115 c^6-26031 c^5+210207 c^4-137781 c^3-980188 c^2+1750324 c-850320) 
\nonu \\
&& \left. \times \pa^4 G^{+} J
       \right](w)
\nonu \\
&& + \cdots,
\nonu \\
&& W_{  \frac{1}{3}}^{(\frac{7}{2})}(z) \, W_{  -\frac{4}{3}}^{(4)}(w) = 
-\frac{1}{(z-w)^6}\, \frac{20}{7} G^{-}(w)
+ \frac{1}{(z-w)^5}\, \left[ 
-\frac{40
 }{7 (c-1)}\,
J G^{-}
-\frac{20 (c+1) 
}{21 (c-1)}\,
\pa G^{-}
\right](w) \nonu \\
&& + \frac{1}{(z-w)^4} \, \frac{1}{ (c-1) (c+6) (2 c-3)}\, \left[ 
-\frac{10}{7} (3 c^2+70 c-104) 
\,
\pa J G^{-}
-\frac{20}{21} (5 c^2-70 c+24) 
\,
\pa G^{-} J
\right. 
\nonu \\
&& \left. -\frac{20}{21} (5 c-6) (9 c+2) 
\,
G^{-} T
+\frac{40}{7} (9 c+2) 
\,
J J G^{-} -\frac{5}{21} (2 c^3-19 c^2+118 c-48) 
\,
\pa^2 G^{-}
\right](w) \nonu \\
&& + \frac{1}{(z-w)^3} \, \frac{1}{ (c-1) (c+1) (c+6) (2 c-3) (5 c-9)} \, 
\left[ 
\frac{80}{21} (157 c^2+57 c+98) 
\,
J J J G^{-}
\right.
\nonu \\
&& -\frac{160}{21} (57 c^3-91 c^2+104 c-12) 
\,
J G^{-} T
 +\frac{40}{21} (41 c^3+504 c^2-131 c+66) 
\,
\pa G^{-} J J 
\nonu \\
&&
+\frac{40}{7} (47 c^3-134 c^2-11 c+38) 
\,
\pa J J G^{-}
-\frac{10}{21} (c+6)(19 c^3-124 c^2+257 c-128) 
\,
\pa G^{-} \pa J
\nonu \\
&&
-\frac{20}{63} (231 c^4-328 c^3+1121 c^2-540 c-108) 
\,
\pa G^{-} T
\nonu \\
&& -\frac{20}{21} (111 c^4-104 c^3-679 c^2+676 c+84) 
\,
\pa T G^{-}
\nonu \\
&& -\frac{20}{21} (7 c^4-133 c^3+1041 c^2-277 c+126) 
\,
\pa^2 G^{-} J
\nonu \\
&& -\frac{10}{21} (13 c^4+418 c^3-729 c^2-1006 c+1712) 
\,
\pa^2 J G^{-}
\nonu \\
&&
\left.
-\frac{10}{63} (c+1)(3 c^4-84 c^3+365 c^2-860 c+228) 
\,
\pa^3 G^{-}
\right](w) \nonu \\
&& + \frac{1}{(z-w)^2}\, \frac{(c+18)(4c-9)(7c-15)}{d(c)} \, \left[ 
\right. \nonu \\
&& -\frac{40}{63} (c+12)(57 c^5-156 c^4+720 c^3-1751 c^2+788 c+84) 
\,
\pa G^{-} \pa T
 \nonu \\
&&
-\frac{40}{63} (291 c^5-3488 c^4-3415 c^3+27448 c^2-7452 c-4176) 
\,
\pa G^{-} J T
\nonu \\
&&
+\frac{40}{63} (389 c^4-3003 c^3+2976 c^2+16652 c-4368) 
\,
\pa G^{-} J J J
\nonu \\
&&
+\frac{40}{21} (16 c^5-1065 c^4+9163 c^3-20196 c^2+10588 c+1296) 
\,
\pa G^{-} G^{-} G^{+}
\nonu \\
&&
+\frac{40}{21} (44 c^5+693 c^4+5645 c^3-17696 c^2+8380 c-4656) 
\,
\pa G^{-} \pa J J
\nonu \\
&&
-\frac{10}{63} (117 c^6-799 c^5-1949 c^4+49559 c^3-119228 c^2+47724 c+12240)
\,
\pa^2 G^{-} T  
\nonu \\
&&
+\frac{10}{21} (38 c^5+13 c^4+7023 c^3-10384 c^2-28428 c+9552) 
\,
\pa^2 G^{-} J J 
\nonu \\
&&
-\frac{20}{21} (c+12)(225 c^4-638 c^3+969 c^2-1468 c+308) 
\,
\pa T J G^{-}
\nonu \\
&&
-\frac{20}{21} (33 c^6+52 c^5-1034 c^4-1057 c^3+9216 c^2-7396 c-336) 
\,
\pa^2 T G^{-}
\nonu \\
&&
+\frac{20}{7} (289 c^4+4497 c^3-8548 c^2-1532 c-4880) 
\,
\pa J J J G^{-}
\nonu \\
&&
-\frac{20}{21} (171 c^5+4874 c^4-16741 c^3+15880 c^2-10500 c+7440) 
\,
\pa J G^{-} T
\nonu \\
&&
+\frac{5}{7} (119 c^5-206 c^4-12753 c^3+38952 c^2-21364 c-11056) 
\,
\pa J \pa J G^{-}
\nonu \\
&&
+\frac{10}{21} (169 c^5+2086 c^4-9375 c^3+4496 c^2+6692 c+3312) 
\,
\pa^2 J J G^{-}
\nonu \\
&&
-\frac{5}{21} (c+12)(11 c^5-97 c^4+2225 c^3-5335 c^2+3016 c-1548) 
\,
\pa^2 G^{-} \pa J
\nonu \\
&&
-\frac{40}{21} (632 c^3-533 c^2-3790 c+48) 
\,
J J  J J G^{-}
\nonu \\
&&
+\frac{40}{21} (1077 c^4-627 c^3-6128 c^2+5356 c-3024) 
\,
J J G^{-} T
\nonu \\
&&
-\frac{20}{63} (2673 c^5-2826 c^4-24367 c^3+51008 c^2-23964 c-3024) 
\,
G^{-} T T
\nonu \\
&&
-\frac{10}{63} (9 c^6-145 c^5-2265 c^4+30601 c^3-55972 c^2-1564 c-528) 
\,
\pa^3 G^{-} J
\nonu \\
&&
-\frac{10}{63} (16 c^6+351 c^5-2794 c^4-16877 c^3+90464 c^2-112916 c+36048) \,
\pa^2 J \pa G^{-}
\nonu \\
&&
-\frac{10}{63} (9 c^6+193 c^5+1254 c^4-5078 c^3-24752 c^2+85128 c-54624) 
\,
\pa^3 J G^{-}
\nonu \\
&&
\left.
-\frac{5}{1512} (24 c^7-376 c^6-6065 c^5+46854 c^4-250485 c^3+448368 c^2-165044 c+87312) 
\,
\pa^4 G^{-}
\right](w)\nonu \\
&& + \frac{1}{(z-w)} \, \frac{(c+18)(4c-9)}{d(c)}\, \left[ 
\right. \nonu \\
&&
\frac{40}{21} (343 c^5-22056 c^4+159035 c^3-268298 c^2+52988 c+67120) \,
\pa G^{-} J G^{-} G^{+}
 \nonu \\
&&
-\frac{20}{63} (2013 c^6-26895 c^5+7779 c^4+590171 c^3-1217720 c^2+399084 c+83280) 
\,
\pa G^{-} \pa T J
\nonu \\
&&
-\frac{20}{63} 
(c+12)(1551 c^5-17093 c^4+63725 c^3-87679 c^2+26108 c+11620) 
\,
\pa G^{-} \pa J T
\nonu \\
&&
+\frac{20}{21} (2525 c^5-16444 c^4+82083 c^3-69036 c^2-67316 c-68080) 
\,
\pa G^{-} \pa J J J 
\nonu \\
&&
+\frac{5}{21} (805 c^6+275 c^5+47995 c^4-282227 c^3+459668 c^2-300236 c+189040) 
\,
\pa G^{-} \pa J \pa J
\nonu \\
&&
-\frac{10}{63} (405 c^7-3422 c^6+1990 c^5+246180 c^4-1334775 c^3+2132778 c^2-830868 c-107280) 
\nonu \\
&& \times \pa^2 G^{-} \pa T
\nonu \\
&&
-\frac{40}{63} (576 c^6-10675 c^5+76477 c^4-90019 c^3+57841 c^2-229884 c+157920) 
\,
\pa^2 G^{-} J T
\nonu \\
&&
+\frac{40}{63} (755 c^5-11985 c^4+168910 c^3-124469 c^2-373268 c+318720) 
\,
\pa^2 G^{-} J J J
\nonu \\
&&
+\frac{20}{21} (113 c^6-7842 c^5+87405 c^4-333968 c^3+453000 c^2-121400 c-68640) 
\,
\pa^2 G^{-} G^{-} G^{+}
\nonu \\
&&
+\frac{10}{21} (290 c^6-2145 c^5+52260 c^4-302029 c^3+405772 c^2-52260 c+109200) 
\,
\pa^2 G^{-} \pa J J
\nonu \\
&&
-\frac{10}{189} (495 c^7-12657 c^6+1367 c^5+456081 c^4-1935686 c^3+2502632 c^2-365400 c-560160) 
\nonu \\
&& \times 
\pa^3 G^{-} T
\nonu \\
&&
+\frac{20}{63} (75 c^6-2056 c^5+28153 c^4-171504 c^3+33758 c^2+473512 c-280640) 
\,
\pa^3 G^{-} J J
\nonu \\
&&
-\frac{20}{21} (591 c^6+15943 c^5-76297 c^4+64289 c^3+53510 c^2
+16972 c-88720) 
\,
\pa T \pa J G^{-}
\nonu \\
&&
-\frac{20}{63} (237 c^7-274 c^6+9140 c^5-106920 c^4+397875 c^3-608454 c^2+320956 c+18960) 
\,
\pa^2 T \pa G^{-}
\nonu \\
&&
-\frac{20}{21} (468 c^6-297 c^5-7014 c^4+18319 c^3-57076 c^2+79708 c+7440) 
\,
\pa^2 T J G^{-}
\nonu \\
&&
-\frac{10}{189} (918 c^7-21 c^6-41391 c^5+23333 c^4+557597 c^3-1276856 c^2+723796 c+37680) 
\,
\pa^3 T G^{-}
\nonu \\
&&
+\frac{10}{7} (2001 c^5+30410 c^4-171687 c^3+219444 c^2-30060 c+5360) 
\,
\pa J \pa J J G^{-}
\nonu \\
&&
+\frac{40}{21} (897 c^5+18183 c^4-68416 c^3+51452 c^2-41480 c+111040) 
\,
\pa^2 J J J G^{-}
\nonu \\
&&
-\frac{20}{63} (1035 c^6+38377 c^5-214035 c^4+406107 c^3-422272 c^2+386236 c-198000) 
\,
\pa^2 J G^{-} T
\nonu \\
&&
+\frac{10}{63} (1115 c^6+20031 c^5+165553 c^4-834539 c^3+937820 c^2-446396 c+561840) 
\,
\pa^2 J \pa G^{-} J
\nonu \\
&&
+\frac{20}{21} (c-2)(c+12)(7c-15)(60 c^3-693 c^2+705 c+226) 
\,
\pa^2 J \pa J G^{-}
\nonu \\
&&
+\frac{10}{63} (780 c^6-1459 c^5-45978 c^4+201337 c^3
-71800 c^2-285948 c+34800) 
\,
\pa^3 J J G^{-}
\nonu \\
&&
-\frac{10}{189} (c+12)(75 c^6-570 c^5-2592 c^4+24584 c^3-103835 c^2+166110 c-67140) \,
\pa^3 J \pa G^{-}
\nonu \\
&&
-\frac{80}{21} (6096 c^3-2809 c^2-17966 c+13808) 
\,
J J J J J G^{-}
\nonu \\
&&
+\frac{80}{21} (1860 c^5-3491 c^4-24593 c^3+46968 c^2+4520 c-12080) 
\,
\pa T J J G^{-}
\nonu \\
&&
-\frac{80}{21} (4653 c^4-25375 c^3+2955 c^2+59364 c-5200) 
\,
\pa J J J J G^{-}
\nonu \\
&&
+\frac{80}{21} (3930 c^5-19209 c^4+16246 c^3+24511 c^2-29344 c+13360) 
\,
\pa J J G^{-} T
\nonu \\
&&
+\frac{160}{63} (13257 c^4-27821 c^3-7135 c^2+53020 c-38640) 
\,
J J J G^{-} T
\nonu \\
&&
-\frac{40}{63} (19197 c^5-72906 c^4+75965 c^3-31032 c^2+80596 c-80400) 
\,
J G^{-} T T
\nonu \\
&&
+\frac{40}{63} (7095 c^5+55072 c^4-146979 c^3+144732 c^2-214244 c
+115440) 
\,
\pa G^{-} J J T
\nonu \\
&&
-\frac{40}{63} (3508 c^4+96405 c^3-132585 c^2-131186 c+155280)
\,
\pa G^{-} J J J J 
\nonu \\
&&
-\frac{80}{63} (54 c^6-4068 c^5+58564 c^4-267659 c^3+439087 c^2-211052 c-12000) 
\,
\pa G^{-} \pa G^{+} G^{-}
\nonu \\
&&
-\frac{5}{63} (65 c^7+747 c^6+1133 c^5+134637 c^4-502298 c^3+527916 c^2-420584 c+566880) 
\,
\pa^2 G^{-} \pa^2 J
\nonu \\
&&
-\frac{5}{63} (51 c^7-299 c^6-9185 c^5+150163 c^4-624670 c^3+860620 c^2-450328 c+351520) 
\,
\pa^3 G^{-} \pa J
\nonu \\
&&
-\frac{80}{63} (4617 c^6-10941 c^5
-66676 c^4+284335 c^3-363779 c^2+146984 c+6960) 
\,
\pa T G^{-} T
\nonu \\
&&
-\frac{5}{63} (24 c^7+35 c^6-5698 c^5+36241 c^4-177568 c^3+637562 c^2-1033336 c+539264) 
\,
\pa^4 J G^{-}
\nonu \\
&&
-\frac{20}{189} (19683 c^6-85935 c^5+280109 c^4-784625 c^3+1094060 c^2-499092 c-45360) 
\,
\pa G^{-} T T
\nonu \\
&&
-\frac{1}{756} (60 c^8-1820 c^7-22026 c^6+264841 c^5-1045468 c^4
 +1078521 c^3+958948 c^2\nonu \\
&& -597660 c-1544400) 
 \,
\pa^5 G^{-}
\nonu \\
&& 
-\frac{5}{756} (264 c^7-6566 c^6-159073 c^5+1307448 c^4
 -4028653 c^3+2944036 c^2+2837548 c-1796400) 
\nonu \\
&& \left. \times \,
\pa^4 G^{-} J
\right](w)+ \cdots,
\nonu \\
&& W_{  \frac{1}{3}}^{(\frac{7}{2})}(z) \, W_{  -\frac{1}{3}}^{(\frac{9}{2})}(w) = 
-\frac{1}{(z-w)^8} \, \frac{c}{21} +\frac{1}{(z-w)^7} \, 
\frac{62}{21} \, J(w) \nonu \\
&& + \frac{1}{(z-w)^6} \, \left[ 
\frac{10 }{3 (c-1)}\,
J J 
-\frac{5 (3 c+25) }{63 (c-1)}\,
T
-\frac{1}{42} \, \pa J
\right](w) \nonu \\
&& + \frac{1}{(z-w)^5}\, \frac{1}{ (c-1) (c+6) (2 c-3)} \, \left[ 
-\frac{4}{21}  (377 c^2-981 c+94) 
\,
G^{-} G^{+}
\right.
\nonu \\
&& +\frac{8}{63} (327 c^2-272 c+1260) 
\,
J T 
\nonu \\
&& +\frac{40 }{21} (c+6)(2c-3)\,
\pa J J  
-\frac{8}{63} (467 c+358) 
\,
J J J
-\frac{1}{63} (18 c^3-2119 c^2+6003 c-1122) 
\,
\pa T
\nonu \\
&& \left. -\frac{1}{126} (2 c^3+1515 c^2-3951 c+394) 
\,
\pa^2 J
\right](w) \nonu \\
&& + \frac{1}{(z-w)^4} \, \frac{1}{(c-1) (c+1) (c+6) (2 c-3) (5 c-9)}\, \left[ 
-\frac{5}{63}(c+1)(5c-9)(3 c^2+943 c-2250) 
\,
\pa J T
\right. \nonu \\
&& -\frac{2}{63} (10084 c^2+10239 c+9032) 
\,
J J J J
 \nonu \\
&&
-\frac{2}{21} (3941 c^3-15760 c^2-5735 c-9706) 
\,
J G^{-} G^{+}
+\frac{2}{63} (7077 c^3+2290 c^2+36671 c+29622) 
\,
J J T
\nonu \\
&&
+\frac{1}{63} (7524 c^4-32607 c^3+19584 c^2+77753 c-5634) 
\,
\pa G^{+} G^{-}
\nonu \\
&&
-\frac{1}{63} (7551 c^4-18378 c^3-11829 c^2-10688 c-1116) 
\,
\pa G^{-} G^{+}
\nonu \\
&&
+\frac{1}{63} (6525 c^4+12728 c^3-67075 c^2-25710 c-23448) 
\,
\pa T J
\nonu \\
&&
-\frac{1}{188} (999 c^4+27978 c^3-80293 c^2+91140 c+9036) 
\,
T T
\nonu \\
&&
+\frac{1}{126} (791 c^4-4804 c^3+31815 c^2-20886 c+60064) 
\,
\pa^2 J J
\nonu \\
&&
+\frac{1}{84} (109 c^4+252 c^3-13255 c^2+31546 c-26072) 
\,
\pa J \pa J
\nonu \\
&&
-\frac{65}{21} (c+1)(5c-9)(19 c+14) 
\,
\pa J J J
-\frac{1}{126} (57 c^5+126 c^4-10230 c^3+23660 c^2+9117 c-1170) 
\nonu \\
&& \times
\pa^2 T
\nonu \\
&& \left.
-\frac{1}{756} (15 c^5+20628 c^4-308 c^3-130070 c^2-17631 c-50514) 
\,
\pa^3 J
\right](w) 
\nonu \\
&& + \frac{1}{(z-w)^3} \, \frac{(c+18)(4c-9)(7c-15)}{d(c)} \, \left[ 
-\frac{1}{1134} (117 c^7-2535 c^6-307215 c^5+392675 c^4
\right. \nonu \\
&& +2562058 c^3-4588420 c^2+1245000 c-15840) 
\,
\pa^3 T
\nonu \\
&& -\frac{16}{189} (26112 c^4+4477 c^3-35377 c^2-36338 c+19776) 
\,
J J J T
\nonu \\
&& +\frac{4}{63} (24896 c^3+40675 c^2-153502 c+48816)
\,
J J J J J 
\nonu \\
&&
+\frac{8}{21} (10823 c^4-19437 c^3-135958 c^2+266168 c-92576) 
\,
J J G^{-} G^{+}
\nonu \\
&&
+\frac{2}{189} (72927 c^5-94098 c^4+365507 c^3-2601712 c^2+4749420 c-2325744)
\,
J T T
\nonu \\
&&
-\frac{4}{63} (45249 c^5-147298 c^4-67143 c^3+578128 c^2-465852 c+7536) 
\,
G^{-} T G^{+}
\nonu \\
&&
-\frac{1}{63} (45 c^6+4508 c^5-185603 c^4+530914 c^3+567876 c^2-2268104 c+897504)
 \,
\pa T \pa J
\nonu \\
&&
+\frac{8}{63} (783 c^5+36555 c^4-102040 c^3-204380 c^2+391712 c+82080) 
\,
\pa G^{+} J G^{-}
\nonu \\
&&
+\frac{1}{63} (1875 c^6-53076 c^5+93907 c^4+432438 c^3-1657988 c^2+1240424 c+364320) 
\,
\pa^2 G^{+} G^{-}
\nonu \\
&&
-\frac{1}{189} (75 c^6+107221 c^5+224505 c^4-2633553 c^3+2667788 c^2+4859116 c-6579312) 
\,
\pa^2 J T
\nonu \\
&& +\frac{1}{63} (c+12)(56 c^4-1861 c^3-1170 c^2+17169 c-2234) 
\,
\pa^2 J \pa J
\nonu \\
&&
+\frac{1}{189} (291 c^6-5004 c^5+46097 c^4-54132 c^3-2010044 c^2+4221312 c-1602240) 
\,
\pa^3 J J
\nonu \\
&&
-\frac{4}{21} (c+12)(947 c^3-1990 c^2-2315 c-4042) 
\,
\pa J G^{-} G^{+}
\nonu \\
&&
+\frac{16}{63} (c+12)(492 c^3-865 c^2+4801 c+4992) 
\,
\pa J J T
 -\frac{8}{63} (c+12)(3881 c^2+3801 c+3418) 
\,
\pa J J J J
\nonu \\
&&
-\frac{2}{63} (c+12)(2759 c^4-10041 c^3+35762 c^2-76234 c+60004) 
\,
\pa^2 J J J 
\nonu \\
&&
-\frac{1}{42} (3667 c^5+1566 c^4-3537 c^3-370328 c^2+1049516 c-789424) 
\,
\pa J \pa J J
\nonu \\
&&
-\frac{8}{63} (1233 c^5-26655 c^4-50444 c^3+432404 c^2-387872 c-4896) 
\,
\pa G^{-} J G^{+}
\nonu \\
&&
-\frac{1}{63} (1893 c^6-10845 c^5-56025 c^4+712753 c^3-2050716 c^2+1569908 c+22512) 
\,
\pa^2 G^{-} G^{+}
\nonu \\
&&
+\frac{4}{63} (1809 c^5-11819 c^4+44652 c^3+389078 c^2-887532 c+251952) 
\,
\pa T J J
\nonu \\
&&
+\frac{2}{63} (978 c^6+3849 c^5-67740 c^4-13155 c^3+771232 c^2-1122284 c+306960) 
\,
\pa^2 T J
\nonu \\
&&
-\frac{1}{189} (7533 c^6+36543 c^5+746959 c^4-2832891 c^3+738668 c^2+2463636 c+26352) 
\,
\pa G^{-} \pa G^{+}
\nonu \\
&&
-\frac{2}{189} (666 c^6-118575 c^5+498988 c^4-335159 c^3-600792 c^2+564516 c-103824) 
\,
\pa T T
\nonu \\
&&
-\frac{1}{756} (3 c^7+3456 c^6+75823 c^5-184006 c^4
 -689296 c^3+1856264 c^2-631952 c-258432) 
\nonu \\
&& \left. 
\times \pa^4 J
\right](w) \nonu \\
&& +\frac{1}{(z-w)^2} \, \frac{(c+18)(4c-9)}{d(c)}\,\left[ 
- \frac{1}{6804} (891 c^8+4005 c^7-1076403 c^6+3751935 c^5+4850454 c^4
\right.
\nonu \\
&& -6470230 c^3-95422492 c^2+119102160 c-14766720) 
\,
\pa^4 T
 \nonu \\
&&
-\frac{2}{189} (1534128 c^4-420433 c^3+2987923 c^2-6448618 c+7860240) 
\,
J J J J T
\nonu \\
&&
+\frac{16}{189} (141040 c^3+135667 c^2-658666 c+449864) 
\,
J J J J J J
\nonu \\
&&
+\frac{8}{63} (243776 c^4-963011 c^3-2649879 c^2+7047194 c-4592960) 
\,
J J J G^{-} G^{+}
\nonu \\
&&
+\frac{4}{189} (262953 c^5-103656 c^4+1139981 c^3-11673926 c^2+24972848 c-13112400) 
\,
J J T T
\nonu \\
&&
+\frac{8}{63} (82419 c^5-472936 c^4+592296 c^3+532459 c^2-1667198 c+691120)
\, 
J G^{-} T G^{+}
\nonu \\
&&
+\frac{4}{63} (154062 c^5-260355 c^4-1881035 c^3+934900 c^2+1785868 c-2216480) 
\,
\pa G^{-} J J G^{+}
\nonu \\
&&
-\frac{8}{189} (162387 c^6-959667 c^5+2759280 c^4-6246750 c^3+8914498 c^2-5511908 c-33840) 
\nonu \\
&& \times
\pa G^{-} T G^{+}
\nonu \\
&&
-\frac{1}{1701} (54675 c^6+6848379 c^5-43308495 c^4
\nonu \\
&&
+113526145 c^3-170739620 c^2+157450836 c-63193680) 
\,
T T T
\nonu \\
&&
+\frac{1}{189} (1018647 c^6+452583 c^5-16567187 c^4 \nonu \\
&& +1693357 c^3+105479556 c^2-164808876 c+61845360) 
\,
\pa T J T
\nonu \\
&&
-\frac{4}{189} (364707 c^5+699485 c^4-6863090 c^3-2476570 c^2+19905948 c-12966000) \,
\pa T J J J 
\nonu \\
&&
-\frac{1}{63} (7c-15)(88227 c^5-147314 c^4-1119489 c^3+3380864 c^2-2567076 c+12048) \,
\pa T G^{-} G^{+}
\nonu \\
&&
-\frac{1}{252} (7c-15)(57 c^6+8226 c^5-251859 c^4+1987048 c^3-4886140 c^2+3480592 c+96576) 
\,
\pa^2 T \pa J
\nonu \\
&&
+\frac{1}{189} (48051 c^6+3477747 c^5-5682831 c^4
\nonu \\
&&
-55539159 c^3+172779708 c^2-136399436 c+7347760) 
\,
\pa G^{+} \pa J G^{-}
\nonu \\
&&
+\frac{1}{1134} (78714 c^7-2554251 c^6+24128973 c^5-85922673 c^4
\nonu \\
&& +164075709 c^3-274309532 c^2+236074180 c-70633680) 
\,
\pa^2 G^{+} \pa G^{-}
\nonu \\
&&
+\frac{1}{189} (31887 c^6-1321581 c^5+10111809 c^4 \nonu \\
&& -428619 c^3-137873628 c^2+237571892 c-115959760) 
 \,
\pa^2 G^{+} J G^{-}
\nonu \\
&&
+\frac{1}{567} (23553 c^7-484050 c^6-1424892 c^5+29078022 c^4
\nonu \\
&& -92638365 c^3+73614152 c^2+49889100 c-65810480) 
\,
\pa^3 G^{+} G^{-} 
\nonu \\
&&
-\frac{4}{63} (7c-15)(34611 c^4-62194 c^3+131689 c^2-136954 c+88368) 
\,
\pa J J J T
\nonu \\
&&
+\frac{5}{63} (7c-15)(39664 c^3+63461 c^2-244466 c+76752) 
\,
\pa J J J J J 
\nonu \\
&&
+\frac{4}{21} (7c-15)(20809 c^4+3069 c^3-309314 c^2+461464 c-175648) 
\,
\pa J J G^{-} G^{+}
\nonu \\
&&
+\frac{1}{504} (7c-15)(337 c^5+40306 c^4-1579987 c^3+6426392 c^2-8368764 c+3281136) 
\,
\pa J \pa J \pa J
\nonu \\
&&
-\frac{1}{189} (315 c^7+323843 c^6-1819335 c^5-1494483 c^4
\nonu \\
&& +29524468 c^3-67134976 c^2+46306088 c+2785760) 
\,
\pa^2 J \pa T
\nonu \\
&&
+\frac{1}{2268} (2568 c^7-147447 c^6-1266435 c^5+11353227 c^4
\nonu \\
&& -25725705 c^3+47372884 c^2-110683412 c+112333200) 
\,
\pa^2 J \pa^2 J
\nonu \\
&&
-\frac{1}{2268} 
(1575 c^7+3588465 c^6+7191555 c^5-107297629 c^4 \nonu \\
&& +108169454 c^3+476070876 c^2-1080325896 c+613164000) 
\,
\pa^3 J T
\nonu \\
&&
+\frac{1}{4536} (6813 c^7-437913 c^6+6426741 c^5-21910947 c^4
\nonu \\
&& -67185186 c^3+417231796 c^2-608810984 c+293459040) 
\,
\pa^3 J \pa J
\nonu \\
&&
+\frac{1}{2268} (4797 c^7-149193 c^6+325857 c^5+11511513 c^4\nonu \\
&& -65327910 c^3+118316680 c^2-96538512 c+34015808) 
\,
\pa^4 J J
\nonu \\
&&
-\frac{1}{252} (7c-15)(14671 c^5-141822 c^4-825473 c^3+4230208 c^2-3725764 c-94320) 
\,
\pa^2 J \pa J J
\nonu \\
&&
-\frac{1}{378} (7c-15)(999 c^5+1150854 c^4-5105581 c^3+8815376 c^2-10048020 c+6438672) 
\,
\pa J T T 
\nonu \\
&&
-\frac{1}{21} (40447 c^5+235146 c^4-1238821 c^3-508264 c^2+4638252 c-3057680) 
\,
\pa J \pa J J J 
\nonu \\
&&
-\frac{4}{63} (160497 c^5-1534281 c^4+1076126 c^3+14267894 c^2-22331836 c+7982240) 
\,
\pa G^{+} J J G^{-} 
\nonu \\
&&
-\frac{1}{189} (58725 c^6-96303 c^5+8140831 c^4-19892901 c^3-47933404 c^2+85041532 c-23553200) 
\nonu \\
&& \times
\pa G^{-} \pa G^{+} J
\nonu \\
&&
-\frac{1}{189} (111681 c^6-2530203 c^5-7955685 c^4 \nonu \\
&& +87259695 c^3-166052556 c^2+69500908 c+30118480) 
\,
\pa G^{-} \pa J G^{+}  
\nonu \\
&&
-\frac{1}{189} (53937 c^6-453135 c^5-963849 c^4+5814459 c^3-44129016 c^2+52102964 c-28536880) 
\nonu \\
&& \times \pa^2 G^{-} J G^{+}
\nonu \\
&&
+\frac{2}{189} (22959 c^6+77121 c^5-2166540 c^4+3106800 c^3+19998036 c^2-42052496 c+24084400) 
\nonu \\
&& \times
\pa^2 T J J
\nonu \\
&&
+\frac{2}{189} (648819 c^6-5104821 c^5+14064021 c^4-15554331 c^3+6539012 c^2+961460 c-406800) 
\nonu \\
&& \times
\pa G^{+} G^{-} T
\nonu \\
&&
-\frac{2}{189} (101567 c^5+769123 c^4-2635318 c^3+10234468 c^2-25947760 c+28292080) 
\,
\pa^2 J J J J 
\nonu \\
&&
-\frac{1}{189} (25536 c^6-341244 c^5-2826253 c^4+15988663 c^3-6292774 c^2-11383008 c+4473840) 
\nonu \\
&& \times
\pa^3 J J J 
\nonu \\
&&
-\frac{1}{378} (159069 c^6+852561 c^5-7118553 c^4
\nonu \\
&&
+9544203 c^3-20148012 c^2+90610972 c-107219120) 
\,
\pa^2 J G^{-} G^{+}
\nonu \\
&&
+\frac{1}{378} (172017 c^6-4011159 c^5+11551539 c^4
\nonu \\
&& +61959291 c^3-319949508 c^2+456597580 c-217487600) 
\,
\pa T \pa J J
\nonu \\
&&
+\frac{1}{378} (109731 c^6-578729 c^5-438543 c^4+23620053 c^3-33031284 c^2-87846428 c+160660400) 
\nonu \\
&& \times 
\pa^2 J J T
\nonu \\
&&
-\frac{1}{1134} (79416 c^7-910083 c^6+14867961 c^5-77902221 c^4
\nonu \\
&& +114370335 c^3+122665892 c^2-248498260 c+70676880) 
\,
\pa^2 G^{-} \pa G^{+}
\nonu \\
&&
-\frac{1}{1134} (47835 c^7-1154514 c^6+99528 c^5+53424738 c^4
\nonu \\
&& -229088823 c^3+376255232 c^2-201197676 c+361840) 
\,
\pa^3 G^{-} G^{+}
\nonu \\
&&
+\frac{1}{756} (34767 c^6-294753 c^5-13078011 c^4+83073621 c^3-154147308 c^2+45225364 c+74299120) 
\nonu \\
&& \times
\pa J \pa J T
\nonu \\
&&
-\frac{1}{567} (8748 c^7-2636469 c^6+10403415 c^5+25981905 c^4\nonu \\
&& -185461383 c^3+295692244 c^2-121685180 c-41186640) 
\,
\pa T \pa T
\nonu \\
&&
+\frac{1}{1134} (54747 c^7+155112 c^6-3254502 c^5+8876512 c^4
\nonu \\
&& +17035511 c^3-117219200 c^2+127823420 c-32416240) 
\,
\pa^3 T J 
\nonu \\
&&
-\frac{1}{1134} (18171 c^7+344223 c^6-5902779 c^5+8262453 c^4
\nonu \\
&& +34520856 c^3-44477740 c^2-75595744 c+110467200) 
\,
\pa^2 T T 
\nonu \\
&& 
-\frac{1}{22680} (105 c^8+96837 c^7+2189421 c^6-8242205 c^5-50622940 c^4
\nonu \\
&& \left. 
+299813218 c^3-529190356 c^2+317863280 c-17794560) 
\,
\pa^5 J
\right](w) \nonu \\
&& + \frac{1}{(z-w)} \, \frac{1}{d(c)} \, \left[
\right. \nonu \\
&&
-\frac{1}{34020} (2700 c^{10}+13005 c^9+130968 c^8-143702742 c^7+500284020 c^6+4713963729 c^5  \nonu \\
&&  -31266014148 c^4+68264908412 c^3-65202679120 c^2+28281314976 c-7297992000) \,
\pa^5 T 
\nonu \\
&&
+\frac{8}{189} (1256832 c^6-21988368 c^5-150949083 c^4\nonu \\
&& +1177579763 c^3-2405225500 c^2+1504296996 c-198944640) 
\,
\pa G^{+} J J J G^{+}
\nonu \\
&&
-\frac{8}{189} (849456 c^7-17026557 c^6-70058022 c^5+937834181 c^4\nonu \\
&& -2803475222 c^3+3380729648 c^2-1521336144 c+48396960) 
\,
\pa G^{-} J T G^{+}
\nonu \\
&&
-\frac{1}{139} (2535948 c^8+27306369 c^7-163802421 c^6-401973987 c^5+3390486509 c^4 \nonu \\
&& -5779572926 c^3+1600358228 c^2+2979266280 c-1167717600) 
 \,
\pa G^{-} \pa T G^{+}
\nonu \\
&&
-\frac{2}{567} (2659068 c^8+15024285 c^7+47512983 c^6+279545857 c^5-6152018439 c^4 \nonu \\
&& +20248988426 c^3-23764255164 c^2+7967357064 c+1809954720) 
\,
\pa G^{-} \pa G^{+} T
\nonu \\
&&
+\frac{4}{189} (668196 c^7+2762433 c^6+62657758 c^5+89088661 c^4
\nonu \\
&& -2795656372 c^3+7925558788 c^2-6869258944 c+1796762880) 
 \,
\pa G^{-} \pa G^{+} J J 
\nonu \\
&&
+\frac{8}{189} (875106 c^7+14188944 c^6-11443461 c^5-412800303 c^4
\nonu \\
&& +1036077102 c^3-406964942 c^2-550212836 c+599577840) 
 \,
\pa G^{-} \pa J J G^{+} 
\nonu \\
&&
+\frac{4}{189} (467748 c^7-1427811 c^6-25796121 c^5+609575883 c^4
\nonu \\
&& -3027712821 c^3+5804492824 c^2-3774941972 c+725987520) 
\,
\pa^2 G^{-} J J G^{+}
\nonu \\
&&
-\frac{2}{189} (658476 c^8-2487525 c^7-84682737 c^6+1032563151 c^5
-4612786331 c^4 \nonu \\
&& +9775940010 c^3-9818091028 c^2+3770880024 c+55049760) 
\,
\pa^2 G^{-} T G^{+}
\nonu \\
&&
+\frac{16}{189} (269496 c^7-1273770 c^6+9899405 c^5+87086908 c^4
\nonu \\
&& -767774059 c^3+1989872268 c^2-2180619948 c+806803200) 
\,
\pa T J J T
\nonu \\
&&
-\frac{2}{189} (3105792 c^6+19774972 c^5-17336203 c^4
\nonu \\
&& +738823903 c^3
-3915256860 c^2+5184776676 c-2240866080) 
\,
\pa T J J J J 
\nonu \\
&&
-\frac{8}{63} 
(c+18)(4c-9)(80523 c^5-349672 c^4-334563 c^3+2949988 c^2-3874496 c+1069840) 
\nonu \\
&& \times \pa T J G^{-} G^{+} 
\nonu \\
&&
-\frac{1}{567} (218700 c^8-81427959 c^7+529933107 c^6-729814723 c^5
-1636669795 c^4 \nonu \\
&& +3742140834 c^3+2503288788 c^2-9603271512 c+5128025760) 
\,
\pa T T T
\nonu \\
&&
+\frac{1}{567} (3060828 c^8+59617701 c^7+18683631 c^6-2486505495 c^5+7709143017 c^4 \nonu \\
&& -5396644846 c^3-6456575356 c^2+8359521960 c-296943840) 
\,
\pa T \pa T J
\nonu \\
&&
-\frac{4}{189} (1454436 c^7+23901699 c^6+26061894 c^5-724114158 c^4
\nonu \\
&& +276116517 c^3+4976578288 c^2-8466311636 c+4340777760) 
\,
\pa T \pa J J J 
\nonu \\
&&
+\frac{1}{567} (3654828 c^8+29483649 c^7-413586165 c^6+219278829 c^5
+6195189765 c^4 \nonu \\
&& -18462511222 c^3+19139061668 c^2-8530109592 c+2566542240) 
\,
\pa^2 T J T
\nonu \\
&&
-\frac{4}{189} (436116 c^7+4115163 c^6-45688737 c^5-27158419 c^4
\nonu \\
&& +1039527233 c^3-3010834532 c^2+3073714396 c-1129569120) 
 \,
\pa^2 T J J J 
\nonu \\
&&
-\frac{1}{63} (728412 c^8+3360285 c^7-66613113 c^6+185815153 c^5+329504409 c^4
\nonu \\
&& -2457472246 c^3+4356379044 c^2-2812096824 c+302482080) 
 \,
\pa^2 T G^{-} G^{+}
\nonu \\
&&
-\frac{1}{756} (1596 c^9+1677177 c^8+2399928 c^7-125161744 c^6+1469240110 c^5
-7616200253 c^4 \nonu \\
&& +18150890398 c^3-18839448532 c^2+5097015720 c+2186157600) 
 \,
\pa^2 T \pa^2 J
\nonu \\
&&
-\frac{1}{2268} (3276 c^9+319557 c^8-15477228 c^7-256634968 c^6+2407016866 c^5-8590189625 c^4 \nonu \\
&& +18242000166 c^3-20911309220 c^2+6003412296 c+3429038880) 
\,
\pa^3 T \pa J
\nonu \\
&&
+\frac{1}{567 } (191988 c^8-12054981 c^7+136930149 c^6+1420894023 c^5-17379411981 c^4 \nonu \\
&& +63644622014 c^3-103746408292 c^2+69844158840 c-16214700960) 
\,
\pa^2 G^{+} \pa G^{-} J
\nonu \\
&&
+\frac{1}{189} (24300 c^8+174771 c^7-33130905 c^6+183814395 c^5+1065343953 c^4
\nonu \\ 
&& -10033024854 c^3+25066284908 c^2-24272226088 c+8413607520) 
\,
\pa^2 G^{+} \pa J G^{-}
\nonu \\
&&
+\frac{2}{1701} (46980 c^9-290565 c^8-14115147 c^7+243838455 c^6-185864595 c^5-4952620746 c^4 \nonu \\
&& +17916629818 c^3-24444493736 c^2+13982959416 c-2838062880) 
\,
\pa^3 G^{+} \pa G^{-}
\nonu \\
&&
-\frac{4}{567} (39816 c^8-1658346 c^7+38728077 c^6+5737170 c^5-691300545 c^4
\nonu \\
&& +2102858080 c^3-4214045876 c^2+4923423984 c-2247773760) 
\,
\pa^3 G^{+} J G^{-} 
\nonu \\
&&
+\frac{1}{2268} (62640 c^9-1560816 c^8-93560949 c^7+333288687 c^6+2042191759 c^5-12366656451 c^4\nonu \\
&& +20729425326 c^3-7055217724 c^2-10735586792 c+7446849120) 
 \,
\pa^4 G^{+} G^{-}
\nonu \\
&&
-\frac{8}{189} (c+18)(4c-9)(587028 c^4-774413 c^3+3188783 c^2-4177778 c+4533840) 
\,
\pa J J J J T
\nonu \\
&&
+\frac{8}{63} (c+18)(4c-9)(237680 c^3+220973 c^2-1100714 c+753136) 
\,
\pa J J J J J J 
\nonu \\
&&
+\frac{8}{21} (c+18)(4c-9)
(117496 c^4-263851 c^3-1574409 c^2+3306754 c-2234560) 
\,
\pa J J J G^{-} G^{+}
\nonu \\
&&
+\frac{4(c+18)}{189} (4c-9)
(141291 c^5-762042 c^4+4618727 c^3-15711872 c^2+26527796 c
\nonu \\ 
&& -14041200) 
\, 
\pa J J T T
\nonu \\
&&
-\frac{16}{63} (c+18)(4c-9)
(39684 c^5-156851 c^4+44001 c^3+504134 c^2-964078 c+475520) 
\nonu \\
&& \times
\pa J G^{-} T G^{+}
\nonu \\
&&
-\frac{2}{189} (1730772 c^7-4145529 c^6-197274204 c^5+1208070807 c^4
\nonu \\
&& -2122442094 c^3-237564164 c^2+3470926552 c-1938169440) 
\,
\pa J \pa J J T
\nonu \\
&&
+\frac{20}{63} (221800 c^6+2500468 c^5-9758719 c^4
\nonu \\
&& -30316995 c^3+156834788 c^2-194225012 c+69170400) 
\,
\pa J \pa J J J J 
\nonu \\
&&
+\frac{2}{21} (341648 c^7+2210359 c^6-26345196 c^5+43371183 c^4
\nonu \\
&& +186352774 c^3-843515996 c^2+1287402488 c-730208160) 
\,
\pa J \pa J G^{-} G^{+}
\nonu \\
&&
-\frac{4}{189} (864168 c^7+8924654 c^6-120045031 c^5+644006368 c^4
\nonu \\
&& -1899786271 c^3+3031168464 c^2-2614023892 c+1026930240) 
\,
\pa^2 J J J T
\nonu \\
&&
+\frac{5}{189} (990208 c^6+17539388 c^5-77591303 c^4
\nonu \\
&& +76702611 c^3-126701740 c^2+376802516 c-378283680) 
\,
\pa^2 J J J J J 
\nonu \\
&&
+\frac{8}{189} 
(772008 c^7+16003269 c^6-78079596 c^5-104109147 c^4\nonu \\
&& +907472394 c^3-1215054586 c^2+198196868 c+281136240) 
\,
\pa^2 J J G^{-} G^{+}
\nonu \\
&&
+\frac{1}{378} (149220 c^8+8184555 c^7+219388677 c^6-381190633 c^5
-6985966733 c^4 \nonu \\
&& +30533641758 c^3-44416514900 c^2+19690229976 c+3023002080) 
\,
\pa^2 J \pa G^{+} G^{-}
\nonu \\
&&
+\frac{1}{504} (9604 c^8+3814691 c^7-7066031 c^6-633932929 c^5+5017729671 c^4
\nonu \\
&& -15398599186 c^3+21681571676 c^2-12151495016 c+712919520) 
\,
\pa^2 J \pa J \pa J
\nonu \\
&&
-\frac{1}{6804} (11340 c^9+20135997 c^8+324094482 c^7-1319520834 c^6-7326722208 c^5 \nonu \\
&& +49791378609 c^4 
 -111506612998 c^3+113034966500 c^2-36354286248 c-9335118240) 
\nonu \\
&& \times 
\pa^3 J \pa T
\nonu \\
&&
+\frac{1}{4536} (10380 c^9-783867 c^8-5584480 c^7+269274060 c^6-974752386 c^5-3081843917 c^4 \nonu \\
&& +23868190814 c^3-52910255284 c^2+54463582760 c-23474774880) 
\,
\pa^3 J \pa^2 J 
\nonu \\
&&
-\frac{1}{6804} (3780 c^9+7294959 c^8+204371370 c^7-681169902 c^6-3378946500 c^5+9778552227 c^4\nonu \\
&& +35990418494 c^3-167939131492 c^2+227549210664 c-104352524640) 
\,
\pa^4 J T
\nonu \\
&&
+\frac{1}{1512} (1716 c^9-92781 c^8+66254 c^7+2832370 c^6-128115680 c^5+833902015 c^4 \nonu \\
&& -1128653898 c^3-4088320628 c^2+12855689736 c-9712276704) 
\,
\pa^4 J \pa J
\nonu \\
&&
+\frac{1}{11340} (15720 c^9-83514 c^8-11345004 c^7+107801173 c^6+841151006 c^5-6440949823 c^4 \nonu \\
&& +9589448846 c^3+5588219440 c^2-17287283424 c+8342304480)\,
\pa^5 J J
\nonu \\
&&
-\frac{1}{63} (c+18) (4 c-9)(25633 c^5+4462 c^4+482213 c^3-6228508 c^2+14794500 c-11239440) 
\nonu \\
&& \times
\pa J \pa J \pa J J
\nonu \\
&&
-\frac{2}{63} (c+18) (4 c-9)(39235 c^5+62993 c^4-1319228 c^3+5725268 c^2-8247968 c+6258320) 
\nonu \\
&& \times
\pa^2 J \pa J J J 
\nonu \\
&&
-\frac{1}{189} (c+18) (4 c-9)(26811 c^6+90411 c^5+3323489 c^4
\nonu \\
&& -2754519 c^3-34087412 c^2+32848980 c+4683120)
 \,
\pa G^{-} \pa G^{+} \pa J
\nonu \\
&&
+\frac{1}{126} (c+18) (4 c-9)(17427 c^6+36759 c^5-964215 c^4
\nonu \\
&&
-6245515 c^3+39946468 c^2-50701884 c+19146160) 
\,
\pa^2 T \pa J J
\nonu \\
&&
+\frac{1}{189} (c+18) (4 c-9)(6015 c^6-353291 c^5-1951923 c^4
\nonu \\
&&
+15747003 c^3-20277168 c^2-15791836 c+33432240) 
\,
\pa^2 J \pa J T
\nonu \\
&&
-\frac{1}{378} (c+18) (4 c-9)(39747 c^6-439473 c^5+1518521 c^4
\nonu \\
&&
+4005429 c^3-35366036 c^2+76713252 c-54966480) 
\,
\pa^3 J G^{-} G^{+}
\nonu \\
&&
-\frac{880}{63} 
(50048 c^5-87950 c^4-1615395 c^3+6639000 c^2-7876228 c+3215520) 
\,
J J J J G^{-} G^{+}
\nonu \\
&&
+\frac{  1760}{63} (34896 c^6-109310 c^5-720956 c^4+4384211 c^3-8814282 c^2+7933696 c-2580480) 
 \nonu \\
&& \times
J J G^{-} T G^{+}
\nonu \\
&&
+\frac{880}{189} (75072 c^5-22400 c^4-949713 c^3+1594868 c^2-1242708 c+182304) 
\,
 J J J J J T
\nonu \\
&&
-\frac{1760}{189} (17920 c^4+18322 c^3-367777 c^2+584066 c-272880) 
\,
J J J J J J J
\nonu \\
&&
-\frac{440}{189} (145800 c^7-659337 c^6-1084556 c^5+11396927 c^4
\nonu \\
&& -24532882 c^3+20544604 c^2-4304856 c-1317600)
\,
G^{-} T T G^{+} 
\nonu \\
&&
-\frac{1}{189} (135900 c^8-505629 c^7-5862309 c^6+187355031 c^5-1057859151 c^4
\nonu \\
&& +1450523610 c^3+613593028 c^2-1893510200 c+1670231520) 
\,
\pa^2 G^{-} \pa J G^{+}
\nonu \\
&&
-\frac{8}{189} (805392 c^6+41147892 c^5-156526743 c^4
\nonu \\
&& -799907177 c^3+3903743980 c^2-4587159804 c+1755077760) 
\,
\pa G^{+} J J J G^{-} 
\nonu \\
&&
+\frac{8}{189} (528696 c^7+27629043 c^6-131000442 c^5-57303479 c^4
\nonu \\
&& +1049337698 c^3-1738328912 c^2+1090764336 c-164492640) 
\,
\pa G^{+} J G^{-} T
\nonu \\
&&
-\frac{8}{189} (924606 c^7+7328217 c^6-132572943 c^5+36015690 c^4
\nonu \\
&&+2712910794 c^3-7873859162 c^2+7366663228 c-2154448080) 
\,
\pa G^{+} \pa J J G^{-}
\nonu \\
&&
-\frac{4}{189} (489528 c^7-12054336 c^6-131541 c^5+535667658 c^4
\nonu \\
&& -3718103061 c^3+10204280224 c^2-11125471532 c+4531181760) 
\,
\pa^2 G^{+} J J G^{-}
\nonu \\
&&
+\frac{8}{189} (163890 c^8-4408800 c^7-3487143 c^6+228738352 c^5-1070678893 c^4
\nonu \\
&& +2136988528 c^3-1976518650 c^2+712964196 c-3473280) 
\,
\pa^2 G^{+} G^{-} T
\nonu \\
&&
+\frac{2}{189} (56880 c^8-856216 c^7-34653531 c^6+209033484 c^5+25977471 c^4
\nonu \\
&& -2229375140 c^3+5064950012 c^2-4170576640 c+1038300480) 
\,
\pa^2 J \pa T J
\nonu \\
&&
+\frac{1}{189} (2530116 c^8+22465611 c^7-233016975 c^6-43696737 c^5
+4236817959 c^4 \nonu \\
&& -12458812666 c^3+13519170620 c^2-4242153288 c-1131092640) 
\,
\pa T \pa G^{+} G^{-}
\nonu \\
&&
-\frac{2}{189} (9324 c^8+5562585 c^7-27531661 c^6-267045139 c^5+2601357233 c^4
\nonu \\
&& -9011736162 c^3+16116933028 c^2-15167286408 c+5841309600) 
\,
\pa T \pa J T
\nonu \\
&&
-\frac{1}{378} (276660 c^8+881415 c^7-228416607 c^6-116799917 c^5
+9029380823 c^4 \nonu \\
&& -32234080698 c^3+41114387900 c^2-14426908296 c-4188235680) 
\,
\pa^2 J \pa G^{-} G^{+}
\nonu \\
&&
+\frac{1}{252} (28620 c^8-3857591 c^7-28006981 c^6+384892509 c^5-558492899 c^4
\nonu \\
&& -4621672870 c^3+19529757972 c^2-28925192760 c+15743440800) 
\,
\pa T \pa J \pa J
\nonu \\
&&
-\frac{ 1760}{567} (78516 c^6-128676 c^5-272257 c^4-2046118 c^3+10495212 c^2-14684472 c+5922720) 
\nonu \\
&& \times
J J J T T
\nonu \\
&&
-\frac{1}{567} (169308 c^8+7476609 c^7-48167547 c^6+288089697 c^5+4730417163 c^4
\nonu \\
&& -39733102610 c^3+97545097348 c^2-84318805848 c+23973878880) 
\,
\pa^2 G^{-} \pa G^{+} J
\nonu \\
&&
-\frac{2}{567} (64116 c^8+784527 c^7-30905952 c^6+140768751 c^5+727952880 c^4 \nonu \\
&& -6375760138 c^3+14498198648 c^2-11281348392 c+3385640160) 
\,
\pa^3 G^{-} J G^{+}
\nonu \\
&&
-\frac{2}{567} (282372 c^7+1832811 c^6-51929 c^5-183969763 c^4
\nonu \\
&& -1057024799 c^3+5972757636 c^2-5459361108 c+247272480) 
\,
\pa^3 J J J J 
\nonu \\
&&
-\frac{1}{1134} (79020 c^9-2558907 c^8+20295987 c^7+477492177 c^6-6769121979 c^5+34420832226 c^4 \nonu \\
&& -88945752796 c^3+121576254752 c^2-76450845360 c+17710375680) 
\,
\pa^2 G^{-} \pa^2 G^{+}
\nonu \\
&&
+\frac{1}{567} (122868 c^8+1706703 c^7-59540139 c^6-101294885 c^5+1967154123 c^4
\nonu \\
&& -3761692914 c^3-1008545380 c^2+5491603864 c-3341391840) 
\,
\pa^3 T J J 
\nonu \\
&&
-\frac{1}{1134} (108468 c^8-1487385 c^7-72427635 c^6+78196059 c^5
+2451552783 c^4 \nonu \\
&& -10228586846 c^3+17438203276 c^2-15069737880 c+5415815520) 
\,
\pa^4 J J J
\nonu \\
&&
-\frac{1}{378} (14940 c^9-1780647 c^8-12101562 c^7+146091846 c^6-136796020 c^5-2014957467 c^4 \nonu \\
&& +7763992546 c^3-10888217580 c^2+5058841464 c+516106080) 
\,
\pa^2 T \pa T
\nonu \\
&&
+\frac{ 440}{1701} (218700 c^7-654759 c^6+1552032 c^5-24397223 c^4
\nonu \\
&& +119501698 c^3-247092684 c^2+236806776 c-87726240) 
\,
J T T T
\nonu \\
&&
-\frac{1}{2268} (663324 c^8-4225851 c^7-314167077 c^6-23872215 c^5+13216047789 c^4 \nonu \\
&& -51958623502 c^3+76280987540 c^2-42364422168 c+4496182560) 
\,
\pa^2 J \pa^2 J J
\nonu \\
&&
+\frac{1}{1134} (324900 c^8-3789141 c^7+15884877 c^6-358419433 c^5
-1463814821 c^4 \nonu \\
&& +22363345502 c^3-76141527108 c^2+110566594104 c-59261928480) 
\,
\pa^3 J  J T
\nonu \\
&&
-\frac{1}{2268} (878076 c^8-13769811 c^7-611873025 c^6-206858007 c^5+17525367993 c^4\nonu \\
&& -28669193798 c^3-46241240236 c^2+124926567208 c-76100940000) \,
\pa^3 J \pa J J 
\nonu \\
&&
-\frac{1}{3402} (191484 c^9-2320119 c^8-49491960 c^7+929777388 c^6
-4437518802 c^5-617618961 c^4 \nonu \\
&& +55963744870 c^3-137773727348 c^2+107590940328 c-22382650080) 
\,
\pa^3 G^{-} \pa G^{+}
\nonu \\
&&
-\frac{1}{4536} (127872 c^9-868068 c^8-74651247 c^7+187583217 c^6+2199813789 c^5-14209849037 c^4 \nonu \\
&& +38169725622 c^3-52608313412 c^2+32376970424 c-8013814560) 
\,
\pa^4 G^{-} G^{+}
\nonu \\
&&
-\frac{1}{1134} (27972 c^8+86682291 c^7+485749025 c^6-7037482529 c^5+22528567935 c^4 \nonu \\
&& -26413043018 c^3+4953558972 c^2+7535542392 c-919499040) 
\,
\pa^2 J T T
\nonu \\
&&
-\frac{1}{3402} (50220 c^9-1057923 c^8-208978824 c^7+592836684 c^6+1724525830 c^5-6436904581 c^4\nonu \\
&& +4152483838 c^3-5104890708 c^2+20264031144 c-17806301280) 
\,
\pa^3 T T
\nonu \\
&&
+\frac{1}{6804} (234576 c^9+3430152 c^8-38975847 c^7-410863911 c^6+2416398513 c^5+7373924155 c^4 \nonu \\
&& -67920494810 c^3+149424137268 c^2-120241255896 c+31105749600) 
\,
\pa^4 T J
\label{j2ope} \\
&&
-\frac{1}{204120} (540 c^{10}+439965 c^9+15257133 c^8+492920787 c^7-1947291627 c^6-14544218028 c^5 \nonu \\
&&  
+105271986982 c^4-258277861072 c^3+286385121712 c^2-119379586992 c+3559550400) 
\nonu \\
&& \left. \times 
\pa^6 J
\right](w)  + \cdots.
\nonu 
\eea
}
Here $d(c)$ was given in Appendix (\ref{ddefinition}).
In the large $c$ limit,
all the nonlinear terms in Appendix (\ref{j2ope}) disappear.
One can also analyze the $\frac{1}{c}, \cdots, \frac{1}{c^6}$-terms.

\subsection{The component OPEs  in the OPE 
${\bf W}_{  -\frac{1}{3}}^{(\frac{7}{2})}(Z_1) \, 
{\bf W}_{  -\frac{1}{3}}^{(\frac{7}{2})}(Z_2)$ }

The $4$ component OPEs corresponding to Appendix (\ref{44OPE})
can be summarized by
{\small
\bea
&& 
\frac{1}{C_{(\frac{7}{2})(\frac{7}{2})}^{(\frac{7}{2})-}} \,
W_{  -\frac{1}{3}}^{(\frac{7}{2})}(z) \, W_{  -\frac{1}{3}}^{(\frac{7}{2})}(w) = 
-\frac{1}{(z-w)^3} \,  \frac{3}{11} \, W_{  -\frac{2}{3}}^{(4)}(w) 
-\frac{1}{(z-w)^2} \, \frac{3}{22} \, \pa   W_{  -\frac{2}{3}}^{(4)}(w) 
\nonu \\
&& + \frac{1}{(z-w)} \, 
\frac{1}{ (c+9) (3 c+4) (27 c-46) (3 c^2+90 c-265)} \, \left[ 
\right. \nonu \\
&& \frac{216}{11} (42 c^3+614 c^2-1651 c+1385) 
 \,
J J W_{  -\frac{2}{3}}^{(4)}
\nonu \\
&& +36 (9 c^3-372 c^2+3333 c-4030) 
 \,
J G^{-} W_{  \frac{1}{3}}^{(\frac{7}{2})}
\nonu \\
&&
-\frac{18}{11} (27 c^4-1575 c^3+507 c^2+16059 c-10778) 
 \,
G^{-} W_{  \frac{1}{3}}^{(\frac{9}{2})}
\nonu \\
&& +\frac{3}{22} (2025 c^4+27756 c^3-38283 c^2-269878 c+353640) 
 \,
T W_{  -\frac{2}{3}}^{(4)} 
\nonu \\
&&
+\frac{18}{11} (9 c^4-390 c^3+4957 c^2+17064 c-78580) 
 \,
\pa J  W_{  -\frac{2}{3}}^{(4)} 
\nonu \\
&&
-\frac{9}{2} (9 c^4-884 c^3-7131 c^2+26506 c-11080) 
\,
\pa G^{-}  W_{  \frac{1}{3}}^{(\frac{7}{2})}
\nonu \\
&&
+\frac{9}{22} (45 c^4-4242 c^3-36687 c^2+154420 c-87036) 
 \,
G^{-} \pa  W_{  \frac{1}{3}}^{(\frac{7}{2})}
\nonu \\
&& 
-\frac{9}{22} (9 c^4-528 c^3-8139 c^2+103598 c-178320) 
 \,
J  \pa  W_{  -\frac{2}{3}}^{(4)} 
\nonu \\
&&
\left.
-\frac{3}{44} (3 c-1) (45 c^4+1302 c^3+1407 c^2-42794 c+75240) 
 \,
\pa^2  W_{  -\frac{2}{3}}^{(4)}
\right](w) + \cdots,
\nonu \\
&& 
\frac{1}{C_{(\frac{7}{2})(\frac{7}{2})}^{(\frac{7}{2})-}} \,
W_{  -\frac{1}{3}}^{(\frac{7}{2})}(z) \, W_{  \frac{2}{3}}^{(4)}(w) = 
-\frac{1}{(z-w)^4} \, W_{  \frac{1}{3}}^{(\frac{7}{2})}(w) +
\frac{1}{(z-w)^3} \left[ 
\frac{3}{22}  
W_{  \frac{1}{3}}^{(\frac{9}{2})} 
-\frac{19}{44} \, \pa  W_{  \frac{1}{3}}^{(\frac{7}{2})} 
\right](w) \nonu \\
&& + \frac{1}{(z-w)^2} \, 
\frac{1}{ (c+9) (27 c-46) (3 c^2+90 c-265)}
\,\left[ 
-\frac{1}{2} (27c-46) (51 c^2+614 c-1725) 
 \,
T W_{  \frac{1}{3}}^{(\frac{7}{2})}
\right. \nonu \\
&& +\frac{36}{27} (c+9)(31 c-35) 
\,
J J W_{  \frac{1}{3}}^{(\frac{7}{2})}
 \nonu \\
&&
+\frac{3}{2} (45 c^3-2904 c^2+4281 c+5998) 
 \,
G^{-}  W_{  \frac{4}{3}}^{(4)} 
-\frac{9}{22} (687 c^3+7144 c^2-25501 c-6790) 
 \,
G^{+}  W_{  -\frac{2}{3}}^{(4)} 
\nonu \\
&& 
-\frac{9}{22} (39 c^3-984 c^2+18923 c-22118) 
 \,
J W_{  \frac{1}{3}}^{(\frac{9}{2})} 
-\frac{3}{4} (3 c^3-1544 c^2-35529 c+108050) 
 \,
\pa J  W_{  \frac{1}{3}}^{(\frac{7}{2})}
\nonu \\
&&
+\frac{3}{44} (15 c^3-6552 c^2-121229 c+352266) 
 \,
J \pa  W_{  \frac{1}{3}}^{(\frac{7}{2})}
\nonu \\
&& +\frac{3}{22} (36 c^4+1695 c^3+15900 c^2-65125 c+27114) 
 \,
\pa  W_{  \frac{1}{3}}^{(\frac{9}{2})} 
\nonu \\
&& \left.
-\frac{1}{44} (387 c^4+7818 c^3-9849 c^2-116948 c+200412) 
 \,
\pa^2  W_{  \frac{1}{3}}^{(\frac{7}{2})}
\right](w) \nonu \\
&& +\frac{1}{(z-w)} 
\, 
\frac{1}{ (c+9) (3c+4)(27 c-46) (3 c^2+90 c-265)}
\,
\left[ 
\frac{3}{22} (27c-46) (81 c^2+87 c-698) 
 \,
T W_{  \frac{1}{3}}^{(\frac{9}{2})} 
\right.
\nonu \\
&& +\frac{9}{22} (27 c^4-567 c^3+15243 c^2-23565 c+22462) 
 \,
J \pa W_{  \frac{1}{3}}^{(\frac{9}{2})} 
\nonu \\
&&
+\frac{54}{11} (42 c^3+614 c^2-1651 c+1385) 
 \,
J J  W_{  \frac{1}{3}}^{(\frac{9}{2})} 
\nonu \\
&&
+\frac{18}{11} (897 c^3+7574 c^2+5294 c-18015) 
\,
J J \pa   W_{  \frac{1}{3}}^{(\frac{7}{2})}
\nonu \\
&&
- 18 (9 c^3-372 c^2+3333 c-4030) 
 \,
J T W_{  \frac{1}{3}}^{(\frac{7}{2})}
\nonu \\
&&
+\frac{216}{11} (42 c^3+614 c^2-1651 c+1385) 
\,
J G^{+} W_{  -\frac{2}{3}}^{(4)} 
\nonu \\
&&
+\frac{3}{22} (18 c^4-5295 c^3-103578 c^2+220669 c+218526) 
 \,
J \pa^2 W_{  \frac{1}{3}}^{(\frac{7}{2})}
\nonu \\
&&
+3 (27 c^4-1629 c^3+2739 c^2-3939 c+13402) 
 \,
G^{-} \pa W_{  \frac{4}{3}}^{(4)} 
\nonu \\
&&
+\frac{3}{2} (81 c^4-5592 c^3-10083 c^2+57318 c-4624) 
 \,
\pa G^{-} W_{  \frac{4}{3}}^{(4)} 
\nonu \\
&&
+\frac{1}{88}
(39771 c^4+408798 c^3-2058153 c^2+1525928 c+1647156) 
 \,
T \pa  W_{  \frac{1}{3}}^{(\frac{7}{2})}
\nonu \\
&&
+\frac{1}{4}
(2025 c^4+27756 c^3-38283 c^2-269878 c+353640) 
 \,
\pa T W_{  \frac{1}{3}}^{(\frac{7}{2})}
\nonu \\
&&
-\frac{9}{44} 
(99 c^4-1206 c^3+71647 c^2+45476 c-381236) 
 \,
\pa J  W_{  \frac{1}{3}}^{(\frac{9}{2})}
\nonu \\
&&
+9 (363 c^3+3796 c^2-3209 c-1010) 
\,
\pa J J W_{  \frac{1}{3}}^{(\frac{7}{2})}
\nonu \\
&&
+\frac{3}{4} (9 c^4+984 c^3+45685 c^2-51258 c-232720) 
 \,
\pa^2 J  W_{  \frac{1}{3}}^{(\frac{7}{2})}
\nonu \\
&&
-18 (9 c^3-372 c^2+3333 c-4030) 
\,
J G^{-}  W_{  \frac{4}{3}}^{(4)} 
\nonu \\
&&
-18 (9 c^3-372 c^2+3333 c-4030) 
\,
G^{-} G^{+}  W_{  \frac{1}{3}}^{(\frac{7}{2})}
\nonu \\
&&
-\frac{9}{22} (1035 c^4+11826 c^3-28033 c^2-9388 c-102740) 
 \,
G^{+} \pa W_{  -\frac{2}{3}}^{(4)} 
\nonu \\
&&
-\frac{3}{22} (2025 c^4+27756 c^3-38283 c^2-269878 c+353640) 
 \,
\pa G^{+} W_{  -\frac{2}{3}}^{(4)}
\nonu \\
&&
+\frac{3}{22} (27 c^5+1233 c^4+17949 c^3-40569 c^2-15268 c-74292) 
\,
\pa^2 W_{  \frac{1}{3}}^{(\frac{9}{2})}
\nonu \\
&&
-\frac{1}{66}
(324 c^5-846 c^4-67125 c^3+228810 c^2-155309 c-27234) 
 \,
\pa^3  W_{  \frac{1}{3}}^{(\frac{7}{2})}
\nonu \\
&&
\left.
-\frac{3}{88} (297 c^4-48630 c^3-901075 c^2+2098520 c+2141068) 
 \,
\pa J \pa W_{  \frac{1}{3}}^{(\frac{7}{2})}
\right](w)+ \cdots,
\nonu \\
&& 
\frac{1}{C_{(\frac{7}{2})(\frac{7}{2})}^{(\frac{7}{2})-}} \,
W_{  -\frac{1}{3}}^{(\frac{7}{2})}(z) \, W_{  -\frac{4}{3}}^{(4)}(w) = 
\frac{1}{(z-w)^2} \, \frac{30 }{11 (c+9)} \,
G^{-} W_{  -\frac{2}{3}}^{(4)}(w)
\nonu \\
&&  + \frac{1}{(z-w)} \, \frac{1}{(c+9) (3 c+4)} \, \left[ 
\frac{30}{11} \, (c-1) \,
\pa G^{-} W_{  -\frac{2}{3}}^{(4)}
+\frac{15}{11} \, (3 c+7)  \,
G^{-} \pa  W_{  -\frac{2}{3}}^{(4)}
+\frac{90 }{11}  \,
J G^{-}  W_{  -\frac{2}{3}}^{(4)}
\right](w) \nonu \\
&& + \cdots,
\nonu \\
&& 
\frac{1}{C_{(\frac{7}{2})(\frac{7}{2})}^{(\frac{7}{2})-}} \,
W_{  -\frac{1}{3}}^{(\frac{7}{2})}(z) \, W_{  -\frac{1}{3}}^{(\frac{9}{2})}(w) = 
-\frac{1}{(z-w)^4} \, \frac{1}{2} \, W_{  -\frac{2}{3}}^{(4)}(w) 
\nonu \\
& & +
\frac{1}{(z-w)^3} \, \frac{1}{(c+9)} \, \left[ 
10  \,
G^{-} W_{  \frac{1}{3}}^{(\frac{7}{2})}
-\frac{30}{11} \, 
J  W_{  -\frac{2}{3}}^{(4)}
-\frac{2 (c+24) }{11} \,
\pa  W_{  -\frac{2}{3}}^{(4)}
\right](w)  \nonu \\
&& 
+ \frac{1}{(z-w)^2} \, 
\frac{1}{(c+9) (27 c-46) (3 c^2+90 c-265)}
\,\left[ 
18 (c+9) (31 c-35) \,
J J  W_{  -\frac{2}{3}}^{(4)}
\right. \nonu \\
&&
+36 (c+9) (31 c-35) \,
J G^{-}  W_{  \frac{1}{3}}^{(\frac{7}{2})}
-\frac{6}{11} (279 c^3+1368 c^2-15255 c+30428) 
\,
G^{-} W_{  \frac{1}{3}}^{(\frac{9}{2})}
\nonu \\
&&
+\frac{1}{11} (8235 c^3+41664 c^2-187409 c+40170) 
 \,
T  W_{  -\frac{2}{3}}^{(4)} 
\nonu \\
&& +\frac{1}{4} (927 c^3+43272 c^2-117629 c+19370) 
\,
\pa G^{-}  W_{  \frac{1}{3}}^{(\frac{7}{2})}
\nonu \\
&& 
+\frac{3}{11} (51 c^3-928 c^2+67527 c-181590) 
\,
\pa J   W_{  -\frac{2}{3}}^{(4)}
\nonu \\
&& +\frac{1}{44} (16155 c^3+377736 c^2-2257257 c+2943466) 
\,
G^{-} \pa W_{  \frac{1}{3}}^{(\frac{7}{2})}
\nonu \\
&&
-\frac{3}{44} (1671 c^3+47640 c^2-136901 c+34490) 
 \,
J  \pa  W_{  -\frac{2}{3}}^{(4)} 
\nonu \\
&& 
\left.
-\frac{1}{88}
(279 c^4+12453 c^3+217491 c^2-1276973 c+1581870) 
\,
\pa^2  W_{  -\frac{2}{3}}^{(4)}
\right](w)  \nonu \\
&& + \frac{1}{(z-w)} \, 
\frac{1}{ (c+9) (3 c-2) (3 c+4) (27 c-46) (3 c^2+90 c-265)}
\, 
 \nonu \\
&& \times \left[ 
18 (99 c^4+2487 c^3+1256 c^2-28072 c+35600) 
\,
J J  \pa  W_{  -\frac{2}{3}}^{(4)}
\right.
 \nonu \\
&&
+\frac{1080}{11} (252 c^3+141 c^2-3017 c+3694) 
\,
J J J  W_{  -\frac{2}{3}}^{(4)}
\nonu \\
&&
-\frac{90}{11} (108 c^4-1809 c^3-23475 c^2+85134 c-71288) 
\, J G^{-} W_{  \frac{1}{3}}^{(\frac{9}{2})}
\nonu \\
&&
+\frac{3}{11} (17388 c^4+124749 c^3-15033 c^2+942306 c-2281720) 
\,
J G^{-} \pa  W_{  \frac{1}{3}}^{(\frac{7}{2})}
\nonu \\
&&
-\frac{30}{11} (6075 c^4-11664 c^3+56673 c^2-338112 c+406316) 
\,
J T  W_{  -\frac{2}{3}}^{(4)} 
\nonu \\
&&
-\frac{3}{11} 
(2187 c^5+39042 c^4-104463 c^3-321174 c^2+1301856 c-1029808) 
 \,
G^{-} \pa  W_{  \frac{1}{3}}^{(\frac{9}{2})}
\nonu \\
&&
+\frac{90}{11} (3456 c^4-6345 c^3-29889 c^2+71576 c-55076) 
\,
G^{-} G^{+}  W_{  -\frac{2}{3}}^{(4)} 
\nonu \\
&&
+\frac{1}{44}
(37665 c^5+192834 c^4-4184127 c^3+7278324 c^2+7100772 c-9611168) 
 \,
G^{-} \pa^2  W_{  \frac{1}{3}}^{(\frac{7}{2})}
\nonu \\
&&
-\frac{3}{11} 
(1782 c^5+927 c^4-264858 c^3+764781 c^2-869644 c+694052) 
 \,
\pa G^{-} W_{  \frac{1}{3}}^{(\frac{9}{2})}
\nonu \\
&&
+ 6 (594 c^4+4707 c^3-17979 c^2-95292 c+154940) 
 \,
\pa G^{-} J W_{  \frac{1}{3}}^{(\frac{7}{2})}
\nonu \\
&&
+\frac{1}{22}
(21384 c^5+791397 c^4+367884 c^3-6884313 c^2+812852 c+4057516)\,
\pa G^{-} \pa  W_{  \frac{1}{3}}^{(\frac{7}{2})}
\nonu \\
&&
+\frac{1}{4}
(1863 c^5+81252 c^4-536739 c^3+843948 c^2+1257316 c-2308160) 
 \,
\pa^2 G^{-} W_{  \frac{1}{3}}^{(\frac{7}{2})}
\nonu \\
&&
-\frac{1}{22}
(27945 c^5+465156 c^4-1091367 c^3+47394 c^2-6488788 c+9919800)  \,
T \pa  W_{  -\frac{2}{3}}^{(4)} 
\nonu \\
&&
-\frac{1}{22}
(35235 c^5+673866 c^4-2238417 c^3-4832676 c^2+21662612 c-20067600) 
 \,
\pa T W_{  -\frac{2}{3}}^{(4)} 
\nonu \\
&&
+\frac{3}{44} (567 c^5-180 c^4+367575 c^3-1038530 c^2+86068 c-2907960) 
 \,
\pa J \pa W_{  -\frac{2}{3}}^{(4)}
\nonu \\
&&
+\frac{9}{11} (6381 c^4+49758 c^3-108251 c^2+554592 c-920020) 
\,
\pa J J W_{  -\frac{2}{3}}^{(4)}
\nonu \\
&&
+ 3 (1053 c^4-1791 c^3-17208 c^2+150256 c+17000) 
 \,
\pa J G^{-} W_{  \frac{1}{3}}^{(\frac{7}{2})}
\nonu \\
&&
+\frac{3}{11} 
(189 c^5+9099 c^4+367767 c^3-801239 c^2-1350036 c+1509340) 
 \,
\pa^2 J W_{  -\frac{2}{3}}^{(4)}
\nonu \\
&&
+ 10 (3c+4) (27c-46) (81 c^2-183 c-44) 
\,
G^{-} T W_{  \frac{1}{3}}^{(\frac{7}{2})}
\nonu \\
&&
-360 (252 c^3+141 c^2-3017 c+3694) 
 \,
J J G^{-} W_{  \frac{1}{3}}^{(\frac{7}{2})}
\nonu \\
&&
-\frac{3}{44} 
(4239 c^5+86526 c^4-127797 c^3-633936 c^2+3066788 c-4363360)  \,
J \pa^2  W_{  -\frac{2}{3}}^{(4)}
\nonu \\
&&
-\frac{1}{264} 
(1215 c^6+33723 c^5+474705 c^4-5268615 c^3+5682720 c^2+18658892 c
-21364560) 
\nonu \\
&& \left. 
\pa^3  W_{  -\frac{2}{3}}^{(4)}
\right](w) \nonu \\
&& + \cdots.
\label{j3ope}
\eea}
In the large $c$ limit,
all the nonlinear terms in Appendix (\ref{j3ope}) disappear.
One can also analyze the $\frac{1}{c}, \cdots, \frac{1}{c^3}$-terms. 


\end{document}